\documentclass[12pt]{article}
\usepackage{graphicx} 
\def\laq{\raise 0.4ex\hbox{$<$}\kern -0.8em\lower 0.62
ex\hbox{$\sim$}}
\def\gaq{\raise 0.4ex\hbox{$>$}\kern -0.7em\lower 0.62
ex\hbox{$\sim$}}
\def\vk{\vec{k}}
\def\vp{\vec{p}}

\begin{document}

\begin{titlepage}
\begin{flushright}
CERN-PH-TH/2007-048
\end{flushright}
\vspace*{1cm}

\begin{center}
{\LARGE {\bf Why CMB physics?}}
\vskip2.cm
\large{Massimo Giovannini \footnote{e-mail address: massimo.giovannini@cern.ch}}
\vskip 1.cm
{\it Centro ``Enrico Fermi",  Via Panisperna 89/A, 00184 Rome, Italy}
\vskip 0.5cm
{\it Department of Physics, Theory Division, CERN, 1211 Geneva 23, Switzerland}

\end{center}
\begin{abstract}
The aim of these lectures is to introduce some basic problems arising in gravitation and modern cosmology.
All along the discussion the guiding theme is provided by the phenomenological and theoretical 
properties of the Cosmic Microwave Background (CMB).
These lectures have been prepared for a regular Phd course of the University of Milan-Bicocca.
\end{abstract}
\end{titlepage}

\pagenumbering{arabic}

\tableofcontents

\newpage
\renewcommand{\theequation}{1.\arabic{equation}}
\setcounter{equation}{0}
\section{Electromagnetic emission of the observable Universe}
\label{sec1}

\subsection{Motivations and credits}
The lectures collected in the present paper have been written on the on the occasion 
of a course prepared for the Phd program of the University of Milan-Bicocca. 
Following the kind invitation 
of   G. Marchesini and C. Destri I have been very glad of presenting 
a logical collection of topics selected among  
the physics of Cosmic Microwave Background (CMB in what follows).
While preparing these lectures I have been faced with the problems 
usually encountered  in trying to introduce the conceptual foundations 
of a rapidly growing field. To this difficulty one must add, as usual, that 
the cultural background of Phd students is often diverse: not all Phd
students are supposed to take undergraduate courses 
in field theory, general relativity or cosmology.  Last year, during the whole
summer semester, I used to teach a cosmology course 
at the technical school of Lausanne (EPFL). Using a portion of the material prepared 
for that course, I therefore summarized the essentials  
needed for a reasonably self-contained presentation of CMB physics. While 
I hope that this compromise will be appreciated, it is also 
clear that a Phd course, for its own nature, imposes  a necessary selection 
among the possible topics.

In commencing this script I wish also to express my very special 
an sincere gratitude to G. Cocconi and E. Picasso.
I am indebted to G. Cocconi for his advices in the preparation 
of the first section. I indebted to E. Picasso for 
sharing his deep knowledge of general relativity and gravitational 
physics and for delightful discussions which have been extremely relevant 
both for the selection of topics and for the overall spirit of the course. The 
encouragement and comments of L. Alvarez-Gaum\'e have been also 
greatly appreciated.

While lecturing in Milan I had many stimulating questions and comments 
on my presentations from L. Girardello, G. Marchesini, 
 P. Nason, S. Penati and C. Oleari. These interesting comments  
led to an improvement of the original plan of various lectures. I also wish to express 
my gratitude for the discussions with the members of the astrophysics group. In particular I acknowledge 
very interesting discussions with  G. Sironi on the low-frequency measurements of CMB distorsions and 
on the prospects of polarization experiments. 
I also thank S. Bonometto and L. Colombo for stimulating questions and advices.
Last but not least, I really appreciated the lively and pertinent questions of the Phd students 
attending the course. In particular, it is a pleasure to thank A. Amariti, S. Alioli, C.-A. Ratti, E. Re and S. Spinelli.

\subsection{Electromagnetic emission of the Universe}

\begin{figure}
\centering
\includegraphics[height=9cm]{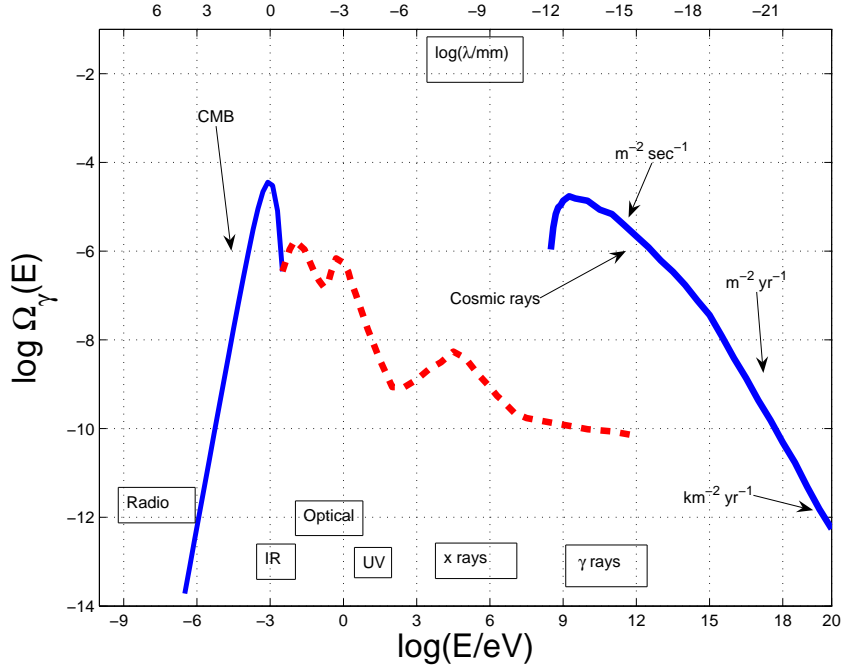}
\caption{The (extragalactic) electromagnetic emission is illustrated.
 On the vertical axis the logarithm (to base $10$)  of the 
  emitted energy density is reported in units of $\rho_{\mathrm{crit}}$ (see Eq. (\ref{CRIT})). 
The logarithm of energy of the photons is instead reported on the horizontal axis.
 The wavelength scale is inserted at the top of the plot. The 
 cosmic ray spectrum is included for comparison and 
 in the same units used to describe the electromagnetic 
 contribution.}
\label{F1}      
\end{figure}

In the present section, after a general introduction to black body emission, 
the question reported in the title of this paper will be partially answered. 
The whole observable Universe will therefore be approached, in the first approximation, 
as a system emitting electromagnetic radiation. 
The topics to be treated in the present section are therefore the following:
\begin{itemize}
\item{} electromagnetic emission of the Universe;
\item{} the black-body spectrum and its physical implications;
\item{} a bit of history of the CMB observations;
\item{} the entropy of the CMB and its implications;
\item{} the time evolution of the CMB temperature.
\end{itemize}
All along these lectures the natural system of units will be adopted.
In this system $\hbar = c = \kappa_{\mathrm{B}} =1$.
In order to pass from one system of units to the other it is useful to recall that 
\begin{itemize}
\item{} $\hbar c= 197.327 \, \mathrm{MeV}\, \mathrm{fm}$;
\item{} $\mathrm{K} = 8.617\times 10^{-5} \,\mathrm{eV}$;
\item{} $(\hbar c)^2 = 0.389\, \mathrm{GeV}^2 \,\mathrm{mbarn}$;
\item{} $c= 2.99792\times 10^{10} \,\,\mathrm{cm}/\mathrm{sec}$.
\end{itemize}
In Fig. \ref{F1} a rather intriguing plot summarizes the electromagnetic emission of our own Universe.
Only the extra-galactic emissions are reported.
On the horizontal axis we have the logarithm of the energy of the photons (expressed in 
eV). On the 
vertical axis we reported the logarithm (to base $10$) of $\Omega_{\gamma}(E)$ which is 
 the energy density of the emitted radiation 
in critical units and per logarithmic interval of photon momentum (see, for instance, Eq. (\ref{critsp})).
For comparison also the associated wavelength of the emitted 
radiation is illustrated (see the top of the figure) in units of  mm.
Figure  \ref{F1} motivates the choice 
of studying accurately the properties of CMB. Moreover, it can be also 
argued that the properties of CMB encode, amusingly enough,
not only the successes of the standard cosmological model
but also its potential drawbacks.

In Fig. \ref{F1} the maximum of $\Omega_{\gamma}(E)$ is located for a wavelength of the $\mathrm{mm}$ (see 
the scale of wavelengths at the top of  Fig. \ref{F1}) corresponding 
to  typical energies of the order\footnote{In natural units $\hbar = c = k_{\mathrm{B}}=1$ we have 
 $E_{k} = \hbar \omega = \hbar c \,k$ and 
that $k = 2\pi/\lambda$. So $E_{k} = k$ and $\omega = 2\pi\nu$.}  of
$10^{-3}\,\mathrm{eV}$. In the optical and ultraviolet range of wavelengths
the energy density drops of almost two orders of magnitude.
In the $x$-rays (i.e. $10^{-6} \mathrm{mm}<\lambda < 10^{-9} \mathrm{mm}$)
the energy density of the emitted radiation drops of more than three  orders 
of magnitude in comparison with the maximum. The $x$-ray range 
corresponds to photon energies $E > \mathrm{keV}$.
In the $\gamma$-rays (i.e. $10^{-9} \mathrm{mm} < \lambda< 10^{-12} \mathrm{mm}$) 
the spectral amplitude is roughly $5$ orders of 
magnitude smaller than in the case of the millimeter maximum.
The range of $\gamma$-rays occurs for photon energies 
$E > \mathrm{GeV}$.  

While the CMB represents $0.93$ of the 
extragalactic emission, the infra-red and visible part give, respectively, $0.05$ and $0.02$. The $x$-ray and $\gamma$-ray 
branches contribute, respectively, by $2.5\times 10^{-4}$ and $2.5\times 10^{-5}$. The CMB is therefore the 93 \% of the 
total extragalactic emission.
The CMB spectrum has been discovered 
by Penzias and Wilson \cite{PW} (see also \cite{PB}) 
and predicted, on the basis of the hot big-bang model, by Gamow, Alpher and Herman (see, for 
instance,\cite{GAM}).
Wavelengths as large as $\lambda \sim \mathrm{m}$ lead to an 
emission which is highly anisotropic and will not be treated here as a cosmological probe. In any case, for $\lambda \geq \mathrm{m}$ we are in the domain of the radio-waves whose analysis is of upmost importance for a variety of problems including, for instance,
large scale magnetic fields (both in galaxies and in clusters) \cite{GOV}, pulsar astronomy \cite{PUL} and, last but not least, extraction of CMB foregrounds. In fact it should be mentioned that also our own galaxy is 
an efficient emitter of electromagnetic radiation. Since our galaxy possess a magnetic field, it emits synchrotron 
radiation as well as thermal bremsstrahlung. A very daring project that will probably be at the forefront 
of radio-astronomical investigations during the next 10 years is SKA (Square Kilometer Array) \cite{SKA1}.
While the technical features of the instrument cannot be thoroughly 
discussed in the present contribution, it suffices to notice that the collecting 
area of the instrument, as the name suggest, will be of $10^{6}\, {\rm m}^2$. The design of SKA will 
 probably allow full sky surveys of Faraday Rotation and better understanding of galactic emission \footnote{We will not enter
 here in the vast subject of CMB foregrounds. It suffices to appreciate that while the spectrum of synchrotron 
 increases with frequency, for wavelengths shorter than the mm the emission is dominated by thermal dust 
 emission whose typical spectrum decreases with frequency. It is opinion of the author that a better 
 understanding of the spectral slope of the synchrotron would be really needed (not only from extrapolation). This 
 seems important especially in the light of forthcoming satellite missions.}

In Fig. \ref{F1} the spectrum 
of the cosmic rays  is also reported, for comparison. This inclusion is somehow arbitrary since the cosmic rays 
of moderate energy are known to come from within the galaxy. However, it is useful 
to plot also this quantity to compare the energy density of the cosmic rays to the energy density 
of the CMB. The energy density 
of the cosmic rays is, roughly,  of the same order of  the energy density of the CMB. For energies 
smaller than $10^{15}$ eV the rate is approximately of 
one particle per $\mathrm{m^2}$ and per second.
For energies larger than $10^{15}$ eV the rate is approximately of 
one particle per $\mathrm{m^2}$ and per year.  The difference in these 
two rates corresponds to a slightly different spectral behaviour 
of the cosmic ray spectrum, the so-called knee.
Finally, for energies larger than $10^{18}$ eV, the rate of the so-called 
ultra-high-energy cosmic rays (UHECR) is even smaller and of the 
order of one particle per $\mathrm{km^2}$ and per year.
The sudden drop in the flux corresponds to another small change 
in the spectral behaviour, the so-called ankle. In the forthcoming years 
the spectrum above the ankle will be scrutinized by the AUGER experiment \cite{auger1,auger2}.

In the parametrization chosen in Fig. \ref{F1} the cosmic ray spectrum 
does not decrease as $E^{-3}$ but rather as $E^{-2}$.
The rationale for this difference stems from the fact that, in the parametrization
of Fig. \ref{F1} we plot the energy density of cosmic rays per logarithmic 
interval of $E$ while, in the standard parametrization the plot is in terms 
of $d\rho_{\mathrm{crays}}/dE$. 
It is important to stress that while the CMB represents the 93 \% of the extragalactic emission, the diffuse $x$-ray 
and $\gamma$-ray backgrounds are also of upmost importance for cosmology.
Various experiments have been  dedicated to the study of the $x$-ray background such as ARIEL, EINSTEIN, GINGA, ROSAT and, last but not least, BEPPO-SAX, an $x$-ray satellite named after  Giuseppe (Beppo) 
Occhialini. Among $\gamma$-ray satellites we shall just mention COMPTON, EGRET and the forthcoming GLAST.

\subsection{The black-body spectrum ad its physical implications}
According to Fig. \ref{F1},  in the mm range  
the electromagnetic spectrum of the Universe is very well fitted by a 
black-body spectrum: if we would plot the error bars magnified 400 times they would still be hardly distinguishable from the thickness of the curve.  Starting from the discovery of Penzias and Wilson \cite{PW} various groups
confirmed, independently, the black-body nature of this emission (see below, in this section, for an oversimplified 
account of the intriguing history of CMB observations). 
As it is well known the black-body has the property of depending 
only upon one single parameter which is the temperature $T_{\gamma}$ of the photon 
gas at the thermodynamic equilibrium. Such a temperature is given by 
\begin{equation}
T_{\gamma} = 2.725 \pm 0.001\,\, \mathrm{K}.
\label{Tgamma}
\end{equation}
According to Wien's law $\lambda T_{\gamma} = 2.897\times 10^{-3}\,\, \mathrm{m\,K}$.
Thus, as already remarked the wavelength of the maximum will be $\lambda \simeq \mathrm{mm}$.
For a photon gas in thermodynamic equilibrium the energy density of the emitted 
radiation is given by 
\begin{equation}
d\rho_{\gamma} = g\times\omega\times \frac{d^3 \omega}{(2\pi)^3}\times 
\overline{n}_{\omega},
\label{endens}
\end{equation}
where $g$ is the number of  intrinsic degrees of freedom ($g = 2$ in the case of photons) and $n_{\omega}$ is the Bose-Einstein occupation number:
\begin{equation}
\overline{n}_{\omega} = \frac{1}{e^{\omega/T_{\gamma}} -1}.
\label{BE}
\end{equation}
Since  in natural units, $E_{k} = k = \omega$, the energy density of the emitted 
radiation per logarithmic interval of frequency is given by:
\begin{equation}
\frac{d\rho_{\gamma}}{d \ln{k}} = \frac{1}{\pi^2} \frac{k^4}{e^{k/T_{\gamma}} -1}.
\label{DIFFFSP}
\end{equation}
Equation (\ref{DIFFFSP}) allows also to compute the total (i.e. integrated) energy density 
$\rho_{\gamma}$. The differential spectrum (\ref{DIFFFSP}) can then be referred to the integrated energy density 
expressed, in turn, in units of the critical energy density.
From Eq. (\ref{DIFFFSP}) the integrated energy density of photons is simply 
given by 
\begin{equation} 
\rho_{\gamma}(t_{0}) = \frac{T_{\gamma}^4}{\pi^2} \int_{0}^{\infty} \frac{x^3}{e^{x} -1} = \frac{\pi^2}{15} T_{\gamma}^4,
\label{rhogamma}
\end{equation}
where the ratio $x = k/T_{\gamma}$ has been defined and where the integral in the second equality is given by 
$\pi^4/15$.

A useful way of measuring energy densities is to refer them to the {\em critical 
energy density} of the Universe (see section \ref{sec2} for a more detailed discussion 
of this important quantity). According to the present data 
it seems that the critical energy density indeed coincides with the 
{\em total} energy density of the Universe. This is just because 
experimental data seem to favour a spatially flat Universe.
The critical energy density today is given by: 
\begin{equation}
\rho_{\mathrm{crit}} = \frac{3 H_{0}^2}{8\pi G} = 1.88\times 10^{-29} \,\,h_{0}^2 \,
\mathrm{g\,cm^{-3}} = 1.05 \times 10^{-5}\,\, h_{0}^2 \, \mathrm{GeV\,cm^{-3}},
\label{CRIT}
\end{equation}
where $h_{0}$ (often assumed to be $\sim 0.7$ for the purpose of numerical estimates along these lectures) measures the indetermination on the present 
value of the Hubble parameter $H_{0} = 100\,\, \mathrm{km \,sec^{-1} \, Mpc^{-1}} h_0$.
From the second equality appearing in Eq. (\ref{CRIT}), recalling that the proton mass is 
$m_{\mathrm{p}} = 0.938\, \mathrm{GeV}$, it is also possible to deduce 
\begin{equation}
\rho_{\mathrm{crit}} = 5.48 \,\biggl(\frac{h_{0}}{0.7}\biggr)^{2} \,\, \frac{m_{\mathrm{p}}}{\mathrm{m}^3},
\label{critp}
\end{equation}
showing that, the critical density is, grossly speaking, the equivalent of $6$ proton masses per cubic meter.

From Eqs. (\ref{DIFFFSP}) and (\ref{CRIT}) we can obtain the energy density 
of photons per logarithmic interval of energy and in critical units, i.e. 
\begin{equation}
\Omega_{\gamma}(k) = \frac{1}{\rho_{\mathrm{crit}}} \frac{d \rho_{\gamma}}{d\ln{k}}.
\label{critsp}
\end{equation}
Recalling that $E_{k} = k$ (and neglecting the subscript) we have that 
\begin{equation}
\Omega_{\gamma}(E) = \frac{15}{\pi^4} \Omega_{\gamma_{0}} \frac{x^4}{e^{x} -1},
\label{critsp1}
\end{equation}
where
\begin{eqnarray}
&&x = \frac{E}{T_{\gamma}} = 4.26\times 10^{3} \biggl(\frac{E}{\mathrm{eV}}\biggr),
\nonumber\\
&&\Omega_{\gamma0} =\frac{\rho_{\gamma}(t_0)}{\rho_{\mathrm{crit}}}
= 2.471 \times 10^{-5}\,\,h_{0}^{-2}.
\label{omegagamma}
\end{eqnarray}
The quantities $\Omega_{\gamma}(E)$ and $\Omega_{\gamma0}$ 
are physically different: $\Omega_{\gamma0}$ is 
the ratio between the total (present) energy density 
of CMB photons and the critical energy density and it is independent on the frequency.
It can be explicitly verified that, inserting the numerical value of $T_{\gamma}$
and $\rho_{\rm c}$  (i.e. Eqs. (\ref{Tgamma}) and (\ref{CRIT})), the figure
of Eq. (\ref{omegagamma}) is swiftly reproduced. 

\begin{figure}
\centering
\includegraphics[height=7cm]{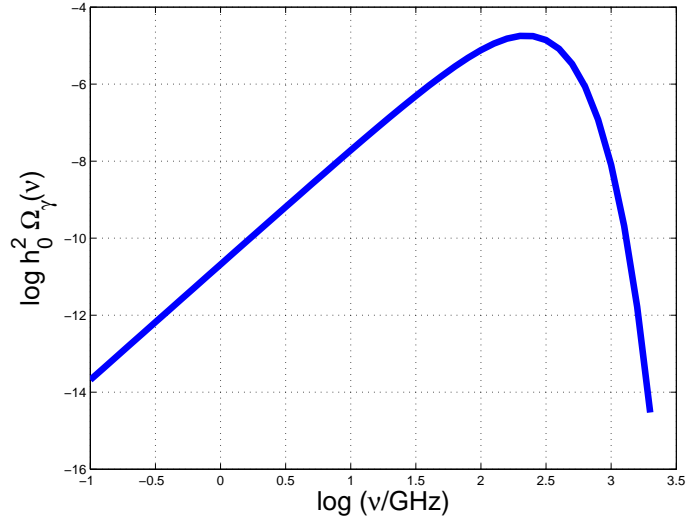}
\caption{The CMB logarithmic energy spectrum here illustrated in terms of the frequency.}
\label{F3}      
\end{figure}

The spectrum of Eq. (\ref{critsp1}) can be also plotted in terms of the frequency.
Recalling that, in natural units, $\nu = 2\pi k$ and that 
\begin{equation}
x = \frac{k}{T_{\gamma}} = 0.01765\,\, \biggl(\frac{\nu}{\mathrm{GHz}}\biggr),
\label{xtonu}
\end{equation}
the spectrum $\Omega_{\gamma}(\nu)$  is 
reported in Fig. \ref{F3}.
It should be borne in mind that the CMB spectrum could be distorted by several energy-releasing 
processes. These distorsions have not been observed so far. In particular we could wander if 
a sizable chemical potential is allowed.
The presence of a chemical potential will affect the Bose-Einstein occupation number 
which will become, in our rescaled notations $\overline{n}_{k}^{{\rm B}} = (e^{x +\mu_{0}} -1)^{-1}$.
Now the experimental data imply that $|\mu_{0}| < 9 \times 10^{-5}$ (95\% C.L.).

It is useful to mention, at this point, the energy density of the CMB in different units and to compare it 
directly with the cosmic ray spectrum as well as with the energy density of the galactic magnetic field.
In particular we will have that 
\begin{eqnarray}
&& \rho_{\gamma} = \frac{\pi^2}{15} T_{\gamma}^4 = 2\times 10^{-51} \,\,\biggl(\frac{T_{\gamma}}{2.725}\biggr)^4\,\, 
\mathrm{GeV}^4,
\label{cmbex}\\
&& \rho_{\mathrm{B}} = \frac{B^2}{8\pi} = 1.36\times10^{-52} \biggl(\frac{B}{3\mu \mathrm{G}}\biggr)^2\,\,\mathrm{GeV}^4.
\label{Bex}
\end{eqnarray}
From Eqs. (\ref{cmbex}) and (\ref{Bex}) it follows that the CMB energy density 
is roughly comparable  with the magnetic energy density of the galaxy. Furthermore $\rho_{\mathrm{crays}} \simeq \rho_{\mathrm{B}}$.

\subsection{A bit of history of CMB observations}

The black-body nature of CMB emission is one of the cornerstones of the Standard cosmological model 
whose essential features will be introduced in section \ref{sec2}.
The first measurement of the CMB spectrum goes back to the work of Penzias and Wilson 
\cite{PW}. The Penzias and Wilson measurement referred to a wavelength of $7.35$ cm (corresponding 
to $4.08$ GHz).
They estimated a temperature of $3.5\,\,^{0} \mathrm{K}$.  Since the Penzias 
and Wilson measurement the black-body nature of the CMB spectrum has been 
investigated and confirmed for a wide range of frequencies extending from 
$0.6$ GHz \cite{sironi} (see also \cite{HS})  up to  300 GHz. 
The history of the measurements of the CMB temperature is a subject by itself 
which has been reviewed in the excellent book of B. Patridge \cite{patridge}.
Before 1990 the measurements of CMB properties have been conducted 
always through terrestrial antennas or even by means of balloon borne 
experiments. In the nineties the COBE satellite \cite{mat0,wright,smoot,kogut1,Bennett1,mat1,fixsen1,Bennett2} allowed to measure 
the properties of the CMB spectrum in a wide range of frequencies 
including the maximum (see Fig. \ref{F3}).
The COBE satellite had two instruments: FIRAS and DMR.

The DMR was able to probe the angular power spectrum\footnote{While the precise definition of angular power 
spectrum will be given later on, here it suffices to recall that $\ell (\ell +1) C_{\ell}/(2\pi)$ measures the 
degree of inhomogeneity in the temperature distribution per logarithmic interval of $\ell$. Consequently,
a given multipole $\ell$ can be related to a given spatial structure in the microwave sky: small $\ell$ will correspond 
to low wavenumbers, high $\ell$ will correspond to larger wave-numbers.}
up to $\ell \simeq 26$.  As the name says, DMR 
was a differential instrument measuring temperature differences in the microwave 
sky. The angular resolution of  a given instrument, i.e. $\vartheta$, is related 
to the maximal multipole probed in the sky according to the approximate relation 
$\vartheta \simeq  \pi/\ell$. Consequently, since the angular resolution of COBE was $7^{0}$, the maximal 
$\ell$ accessible to that experiment was $\ell \simeq 180^{0}/7^{0} \sim 26$.  Since the angular 
resolution of WMAP is $0.23^{0}$, the corresponding maximal harmonic 
probed by WMAP will be $\ell \simeq 180^{0}/0.23^{0} \sim 783$. Finally, the Planck experiment, to be soon launched
will achieve an angular resolution of $5'$, implying $\ell \simeq 180^{0}/5' \sim 2160$.

After the COBE mission, various experiments attempted the exploration of smaller 
angular separation, i. e. larger multipoles.  
A definite convincing evidence of the existence and location of the first peak 
in the $C_{\ell}$ spectrum came from the Boomerang \cite{boom1,boom2} , Dasi \cite{dasi}
and Maxima \cite{maxima} experiments.  Both Boomerang and Maxima were balloon borne (bolometric) 
experiments. Dasi was a ground based interferometer. The data points of these last 
three experiments explored multipoles up to $1000$, determining the first acoustic oscillation 
(in the jargon the first Doppler peak) for $\ell \simeq 220$. 
Another important balloon borne experiments was Archeops \cite{archeops} providing 
interesting data for the region characterizing the first rise of the $C_{\ell}$ spectrum.
Some other useful references on earlier CMB experiments can be found in \cite{silk}.

\begin{figure}[t!]
\centering
\includegraphics[height=11cm,angle=90]{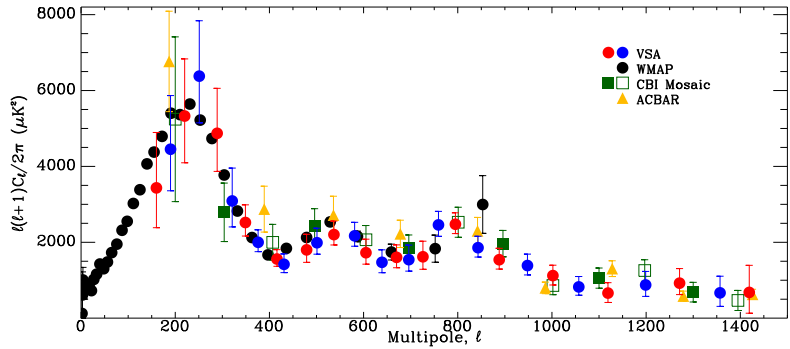}
\caption{Some CMB anisotropy data are reported (figure adapted from \cite{THTH}): 
 WMAP data (filled circles);
VSA data (shaded circles) \cite{DICK}; 
CBI data  (squares) \cite{mason,pearson};
ACBAR data (triangles) \cite{kuo}.}
\label{F4} 
\end{figure}

The $C_{\ell}$ spectrum, as measured by different recent experiments is reported in 
Fig. \ref{F4} (adapted from Ref. \cite{THTH}).
At the moment the most accurate determinations of CMB observables are derived from 
the data of WMAP (Wilkinson Microwave Anisotropy Probe). 
The first release of WMAP data are the subject of Refs. \cite{WMAP01,WMAP02,map1,map2}.
The three-years release of WMAP data is discussed in Refs. \cite{map3,map4}.
The WMAP   data (filled circles in Fig. \ref{F4}) 
provided, among other important pieces of information the precise determination of the position 
of the first peak (i.e. $\ell = 220.1 \pm 0.8$ \cite{WMAP02}) the evidence  of the second
peak. The WMAP experiment also 
measured temperature-polarization correlations providing a  distinctive 
signature (the so-called anticorrelation peak in the temperature-polarization 
power spectrum for $\ell \sim 150$) of primordial adiabatic fluctuations (see sections \ref{sec8} and \ref{sec9}
and, in particular Fig. \ref{POL}).  
To have a more detailed picture of the evolution and relevance 
of CMB experiments we refer the reader to  Ref. \cite{white} (for review of the pre-1994 status of the art) 
 and Ref.  \cite{dodelson} for a review of the pre-2002 situation). The rather broad set of lectures by 
Bond \cite{bond} may also be usefully consulted.

In recent years, thanks to combined observations 
of CMB anisotropies \cite{WMAP01,WMAP02}, 
Large scale structure  \cite{LSS01,CL01}, supernovae of type Ia \cite{riess}, 
big-bang nucleosyntheis \cite{BBN01},
some kind of paradigm for the evolution of the late time (or even present)
Universe emerged. 
It is  normally called by practitioners $\Lambda$CDM 
model or even, sometimes, ``concordance model". The terminology 
of $\Lambda$CDM refers to the fact that, in this model, the dominant 
(present) component of the energy density of the Universe 
is given by a cosmological constant $\Lambda$ and a fluid of cold dark matter 
particles interacting only gravitationally with the other (known) 
particle species such as baryons, leptons, photons.
According to this paradigm, our understanding of the Universe can be summarized 
in two sets of cosmological parameters: the first set of parameters refers to the homogeneous 
background, the second set of parameters to the inhomogeneities.
So, on top of the indetermination 
on the (present) Hubble expansion rate, 
i.e. $h_{0}$, there are various other parameters such as:
\begin{itemize}
\item{} the (present) dark energy density in critical units\footnote{Instead of giving the critical fraction of the 
total energy density alone, it is common practice to multiply this figure by $h_{0}^2$ so that the final 
number will be independent of $h_{0}$.}, i.e. $h_{0}^2 \Omega_{\Lambda 0}$;
\item{} the (present) cold dark matter (CDM in what follows) energy density, i.e. $h_{0}^2 \Omega_{\mathrm{c}0}$;
\item{} the (present) baryon energy density, i.e. $h_{0}^2 \Omega_{\mathrm{b}0}$;
\item{} the (present) photon energy density (already introduced) $h_{0}^2 \Omega_{\gamma0}$;
\item{} the (present) neutrino energy density, i.e.  $h_{0}^2 \Omega_{\nu0}$;
\item{} the optical depth at reionization (denoted by $\epsilon$ but commonly 
named $\tau$ which denotes instead, in the present lectures, the conformal time coordinate, see section \ref{sec2});
\item{} the spectral index of the primordial (adiabatic) mode for the scalar fluctuations $n_{\mathrm{r}}$;
\item{} the amplitude of the curvature perturbations $A_{\mathrm{ad}}$;
\item{} the bias parameter (related to large scale structure).
\end{itemize}
To this more or less standard set of parameters one can also add other parameters reflecting 
a finer description of pre-decoupling physics:
\begin{itemize}
\item{} the neutrinos are, strictly speaking, massive and their masses can then 
constitute an additional set of parameters;
\item{} the dark energy may not be exactly a cosmological constant and, therefore, the 
barotropic index of dark energy may be introduced as the ratio between the pressure 
of dark energy and its energy density (similar argument can entail also the introduction 
of  the sound speed off dark energy);
\item{} the spectral index may not be constant as a function of the wave-number and this 
consideration implies a further parameter;
\item{} in the commonly considered  inflationary scenarios there are not only scalar (adiabatic) 
modes but also tensor modes and this evidence suggests the addition of the relative 
amplitude and spectral index of tensor perturbations, i.e., respectively, $r$ and $n_{\mathrm{T}}$.
\end{itemize}
Different parameters can be introduced in order to account for even more 
daring departures from the standard cosmological lore. These 
parameters include
\begin{itemize}
\item{} the amplitude and spectral index of primordial non-adiabatic perturbations;
\item{} the amplitude and spectral index of the correlation between adiabatic 
and non-adiabatic modes;
\item{} a primordial magnetic field which is fully inhomogeneous and characterized, again, by a 
given spectrum and an amplitude.
\end{itemize}
This list can be easily completed by other possible (and physically reasonable) parameters.
We just want to remark that the non-adiabatic modes represent a whole set of physical 
parameters since, as it will be swiftly discussed, there are $4$ non-adiabatic modes. Consequently, already 
a thorough parametrization of the non-adiabatic sector will entail, in its most general incarnation, 
4 spectral indices, 4 spectral amplitudes and the mutual correlations of each non-adiabatic mode 
with the adiabatic one.
Having said this it is important to stress that the lectures will not deal with the problem 
 of data analysis (or parameter extraction from the CMB data). The purpose of the 
 present lectures, as underlined before in this introduction, will be to use CMB as a guiding theme 
 for the formulation of a consistent cosmological framework which might be in sight but which 
 is certainly not yet present.
 
\subsection{The entropy of the CMB and its implications}
The pressure of black-body photons is simply 
$p_{\gamma} = \rho_{\gamma}/3$. Since the chemical 
potential exactly vanishes in the case of a photon gas at the thermodynamic 
equilibrium, the entropy density 
of the black-body is given, through the fundamental identity 
of thermodynamics (see Appendix \ref{APPB}),  by 
\begin{equation}
s_{\gamma} =\frac{S_{\gamma}}{V}= \frac{\rho_{\gamma} + p_{\gamma}}{T_{\gamma}} = 
\frac{4}{45} \pi^2 T_{\gamma}^3,
\label{sgamma}
\end{equation}
where $S_{\gamma}$ is the entropy and $V$ is a fiducial volume.
Equation (\ref{sgamma}) implies 
that the entropic content of the present Universe is dominated 
by the species that are relativistic today (i.e. photons) and that
the total entropy contained in the Hubble volume, i.e. $S_{\gamma}$ is 
{\em huge}.  The Hubble volume can be thought as 
the present size of our observable Universe and it is 
roughly given by $ V_{H} = 4\pi H_{0}^{-3}/3$.  Thus, we will have 
that 
\begin{equation}
S_{\gamma} = \frac{4}{3} \pi s_{\gamma} H_{0}^{-3} \simeq 
1.43 \times 10^{88} \,\, \biggl(\frac{h_{0}}{0.7}\biggr)^{-3}.
\label{Sg}
\end{equation}
The figure provided by Eq. (\ref{Sg}) is still one of the 
major problems of the standard cosmological model.
Why is the entropy of the observable Universe so large?
Note for the estimate of Eq. (\ref{Sg}) it is practical to express 
both $T_{\gamma}$ and $H_{0}$ in Planck units, namely:
\begin{equation}
T_{\gamma} = 1.923\times 10^{-32}\,\, M_{\mathrm P},
\qquad 
H_{0} = 1.22\times 10^{-61} \biggl(\frac{h_{0}}{0.7}\biggr) \,\, M_{\mathrm{P}}.
\label{num}
\end{equation}
It is clear that the huge value of the present entropy is a direct 
consequence of the smallness of $H_{0}$ in Planck units. This implies
that $T_{\gamma}/H_{0} \simeq 1.57\times 10^{29}$. Let us just 
remark that the present estimate only concerns the usual entropy, i.e. 
the thermodynamic entropy. Considerations related with the validity 
(also in the early Universe) of the second law of thermodynamics 
seem to suggest that also the entropy of the gravitational field itself
may play a decisive r\^ole. While some motivations seem to be 
compelling there is no consensus, at the moment, on what should be 
the precise mathematical definition of the entropy of the gravitational field.
This remark is necessary since we should keep our minds open. It may well 
be that the true entropy of the Universe (i.e. the entropy of the sources and of the 
gravitational field) is larger than the one computed in Eq. (\ref{Sg}).
Along this direction it is possible to think that the maximal entropy that can be 
stored inside the Hubble radius $r_{\mathrm{H}}$ is of the order of a black-hole with  radius $r_{\mathrm{H}}$ 
which would give 
\begin{equation}
r_{\mathrm{H}}^2 M_{\mathrm{P}}^2 \simeq 10^{122}.
\label{GrEn}
\end{equation}

In connection with Eq. (\ref{num}), it is also useful to point out  that the critical 
density can be expressed directly in terms of the fourth power of the Planck mass, i.e. :
\begin{equation}
\rho_{\mathrm{crit}} = \frac{3}{8\pi} H_{0}^2 M_{\mathrm{P}}^2 = 
1.785\times 10^{-123} \biggl(\frac{h_{0}}{0.7}\biggr)^{2} 
\,\, M_{\mathrm P}^{4}.
\label{CRIT2}
\end{equation}
The huge hierarchy between the critical energy density of the 
present Universe and the Planckian energy density 
is, again, a direct reflection of the hierarchy between 
the Hubble parameter and the Planck mass.
Such a hierarchy would not be, by itself, problematic.
The rationale for such a statement is connected to the fact 
that in the SCM the energy densities as well as the 
related pressures decrease as the Universe expand. 
However, it turns out that, today, the largest 
portion of the energy density of the Universe is determined 
by a component called {\em dark energy}. The term {\em dark} 
is a coded word of astronomy. It means that a given 
form of matter neither absorbs nor emit radiation. Furthermore 
the dark energy is homogeneously distributed and, unlike 
{\em dark matter}, is not concentrated in the galactic halos and in the 
clusters of galaxies. 
Now, one of the chief properties of dark energy is that it  is not affected 
by the Universe expansion and this is the reason why 
it is usually parametrized in terms of a cosmological constant.
measurements tell us that $\rho_{\Lambda} \simeq 0.7 \rho_{\mathrm{crit}}$ which implies, from Eq. (\ref{CRIT2}) that 
\begin{equation}
\rho_{\Lambda} \simeq 1.24 \times 10^{-123}\,\, M_{\mathrm{P}}^{4}.
\label{rholambda}
\end{equation}
Since $\rho_{\Lambda}$ {\em does not} decreases with the expansion 
of the Universe, we have also to admit that Eq. (\ref{rholambda}) 
was enforced at any moment in the life of the Universe and, in particular 
at the moment when  the initial conditions of the SCM were set. 
A related way of phrasing this impasse relies on the field theoretical 
interpretation of the cosmological constant. In field theory 
we do know that the zero-point (vacuum) fluctuations have 
an energy density (per logarithmic interval of frequency) that goes as 
$k^4$. Now, adopting the Planck mass as the ultraviolet cut-off
we would be led to conclude that the total energy density of the zero-point 
vacuum fluctuations would be of the order of $M_{\mathrm{P}}^4$. 
On the contrary, the result of the measurements 
simply gives us a figure which is $122$ orders of magnitude smaller. 

The expression of the black-body spectrum also allows the calculation of the 
photon concentration.  Recalling that, in the case of photons,
$d n = (k^3 n_{k}/\pi^2) d\log{k}$ we have, after integration 
over $k$ that the concentration of photons is given by
\begin{equation}
n = \frac{2 \zeta(3)}{\pi^2} T_{\gamma}^{3} \simeq 411 \,\,\mathrm{cm}^{-3}
\end{equation}
where $\zeta(r)$ is the Riemann zeta function with argument $r$.

\subsection{The time evolution of the CMB temperature}

In summary we can therefore answer, in the first approximation, to the 
title of this lecture series:
\begin{itemize}
\item{} in the electromagnetic spectrum the contribution of the CMB is by far 
larger than the other branches and constitutes, roughly, the 93 \% of the whole 
emission;
\item{} the CMB energy density is comparable with (but larger than) the 
energy density of cosmic rays;
\item{} the CMB energy density is a tiny fraction of the total energy density 
of the Universe (more precisely 24 millionth of the critical energy density);
\item{} the CMB dominates the total entropy of the present Hubble patch: 
$S_{\gamma} \simeq 10^{88}$.
\end{itemize}

The fact that we observe a CMB seems implies that CMB photons 
are in thermal equilibrium at the temperature $T_{\gamma}$. This 
occurrence strongly suggests that the evolution of the whole Universe 
must be somehow adiabatic. This observation is one of the cornerstones 
of the standard cosmological model (SCM) whose precise formulation 
will be given in the following section.

In a preliminary perspective, the following naive observation 
is rather important. Suppose that the spatial coordinates expand thanks to a 
time-dependent rescaling. Consequently the wave-numbers 
will be also rescaled accordingly, i.e. 
\begin{equation}
\vec{x}_{0} \to \vec{x} = a(t) \vec{x}_{0},\qquad  \vec{k}_{0} \to \vec{k} = \frac{\vec{k}_{0}}{a(t)}.
\label{rescaling}
\end{equation} 
In the jargon $\vec{k}_{0}$ is commonly referred to as the comoving 
wave-number (which is insensitive to the expansion), while $\vec{k}$ 
is the physical wave-number.
Consider then the number of photons contained in an infinitesimal 
element of the phase-space and suppose that the whole Universe 
expands according to Eq. (\ref{rescaling}). 
At a generic time $t_{1}$ we will then have 
\begin{equation}
d n_{k}(t_1) = \overline{n}_{k}(t_1) d^{3} k_1 d^{3} x_1.
\label{timet1}
\end{equation}
At a generic time $t_{2} > t_1$ we will have, similarly,
\begin{equation}
d n_{k}(t_2) = \overline{n}_{k}(t_2) d^{3} k_2 d^{3} x_2.
\label{timet2}
\end{equation}
By looking at Eqs. (\ref{timet1}) and (\ref{timet2}) it is rather 
easy to argue that $d n_{k}(t_1) = d n_{k}(t_2)$ {\em provided} 
$\overline{n}_{k}(t_1) = \overline{n}_{k}(t_2)$.
By looking at the specific form of the Bose-Einstein 
occupation number it is clear that the latter occurrence 
is verified provided $k(t_1)/T_{\gamma}(t_1) = k(t_2)/T_{\gamma}(t_2)$.
From this simple argument we can already argue an important 
fact: the black-body distribution is preserved under the rescaling 
(\ref{rescaling}) provided the black-body temperature scales as the 
inverse of the scale factor $a(t)$, i.e. 
\begin{equation}
T_{\gamma 0} \to T_{\gamma} = \frac{T_{\gamma0}}{a(t)}.
\label{scT}
\end{equation}
The property summarized in Eq. (\ref{scT}) holds also 
in the context of the SCM where $a(t)$ will be correctly 
defined as the time-dependent scale factor of a Friedmann-Robertson-Walker 
(FRW) Universe. The physical consequence of Eq. (\ref{scT}) is that 
the temperature of CMB photons is higher at higher redshifts (see Appendix A for a 
definition of redshift). More precisely:
\begin{equation}
T_{\gamma} = (1 + z) T_{\gamma0}.
\label{rstemp}
\end{equation}
This consequence of the theory can  be tested experimentally 
\cite{songaila}.  In short, the argument goes as follows. The CMB will populate 
excited levels of atomic and molecular species when the energy separations involved are 
not too different from the peak of the CMB emission. The first measurement of the 
local CMB temperature was actually made with this method 
by using the fine structure lines of CN (cyanogen) \cite{CN}. Using the same philosophy 
it is reasonable to expect that clouds of other chemical elements (like Carbon, in Ref. \cite{songaila}) 
may be sensitive to CMB photons also at higher redshifts. For instance  in \cite{songaila}
measurements were performed at $z = 1.776$ and the estimated temperature 
was found to be of the order of $T_{\gamma}(z) \simeq 7.5$ $^{0}K$.
These measurements are potentially very instructive but have been a bit neglected, in the recent 
past, since the attention of the community focused more on the properties of CMB anisotropies.

For the limitations imposed to the present script it is not possible to treat in detail the very interesting 
physics of another important effect that gives us important informations concerning 
the CMB and its primeval origin. This effect should be anyway mentioned and it is called 
Sunyaev-Zeldovich effect  \cite{SZ1,SZ2,SZ3}. The physics of this effect is, in a sense, rather simple. 
If you have a cluster of galaxies, that cluster o galaxies has a deep potential well and on the average, by the 
virial theorem, its kinetic motion is of the order o few keV. So some fraction of the hot gas 
can get ionized and we will have ionized plasma around. That plasma emits 
$x$-rays that, for instance, the ROSAT satellite has seen \footnote{It is actually interesting, incidentally, that from the ROSAT 
full sky survey (allowing to determine the surface brightness o various clusters in the $x$-rays), 
the average electron density has been determined can be determined \cite{roscat1} and this 
allowed interesting measurements of magnetic fields inside a sample of Abell clusters.}. 
Now the CMB will sweep the whole space. By looking at a direction where 
there is nothing between the observer and the last scattering surface the radiation arrives basically 
unchanged except for the effect due to the expansion of the Universe. But if the observation is now made along a direction 
passing through a cluster of galaxies, some small fraction of the CMB photons (roughly one over $1000$ CMB photons)
will be scattered by the hot gas. Because the gas is actually hot, there is more probability that photons will 
be scattered at high energy rather than at low energy. They will also be scattered almost at isotropic angle. 
The bottom line is that the CMB spectrum along a line of sight that crosses a cluster of galaxies will 
have a slight excess of high energy photons and a slight deficiency of low energy photons. 
So if you see this effect (as we do)  it means that the CMB photons come from behind the clusters. Some of these 
clusters are at redshift $0.07 < z< 1.03$. The measurements of the Sunyaev-Zeldovich effect have been attempted for roughly 
two decades but in the last decade a remarkable progress has been made. As already mentioned, the Sunyaev-Zeldovich 
effect tells that the CMB is really an extra-galactic radiation.
\newpage
\renewcommand{\theequation}{2.\arabic{equation}}
\setcounter{equation}{0}
\section{From CMB to the standard cosmological model}
\label{sec2}
Various excellent publications treat the essential elements of the Standard Cosmological Model 
(SCM in what follows)
within different perspectives (see, for instance, \cite{weinberg,peebles,peebles2,KT}).
The purpose here will not be to present the conceptual foundations SCM but to 
introduce its main assumptions and its most relevant consequences with 
particular attention to those aspects and technicalities that are germane to our theme, i.e. CMB physics.

It should be also mentioned that there are a number of relatively ancient papers that can  be usefully
consulted to dig out both the historical and conceptual foundations of the SCM.
In the he issue number 81 of the ``Uspekhi Fizicheskikh
Nauk'' , on the occasion of the seventy-fifth 
anniversary of the birth of A. A. Friedmann, a number of rather 
interesting papers were published. Among them there is a 
review article of the development of Friedmannian 
cosmology by Ya. B. Zeldovich \cite{zel1} and  
the inspiring paper of Lifshitz and Khalatnikov \cite{LK} on the 
relativistic treatment of cosmological perturbations.

Reference \cite{zel1} describes mainly Friedmann's contributions \cite{friedmann}.
Due attention should also be paid 
to the work of G. Lema\^itre \cite{lemaitre1,lemaitre12,lemaitre2} that 
was also partially motivated by the debate with A. Eddington
\cite{eddington}. According to the idea of Eddington 
the world evolved from an Einstein static Universe and so developed 
``infinitely slowly from a primitive uniform distribution in 
unstable equilibrium'' \cite{eddington}. The point of view of Lema\^itre 
was, in a sense, more radical since he suggested, in 1931, that the expansion 
really did start with the beginning of the entire Universe. Unlike the 
Universe of some modern big-bang cosmologies, the description of Lema\^itre 
did not evolve from a true singularity but from a material 
pre-Universe, what Lema\^itre liked to call ``primeval atom'' \cite{lemaitre2}.
The primeval atom was a unique atom whose atomic weight was the total 
mass of the Universe. This highly unstable atom would have 
experienced some type of fission and would have divided into smaller 
and smaller atoms by some kind of super-radioactive processes. 
The perspective  of Lema\^itre was that the early expansion of the Universe 
could be a well defined object of study for natural sciences even in the 
absence of a proper understanding of the initial singularity.
The discussion of the present section follows four main lines:
\begin{itemize}
\item{} firstly the SCM will  be formulated 
in his essential elements;
\item{} then the matter content of the present Universe 
will be introduced as it emerges in the concordance model;
\item{} the (probably cold) future of our own Universe will be swiftly
discussed;
\item{} finally the (hot) past of the Universe will be scrutinized 
in connection with the properties of the CMB.
\end{itemize}
Complementary discussions on the concept of distance in cosmology and on the 
kinetic description of hot plasmas are collected, respectively,  
in Appendix A and in Appendix B.

\subsection{The standard cosmological model (SCM)}
\label{sub21}
The standard cosmological model (SCM) rests on the following three 
important assumptions:
\begin{itemize}
\item{} for typical length-scales larger than $50$ Mpc the Universe 
is homogeneous and isotropic;
\item{} the matter content of the Universe 
can be parametrized in terms of perfect barotropic fluids;
\item{} the dynamical law connecting the evolution of the sources 
to the evolution of the geometry is provided by General Relativity 
(GR).
\end{itemize}

\subsubsection{Homogeneity and isotropy}
The assumption of homogeneity and isotropy implies that the geometry of the Universe is invariant
for spatial roto-translations.  In four space-time dimensions 
the metric tensor will have 10 independent components.
Using homogeneity and isotropy the ten independent components can be 
reduced from 10 to 4 (having taken into account the 3
spatial rotation and the 3 spatial translations). The most general form 
of a line element which is invariant under spatial rotations and spatial translations 
can then be written as 
\begin{equation}
ds^2 =  e^{\nu} dt^2 - e^{\lambda} dr^2 - e^{\mu}( r^2 d\vartheta^2 + r^2 \sin^2{\vartheta} d\varphi^2)
+ 2 e^{\sigma} d r d t.
\label{genFRW}
\end{equation}
The freedom of choosing a gauge can the be exploited and the 
metric can be reduced to its canonical Friedmann-Robertson-Walker
(FRW) form\footnote{The transition from Eq. (\ref{genFRW}) to Eq. (\ref{FRW}) by successive gauge 
choices can be followed in the book of Tolman \cite{tolman}. } :
\begin{equation}
ds^2 = g_{\mu\nu} dx^{\mu} dx^{\nu} = dt^2 - a^2(t) \biggl[ \frac{dr^2}{1 - k r^2} + r^2 (d\vartheta^2 +
\sin^2{\vartheta} d\varphi^2)\biggr],
\label{FRW}
\end{equation}
where $g_{\mu\nu}$ is the metric tensor of the FRW geometry and $a(t)$ is the scale factor.
In the parametrization of Eq. (\ref{FRW}), $k=0$ corresponds to a spatially flat 
Universe; if  $k>0$ the Universe is spatially closed and, finally,   $k<0$ corresponds to a spatially open Universe.
The line element (\ref{FRW}) is invariant under the following transformation
\begin{eqnarray}
&& r \to \tilde{r} = \frac{r}{r_{0}},
\nonumber\\
&& a(t) \to \tilde{a} = a(t)\, r_{0},
\nonumber\\
&& k\to \tilde{k} =  k \,r_{0}^2,
\label{param2}
\end{eqnarray}
where $r_{0}$ is a dimensionfull constant. In the parametrization 
(\ref{param2}) the scale factor is dimensionfull and $\tilde{k}$ is $0$, $+1$ 
or $-1$ depending on the spatial curvature of the internal space. 
Throughout these lectures, the parametrization where the scale 
factor is dimensionless will be consistently employed.
In Eq. (\ref{FRW}) the time $t$ is the {\em cosmic} time coordinate. 
Depending upon the physical problem 
at hand, different time parametrizations can be also
adopted. A particularly useful one (especially in the study of cosmological 
inhomogeneities)  is the so-called 
conformal time parametrization. In the conformal time coordinate 
$\tau$ the line element of Eq. (\ref{FRW}) can be written as
\begin{equation}
ds^2 =g_{\mu\nu} dx^{\mu} d x^{\nu} = 
a^2(\tau)\biggl\{ d\tau^2 - \biggl[ \frac{dr^2}{1 - k r^2} + r^2 (d\vartheta^2 +
\sin^{2}\vartheta d\varphi^2)\biggr]\biggr\}.
\label{FRW2}
\end{equation}
The line element (\ref{FRW}) describes a situation where 
the space-time is homogeneous and isotropic. It is 
possible to construct geometries that are homogeneous 
but {\em not} isotropic. The Bianchi geometries are, indeed,
homogeneous but not isotropic. For instance, the Bianchi type-I
metric can be written, in Cartesian coordinates, as  
\begin{equation}
ds^2 = dt^{2} - a^2(t) dx^2 - b^2(t) dy^2 - c^2(t) dz^2. 
\label{bianchiI}
\end{equation}
Equation (\ref{bianchiI}) leads to a Ricci 
tensor that depends only on time and not on the spatial coordinates.
Another, less obvious, example is given by the following line element:
\begin{equation}
ds^2 = dt^2 - a^2(t) dx^2 - e^{2 \alpha x} b^2(t) dy^2 - c^2(t) dz^2.
\label{otherbianchi}
\end{equation}
For $\alpha = -1$ we get the Bianchi III line element while, for $\alpha = -2$
we obtain the Bianchi $\mathrm{VI}_{-1}$ line element \cite{ryan}. In both cases 
the geometry is homogeneous but not isotropic. This example 
shows that it is a bit dangerous to infer the homogeneity 
properties of a given background only by looking at the form of the 
line element. A more efficient strategy is to scrutinize 
the properties of curvature invariants.

\subsubsection{Perfect barotropic fluids}
The material content of the Universe 
is often described in terms of perfect fluids (i.e. fluids that are not viscous) 
which are also barotropic (i.e. with a definite relation between 
pressure and energy-density). 
An example of (perfect) barotropic fluid has been already provided: the gas 
of photons in thermal equilibrium  introduced in section \ref{sec1}. 
The energy-momentum tensor of a gas of photons can be 
indeed written as 
\begin{equation}
T_{\mu}^{\nu} = (p_{\gamma} + \rho_{\gamma}) u_{\mu} u^{\nu} - p_{\gamma} \delta_{\mu}^{\nu}, 
\label{Tmunugamma}
\end{equation}
where 
\begin{equation}
u^{\mu} = \frac{dx^{\mu}}{ds},\qquad p_{\gamma} = \frac{\rho_{\gamma}}{3},\qquad g_{\mu\nu} u^{\mu} u^{\nu} =1.
\label{def1}
\end{equation}
Equations (\ref{Tmunugamma}) and (\ref{def1}) are the first example 
of a radiation fluid. The covariant conservation of the energy-momentum tensor 
can be written as 
\begin{equation}
\nabla_{\mu} T^{\mu}_{\nu} =0, 
\label{cov1}
\end{equation}
where $\nabla_{\mu}$ denotes the covariant derivative with respect 
to the metric $g_{\mu\nu}$ of Eq. (\ref{FRW}).  Using the definition of covariant derivative, Eq. (\ref{cov1}) 
can be also written, in more explicit terms, as 
\begin{equation}
\partial_{\mu} T^{\mu}_{\nu} + \Gamma_{\alpha\mu}^{\mu}T^{\alpha}_{\nu} - 
\Gamma_{\nu\alpha}^{\beta} T^{\alpha}_{\beta} =0,
\label{cov2}
\end{equation}
where 
\begin{equation}
\Gamma_{\mu\nu}^{\alpha} = \frac{1}{2} g^{\alpha\beta} ( - \partial_{\beta} 
g_{\mu\nu} + \partial_{\nu} g_{\beta\mu} + 
\partial_{\mu} g_{\nu\beta}),
\label{christconn}
\end{equation}
are the Christoffel connections computed from the metric tensor.
In the FRW metric, Eq. (\ref{cov2}) implies 
\begin{equation}
\dot{\rho}_{\gamma} + 3 H ( \rho_{\gamma}+ p_{\gamma}) =0,
\label{cov3}
\end{equation}
where the overdot denotes a derivation with respect to the cosmic time 
coordinate $t$ and 
\begin{equation}
H = \frac{\dot{a}}{a}, 
\label{hubbpar}
\end{equation}
is the Hubble parameter.
The covariant conservation of the energy-momentum 
tensor implies that the evolution is adiabatic, i.e. $\dot{S} =0$, where 
$S$ is the entropy which, today, is effectively dominated by photons (i.e., today, $S = S_{\gamma}$).
Recall, to begin with, that  the fundamental thermodynamic identity 
and the first law of thermodynamics, stipulate, respectively \footnote{See Appendix B for more details 
on the kinetic and chemical description of hot plasmas.}, 
\begin{eqnarray}
&& {\mathcal E} = T S - pV + \mu N,
\label{thid}\\
&&d {\mathcal E} = T dS - p dV + \mu dN.
\label{firstlaw}
\end{eqnarray}
In the case of the photons the chemical potential is zero.
The volume $V$ will be then given by a fiducial volume 
(for instance the Hubble volume) rescaled through the third 
power of the scale factor. In analog terms one can write 
the energy. In formulas
\begin{equation}
V(t) = V_{0} \biggl(\frac{a}{a_{0}}\biggr)^3, \qquad 
{\mathcal E} = V(t) \rho_{\gamma}.
\label{def2}
\end{equation}
Thus, using Eq. (\ref{def2}) into Eq. (\ref{firstlaw}), we do get 
\begin{equation}
T \frac{d S}{d t} = V_0\biggl(\frac{a}{a_0}\biggr)^3 
[ \dot{\rho}_{\gamma} + 3 H (\rho_{\gamma} + p_{\gamma})].
\label{def3}
\end{equation} 
Equation (\ref{def3}) shows that $\dot{S}=0$ provided 
the covariant conservation (i.e. Eq. (\ref{cov3})) is enforced.
Different physical fluids will also imply different equations 
of state. Still, as long as the total fluid is not viscous, it is 
possible to write, in general terms, the total energy-momentum 
tensor as 
\begin{equation}
T_{\mu}^{\nu} = (p_{\mathrm{t}} + \rho_{\mathrm{t}}) u_{\mu}u^{\nu} 
- p_{\mathrm{t}} \delta_{\mu}^{\nu},
\label{totenmom}
\end{equation}
where $u_{\mu}$ is the peculiar velocity field of the (total) fluid still 
satisfying $g_{\mu\nu} u^{\mu} u^{\nu} = 1$. For instance (see Appendix B) non-relativistic 
matter (i.e. bosons or fermions in  equilibrium at a temperature that is far below the 
threshold of pair production) leads naturally to an equation of state $p=0$ (often called 
dusty equation of state). Another example could be a homogeneous scalar field whose potential 
vanishes exactly (see section \ref{sec5}). In this case the equation of state is $p = \rho$ (also called stiff 
equation of state, since, in this case, the sound speed coincides with the speed of light).

Viscous effects, when included, may spoil 
the homogeneity off the background. This is the case, for instance, of shear 
viscosity \cite{weinberg}. It is however possible to include viscous effects that 
do not spoil the homogeneity of the background. Two examples 
along this direction are the bulk viscosity effects (see \cite{LLF} for the notion of bulk and shear 
viscosity), and the possible 
transfer of energy (and momentum) between different fluids of the mixture.
For a  single fluid, the total energy-momentum 
${\cal T}_{\mu}^{\nu}$ tensor can then be split
into a perfect contribution, denoted in the following by $T_{\mu}^{\nu}$,
and into an imperfect contribution, denoted by $\Delta T_{\mu}^{\nu}$, i.e. 
\begin{equation}
{\cal T} _{\mu}^{\nu} = T_{\mu}^{\nu} + \Delta T_{\mu}^{\nu},
\label{BV1}
\end{equation}
In general coordinates, and within our set of conventions, the 
contribution of bulk viscous stresses can be written, in turn, as \cite{weinberg}
\begin{equation}
\Delta T_{\mu}^{\nu} = \xi \biggl( \delta_{\mu}^{\nu} - u_{\mu}u^{\nu}\biggr) \nabla_{\alpha} u^{\alpha},
\label{BV2}
\end{equation}
where $\xi$ represents the bulk viscosity coefficient \cite{weinberg}. 
The presence of bulk viscosity 
can also be interpreted, at the level of the background, 
 as an effective redefinition of the pressure (or of the enthalpy).
This discussion follows the spirit of the Eckart approach \cite{EK1}. 
 It must be mentioned, that this approach is 
 phenomenological in the sense that the bulk viscosity is not 
 modeled on the basis of a suitable microscopic theory.  For  
 caveats concerning the Eckart approach see \cite{isr} (see also 
 \cite{maartens2} and references therein).
 The Eckart approach, however, fits with the phenomenological inclusion 
 of a fluid decay rate that has been also considered 
 recently for related applications to cosmological perturbation theory \cite{mg1,mg2}.
 It is interesting to mention that bulk viscous effects have been used in the past in order 
 to provide an early completion of the SCM \cite{bel1,murphy1,barrow1}. According 
 to Eq. (\ref{BV2}), bulk viscosity modifies the pressure so that the spatial components 
 of the energy-momentum tensor can be written as 
 \begin{equation}
{\mathcal  T}_{i}^{j} = - {\mathcal P} \delta_{i}^{j}, \qquad {\mathcal P} = p - 3 H \xi.
\label{BVpar1}
\end{equation}
 The bulk viscosity coefficient $\xi$ may depend on the energy density and it can 
 be parametrized as $\xi(\rho) = (\rho/\rho_{1})^{\nu}$ where different values of 
 $\nu$ will give rise to different cosmological solutions \cite{barrow1}. This 
 parametrization allows for various kinds of unstable quasi-de Sitter solutions \cite{murphy1,bel1,barrow1}.
 Notice that, according to the parametrization of Eq. (\ref{BVpar1}), the covariant conservation 
 of the energy-momentum tensor implies that
 \begin{equation}
 \dot{\rho} + 3 H (\rho + p) = 9 H^2 \xi(\rho).
 \label{BVpar2}
 \end{equation}
 So far it has been  assumed that the energy and momentum exchanges 
 between the different fluids of the plasma are negligible. However, there 
 are situations (rather relevant for CMB physics) where the coupling between 
 different fluids cannot be neglected. An example is the tight-coupling between 
 photons and the lepton-baryon fluid  which is exact well before hydrogen recombination
 (see sections \ref{sec8} and \ref{sec9}). There are also situations, in the early Universe, where 
it is mandatory to consider the decay of a given species into another species. For instance, 
massive particles decaying into massless particles.
Consider, for this purpose,  the situation where the plasma is a mixture of two species 
whose associated energy-momentum tensors can be written as 
can be written as
\begin{equation}
T_{{\rm a}}^{\mu\nu} = ( p_{\rm a} + \rho_{\rm a}) u_{\rm a}^{\mu} u_{\rm a}^{\nu} - p_{\rm a} g^{\mu\nu},
\qquad  T_{{\rm b}}^{\mu\nu} = ( p_{\rm b} + \rho_{\rm b}) u_{\rm b}^{\mu} u_{\rm b}^{\nu} - p_{\rm b} g^{\mu\nu}.
\label{tmnab}
\end{equation}
If the fluids are decaying one into the other (for instance the a-fluid  decays into the b-fluid),  the covariant conservation equation only applies to the global relativistic plasma, while the energy-momentum tensors of the single species 
are not covariantly conserved and their specific form accounts for the 
transfer of energy between the a-fluid and the b-fluid:
\begin{equation}
 \nabla_{\mu} T^{\mu\nu}_{\rm a} = - \Gamma g^{\nu\alpha} u_{\alpha} ( p_{\rm a} 
+ \rho_{\rm a}),\qquad \nabla_{\mu} T^{\mu\nu}_{\rm b} = \Gamma g^{\nu\alpha} u_{\alpha} ( p_{\rm a} 
+ \rho_{\rm a}),
\label{atob}
\end{equation}
where the term $\Gamma$ is the decay rate that can be both space- and 
time-dependent; in Eqs. (\ref{atob}) 
$u_{\alpha}$ represents the (total) peculiar velocity field.
Owing to the form of Eqs. (\ref{atob}), it is clear that the total energy-momentum 
tensor of the two fluids, i.e. $T_{\rm tot}^{\mu\nu} = T^{\mu\nu}_{\rm a} + T^{\mu\nu}_{\rm b}$ is indeed covariantly conserved.
Equations (\ref{atob}) can be easily generalized to the description
of more complicated dynamical frameworks, where the relativistic 
mixture is characterized by more than two fluids. 
Consider the situation where the a-fluid 
decays as ${\rm a} \to {\rm b} + {\rm c}$. Then, if a fraction $f$ of the a-fluid 
decays into the b-fluid and a fraction $(1-f)$  into the c-fluid, 
Eqs. (\ref{atob}) can be generalized as 
\begin{eqnarray}
&& \nabla_{\mu} T^{\mu\nu}_{\rm a} = - \Gamma \,g^{\nu\alpha} \,u_{\alpha} ( p_{\rm a} 
+ \rho_{\rm a}),
\nonumber\\
&& \nabla_{\mu} T^{\mu\nu}_{\rm b} = f \Gamma \, g^{\nu\alpha}\, u_{\alpha} ( p_{\rm a} 
+ \rho_{\rm a}),
\nonumber\\
&&\nabla_{\mu} T^{\mu\nu}_{\rm c} = (1 -f) \Gamma\, g^{\nu\alpha} \, u_{\alpha} ( p_{\rm a} + \rho_{\rm a}),
\label{atobc}
\end{eqnarray}
and so on.  In the case of a FRW metric, Eq. (\ref{atob}) can be written in explicit terms as:
\begin{eqnarray}
&& 
\dot{\rho}_{\rm a} + 3 H ( \rho_{a} + p_{\rm a}) + \overline{\Gamma} ( \rho_{\rm a} + p_{\rm a}) =0
\label{Ac1}\\
&& \dot{\rho}_{\rm b} + 3 H ( \rho_{b} + p_{\rm b}) -
\overline{\Gamma} ( \rho_{\rm a} + p_{\rm a}) =0.
\label{Bc1}
 \end{eqnarray}
If the a-fluid is identified with dusty matter and the b-fluid with radiation we will have 
\begin{eqnarray}
&& \dot{\rho}_{\mathrm{m}} + ( 3 H + \overline{\Gamma}) \rho_{\mathrm{m}}=0,
\nonumber\\
&& \dot{\rho}_{\gamma} + 4 H \rho_{\gamma} - \overline{\Gamma} \rho_{\mathrm{m}} =0.
\end{eqnarray}
Note that $\overline{\Gamma}$ is the homogeneous part of the decay rate. 
To first-order, the decay rate may be spatially inhomogeneous and this 
entails various interesting consequences which will be only marginally 
discussed in these lectures (see, however, \cite{mg1,mg2} and references therein).
It is relevant to stress that, owing to the form of the FRW metric, the homogeneous 
decay rate entails only exchange of energy between the fluids of the mixture. 
To first-order, the peculiar velocity fields will also be affected and the exchange 
of momentum is explicit.
 
\subsubsection{Friedmann-Lema\^itre equations}

The third assumption of the SCM is that the evolution of the geometry 
is connected to the evolution of the sources through the GR equations, i.e. 
\begin{equation}
R_{\mu}^{\nu} - \frac{1}{2} \delta_{\mu}^{\nu} R = 8\pi G T_{\mu}^{\nu},
\label{EEQ}
\end{equation}
where  $R_{\mu\nu}$ is the Ricci tensor, i.e.
\begin{equation} 
R_{\mu\nu} = \partial_{\alpha} \Gamma^{\alpha}_{\mu\nu}  - \partial_{\nu} \Gamma_{\mu\alpha}^{\alpha}  + \Gamma_{\mu\nu}^{\alpha} \Gamma_{\alpha\beta}^{\beta} - \Gamma_{\nu\alpha}^{\beta} \Gamma_{\beta\mu}^{\alpha},
\label{riccit}
\end{equation}
and $R=R_{\mu}^{\mu}$ is the Ricci scalar. By computing the Christoffel connections 
Eq. (\ref{FRW}) allows to determine the components of the Ricci tensor and the the Ricci scalar:
\begin{eqnarray}
&& R_{0}^{0} = - 3 ( H^2 + \dot{H}), 
\nonumber\\
&& R_{i}^{j} = - \biggl( \dot{H} + 3 H^2 + \frac{2 k}{a^2}\biggr) \delta_{i}^{j},
\nonumber\\
&& R= - 6 \biggl(\dot{H} + 2 H^2 + \frac{k}{a^2}\biggr).
\label{Riccicosmic}
\end{eqnarray}
Equation (\ref{EEQ}) supplemented by the covariant conservation 
equations form a closed set of equations that can be consistently 
solved once the equation of state is specified. Using Eq. (\ref{Riccicosmic}) into Eq. (\ref{EEQ}), the 
Friedmann-Lema\^itre equations (often quoted as FL equations in what follows) are 
\begin{eqnarray}
&& H^2 = \frac{8\pi G}{3} \rho_{\mathrm{t}} - \frac{k}{a^2},
\label{FL1}\\
&& \dot{H} = - 4\pi G( \rho_{\mathrm{t}} + p_{\mathrm{t}}) + \frac{k}{a^2},
\label{FL2}\\
&& \dot{\rho}_{\mathrm{t}} + 3 H (\rho_{\mathrm{t}} + p_{\mathrm{t}}) =0.
\label{FL3}
\end{eqnarray}
While Eq. (\ref{FL1}) follows from the $(00)$ component of Eq. (\ref{EEQ}), Eq. (\ref{FL2}) 
is a linear combination of the the $(ij)$ and $(00)$ components of Eq. (\ref{EEQ}). Eq. (\ref{FL3})
follows, as already discussed, from the covariant conservation of the energy-momentum tensor.
Equations (\ref{FL1}), (\ref{FL2}) and (\ref{FL3})  are not all independent 
once the equation of state is specified. 
Sometimes a cosmological term is directly introduced in Eq. (\ref{EEQ}).
The addition of a cosmological term entails the presence of a 
term $\Lambda \delta_{\mu}^{\nu}$ at the right hand side of Eq. (\ref{EEQ}).
In the light of forthcoming applications, it is preferable to think about the $\Lambda$ term as to 
a component of the total energy-momentum tensor of the Universe.
Such a component will contribute to $\rho_{\mathrm{t}}$ and to 
$p_{\mathrm{t}}$ with 
\begin{equation}
\rho_{\Lambda} = \frac{\Lambda}{8\pi G}, \qquad p_{\Lambda} = - \rho_{\Lambda}.
\label{lambda}
\end{equation}
If the evolution of the SCM takes place for positive cosmic times (i.e. $t>0$), the Universe 
expands when $\dot{a} >0$ and contracts when $\dot{a} <0$. If $\ddot{a} >0$, the Universe 
accelerates while if $\ddot{a} <0$ the Universe is said to be decelerating. In the SCM the evolution 
of the Universe can be parametrized as $a(t) \simeq t^{\alpha}$ where $0< \alpha < 1$ and $t>0$.
The power $\alpha$ changes depending upon the different stages of the evolution.
As a complementary remark it is useful to mention that, recently, cosmological models inspired 
by string theory try also to give a meaning to the evolution of the Universe 
when the cosmic time coordinate is negative, i.e. $t<0$. What sets the origin of the time coordinate 
is, in this context, the presence of a curvature singularity that can be eventually 
resolved into a stage of maximal curvature \cite{PBB1,PBB2,PBB3} (see also \cite{reg1,reg2,reg3} and references 
therein for some recent progress on the evolution of the fluctuations in a regularized pre-big bang background
with T-duality invariant dilaton potential). In pre-big bang models 
it is important to extend the cosmic time coordinate also for negative values so, for 
instance parametrizations as $a(t) \sim (- t)^{-\gamma}$ are meaningful.
Classically there are a number of reasonable conditions to be required on the 
components of the energy-momentum tensor of a perfect relativistic fluid.
These conditions go under the name of {\em energy conditions} and play 
an important r\^ole in the context of the singularity theorems that can be 
proved in General Relativity \cite{hawkingellis1,hawkingellis2}. Some of these energy conditions 
may be violated once the components of the energy-momentum 
tensor are regarded as the expectation value of the energy density and of the 
pressure of a quantum field  \cite{fordineq}.

The first condition, called weak energy condition (WEC) stipulates that the energy 
density is positive semi-definite, i.e. $\rho_{\mathrm{t}}\geq0$. 
The dominant energy condition (DOC) implies, instead, that the enthalpy of the 
fluid is positive semi-definite, i.e. $\rho_{\mathrm{t}} + p_{\mathrm{t}} \geq 0$.
Finally the strong energy condition (SEC) demands that $\rho_{\mathrm{t}} + 3 p_{\mathrm{t}} \geq 0$.
According to the Hawking-Penrose theorems \cite{hawkingellis1,hawkingellis2}, if the energy conditions are enforced the 
geometry will develop, in the far past, a singularity where the curvature invariants (i.e. 
$R^2$, $R_{\mu\nu}R^{\mu\nu}$, $R_{\mu\nu\alpha\beta} R^{\mu\nu\alpha\beta}$) will all diverge.
If some of the energy conditions are not enforced, the geometry may still be singular 
is the causal geodesics (i.e. null or time-like) are past-incomplete, i.e. if 
they diverge at a finite value of the affine parameter. A typical example of this 
phenomenon is the expanding branch of de Sitter space which will be later 
scrutinized in the context of inflationary cosmology.

The enforcement of the energy conditions (or their consistent violation) implies interesting 
consequences at the level of the FL equations. 
For instance, if the DOC is enforced, Eq. (\ref{FL2}) demands that $\dot{H} <0$ 
when the Universe is spatially flat (i.e. $k =0$). This means that, in such a case 
the Hubble parameter is always decreasing for $t>0$.
If the SEC is enforced the Universe always decelerates {\em in spite of the value of the spatial 
curvature}. Consider, indeed, the sum of Eqs. (\ref{FL1}) and (\ref{FL2}) and recall that 
$\ddot{a} = ( H^2 + \dot{H}) a$. The result of this manipulation will be \footnote{In section 
\ref{sec5} we will see how to generalize this result to the case when the spatial gradients 
are consistently included in the treatment.}
\begin{equation}
\frac{\ddot{a}}{a} = - \frac{4\pi G}{3} ( \rho_{\mathrm{t}} + 3 p_{\mathrm{t}}),
\label{acc}
\end{equation}
showing that, as long as the SEC is enforced $\ddot{a}<0$. This conclusion 
can be intuitively understood since, under normal conditions, gravity is 
an attractive force: two bodies slow down as they move apart.
The second interesting aspect of Eq. (\ref{acc}) is that the spatial 
curvature drops out completely. Again this aspect suggest that inhomogeneities 
cannot make gravity repulsive. Such a conclusion can be generalized to the situation 
where the FL equations are written without assuming the homogeneity of the background geometry 
(see section \ref{sec5} for the first rudiments on this approach).
A radiation-dominated fluid (or a matter-dominated fluid) respect both the  SEC and the DOC.
Hence, in spite of the presence of inhomogeneities, the Universe will always expand (for $t>0$), it will 
always decelerate (i.e. $\ddot{a} <0$) and the Hubble parameter will always decrease (i.e. $\dot{H} <0$).
To have an accelerated Universe the SEC must be violated. By parametrizing 
the equation of state of the fluid as $ p_{\mathrm{t}} = w_{\mathrm{t}} \rho_{\mathrm{t}}$ the SEC 
will be violated provided $w_{\mathrm{t}} < -1/3$.

Introducing the critical density and the critical parameter at a given cosmic time $t$
\begin{equation}
\rho_{\mathrm{crit}} = \frac{3 H^2}{8\pi G},\qquad \Omega_{\mathrm{t}} = \frac{\rho_{\mathrm{t}}}{\rho_{\mathrm{crit}}}
\label{critgen}
\end{equation}
already mentioned in section \ref{sec1}, Eq. (\ref{FL1}) can be written as 
\begin{equation}
\Omega_{\mathrm{t}} = 1 + \frac{k}{a^2 H^2}.
\label{criteq}
\end{equation}
Equation (\ref{criteq}) has the following three direct consequences:
\begin{itemize}
\item{} if $k=0$ (spatially flat Universe), $\Omega_{\mathrm{t}}=1$ (i.e. 
$\rho_{\mathrm{t}} = \rho_{\mathrm{crit}}$);
\item{} if $k <0 $ (spatially open Universe), $\Omega_{\mathrm{t}}<1$ (i.e. 
$\rho_{\mathrm{t}} < \rho_{\mathrm{crit}}$);
\item{}  if $k >0$ (spatially closed Universe), $\Omega_{\mathrm{t}}>1$ (i.e. 
$\rho_{\mathrm{t}} > \rho_{\mathrm{crit}}$).
\end{itemize}

Always at the level of the terminology, the {\em deceleration parameter}
is customarily introduced:
\begin{equation}
q(t) = - \frac{\ddot{a}}{a\,H^2}.
\label{decpar}
\end{equation}
Notice the minus sign in the convention of Eq. (\ref{decpar}): if $q <0$ the Universe 
accelerates (and if $q>0$ the Universe decelerates). In different 
applications, it is important to write, solve and discuss the analog of Eqs. (\ref{FL1}),
(\ref{FL2}) and (\ref{FL3}) in the conformal time parametrization 
already introduced in Eq. (\ref{FRW2}).  From Eq. (\ref{Riccicosmic}) 
\begin{eqnarray}
&& R_{0}^{0} = - \frac{3}{a^2} {\mathcal H}', 
\nonumber\\
&& R_{i}^{j} = - \frac{1}{a^2} \biggl( {\mathcal H}' + 2 {\mathcal H}^2 + 2k\biggr) \delta_{i}^{j},
\nonumber\\
&& R= - \frac{6}{a^2} \biggl({\mathcal H}' + 2 {\mathcal H}^2 + k \biggr).
\label{Ricciconformal}
\end{eqnarray}
Using Eq. (\ref{Ricciconformal}) in Eq. (\ref{EEQ}) the conformal time counterpart of 
Eqs. (\ref{FL1}), (\ref{FL2}) and (\ref{FL3}) 
become, respectively,
\begin{eqnarray}
&& {\mathcal H}^2 = \frac{8\pi G}{3} a^2 \rho_{\mathrm{t}} - k,
\label{FL1C}\\
&& {\mathcal H}^2 - {\mathcal H}' = 4\pi G a^2 (\rho_{\mathrm{t}} + p_{\mathrm{t}}) - k,
\label{FL2C}\\
&& \rho_{\mathrm{t}}' + 3 {\mathcal H}(\rho_{\mathrm{t}} + p_{\mathrm{t}})=0,
\label{FL3C}
\end{eqnarray}
where the prime denotes a derivation with respect to the conformal 
time coordinate $\tau$ and ${\mathcal H} = a'/a$.
Note that Eqs. (\ref{FL1C}), (\ref{FL2C}) and (\ref{FL3C}) 
can be swiftly obtained from Eqs. (\ref{FL1}), (\ref{FL2}) and (\ref{FL3}) by bearing in mind 
the following (simple) dictionnary:
\begin{equation}
H = \frac{{\mathcal H}}{a}, \qquad \dot{H} = \frac{1}{a^2}({\mathcal H}' - {\mathcal H}^2).
\label{HtoHcal}
\end{equation}

\subsection{Matter content of the SCM}
\label{sub22}
According to the present experimental understanding 
 our own Universe is to a good approximation 
spatially flat. Furthermore, the total 
energy density receives contribution from three (physically different) components:
\begin{equation}
\rho_{\mathrm{t}} = \rho_{\mathrm{M}} + \rho_{\mathrm{R}} + \rho_{\Lambda}.
\label{mattercontent}
\end{equation}
Using the definition of the critical density parameter, a critical fraction
is customarily introduced for every fluid of the mixture, i.e. 
\begin{equation}
\Omega_{\mathrm{t}} = \Omega_{\mathrm{M}} + \Omega_{\mathrm{R}} + \Omega_{\Lambda}.
\label{OMs}
\end{equation}
In Eq. (\ref{mattercontent}), $\rho_{\mathrm{M}}$ parametrizes the contribution
of non-relativistic species which are today stable and, in particular, 
$\rho_{\mathrm{M}0} = \rho_{\mathrm{c}0} + \rho_{\mathrm{b}0}$, i.e. $\rho_{\mathrm{M}0}$
(the present matter density) receives contribution from a cold dark matter component 
(CDM) and from a baryonic component\footnote{The subscript $0$, when not otherwise stated, denotes 
the present value of the corresponding quantity.}. Both components have the equation of state 
of non-relativistic matter, i.e. $p_{\mathrm{c}} =0$  and $p_{\mathrm{b}} =0$ and 
the covariant conservation of each species (see Eqs. (\ref{cov1}) and (\ref{cov2})) implies 
\begin{equation}
\dot{\rho}_{b} + 3 H \rho_{\mathrm{b}} =0, \qquad \dot{\rho}_{\mathrm{c}} + 3 H \rho_{\mathrm{c}}=0.
\label{covcons}
\end{equation}
The term {\em cold dark matter} simply means that this component is non-relativistic 
today and it is dark, i.e. it does not emit light and it does not absorb light.
CDM particles are {\em inhomogeneously} distributed. Dark matter may also be hot. However, in this 
case it is more difficult to form structures because of the higher velocities 
of the particles. The present abundance of non-relativistic matter can be appreciated 
by the following illustrative values:
\begin{equation}
h_{0}^2 \Omega_{\mathrm{M}0} = 0.134, \qquad h_{0}^2 \Omega_{\mathrm{c}0} = 0.111,
\qquad h_{0}^2 \Omega_{\mathrm{b}0} = 0.023.
\label{valuesmatter}
\end{equation}
Notice that, in Eq. (\ref{valuesmatter}) the abundances are independent on the indetermination 
of the Hubble parameter $h_{0}$. For numerical estimates $h_{0}$ will be taken between $0.7$ and $0.73$.
Concerning the set of parameters adopted in Eq. (\ref{valuesmatter}) few comments are 
in order. The recent WMAP three year 
data \cite{map3}, when combined with ``gold" sample of SNIa \cite{riess}  give values compatible with the ones adopted here.  However, if WMAP data alone 
are used we would have, instead, a slightly smaller value for the
critical fraction of matter, i.e. $h_0^2 \Omega_{\mathrm{M}0} \simeq 1.268$.
The values of the parameters chosen here are just illustrative 
in order to model, in a realistic fashion, the radiation-matter 
transition (see subsection \ref{sub24}) which is crucial 
for CMB physics. 

For CDM the main observational evidences come from the rotation curves of spiral galaxies, 
from the mass to light ratio in clusters and from CMB physics. For baryonic 
matter an indirect evidence stems from big-bang nucleosynthesis (BBN) (see \cite{bernstein} for 
a self-contained introduction to BBN). Indeed 
for temperatures smaller than $1$ MeV weak interactions fall out of thermal equilibrium 
and the neutron to proton ratio decreases via free neutron decay. A bit later the abundances 
of the light nuclear elements (i.e. $^{4}\mathrm{He}$, $^{3}\mathrm{He}$, $^{7}\mathrm{Li}$ and
D) start being formed. While the discussion of BBN is rather interesting in its own right (see Appendix B for 
some further details), it suffices 
to note here that, within the SCM, the homogeneous BBN only depends, in principle upon two parameters: one is 
the temperature of the plasma (or, equivalently, the expansion rate) and the other is the ratio 
between the concentration of baryons and the concentration of photons. 
Recalling the expression for the concentration of CMB photons, 
and recalling that $\rho_{\mathrm b} = m_{\mathrm{b}} n_{\mathrm{b}}$ we have 
\begin{equation}
\eta_{\mathrm{b}0} = \frac{n_{\mathrm{b}0}}{n_{\gamma0}} = 6.27 \times 10^{-10} 
\biggl(\frac{h_{0}^2 \Omega_{\mathrm{b}0}}{0.023}\biggr),
\label{etab}
\end{equation}
having taken the typical baryon mass of the order of the proton mass.

On top of the dark matter component, the present Universe seems to contain also 
another dark component which is, however, much more homogeneously 
distributed than dark matter.  It is therefore named dark energy and satisfies 
an equation of state with barotropic index smaller than $-1/3$. In particular, a viable 
and current model of dark energy is the one of a simple cosmological term 
with 
\begin{equation}
h_{0}^2 \Omega_{\Lambda 0} = 0.357.
\label{Lambda2}
\end{equation}
Notice that for a fiducial value of $h_{0} \simeq 0.7$
\begin{equation}
\Omega_{\mathrm{M}0} \simeq 0.27, \qquad \Omega_{\mathrm{c}0} \simeq 0.22,\qquad
\Omega_{\mathrm{b}0} \simeq 0.046,\qquad  \Omega_{\Lambda 0} \simeq 0.73.
\label{omegaML}
\end{equation}

Finally, in the present Universe, as discussed in section \ref{sec1} there is also radiation.
In particular we will have that 
\begin{equation}
h_{0}^2 \Omega_{\mathrm{R}0} = h_{0}^2 \Omega_{\gamma0} + h_{0}^2 \Omega_{\nu0} + h_{0}^2
\Omega_{\mathrm{gw}0},
\label{omegaR}
\end{equation}
where $\Omega_{\nu0}$ denotes the contribution of neutrinos and $\Omega_{\mathrm{g}0}$ the contribution 
of relic gravitons.
In Eq. (\ref{omegaR}) we will have 
\begin{equation}
h_{0}^2 \Omega_{\gamma 0} = 2.47 \,\,10^{-5},\qquad h_{0}^2 \Omega_{\nu 0} = 1.68 \,\,10^{-5},
\qquad h_{0}^2 \Omega_{\mathrm{gw} 0} < 10^{-11}.
\label{omegaRval}
\end{equation}
The contribution of relic gravitons is, today, smaller 
than $10^{-11}$ this bound stems from the analysis of the integrated Sachs-Wolfe contribution which 
will be discussed later in this set of lectures (see section \ref{sec7}).
If neutrinos have masses smaller than the $\mathrm{meV}$ they are today non-relativistic 
and, in principle, should not be counted as radiation. However, since the temperature 
of CMB was, in the past, much larger (as it will be discussed below), they will be effectively 
relativistic at the moment when matter decouples from radiation. Since, along these lectures, we will 
be primarily interested in the physics of CMB we will assume that neutrinos 
are effectively massless.  To be more precise, we can say that current oscillation data require at least 
one neutrino eigenstate to have a mass exceeding $0.05$ eV. 
In this minimal case $h_{0}^2 \Omega_{\nu0} \simeq 5\times 10^{-4}$ so that the neutrino contribution to the matter 
budget will be negligibly small.  However, a nearly degenerate pattern of mass eigenstates could allow larger 
densities, since oscillation experiments only measure differences in the values of the squared masses. 

\subsection{The future of the Universe}
\label{sub23}

From the analysis of the luminosity distance (versus the redshift) it appears 
that type-Ia supernovae are dimmer than expected and this suggests that 
at high redshifts (i.e. $z\geq 1$) the Universe is effectively accelerating \cite{riess}.
The 
redshift $z$ is defined (see Appendix \ref{APPA} for further details) as 
\begin{equation}
1 + z = \frac{a_{0}}{a},
\end{equation}
where $a_{0}$ is the present value of the scale factor and $a$ denotes a generic stage of expansion 
preceding the present epoch (i.e. $a< a_{0}$). The concept of redshift (see Appendix \ref{APPA}) is related 
with the observation that, in an expanding Universe, the spectral lines of emitted radiation become 
more red (i.e. they redshift, they become longer) than in the case when the Universe does not expand.
Given the matter content of the present Universe, its destiny can be guessed by using 
the FL equations and by integrating them forward in time. 
From Eq. (\ref{FL1}), with simple algebra, it is possible to obtain the following equation:
\begin{equation}
\frac{d\alpha}{d x} = \sqrt{\frac{\Omega_{\mathrm{M}0}}{\alpha} + \Omega_{\Lambda 0} \alpha^2 + \Omega_{k} +
\frac{\Omega_{\mathrm{R}0}}{\alpha^2}},
\label{Eq1}
\end{equation}
where the following rescaled variables have been defined:
\begin{equation}
\alpha = \frac{a}{a_0},\qquad x = H_{0} t, \qquad \Omega_{k} = - \frac{k}{a_{0}^2 H_{0}^2},
\label{DEFEQ1}
\end{equation}
and the quantity with subscripts $0$ always refer to the present time \footnote{Notice that $k$ and 
$\Omega_{k}$ have opposite sign. While it is useful to define $\Omega_{k}$ as a critical fraction, it may also 
engender unwanted confusions which are related to the fact that, physically, the spatial 
curvature is not a further form of matter. With these caveats the use of $\Omega_{k}$ is rather 
practical.}.
To derive Eq. (\ref{Eq1}) it must also be borne in mind that  a first integration of the covariant conservation 
equations leads to the following relations:
\begin{equation}
\rho_{\mathrm{R}} = \rho_{\mathrm{R}0} \biggl(\frac{a_0}{a}\biggr)^{4},\qquad
\rho_{\mathrm{M}} = \rho_{\mathrm{M}0} \biggl(\frac{a_0}{a}\biggr)^{3},\qquad
\rho_{\Lambda}= \rho_{\Lambda 0}.
\label{DEFEQ2}
\end{equation}
From Eq. (\ref{Eq1}), different possibilities exist for the future dynamics of the Universe. These 
possibilities depend on the relative weight of the various physical components of the present Universe.
In the case $\Omega_{\Lambda0}$  Eq. (\ref{Eq1}) reduces to 
\begin{equation}
\int \frac{\sqrt{\alpha} d\alpha}{\sqrt{\Omega_{\mathrm{M}0} + \Omega_{k} \alpha}} = H_{0}(t -t_0).
\label{onlymatter}
\end{equation}
If $\Omega_{k} =0$, $a(t)$ expands forever with $a(t) \sim t^{2/3}$ (decelerated expansion).
If $\Omega_{k} <0$ (closed Universe) the Universe will collapse in the future an for a 
critical value $\alpha_{\mathrm{coll}} \simeq \Omega_{\mathrm{M}0}/|\Omega_{k}|$.
Finally, if $\Omega_{k} >0$ (open Universe) the geometry will expand forever in a decelerated way.
Notice that, in Eq. (\ref{onlymatter}) the r\^ole of radiation has been neglected 
since radiation is subleading today and it will be even more subleading in the future since 
it decreases faster than matter and faster than the dark energy.

If $\Omega_{\Lambda0} \neq 0$ and $\Omega_{k} =0$ 
Eq. (\ref{Eq1}) can be solved in explicit terms with the result that
\begin{equation}
\frac{a}{a_0} = \biggl(\frac{\Omega_{\mathrm{M}0}}{\Omega_{\Lambda0}}\biggr)^{1/3} \biggl\{
\sinh{\biggl[\frac{3}{2} \sqrt{\Omega_{\Lambda0}} H_{0} (t -t_0)\biggr]}\biggr\}^{2/3}.
\label{solds}
\end{equation}
This solution interpolates between a matter-dominated Universe expanding in a decelerated way 
as $t^{2/3}$ and an exponentially expanding Universe which is also accelerating.
To get to Eq. (\ref{solds}), Eq. (\ref{Eq1}) can be written as 
\begin{equation}
\int\frac{\sqrt{\alpha} d\alpha}{\sqrt{1 + \frac{\Omega_{\Lambda 0}}{
\Omega_{\mathrm{M} 0}} \alpha^3}} = \sqrt{\Omega_{\mathrm{M}0}}\,\, d x.
\end{equation}
By introducing the auxiliary variable 
\begin{equation}
\alpha^{3/2} \sqrt{\frac{\Omega_{\Lambda 0}}{\Omega_{\mathrm{M}0}}} = y,
\end{equation}
we do obtain 
\begin{equation}
\int \frac{d y}{\sqrt{1 + y^2}} = \frac{3}{2} \sqrt{\Omega_{\Lambda 0}} \,\,H_{0} \,( t - t_{0}).
\end{equation}
Finally, by introducing a second auxiliary variable $ y = \sinh{\beta}$ the 
integral can be readily solved and Eq. (\ref{solds}) reproduced. 
While the discussion for $\Omega_{k}\neq 0$, $\Omega_{\Lambda 0} \neq 0$ and $\Omega_{\mathrm{M}0}\neq 0$ 
is more complicated and will not be treated here, it is also clear that given the matter content of the present Universe,
it is reasonable to expect that the, in the future, the Universe will accelerate faster and faster 
while the r\^ole of non relativistic matter (and of radiation) will be progressively negligible.

There are a number of ways in which the kinematical features of the present 
Universe can be observationally accessible. The main tool is represented 
by the various distance concepts used by astronomers. 
The three useful distance measures that could be mentioned 
are (see Appendix \ref{APPA} for further details on the derivation 
of the explicit expressions):
\begin{itemize}
\item{} the distance measure (denoted with $r_{\mathrm{e}}(z)$ in Appendix A and often denoted with $D_{\mathrm{M}}(z)$ in the literature);
\item{} the angular diameter distance $D_{\mathrm{A}}(z)$;
\item{} the luminosity distance.
\end{itemize}
These three distances are all functions of the redshift $z$ and of the (present) 
critical fractions of matter, dark energy, radiation and curvature, i.e., respectively,
$\Omega_{\mathrm{M}0}$, $\Omega_{\Lambda0}$, $\Omega_{\mathrm{R}0}$, and $\Omega_{k}$. In practice, the dependence upon $\Omega_{\mathrm{R}0}$ can be 
dropped and it becomes relevant for {\em very} large redshift, i.e. $z\simeq 10$. 

The three distances introduced in the aforementioned list of items are integrated quantities in the sense that they depend upon the 
integral of the inverse of the Hubble parameter from $0$ to the generic redshift $z$ (see Appendix \ref{APPA} for 
a derivation). 
The angular diameter distance and the luminosity distance are related 
to $r_{\mathrm{e}}(z)$ as 
\begin{equation}
D_{\mathrm{A}}(z) =\frac{a_{0} r_{\mathrm{e}}(z)}{1+z},\qquad 
D_{\mathrm{L}}(z) =a_{0} r_{\mathrm{e}}(z)(1+z),
\end{equation}
where $a_{0}$ is the present value of the scale factor that could be 
conventionally normalized to $1$. The distance measure has been 
denoted by $r_{\mathrm{e}}$ since it represents the coordinate distance 
(defined in the FRW line element) once the origin of the coordinate system is placed
in the Milky way. 
The angular diameter distance gives us the possibility of determining the distance 
of an object by measuring its angular size in the sky. Of course to conduct 
successfully a measurement we must have a set of standard rulers, i.e. a set 
of objects that have, at different redshifts, the same size.

The luminosity distance gives us the possibility of determining the 
distance of an object from its apparent luminosity. Of course, as in the case of the 
angular diameter distance, to complete successfully the measurement we would need 
a set of standard candles , i.e. a set of object with the same absolute 
luminosity.

\begin{figure}
\centering
\includegraphics[height=6cm]{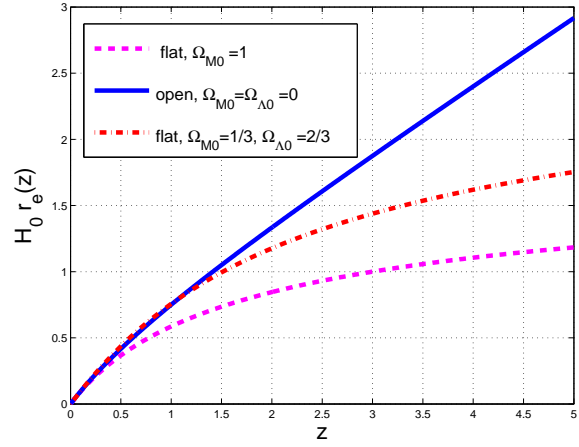}
\caption{The distance measure as a function of the redshift for three 
different models of the Universe.}
\label{DIS1}      
\end{figure}

\begin{figure}
\centering
\includegraphics[height=6cm]{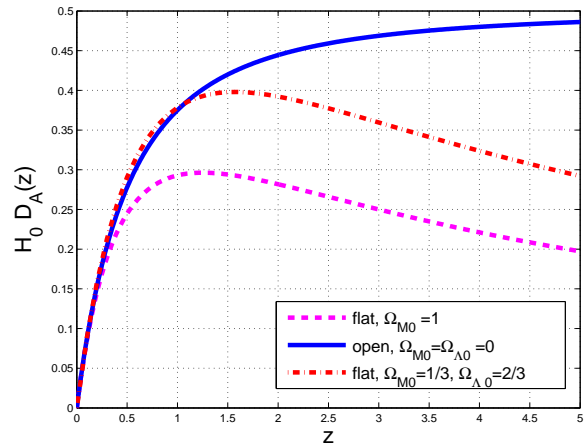}
\caption{The angular diameter distance as a function of the redshift for the same 
sample of models discussed in Fig. \ref{DIS1}.}
\label{DIS2}      
\end{figure}

\begin{figure}
\centering
\includegraphics[height=6cm]{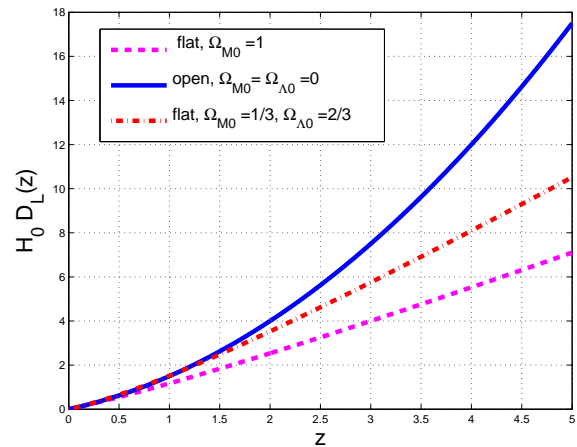}
\caption{The luminosity distance as a function of the redshift for the same 
sample of models discussed in Figs. \ref{DIS1} and \ref{DIS2}.}
\label{DIS3}      
\end{figure}

In Figs. \ref{DIS1}, \ref{DIS2} and \ref{DIS3} 
the three concepts of distance introduced above 
are illustrated. In Fig. \ref{DIS1} the distance measure 
is illustrated in the case of three models. The lowest (dashed)
curve holds in the case of a flat Universe with $\Omega_{\mathrm{M}0}=1$.
The intermediate (dot-dashed) curve holds in the case of a flat Universe
with $\Omega_{\mathrm{M}0}=1/3$ and $\Omega_{\Lambda0}=2/3$. Finally 
the upper curve (full line) holds in the case of an open Universe dominated 
by the spatial curvature (i.e. $\Omega_{\mathrm{M}0} =\Omega_{\Lambda0}=0$ and 
$\Omega_{k} =1$).
The angular diameter distance is reported in Fig. \ref{DIS2} for the same 
sample of models described by Fig. \ref{DIS1}. For large redshift,
the angular diameter distance may well be decreasing, for some models. 
This means that the object that is further away may appear larger in the sky.
Finally, in Fig. \ref{DIS3} the luminosity distance is illustrated.
\subsection{The past of the Universe}
\label{sub24}
Even if, today, $\Omega_{\mathrm{R}0} \ll \Omega_{\mathrm{M}0}$, in the past 
history of the Universe radiation was presumably the dominant component.
By going back in time, the dark-energy does not increase (or it increases 
very slowly) while radiation increases faster than the non-relativistic matter. 
\begin{figure}
\centering
\includegraphics[height=6cm]{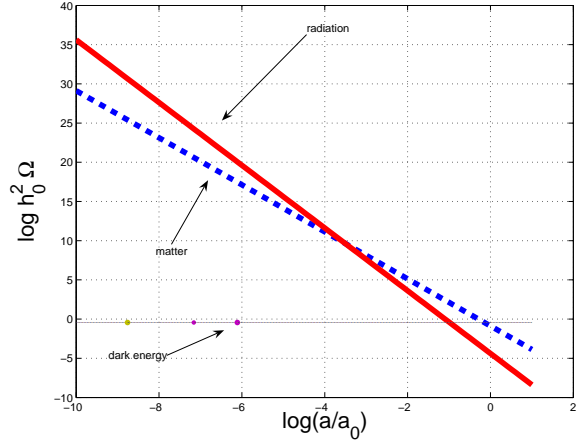}
\caption{The evolution of the critical fractions of matter, radiation and dark energy as a function of the logarithm (to base $10$) of 
$(a/a_{0})$ where 
$a_{0}$ denotes, as usual, the present value of the scale factor.}
\label{HOT}      
\end{figure}
In Fig. \ref{HOT} the evolutions of the critical fractions of matter, radiation and dark energy are reported assuming, as present values 
of the illustrated quantities, the numerical values introduced in the present section (see, for instance, Eq. (\ref{valuesmatter})). Recalling the evolution of the radiation and matter energy densities, 
radiation and matter were equally abundant at a redshift
\begin{equation}
1 + z_{\mathrm{eq}} = \frac{a_{0}}{a_{\mathrm{eq}}} = \frac{h_{0}^2 \Omega_{\mathrm{M}0}}{h_{0}^2 
\Omega_{\mathrm{R}0}}= 3228 \biggl(\frac{h_{0}^2 \Omega_{\mathrm{M}0}}{0.134}\biggr).
\label{zeq}
\end{equation}
For $z> z_{\mathrm{eq}}$ (i.e. $a< a_{\mathrm{eq}}$) the Universe is effectively dominated by radiation. 
For $z< z_{\mathrm{eq}}$   (i.e. $a> a_{\mathrm{eq}}$) the Universe is effectively dominated by non-relativistic 
matter until the moment dark-energy starts being dominant.
Around the equality time, various important phenomena take place in the life 
of the Universe and they are directly related to CMB physics. For this reason it is 
practical to solve the FL equations across the transition between radiation and matter.
Assuming that the only matter content is given by matter and radiation and supposing 
that the Universe is spatially flat, Eq. (\ref{FL1C}) implies the following 
differential equation 
\begin{equation} 
\frac{1}{H_{0}} \frac{d}{d\tau} \biggl(\frac{a}{a_{\mathrm{eq}}}\biggr) = \frac{\Omega_{\mathrm{M}0}}{\sqrt{\Omega_{\mathrm{R}0}}} \biggl[ \biggl(\frac{a}{a_{\mathrm{eq}}}\biggr) +1 \biggr]^{1/2},
\label{eqMR}
\end{equation}
whose solution is simply:
\begin{equation}
a(\tau) = a_{\rm eq} \biggl[ \biggl(\frac{\tau}{\tau_{1}}\biggr)^2 + 2 \biggl( \frac{\tau}{\tau_{1}}\biggr)\biggr],
\label{aint}
\end{equation}
with
\begin{equation}
 \tau_{1} = \frac{2}{H_{0}} \sqrt{\frac{a_{\rm eq}}{\Omega_{\mathrm{M}0}}} \simeq 288.16\, 
\biggl(\frac{h_0^2 \Omega_{\mathrm{M}0}}{0.134} \biggr)^{-1} 
\mathrm{Mpc}.
\label{tau1}
\end{equation}
From Eq. (\ref{aint}) $\tau_{\mathrm{eq}} = (\sqrt{2} -1) \tau_{1}$ and, thus,
\begin{equation}
\tau_{\mathrm{eq}} = 119.35 \,\biggl(\frac{h_0^2 \Omega_{\mathrm{M}0}}{0.134} \biggr)^{-1} 
\mathrm{Mpc},\qquad  \tau_{\mathrm{dec}} = 283.47\, \mathrm{Mpc},
\label{scales1}
\end{equation}
where the second relation holds for $h_{0}^2 \Omega_{\mathrm{M}0} = 0.134$.
Notice that, for $\tau \ll \tau_1$, $a(\tau) \sim \tau$ (which implies $a(t) \sim t^{1/2}$ in cosmic time).
For $\tau \gg \tau_{1}$, $a(\tau)\sim \tau^2$ (which implies $a(t) \sim t^{2/3}$ in cosmic time).
After equality, two important phenomena take place, namely Hydrogen recombination
and the decoupling of radiation from matter. These will be the last two topics treated in the present section.

\subsubsection{Hydrogen recombination}
After electron positron annihilation, the 
concentration of electrons can be written as $n_{\mathrm{e}} = 
x_{\mathrm{e}} n_{\mathrm{B}}$ where $n_{\mathrm{B}}$ is the 
concentration of baryons and $x_{\mathrm{e}}$ is the ionization 
fraction. Before equality, i.e. deep in the radiation-dominated epoch, 
$x_{\mathrm{e}} =1$: the concentration of free electrons exactly 
equals the concentration of protons and the Universe is globally 
neutral. 

After matter-radiation equality, when the temperature drops 
below the eV, protons start combining with free electrons and the 
ionization fraction drops from $1$ to $10^{-4}$--$10^{-5}$. The drop
in the ionization fraction occurs because free electrons are captured by 
protons to form Hydrogen atoms according to the reaction 
$e + p \to H + \gamma$. For sake of simplicity we can think that the Hydrogen is formed in its lowest 
energy level. It would be wrong to guess, however, that this process 
takes place around $13.2$ eV. It takes place, on the contrary, for typical 
temperatures that are of the order of $0.3$ eV. The rationale for this statement 
is that the pre-factor in the equilibrium concentration of free electrons 
is actually small and, therefore, the Hydrogen formation cannot be simply 
estimated from the Boltzmann factor.

The redshift of recombination is defined as the 
moment at which the ionization fraction drops from his equilibrium value 
(i.e. $x_{\rm e} =1$) to $x_{\mathrm e} \simeq 0.1$. The redshift 
of decoupling is the determined by the requirement that the ionization fraction 
decreases even more. At $x_{\mathrm e} \simeq 10^{-4}$ the decoupling is considered 
achieved. Let us go through a more quantitative discussion of these figures. 
When the temperature of the plasma is high enough the reactions 
of recombination and photodissociation of Hydrogen are in thermal equlibrium, i.e.
$e + p \to H + \gamma$ is balanced by $H + \gamma \to e + p$. In this 
situation the concentrations of Hydrogen, of the protons and of the electrons follow, respectively, 
from the equlibrium distribution (see Appendix \ref{APPB}  for further details):
\begin{eqnarray}
&& n_{\rm H} = g_{\mathrm{H}} \biggl(\frac{m_{\mathrm{H}} T}{2\pi} \biggr)^{3/2} 
e^{(\mu_{\mathrm{H}} -m_{\mathrm{H}})/T}, 
\label{nH}\\
&& n_{\rm p} = g_{\mathrm{p}} \biggl(\frac{m_{\mathrm{p}} T}{2\pi} \biggr)^{3/2} 
e^{(\mu_{\mathrm{p}} -m_{\mathrm{p}})/T}, 
\label{np}\\
&& n_{\rm e} = g_{\mathrm{e}} \biggl(\frac{m_{\mathrm{e}} T}{2\pi} \biggr)^{3/2} 
e^{(\mu_{\mathrm{e}} -m_{\mathrm{e}})/T},
\label{ne}
\end{eqnarray}
where $g_{\mathrm{H}}$,  $g_{\mathrm{p}}$ and  $g_{\mathrm{e}}$ are, respectively, 
$4$, $2$ and $2$.
Since we are in a situation of chemical equilibrium (see Appendix \ref{APPB}) we can relate 
the various chemical potentials according to the order of the reaction, i.e. $\mu_{\mathrm{H}} = \mu_{\mathrm{p}}
+\mu_{\mathrm{e}}$.  Eliminating $\mu_{\mathrm{H}}$ from Eq. (\ref{nH}) and using the product of 
 Eqs. (\ref{np}) and (\ref{ne}) to express $\exp{[(\mu_{\mathrm{p}}+ \mu_{\mathrm{e}})/T]}$ in terms 
 of the electron and proton concentrations, the following expression can be obtained:
\begin{equation}
 n_{\mathrm{H}} = n_{\mathrm{e}} n_{\mathrm{p}} \biggl(\frac{m_{\mathrm{e}} T}{2\pi}\biggr)^{-3/2} e^{E_{0}/T},\qquad E_{0} = m_{\mathrm{e}}+ m_{\mathrm{p}} - m_{\mathrm{H}} = 13.26\,\, \mathrm{eV},
 \label{nH2}
 \end{equation}
 where $E_{0}$ is the absolute value of the binding energy of the hydrogen atoms that 
 corresponds to the energy of the lowest energy level since it has been assumed that 
 hydrogen recombines in the fundamental state.
 We now observe that:
 \begin{itemize}
 \item{} the Universe is electrically neutral, hence $n_{\mathrm{p}} = n_{\mathrm{p}}$;
 \item{} the total baryonic concentration of the system is $n_{\mathrm{B}} = n_{\mathrm{H}} + n_{\mathrm{p}}$
 \item{} the concentration of free electrons (or free protons) can be related to the baryonic 
 concentration as $n_{\mathrm{e}} = x_{\mathrm{e}} n_{\mathrm{B}}$ where $x_{\mathrm{e}}$ is the ionization 
 fraction.
 \end{itemize}
 Concerning the second observation, it should be incidentally remarked that  the 
 total baryonic concentration is given, in general terms, by $n_{\mathrm{B}} = n_{\mathrm{N}}- n_\mathrm{\overline{N}}$ 
 (where $n_{\mathrm{N}}$ and $n_{\mathrm{\overline{N}}}$ are, respectively, the concentrations 
 of nucleons and antinucleons). However, for $T< 10\,\,{\mathrm{MeV}}$,  $n_\mathrm{\overline{N}}\ll 1$ and, therefore, $n_{\mathrm{B}} = n_{\mathrm{n}}+ n_{\mathrm{p}}$.  
 The success of big-bang nucleosynthesis implies, furthermore, that approximately one quarter of all
 nucleons form nuclei with atomic mass number $A >1$ (and mostly $^{4}\mathrm{He}$), while 
 the remaining three quarters are free protons. In similar terms we can also say that for temperatures 
 $T< 10\,{\mathrm{keV}}$ the concentration of positrons is negligible in comparison with the concentration 
 of electrons.
 
\begin{figure}
\centering
\includegraphics[height=5cm]{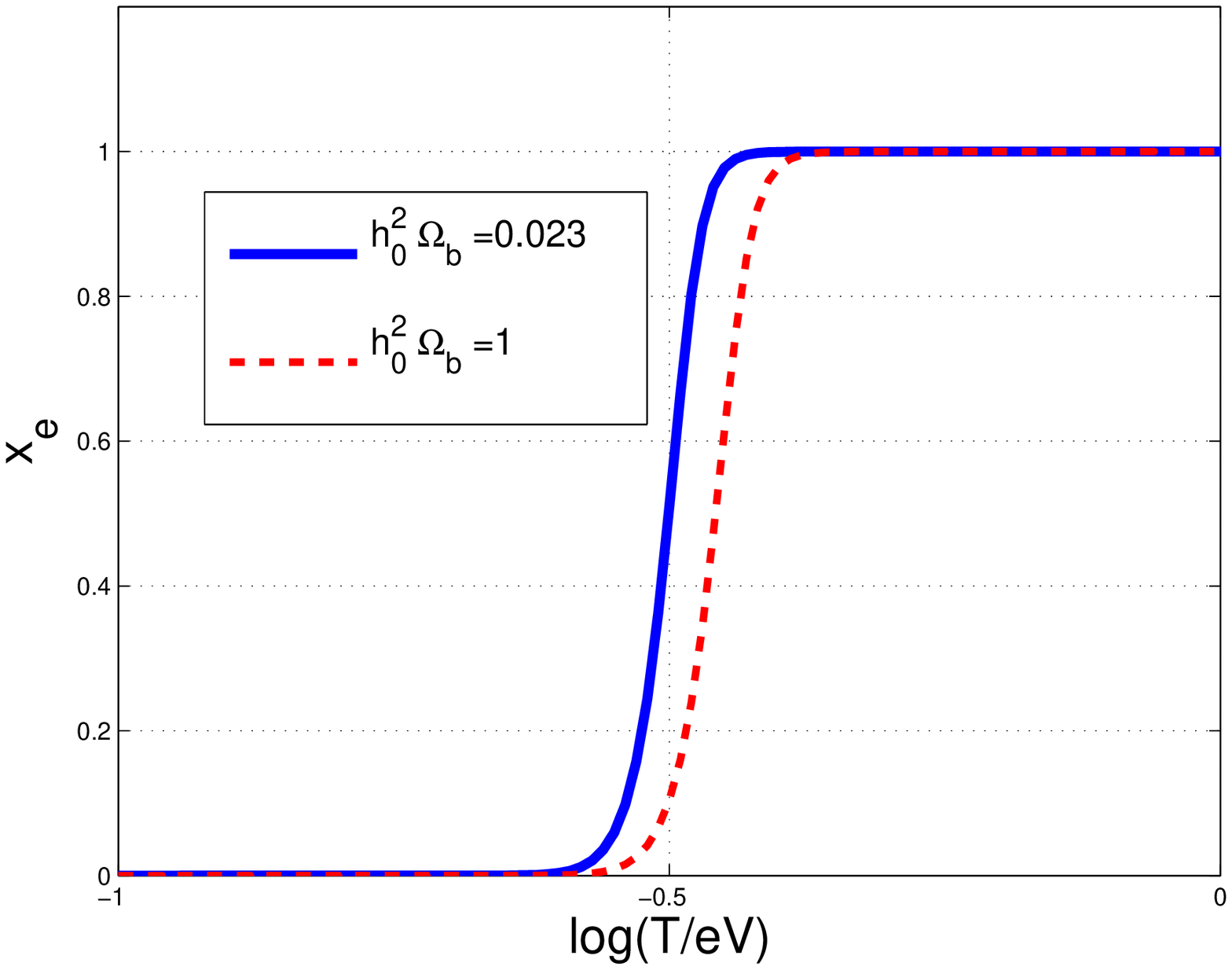}
\includegraphics[height=5cm]{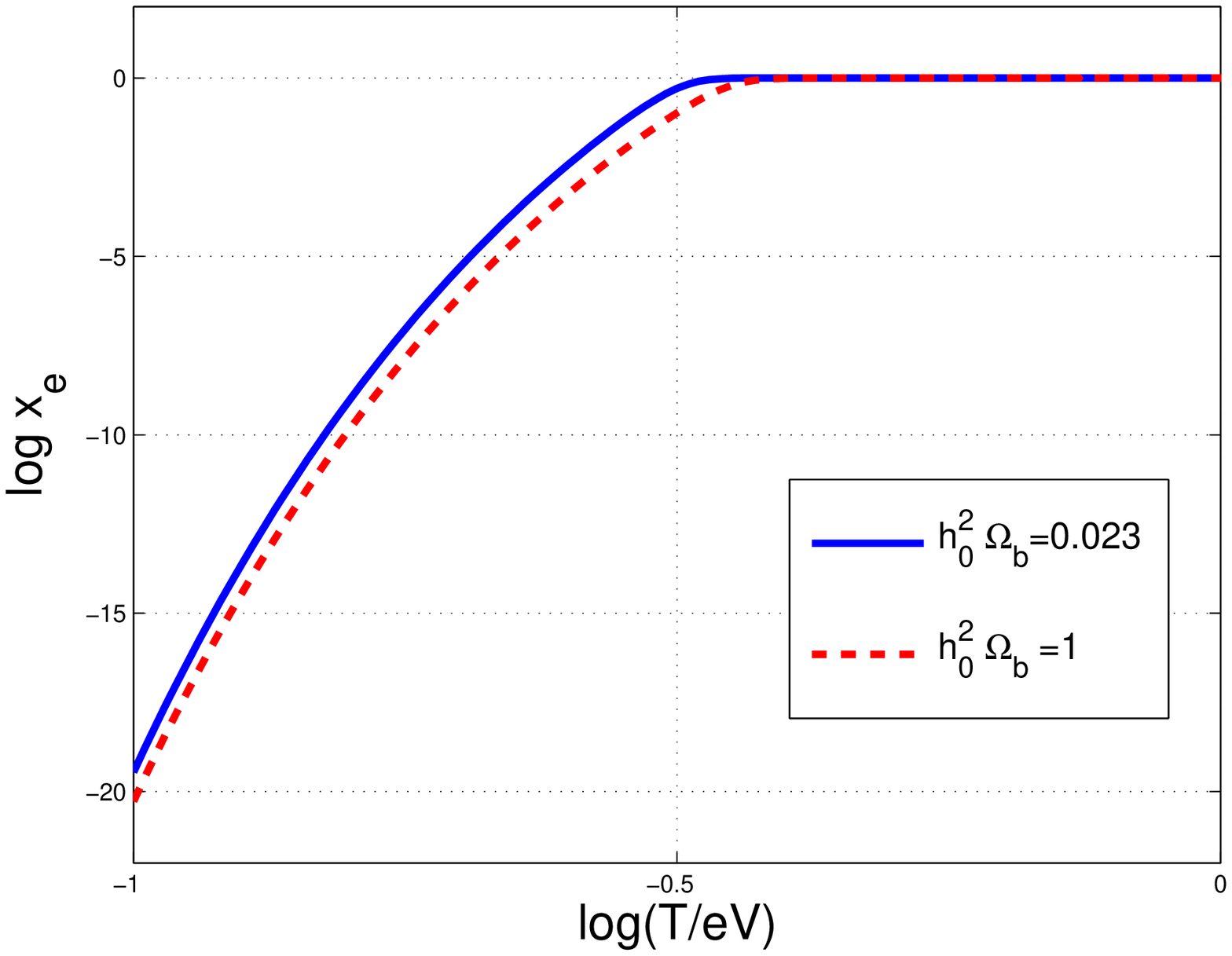}
\caption{The ionization fraction is illustrated as a function of the rescaled temperature 
$y=T/\mathrm{eV}$ for two different scales, i.e. linear (plot at the left) and logarithmic (plot at the right).}
\label{F4a}      
\end{figure}

Using all the aforementioned observations, both sides of Eq. (\ref{nH2}) can be divided by the 
baryonic concentration $n_{\mathrm{B}}$. Then, using of the global charge neutrality of the plasma together with 
Eq. (\ref{etab}), Eq. (\ref{nH2}) can be written as 
\begin{equation}
\frac{1 - x_{\mathrm{e}}}{x_{\mathrm{e}}^2} = \eta_{\mathrm{b}0} \frac{4 \zeta(3) \sqrt{2}}{\sqrt{\pi}} \biggl(\frac{T}{m_{\mathrm{e}}}\biggr)^{3/2} \, e^{E_{0}/T},
\label{saha}
\end{equation} 
which is called the Saha equation for the equlibrium ionization fraction. In Eq. (\ref{saha}) 
the baryonic concentration has been expressed through $\eta_{\mathrm{b}0}$, i.e. the ratio between 
the concentrations of baryons and photons. Introducing now the dimensionless variable $y = T/\mathrm{eV}$ we have that, using the explicit expression of $\eta_{\mathrm{b}}$ (i.e. Eq. (\ref{etab})), 
 Eq. (\ref{saha}) can be written as 
\begin{equation}
\frac{1 - x_{\mathrm{e}}}{x_{\mathrm{e}}^2} = {\mathcal P} y^{3/2} e^{y_{0}/y}, \qquad {\mathcal P} = 
6.530\times10^{-18} \biggl(\frac{h_{0}^2 \Omega_{\mathrm{b}}}{0.023}\biggr).
\label{saha2}
\end{equation}
where:
\begin{equation} 
 \biggl(\frac{T}{m_{\mathrm{e}}}\biggr) = 1.96\times10^{-6} y,\qquad y_{0} = 13.26.
\end{equation}
Equation (\ref{saha2}) stipulates that, when $y\simeq 1$ (corresponding to $T\simeq \mathrm{eV}$)
$\exp{(13.26)}\simeq 10^{5}$: thus we still have  $x_{\mathrm e} \simeq 1$. In fact, the smallness of ${\mathcal P}$ appearing in Eq. (\ref{saha2}) necessarily implies that $1-x_{\mathrm{e}}\simeq 10^{-13} x_{\mathrm{e}}^2$.
This observation shows that atoms do not form neither for $T\sim 10$ eV nor for $T\sim {\mathrm{eV}}$ but only 
when the temperature drops well below the eV. 
Equation (\ref{saha2}) can be made more explicit by solving with respect to $x_{\mathrm{e}}$
\begin{equation}
x_{\mathrm{e}} = 
\biggl(\frac{-1 + \sqrt{1 + 4 {\mathcal P} y^{3/2} e^{y_{0}/y}}}{2 {\mathcal P} y^{3/2}}\biggr)\,e^{y_{0}/y}.
\label{saha3}
\end{equation}
From Fig. \ref{F4a} it appears  that in order to reduce the ionization fraction to an appreciable 
value (i.e. $x_{\mathrm{e}}\simeq 10^{-1}$), $T$ must be as low as $0.3$ eV.
Recalling that $T = T_{\gamma0}(1 + z)$ we can see that\footnote{From now on, without any confusion, we will often drop the subscript $\gamma$ in the temperature.}:
\begin{itemize}
\item{} $x_{\mathrm{e}} \simeq 10^{-1}$ implies $T_{\mathrm{rec}}\simeq 0.3\,{\mathrm{eV}}$ and 
$z_{\mathrm{rec}} \simeq 1300$: this is the moment of hydrogen recombination when photoionization 
reactions are unable to balance hydrogen formation;
\item{}  $x_{\mathrm{e}} \simeq 10^{-4}$ implies $T_{\mathrm{dec}}\simeq 0.2\,{\mathrm{eV}}$ and 
$z_{\mathrm{rec}} \simeq 1100$: this is the moment of decoupling when the photon mean free path 
gets as large as $10^{4} \mathrm{Mpc}$ (see below in this section and, in particular, Eq. (\ref{phex})).
\end{itemize}

Since
the most efficient process that can transfer energy and momentum 
is Thompson scattering, the drop in the ionization fraction entails a dramatic increase of the proton mean free path.  Before decoupling the photon 
mean free path is of the order of the Mpc. After decoupling, the photon mean free path becomes of the order of $10^{4}$ Mpc and the CMB photons 
may reach our detectors and satellites without being scattered. This 
is the moment when the Universe becomes transparent to radiation.

\subsubsection{Coulomb scattering: the baryon-electron fluid}
Before equality electrons and protons are coupled through Coulomb 
scattering while photons scatter protons and electrons with  Thompson 
cross section. Now, the Coulomb rate of interactions is much smaller than the 
Hubble rate at the corresponding epoch. Thus, the protons and electrons form
a single (globally neutral) component where the velocities 
of the electrons and of the protons are approximately equal. This is 
the reason why, baryons and leptons will be described, in the analysis 
of CMB anisotropies by a single set of equations called, somehow 
confusingly, baryon fluid. 

Photons scatter electrons with Thompson cross section and, in principle, 
photons scatter also protons with Thompson cross section.  
However, since the mass of the protons is roughly $2000$ times larger than the mass 
of the electrons, the corresponding cross-section for photon-proton 
scattering will be much smaller than the cross-section for photon-electron 
scattering. This observation implies that the mean free path of photons 
is primarily determined by the photon-electron cross section.

\begin{figure}
\centering
\includegraphics[height=5cm]{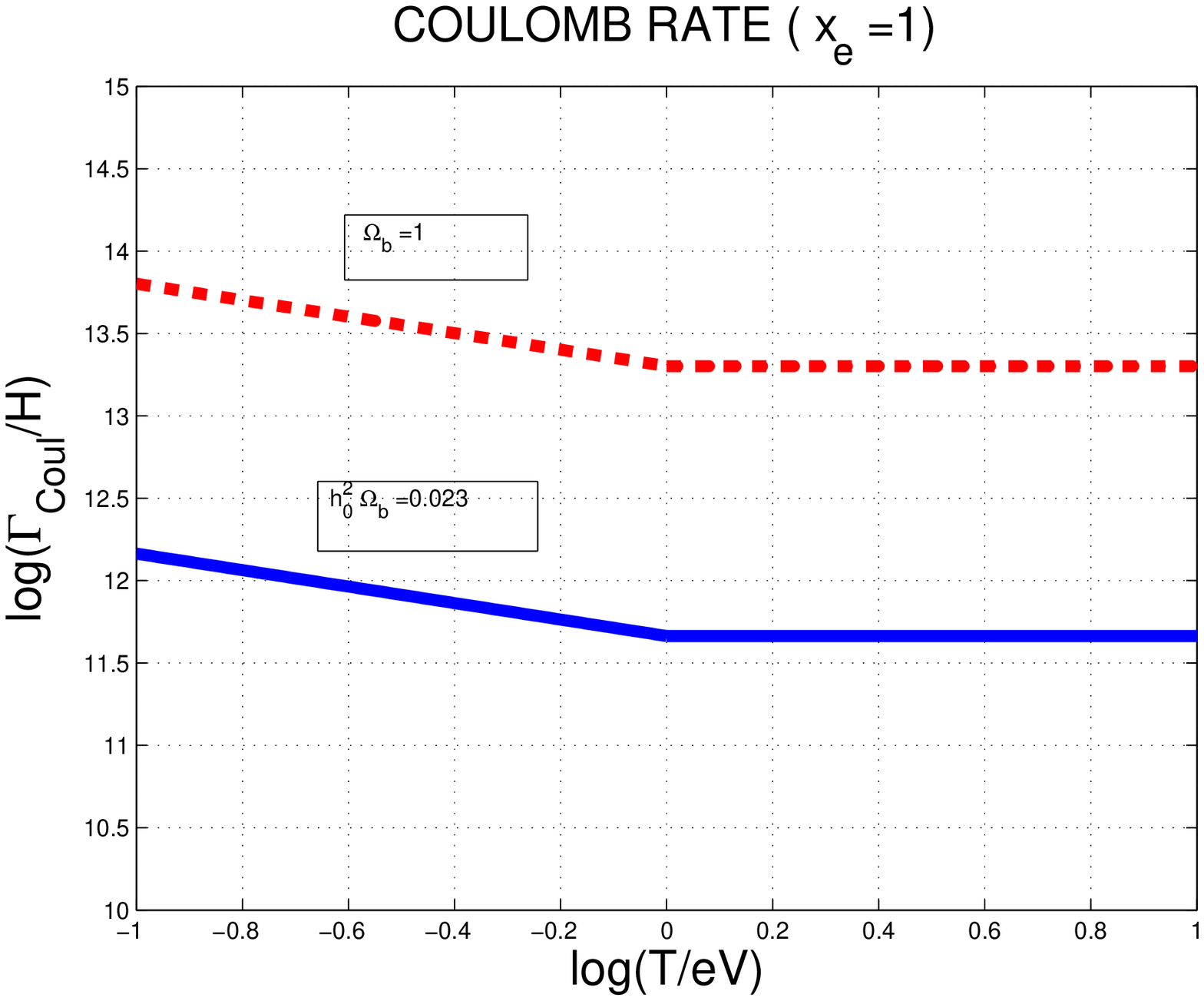}
\includegraphics[height=5cm]{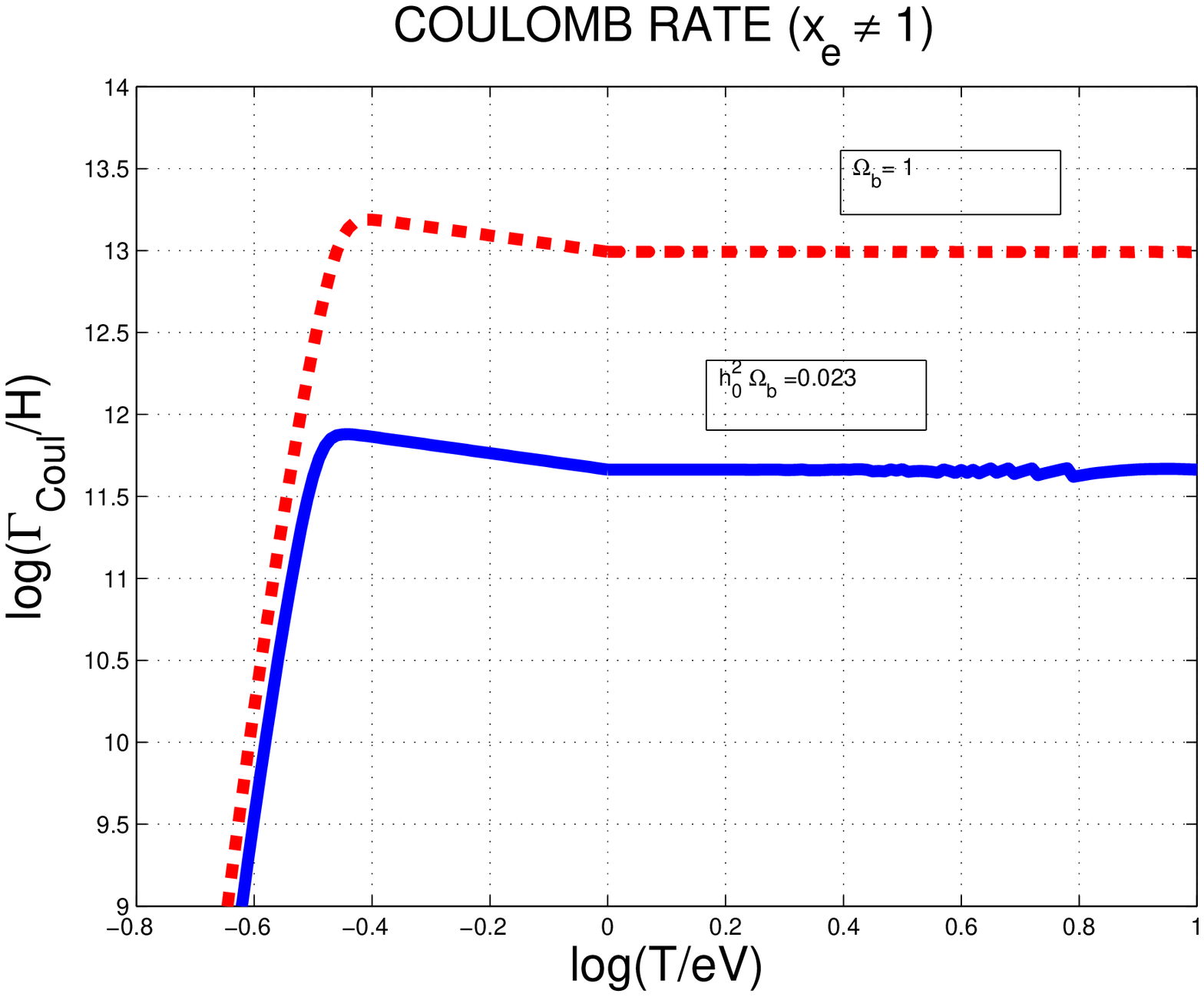}
\caption{The Coulomb rate is illustrated around equality in the case when $x_{\mathrm{e}} =1$ (plot at the left) 
and in the case when $x_{\mathrm{e}}(T)$ is determined from the Saha equation}
\label{COUL}      
\end{figure}

Consider then, for $t< t_{\mathrm{eq}}$, the Coulomb rate of interactions given by:
\begin{equation}
\Gamma_{\mathrm{Coul}}= v_{\mathrm{th}} \sigma_{\mathrm{Coul}} n_{\mathrm{e}},
\label{gammacoul}
\end{equation}
where:
\begin{itemize}
\item{} $v_{\mathrm{th}}\simeq \sqrt{T/m_{\mathrm{e}}}$ is the 
thermal velocity of electrons;
\item{} $\sigma_{\mathrm{Coul}}= (\alpha_{\mathrm{em}}^2/T^2)\ln\Lambda$ is the Coulomb cross section including the Coulomb logarithm;
\item{} $n_{\mathrm{e}} = x_{\mathrm{e}} n_{\mathrm{B}}$ which may also 
be written as 
\begin{equation}
n_{\mathrm{e}} = \frac{ 2 \zeta(3)}{\pi^2}\, T^3\, x_{\mathrm{e}} \eta_{\mathrm{b}0} .
\end{equation}
\end{itemize}
Plugging everything into Eq. (\ref{gammacoul}) we obtain:
\begin{equation}
\Gamma_{\mathrm{Coul}} = 1.15\times10^{-17}\,\,x_{\mathrm{e}}\, \biggl(\frac{T}{\mathrm{eV}}\biggr)^{3/2} \,\,\biggl( \frac{h_{0}^2 \Omega_{\mathrm{b}}}{0.023}
\biggr)\,\,\,\mathrm{eV}.
\label{gammacoul2}
\end{equation}
The Coulomb rate may now be compared with the Hubble rate. Since 
the number of relativistic degrees of feedom is given by $g_{\rho} \simeq 3.36$, according to the general formula (valid for $t<t_{\mathrm{eq}}$ and derived in Eq. (\ref{Hrho}))
\begin{equation}
H = 1.66 \sqrt{g_{\rho}} \frac{T^2}{M_{\mathrm{P}}} =2.49\times10^{-28}\,\, \biggl(\frac{T}{\mathrm{eV}}\biggr)^2\,\,\mathrm{eV}.
\label{HRT}
\end{equation}
For $t> t_{\mathrm{eq}}$ we will have, instead
\begin{equation}
H = H_{\mathrm{eq}} \,\biggl(\frac{T}{\mathrm{eV}}\biggr)^{3/2}\,\, \mathrm{eV}.
\label{HRTmat}
\end{equation}
Therefore, 
\begin{eqnarray}
&& \frac{\Gamma_{\mathrm{Coul}}}{H} = 4.61 \times 10^{11}\,\, \biggl(\frac{T}{\mathrm{eV}}\biggr)^{-1/2} \, x_{\mathrm{e}} 
\,\,\biggl( \frac{h_{0}^2 \Omega_{\mathrm{b}}}{0.023}
\biggr), \qquad T> T_{\mathrm{eq}},
\label{COULRAD}\\
&&  \frac{\Gamma_{\mathrm{Coul}}}{H} = 4.61 \times 10^{11}\,\, \, x_{\mathrm{e}} 
\,\,\biggl( \frac{h_{0}^2 \Omega_{\mathrm{b}}}{0.023}
\biggr), \qquad T< T_{\mathrm{eq}}.
\label{COULMAT}
\end{eqnarray}
Equations (\ref{COULRAD}) and (\ref{COULMAT}) are illustrated in Fig. \ref{COUL}
where the Coulomb rate is plotted in units of the expansion rate.
We can clearly see that $\Gamma_{\mathrm{Coul}} >H$ in the physically 
interesting range of temperatures. This means, as anticipated, that 
charged particles are strongly coupled.

\subsubsection{Thompson scattering: the baryon-photon fluid}

Consider now, always before equality, the Thompson rate of reaction. In this 
case we will have that 
\begin{figure}
\centering
\includegraphics[height=5cm]{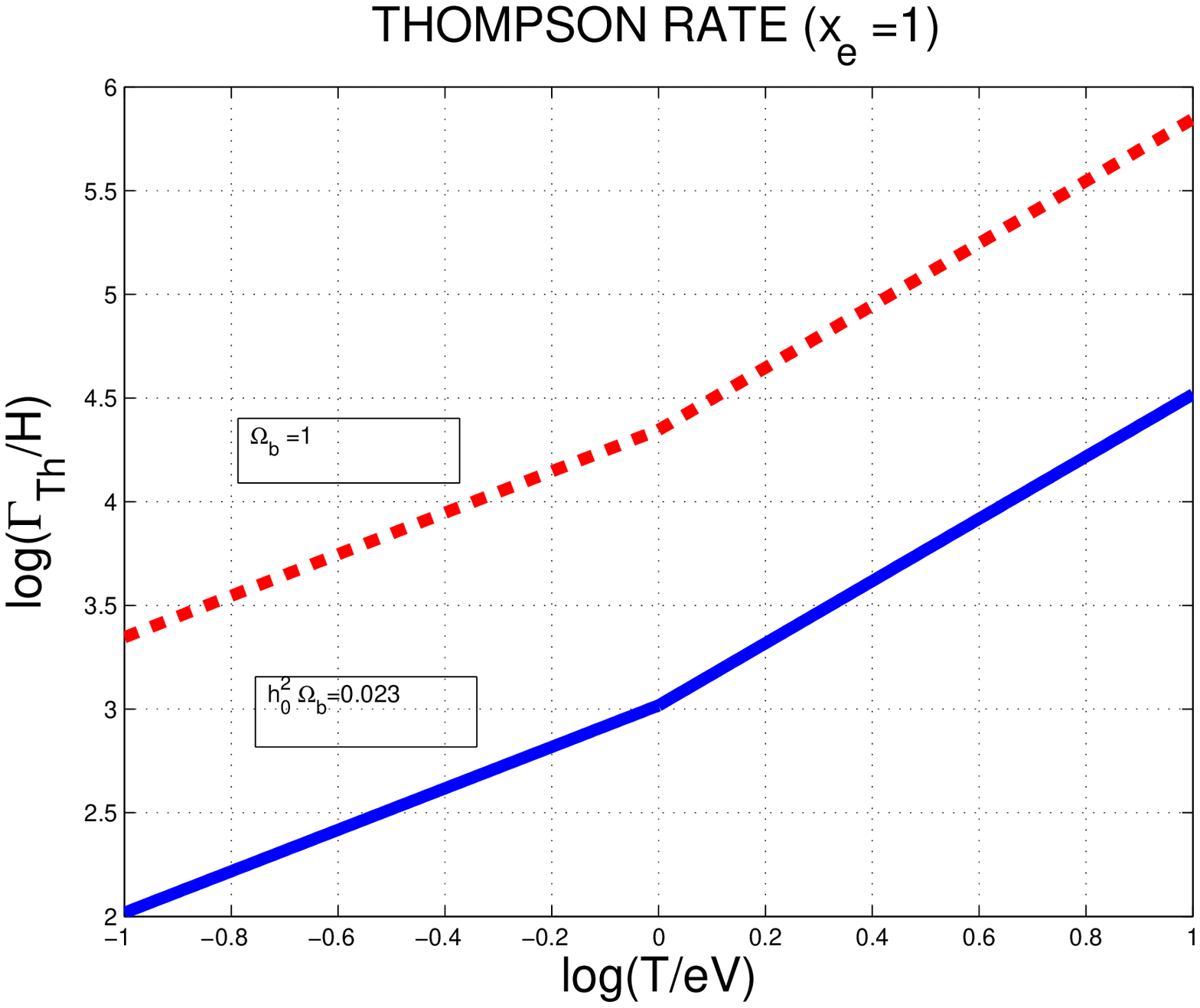}
\includegraphics[height=5cm]{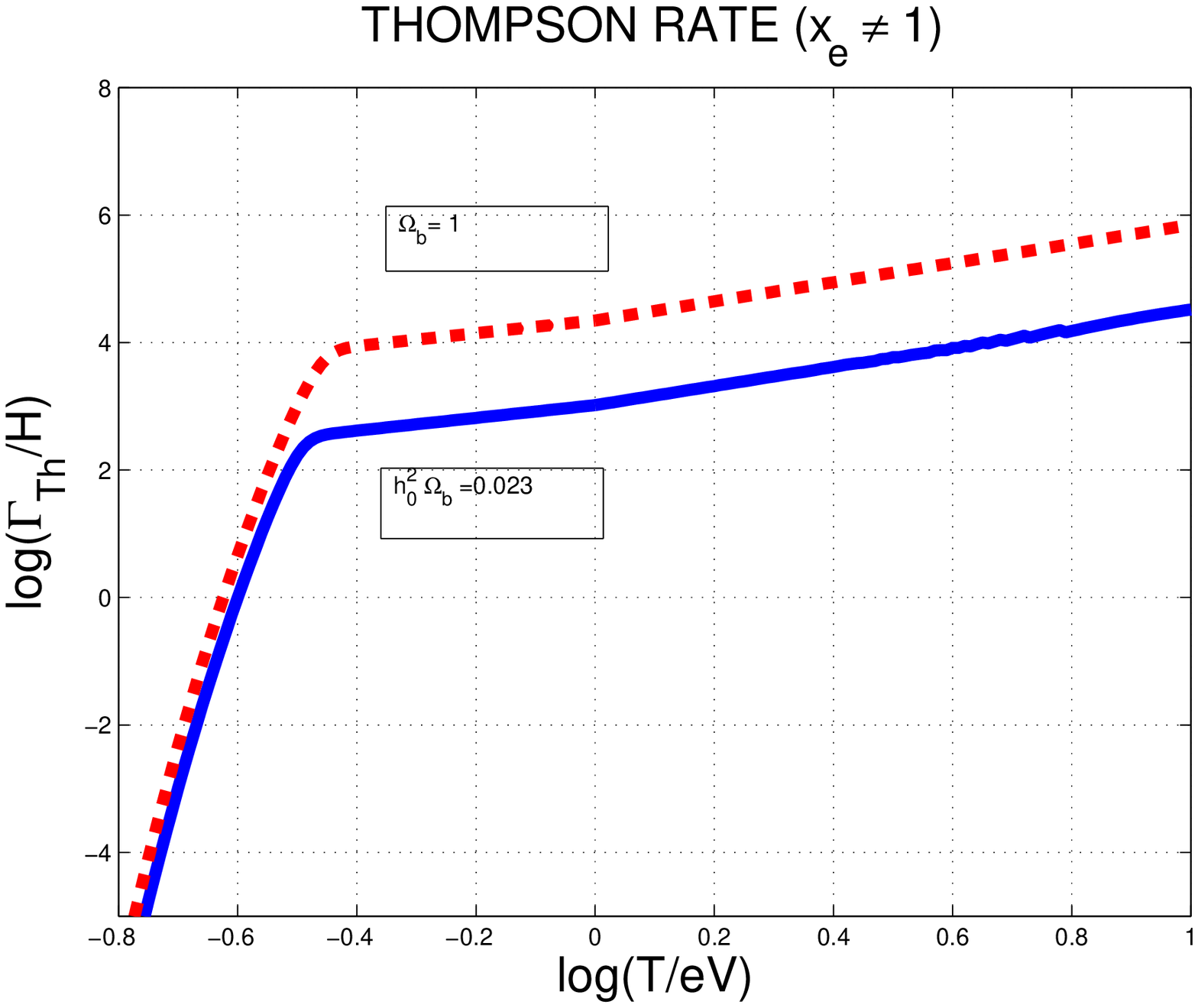}
\caption{The Thompson rate is illustrated around equality in the case when $x_{\mathrm{e}} =1$ (plot at the left) 
and in the case when $x_{\mathrm{e}}(T)$ is taken from the Saha equation}
\label{THOM}      
\end{figure}
\begin{equation}
\Gamma_{\mathrm{Th}} \simeq n_{\mathrm{e}}  \sigma_{\mathrm{T}},
\label{Txs}
\end{equation}
where 
\begin{equation}
\sigma_{\mathrm{T}} = 0.665\,\,\mathrm{barn},\qquad 
1\,\mathrm{barn} = 10^{-24} \mathrm{cm}^2.
\label{Txs2}
\end{equation}
Using Eq. (\ref{Txs2}) into Eq. (\ref{Txs}) we will have 
\begin{equation}
\Gamma_{\mathrm{Th}} = 2.6\times 10^{-25}\,\,x_{\mathrm{e}}\biggl(\frac{T}{\mathrm{eV}}\biggr)^3 \,\,\biggl( \frac{h_{0}^2 \Omega_{\mathrm{b}}}{0.023}
\biggr)\,\,\,\mathrm{eV},
\label{Txs3}
\end{equation}
which shows that $\Gamma_{\mathrm{Coul}}\gg \Gamma_{\mathrm{Th}}$
and also that 
\begin{eqnarray}
&& \frac{\Gamma_{\mathrm{Th}}}{H} = 1.04\times 10^{3} \,\,\biggl(\frac{T}{\mathrm{eV}}\biggr)\,\,x_{\mathrm{e}}\biggl( \frac{h_{0}^2 \Omega_{\mathrm{b}}}{0.023}
\biggr),\qquad T> T_{\mathrm{eq}},
\label{THOMRAD}\\
&& \frac{\Gamma_{\mathrm{Th}}}{H} = 1.04\times 10^{3} \,\,\biggl(\frac{T}{\mathrm{eV}}\biggr)^{3/2}\,\,x_{\mathrm{e}}\biggl( \frac{h_{0}^2 \Omega_{\mathrm{b}}}{0.023}
\biggr),\qquad  T< T_{\mathrm{eq}}.
\label{THOMMAT}
\end{eqnarray}
The previous equations also substantiate the statement that the photon mean 
free path is much larger than the electron mean free path for 
temperatures $T>\mathrm{eV}$. Thus, Thompson scattering is the 
most efficient way of transferring energy and momentum.
Equations (\ref{THOMRAD}) and (\ref{THOMMAT}) are illustrated in Fig. \ref{THOM}.
It is clear that as soon as the ionization fraction drops, the Thompson rate becomes suddenly 
smaller than the expansion rate.
After equality  the 
photon mean free path can be written as 
\begin{equation}
\lambda_{\mathrm{Th}} \simeq \frac{1}{a n_{\mathrm{e}} \sigma_{\mathrm{Th}}},
\label{phmfp}
\end{equation}
which can also be written, in more explicit terms, as 
\begin{equation}
\lambda_{\mathrm{Th}} \simeq 1.8\,\,x_{\mathrm{e}}^{-1}\,\,\biggl(\frac{0.023}{h_{0}^2 \Omega_{\mathrm{b}}}\biggr)\,\,\biggl(\frac{1100}{1 + z_{\mathrm{dec}}}\biggr)^2 \,\,
\biggl(\frac{0.88}{1 - Y_{\mathrm{p}}/2}\biggr)\,\,\, \mathrm{Mpc}.
\label{phex}
\end{equation}
Equation (\ref{phex}) shows clearly that as soon as the ionization fraction drops (at recombinantion) the photon mean free path becomes of the order 
of $10^{4}$--$10^{5}$ Mpc. In Eq. (\ref{phex}) the mass fraction of $^{4}\mathrm{He}$ appears explicitly and it is denoted by $Y_{\mathrm p}$
(typically $Y_{\mathrm{p}} \simeq 0.24$).  This is not a surprise since the Helium nucleus contains four nucleons and the ratio of Helium to the total 
number of nuclei is $Y_{\mathrm{p}}/4$. Each of these absorbs two electrons
(one for each proton). Thus when we count the number of free electrons before recombination  the estimate of the Thompson reaction rate 
must be multiplied by $(1 - Y_{\mathrm{p}}/2)$.
Note, finally, that in the last estimate the recoil energy of the electron
ha been neglected. This is justified since the electron rest mass 
is much larger than the incident photon energy which is, at 
recombination, of the order of the temperature, i.e. $0.3$ eV. 

In summary, it is important to stress that Coulomb scattering 
is rather efficient in keeping rather tight the coupling between 
protons and electrons, at least in the standard treatment. 
This occurrence justifies, at an effective level, to consider 
a single baryon-lepton fluid which is globally neutral but intrinsically charged.
The tight coupling between photons and charged particles (either 
leptons or baryons) is realized before recombination and it is, therefore, 
a very useful analytical tool for the approximate estimate 
of acoustic oscillations arising in the temperature autocorrelations 
which will be discussed in the last three sections of this script. The 
(approximate) tight coupling between photons and charged species 
allows then, in combination with the largeness of the Coulomb rate,
the treatment of a single baryon-lepton-photon fluid or baryon-photon fluid, for short.
This chain of observations will be turn out to be very useful when writing the 
evolution equations for the inhomogeneities prior to decoupling. 
This topic will be discussed in sections \ref{sec8} and \ref{sec9} (see, in particular, before and after Eqs. 
(\ref{pb0}), (\ref{pb1}) and (\ref{pb2}) when talking about the tight coupling appoximation). 

\newpage
\renewcommand{\theequation}{3.\arabic{equation}}
\setcounter{equation}{0}
\section{Problems with the standard cosmological model}
\label{sec3}
The standard cosmological model gives us a rationale for 
two important classes of phenomena that are directly observable in the sky: the recession of galaxies and the 
spectral properties of CMB. In spite of this occurrence, two possible drawbacks 
of the SCM are already understandable: 
\begin{itemize}
\item{}the anisotropies of CMB that are not accounted by the SCM (see Fig. \ref{F4}); 
\item{}  the huge thermodynamic entropy stored in the CMB (see Eq. (\ref{Sg}) and the related discussion) is not so explained within the SCM since evolution of the Universe  was all the time adiabatic (see, for instance, 
 Eqs. (\ref{cov3}) and (\ref{def3})).
\end{itemize}
 The present hierarchy between the matter and radiation 
energy density suggests, furthermore, that the Universe was rather hot in the past.
This conclusion is indirectly tested through the success of big-bang nucleosynthesis (BBN).
As already pointed out, in BBN there are essentially only two free parameters: the temperature and the 
the baryon to photon ratio \footnote{This statement holds, strictly speaking, in the simplest (and also 
most predictive) BBN scenario where the synthesis of light nuclei occurs homogeneously in space 
and in the absence of matter--antimatter fluctuations. In this scenario the antinucleons have almost 
completely disappeared by the time weak interactions fall out of thermal equilibrium.  There are, however, models 
where both assumptions have been relaxed (see, for instance \cite{inBBN1,inBBN2,inBBN3} and references therein). 
In this case the prediction of BBN will also depend upon the typical inhomogeneity scale 
of the baryon to photon ratio.}. After weak interactions fall out of thermal equilibrium the light 
nuclei start being formed. Since the $^{4}\mathrm{He}$ has the largest binding energy 
per nucleon for nuclei with nuclei with atomic number $A <12$, roughly one quarter 
of all the protons will end up in  $^{4}\mathrm{He}$ while the rest will remain in free 
protons. Smaller abundances of other light nuclei (i.e. $D$, $^{3}\mathrm{He}$ and $^{7} \mathrm{Li}$)
can be also successfully computed in the framework of BBN \cite{bernstein}. The synthesis
of light elements is very important since light elements have to turn on the 
thermonuclear reactions taking place in the stars and during supernova explosions.
However, even if the Universe must be sufficiently hot (and probably as hot as several hundreds 
GeV to produce a sizable baryon asymmetry) it cannot be dominated by radiation 
all the way up to the Planck energy scale: this occurrence would lead to logical puzzles 
in the formulation of the SCM. In what follows some of the problems of the SCM will be discussed in a unified 
perspective and, in particular, we shall discuss:
\begin{itemize}
\item{} the horizon (or causality) problem;
\item{} the spatial curvature (or flatness) problem;
\item{} the entropy problem;
\item{} the structure formation problem;
\item{} the singularity problem.
\end{itemize}
The first two problems in the above list of items are often named kinematical problems. 
It is interesting to notice that both the horizon problem as well as the entropy and structure 
formation problems are directly related with CMB physics as it will be stressed below in this section.
\subsection{The horizon problem}
\begin{figure}
\centering
\includegraphics[height=6cm]{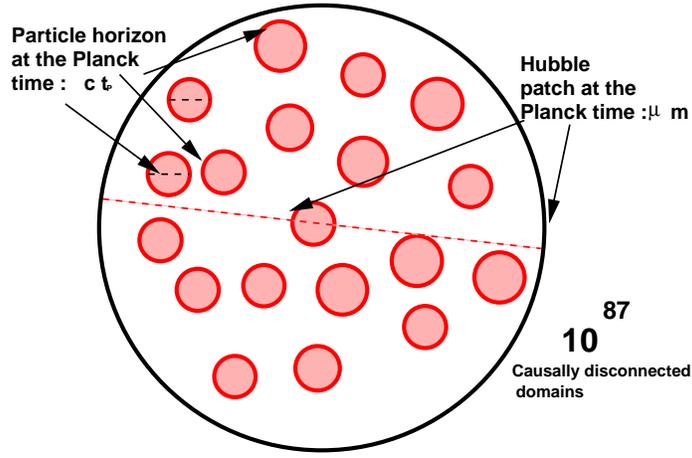}
\caption{A schematic snapshot of the Hubble patch blueshifted at the Planck time (see Eq. (\ref{ratio1})). The filled 
circles represent the typical size of the particle horizon at the corresponding epoch.}
\label{HOR1}      
\end{figure}

Two important concepts appear in the analysis of the causal structure of cosmological models 
\cite{hawkingellis2}, i.e. the proper 
distance of the event horizon:
\begin{equation}
d_{\mathrm{e}}(t) = a(t) \int_{t}^{t_{\mathrm{max}}} \frac{dt'}{a(t')},
\label{evh}
\end{equation}
and the proper distance of the particle horizon 
\begin{equation}
d_{\mathrm{p}}(t) = a(t) \int_{t_{\mathrm{min}}}^{t} \frac{dt'}{a(t')},
\label{ph}
\end{equation}
(see also Appendix \ref{APPA} for further details).
The event horizon measures the size over which we can admit {\em even in the future} a causal 
connection. The particle horizon measures instead the size of causally connected regions at the time $t$.
In the SCM the particle horizon exist while the event horizon does not exist and this 
occurrence is the rationale for a kinematical problem of the standard model. 
According to the SCM, the Universe, in its past expand in a decelerated way as 
\begin{equation}
a(t) \sim t^{\alpha}, \qquad 0 <\alpha < 1, \qquad t>0,
\label{atalpha}
\end{equation}
which implies that $\dot{a}>0$ and $\ddot{a} <0$. Inserting  Eq. (\ref{atalpha}) 
into Eqs. (\ref{evh}) and (\ref{ph}) the following two expressions are swiftly obtained after direct
integration:
\begin{eqnarray}
 d_{\mathrm{e}}(t) &=& \frac{t_{\mathrm{max}}}{1 - \alpha} \biggl[ \biggl(\frac{t}{t_{\mathrm{max}}}\biggr)^{\alpha} - 
 \biggl(\frac{t}{t_{\mathrm{max}}}\biggr)\biggr], 
\label{deex1}\\
d_{\mathrm{p}}(t) &=& \frac{1}{1 -\alpha} \biggl[ t - t_{\mathrm{min}}\biggl(\frac{t}{t_{\mathrm{min}}}\biggr)^{\alpha}\biggr].
\label{dpex1}
\end{eqnarray}
Since $0<\alpha < 1$, Eqs. (\ref{deex1}) and (\ref{dpex1}) lead to the following pair of limits
\begin{eqnarray}
&& \lim_{t_{\mathrm{min}}\to 0} d_{\mathrm{p}}(t) \to \frac{\alpha}{1 - \alpha} H^{-1}(t),
\label{dpex2}\\
&& \lim_{t_{\mathrm{max}}\to \infty} d_{\mathrm{e}}(t) \to \infty,
\label{deex2}
\end{eqnarray}
where both limits are taken while $t$ is kept fixed. Equations (\ref{dpex2}) and (\ref{deex2}) 
show that, in the SCM, the event horizon 
{\em does not exist} while the particle horizon exist and it is finite.
Because of the existence of the particle horizon, for each time in the past history of the Universe 
the typical causal patch will be of the order of the Hubble radius, i.e., restoring for a moment the 
speed of light, $d_{\mathrm{p}}(t) \sim c\,t$. This simple occurrence represents, indeed,  a problem. 
The present extension of the Hubble radius evolves as the scale factor (i.e. faster than the particle horizon).
Let us then see how large was the present Hubble radius at a 
given reference time at which the evolution of the SCM is supposed to start. Such a reference time 
can be taken to be, for instance, the Planck time. The Hubble radius at the Planck time will be of the 
order of the $\mu\mathrm{m}$, i.e., more precisely: 
\begin{equation}
r_{H}(t_{\mathrm{P}}) = 4.08 \, \times 10^{-4}\, \biggl(\frac{0.7}{h_{0}}\biggr) \biggl(\frac{T}{\mathrm{eV}}\biggr)\,\,
\mathrm{cm}.
\label{hrpl}
\end{equation}
The obtained figure can then be measured  in units of the particle horizon at the Planck time, which is 
the relevant scale set by causality at any given time in the life of the SCM:
\begin{equation}
d_{\mathrm{p}}(t_{\mathrm{P}}) \simeq c\,t_{\mathrm{P}} \simeq 10^{-33} \,\,\mathrm{cm}.
\label{dppl}
\end{equation}
Taking the ratio between (\ref{hrpl}) and (\ref{dppl}) 
\begin{equation}
\frac{r_{H}(t_{\mathrm{P}})}{d_{\mathrm{p}}(t_{\mathrm{P}})} \simeq 4.08 \times 10^{29} \,\biggl(\frac{0.7}{h_{0}}\biggr) \biggl(\frac{T_{\mathrm{eq}}}{\mathrm{eV}}\biggr).
\label{ratio1}
\end{equation}
The third power of Eq. (\ref{ratio1}) measures the number of causally disconnected volumes at $t_{\mathrm{P}}$.
This estimate tells that there are, roughly, to $10^{87}$ causally disconnected 
 regions at the Planck time. In Fig. \ref{HOR1} the physics described by Eq. (\ref{ratio1}) is illustrated in pictorial terms.
The Hubble radius at the Planck time has approximate size of the order 
of the $\mu$m and it contains $10^{87}$ causally disconnected volumes each with approximate size 
of the order of the particle horizon at the Planck time.
A drastic change in the reference time at which initial conditions for the evolution are set 
does not alter the essence of the problem.
Suppose that, indeed, the thermal history of the Universe does not extend up to the Planck temperature. Let us take our reference temperature to be of the order of $200$ GeV. 
For such a temperature all the species of the Glashow-Weinberg-Salam (GWS) model are in thermal equilibrium and the particle horizon is given by 
\begin{equation}
d_{\mathrm{p}}(t_{\mathrm{ew}}) \simeq 35 \, \sqrt{\frac{106.75}{g_{\rho}}} \biggl(\frac{T_{\mathrm{ew}}}{200} \biggr)^{-2}\,\,\mathrm{cm}
\end{equation}
where $g_{\rho}(T)$ is the number of relativistic degrees of freedom at the temperature $T$ here 
taken to be of the order of $100$GeV (see Eqs. (\ref{grho}), (\ref{DFGWS})) and (\ref{Hrho}) of Appendix \ref{APPA}). 
The Hubble radius 
blueshifted at the temperature $T_{\mathrm{ew}}\simeq 200$ GeV  will be instead
\begin{equation}
r_{H}(t_{\mathrm{ew}}) \simeq 1.98\times 10^{13}\,\,
 \biggl(\frac{0.7}{h_{0}}\biggr) \biggl(\frac{T_{\mathrm{eq}}}{\mathrm{eV}}\biggr) \,\,\mathrm{cm}.
\label{domain2}
\end{equation}
 Thus, since $r_{H}(t_{\mathrm{ew}})/d_{\mathrm{p}}(t_{\mathrm{ew}})\simeq 10^{12}$, the present Hubble patch 
 will consist, at the temperature of $10^{36}$ causally disconnected regions. Since 
 the temperature fluctuations in the microwave sky are of the order 
 of $\delta T/T \simeq 10^{-5}$, the density contrast in radiation will be of the order of $\delta\rho_{\gamma}/\rho_{\gamma} \sim 10^{-4}$.

 How come that the CMB is so homogeneous, if, in the past history of the Universe there were 
 so many causally disconnected regions. Is there something else than causality 
 that can make our Hubble patch homogenous? The answer to this question seems of course to be 
 negative.
\begin{figure}
\centering
\includegraphics[height=6cm]{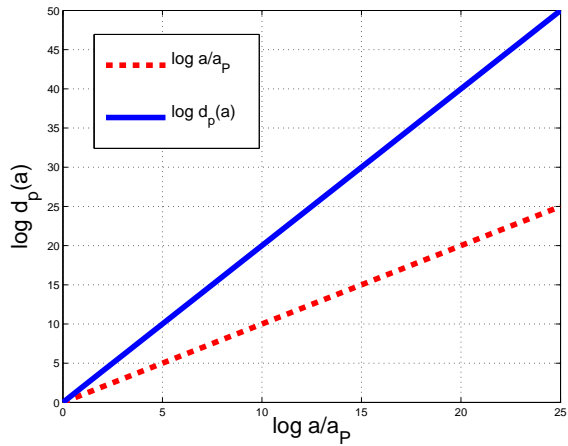}
\caption{The evolution of the particle horizon in the SCM is compared with the evolution of the scale 
factor.  From the plot its clear that, in the SCM, the particle horizon evolves faster than the scale 
factor, since, approximately, $d_{\mathrm{p}}(a) \sim a^{1/\alpha}$ with $0<\alpha <1$. Here, for 
illustration, the case $\alpha =1/2$ (radiation dominance) has been assumed. }
\label{HOR2}      
\end{figure}
The final observation to be borne in mind is that the root of the horizon problem resides in the 
occurrence that, in the SCM, the particle horizon evolves faster than the scale factor. This 
point is summarized in Fig. (\ref{HOR2}) where the evolution of the particle horizon is compared 
with the evolution of the scale factor.
 \subsection{The spatial curvature problem}
 The problem of the spatial curvature can be summarized by the following 
 question: why is the present Universe so close to flat?
 From Eq. (\ref{FL1}), the total energy density in critical units can be 
 written as 
 \begin{equation}
 \Omega_{\mathrm{t}}(a) = 1 + \frac{k}{a^2  H^2}.
 \label{OM1}
 \end{equation}
 Equation (\ref{OM1}) holds at any time where the SCM applies. 
 In particular, at the present time, we will have that 
 \begin{equation}
 \Omega_{\mathrm{t}}(t_0) = 1 + \frac{k}{a_{0}^2\, H_{0}^2}.
 \label{OM2}
 \end{equation}
 According to the experimental data\footnote{Notice that this experimental determination 
 is achieved directly from the position of the first Doppler peak of the CMB temperature autocorrelations.} \cite{WMAP01,map3}
\begin{equation}
\Omega_{\mathrm{t}}(t_0) = 1.02 \pm 0.02.
\label{OM3}
\end{equation}
Equation (\ref{OM3}) implies that the contribution of $|k| (a_{0} H_{0})^{-2}$ is smaller than $1$ (but of order $1$).
The denominator of the second term at the right hand side of 
Eq. (\ref{OM1}) goes as $H^2 a^2 \simeq \dot{a}^2$. So if 
$a(t) \simeq t^{\alpha}$ (with $0<\alpha<1$), $\dot{a}^2$ will be a decreasing 
function of the cosmic time coordinate $t$. But this implies that,
overall, the second term at the right hand side of Eq. (\ref{OM1}) 
will increase dramatically as time goes by.
 
As in the case of the horizon problem, a particular reference 
time may be selected. At this time initial conditions of the SCM 
are ideally set.  Let us take this time to be again, the Planck time
and suppose that, at the Planck time 
\begin{equation}
\frac{|k|}{a_{\mathrm{P}}^2 H_{\mathrm{P}}^2} \simeq {\mathcal O}(1).
\label{OM4}
\end{equation}
If Eq. (\ref{OM4}) holds around  the Planck time, today the same quantity 
will be 
\begin{equation}
\frac{|k|}{a_{0}^2 H_{0}^2} \simeq \frac{|k|}{a_{\mathrm{P}}^2 H_{\mathrm{P}}^2} \biggl(\frac{a_{\mathrm{eq}}}{a_{0}}\biggr)^2 \biggl(\frac{H_{\mathrm{eq}}}{H_{0}}\biggr)^2 \,\biggl(\frac{a_{\mathrm{P}}}{a_{\mathrm{eq}}}\biggr)^2 
\biggl(\frac{H_{\mathrm{P}}}{H_{\mathrm{eq}}}\biggr)^2 \simeq 
10^{60} \frac{|k|}{a_{\mathrm{P}}^2 H_{\mathrm{P}}^2}.
\label{OM5}
\end{equation}
Equation (\ref{OM5}) demands that if the spatial curvature is a contribution of order $1$ 
at the Planck time, today its contribution will be $60$ orders of magnitude larger. By reversing the argument it can be 
argued that if  the 
spatial curvature is (today) smaller than the extrinsic curvature (i.e. $k/a_{0}^2 \leq H_{0}^2$),  at the Planck time we must require an enormous fine-tuning:
\begin{equation}
\frac{|k|}{a_{\mathrm{P}}^2 H_{\mathrm{P}}^2} \simeq 10^{-60}.
\end{equation}
In other words: if the Universe is flat today it must have been even flatter in the past history 
of the Universe and the further we go back in time the flatter the Universe was.

Therefore, in summary, since $|k|/\dot{a}^2$ increases during the radiation 
and matter-dominated epochs, $\Omega_{\mathrm{t}}$ must be fine 
tuned to $1$ with a precision of $10^{-60}$.
By writing the evolution equation of $\Omega_{\mathrm{tot}}$:
\begin{equation}
 \Omega_{\mathrm{t}}(a)  -1 
= \frac{\Omega_{0} -1}{1 - \Omega_{0} + \Omega_{\Lambda 0} \biggl(\frac{a}{a_{0}}\biggr)^2  +\Omega_{\mathrm{M}0}
\biggl(\frac{a_0}{a}\biggr) + 
\Omega_{\mathrm{R}0} \biggl(\frac{a_{0}}{a}\biggr)^2},
\label{OM6}
\end{equation}
where $\Omega_{0} = \Omega_{\mathrm{t}}(t_{0})$. In the limit 
$a\to 0$ , i.e. $a_{0}/a \to \infty$, Eq. (\ref{OM6}) leads to
\begin{equation}
\Omega_{\mathrm{t}}(a) -1 \simeq \frac{\Omega_{0} -1}{
\Omega_{\mathrm{R}0} \biggl(\frac{a_0}{a}\biggr)^2}.
\label{OM7}
\end{equation}
According to Eq. (\ref{OM7}), we need $\Omega(a)\to 1$ with arbitrary 
precision (for the standards of physics) when $a\to 0$ if we want $\Omega_{0}\simeq 1 $ today.

\subsection{The entropy problem}
As discussed in introducing the essential features of the black body 
emission, the total entropy of the present Hubble patch is enormous 
and it is of the order of $10^{88}$ (see, for instance, Eq. (\ref{Sg})). This huge number 
arises since the ratio of $T_{\gamma0}/H_{0} \simeq {\mathcal O}(10^{30})$ (see Eq. (\ref{num})).
The covariant conservation of the energy-momentum 
tensor (see Eq. (\ref{cov3})) implies that the whole Universe is, 
from a thermodynamic point of view (see Eq. (\ref{def3})), 
as an isolated system where the total entropy is conserved.
If the evolution was adiabatic throughout the whole 
evolution of the SCM, why the present Hubble patch has such a 
huge entropy?  Really and truly the entropy problem contains, in itself, various other 
sub-problems that are rarely mentioned. They can be phrased in the following way:
\begin{itemize}
\item{} is the CMB entropy the only entropy that should be included in the formulation 
of the second law of thermodynamics?
\item{} is the second law of thermodynamics valid throughout the history of the Universe?
\item{} is the gravitational field itself a source of entropy?
\item{} how can we associate an entropy to the gravitational field?
\end{itemize}
If the second law of thermodynamics is applied we have two mutually exclusive choices:
either our own observable Universe originated from a Hubble patch with enormous 
entropy or the initial entropy was very small so that the initial state of the Universe 
was highly ordered. The first option applies, of course, if the evolution of the Universe 
was to a good approximation adiabatic.

\subsection{The structure formation problem}
The SCM posits that the geometry is isotropic and fairly homogeneous 
over very large scales. Small deviations from homogeneity are, however, 
observed. For instance, inhomogeneities arise as spatial fluctuations 
of the CMB temperature (see, for instance, Fig. \ref{F4}). These fluctuations will grow during the 
matter-dominated epoch and eventually collapse to form 
gravitationally bound systems such as galaxies and clusters of galaxies.

In what follows a simplistic description of  CMB observables will introduced
 \footnote{The treatment of CMB anisotropies presented here mirrors 
the approach adopted in a recent review \cite{THTH} where the main 
theoretical tools needed for the analysis of CMB anisotropies have been 
discussed within a consistent set of conventions}. By looking at the plots that are customarily introduced 
in the context of CMB anisotropies (like the one reported in Fig. \ref{F4}) 
we will try to understand in more detail what is actually plotted on the vertical as well as on the horizontal axis. 
The logic will be to use various successive expansions with the aim 
of obtaining a reasonably simple parametrization of the CMB temperature 
fluctuations. Consider, therefore the spatial fluctuations in the temperature of the CMB and 
expand them in Fourier integral \footnote{In what follows the subscript $\gamma$ will be dropped from the 
temperature to match with the conventions that customarily employed.}:
\begin{equation}
\Delta_{\mathrm{I}}(\vec{x}, \hat{n},\tau) = \frac{\delta T}{T}(\vec{x}, \hat{n},\tau)= \frac{1}{(2\pi)^{3/2}} \int d^{3} k \,\,
e^{- i \vec{k} \cdot \vec{x}} \Delta_{\mathrm{I}}(\vec{k},\hat{n}, \tau),
\label{firstexp}
\end{equation}
where $\hat{k}$ is the direction of the Fourier wave-number, $\hat{n}$ is the 
direction of the photon momentum. 
Assuming that the observer is located at a time $\tau$ (eventually coinciding with 
the present time $\tau_{0}$) and at $\vec{x}=0$, Eq. (\ref{firstexp}) can be also 
expanded in spherical harmonics, i.e. 
\begin{equation}
\Delta_{\mathrm{I}}(\hat{n}) = \sum_{\ell m} a_{\ell\, m} Y_{\ell\,m}(\hat{n})=
\frac{1}{(2\pi)^{3/2}} \int d^{3} k \,\,
 \Delta_{\mathrm{I}}(\vec{k},\hat{n}, \tau).
\label{sharm}
\end{equation}
Then, the Fourier 
amplitude appearing in Eq. (\ref{firstexp}) can be expanded in 
series of Legendre polynomials according to the well known relation 
\begin{equation}
\Delta_{\mathrm{I}}(\vec{k},\hat{n},\tau) = \sum_{\ell =0}^{\infty} (-i)^{\ell} (2 \ell + 1) 
\Delta_{\mathrm{I}\ell}(\vec{k},\tau) P_{\ell}(\hat{k}\cdot\hat{n}).
\label{LEG}
\end{equation}
Now the Legendre polynomials appearing in Eq. (\ref{LEG}) can be 
expressed via the addition theorem of spherical harmonics stipulating 
that 
\begin{equation}
P_{\ell}(\hat{k}\cdot\hat{n}) = \frac{4\pi}{2\ell + 1} \sum_{m = -\ell}^{\ell} Y_{\ell\,m}^{*}(\hat{k}) Y_{\ell m}(\hat{n}).
\label{SP}
\end{equation}
Inserting now Eq. (\ref{SP}) into Eq. (\ref{LEG}) and recalling the second equality 
of Eq. (\ref{sharm}) the coefficients $a_{\ell m}$ are determined to be 
\begin{equation}
a_{\ell\, m} = \frac{4\pi}{(2\pi)^{3/2}} (-i)^{\ell} \int d^{3}k Y_{\ell\,m}^{*}(\hat{k}) \Delta_{\mathrm{I}\,\ell}(\vec{k},\tau).
\end{equation}

The two-point temperature correlation function on the sky 
between two directions conventionally denoted by $\hat{n}_{1}$ and $\hat{n}_{2}$, can be written as 
\begin{equation}
C(\vartheta) = \langle \Delta_{\rm I}(\hat{n}_{1},\tau_{0})  \Delta_{\rm I}(\hat{n}_{2},\tau_{0}) \rangle,
\end{equation}
where $C(\vartheta)$ does not depend on the azimuthal angle because of isotropy of the background 
space-time and where the angle brackets denote a theoretical ensamble average. 
Since the background space-time is isotropic, the ensamble average of the $a_{\ell m}$ will only depend 
upon $\ell$, not upon $m$, i.e. 
\begin{equation}
\langle a_{\ell m}a^{\ast}_{\ell' m'} \rangle = C_{\ell} \delta_{\ell\ell'} \delta_{m m'},
\label{AVALM}
\end{equation}
where $C_{\ell}$ is the angular power spectrum.
Thus,  the relation (\ref{AVALM}) implies 
\begin{equation}
C(\vartheta) = \langle \Delta_{\rm I}(\hat{n}_{1},\tau_{0})  \Delta_{\rm I}(\hat{n}_{2},\tau_{0}) \rangle \equiv 
\frac{1}{4\pi} \sum_{\ell} (2 \ell + 1) C_{\ell } P_{\ell}(\hat{n}_{1} \cdot \hat{n}_{2}).
\end{equation}
Notice that in Fig. \ref{F4} the quantity $C_{\ell}\,\ell(\ell+1)/(2\pi)$ is directly plotted: as it follows 
from the approximate equality
\begin{equation}
\sum_{\ell} \frac{2\ell + 1}{4\pi} C_{\ell} \simeq \int \frac{\ell(\ell +1)}{2\pi} C_{\ell} d \ln{\ell},
\end{equation}
$C_{\ell}\,\ell(\ell+1)/(2\pi)$ is roughly the power per logarithmic interval of $\ell$.
In Fig. \ref{F4} the angular power spectrum is measured in $(\mu \mathrm{K})^2$. This 
is simply because instead of discussing $\Delta_{\mathrm{I}}$ (which measures the relative temperature fluctuation) one can equally reason in terms   
of $\Delta T = T\, \Delta_{\mathrm{I}}$, i.e. the absolute temperature fluctuation.
\begin{figure}
\centering
\includegraphics[height=8cm]{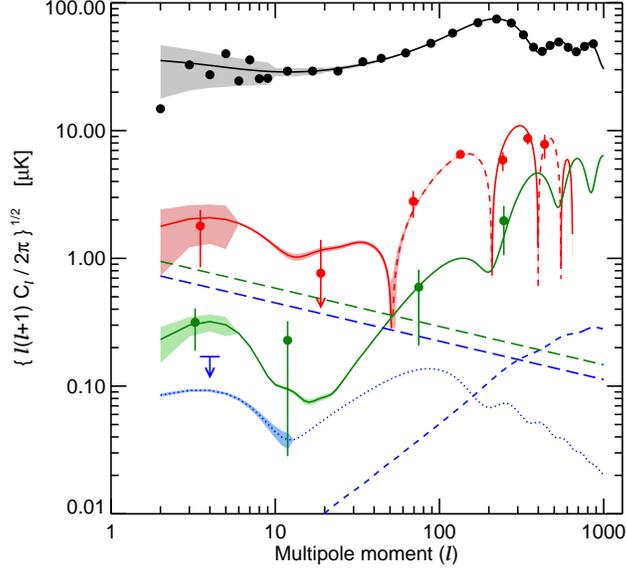}
\caption{From top to bottom the spectra for the $TT$, $TE$ and $EE$ correlations. The dashed lines indicate foregrounds of various nature. This figure 
is adpted from \cite{map4}.}
\label{grand}      
\end{figure}
Similar quantities can be defined for other observables such as, for instance, the degree of polarization. In Fig. \ref{grand} 
the angular power spectra are reported for (from top to bottom): the temperature autocorrelations (i.e. the quantity we just discussed),
the temperature polarization cross-correlation (often indicated as $TE$ spectrum), the polarization 
autocorrelation (often indicated as $EE$ spectrum).
\subsection{The singularity problem}

The singularity problem is probably the most serious fundamental 
drawback of the SCM. While the other problems are certainly 
very important and manifest diverse logical inconsistencies of the SCM,
the singularity problem is fundamental since it is related to the 
structure of the underlying theory of gravitation, i.e. General Relativity.
In the SCM as $t\to 0$ 
\begin{equation}
\rho_{\mathrm{tot}}\simeq \frac{1}{t^2}, \qquad H^2\simeq \frac{1}{t^2}
\end{equation}
and 
\begin{equation}
R^2 \simeq R_{\mu\nu}R^{\mu\nu} \simeq R_{\mu\nu\alpha\beta}R^{\mu\nu\alpha\beta} \simeq \frac{1}{t^4}.
\end{equation}
Thus, in the limit $t\to 0$ the energy density diverges and also 
the relevant curvature invariants diverge. The Weyl invariant 
is automatically vanishing since the geometry is isotropic. 
The singularity problem does not only involve the regularity 
of the curvature invariants but also the possible completeness 
(or incompleteness) of causal (i.e. either time-like or null) 
geodesics. By the Hawking-Penrose theorems \cite{hawkingellis2}
the past-incompleteness of causal geodesics is just another 
diagnostic of a singular space-time.

As already mentioned the singularity problem is deeply rooted in the 
adoption of General Relativity as underlying gravitational theory. In recent years, in the 
context of string-inspired cosmological scenarios (see \cite{PBB2} and references therein)
a lot of work has been done to see if cosmological singularities can be avoided 
(or, even more modestly, addressed) in gravity theories that, at early time are 
different from General Relativity. While the conclusion of these 
endevours is still far, it is certainly plausible that the ability of string theory 
in dealing with gravitational interactions can shed some light on the cosmological singularities 
(and on their possible avoidance). Two key features emerge when string theory is 
studied in a cosmological framework \cite{PBB2}. The first feature is that string theory demands 
the existence of a fundamental length-scale (the string scale which 10 or 100 times larger 
than the Planck length). This occurrence seems to point towards 
the existence of a maximal curvature scale (and of a maximal energy density) 
which is the remnant of the general relativistic singularity. While the resolution 
of cosmological singularities in string theory is still an open problem, there certainly 
exist amusing (toy) models where the singularity are indeed resolved. 
The second key feature of string cosmological scenarios is represented 
by the novelty that gauge couplings are effectively dynamical. This phenomenon
has no counterpart in the standard general relativistic treatment but will not 
be discussed here.

\newpage
\renewcommand{\theequation}{4.\arabic{equation}}
\setcounter{equation}{0}
\section{Beyond the SCM}
\label{sec4}
It is interesting 
to see how these conceptual problems can be reduced or, at least, relaxed in some 
conventional scenarios that can usefully complement the SCM. In spite of the fact that 
some of these scenarios (like the inflationary scenario) can cope with the technical 
problems of the SCM (such as the flatness or the horizon problems) none of these models 
is able to cope with the deepest of all the problems of the SCM, i.e. the singularity problems. 
To this statement it should be added that the inflationary solution of the entropy problem 
relies on the possible decay of inflaton into massless particles with the hope that 
such a process may produce a sufficiently high reheating temperature.
For an introduction to the inflationary paradigm Refs. \cite{olive1,linde1} 
  can be usefully consulted (see also \cite{CONVEN1,CONVEN2,CONVEN3,CONVEN4} for some specific 
  inflationary scenarios). For reasons of space, it will not be possible to treat some of the unconventional 
  approaches to inflation that are rather interesting especially in the light of their connections 
  with string theory. Among them, the pre-big bang scenario (developed in the last fifteen years) 
  represents a rather intriguing option. We refer the reader to the original papers and to some 
  very comprehensive review articles \cite{PBB3} (see also \cite{reg1,reg2}).

\subsection{The horizon and the flatness problems}

The horizon problem in the SCM has to do with the fact that there exist a particle horizon $d_{\mathrm{p}}(t) \simeq 
H^{-1}(t)$. Thus, as we go forward in time  (and for $t>0$) the particle horizon evolves faster than the scale 
factor which goes, in the SCM, as $t^{\alpha}$ with $0<\alpha <1$. This occurrence also implies that at the moment when the initial conditions are ideally set, our observable Hubble volume consisted of a huge amount 
of causally disconnected domains (see, for instance, Eqs. (\ref{ratio1}) and (\ref{domain2})). 
A possible way out of this problem is to consider the completion of the SCM
by means of a phase where not a particle but an {\em event} horizon exist.  Consider, for instance, a scale factor with power-law behaviour going as 
\begin{equation}
a(t) \simeq t^{\beta}, \qquad \beta > 1,\qquad t>0,
\label{tbeta}
\end{equation}
and describing a phase of accelerated expansion (i.e. $\dot{a}>0$, $\ddot{a}>0$).
The particle and event horizons are given, respectively, by 
\begin{eqnarray}
&& d_{\mathrm{p}}(t) = \frac{1}{1-\beta} 
\biggl[ t - t_{\mathrm{min}} \biggl(\frac{t}{t_{\mathrm{min}}}\biggr)^{\beta}\biggr],
\label{dpbeta}\\
&& d_{\mathrm{e}}(t) = \frac{1}{1 -\beta} \biggl[ t_{\mathrm{max}} \biggl(\frac{t}{t_{\mathrm{max}}}\biggr)^{\beta} - t\biggr].
\label{debeta}
\end{eqnarray}
From Eqs. (\ref{dpbeta}) and (\ref{debeta}) it immediately follows that the particle horizon does not exist while 
the event horizon is finite:
\begin{eqnarray}
d_{\mathrm{e}}(t) \simeq \frac{\beta}{\beta -1}\, H^{-1}(t).
\label{debeta2}
\end{eqnarray}
Equation (\ref{debeta2})  follows from Eq. (\ref{debeta}) in the limit $t_{\mathrm{max}} \to +\infty$ while 
in the limit $t_{\mathrm{min}}\to 0$, $d_{\mathrm{p}}(t)$ diverges.
Similar conclusions follow in the case when the phase of accelerated expansion is parametrized in terms of the (expanding) branch of four-dimensional de Sitter space-time, namely
\begin{equation}
a(t) \simeq e^{H_{\mathrm{i}} t},\qquad H_{\mathrm{i}} >0. 
\label{DS}
\end{equation}
In this case, the particle and event horizon are, respectively, 
\begin{eqnarray}
&& d_{\mathrm{p}}(t) = H_{\mathrm{i}}^{-1} \biggl[ e^{H_{\mathrm{i}}(t - t_{\mathrm{min}})} -1\biggr],
\label{dsP}\\
&&  d_{\mathrm{e}}(t) = H_{\mathrm{i}}^{-1} \biggl[1-  e^{H_{\mathrm{i}}(t - t_{\mathrm{max}})}\biggr].
\label{dsE}
\end{eqnarray}
According to Eq. (\ref{DS}) the cosmic time coordinate is allowed to run
from $t_{\mathrm{min}}\to - \infty$ up to $t_{\mathrm{max}}\to + \infty$. Consequently, 
for $t_{\mathrm{min}}\to -\infty$ (at fixed $t$) the particle horizon will diverge and the 
typical size of causally connected regions at time $t$ will scale as 
\begin{equation}
L_{\mathrm{i}}(t) \simeq H_{\mathrm{i}}^{-1} \frac{a(t)}{a(t_{\mathrm{min}})}.
\label{CCregion}
\end{equation}
So while in the SCM the particle horizon increases faster than the scale factor, the 
typical size of causally connected regions scales exactly {\em as the scale factor}.
In the limit $t_{\mathrm{max}}\to \infty$ the event horizon exist and it is given, from Eq.
(\ref{dsE}), by 
\begin{equation}
d_{\mathrm{e}}(t) \simeq H_{\mathrm{i}}^{-1},
\label{dsE2}
\end{equation}
implying that in the case of de Sitter dynamics the event horizon is constant.
Of course, as it will be later pointed out, de Sitter dynamics cannot be exact (see section \ref{sec5}).
In this case, customarily, we talk about quasi-de Sitter stage of expansion where 
$H_{\mathrm{i}}$ is just approximately constant and, more precisely, slightly 
decreasing. 

To summarize, the logic to address the horizon problem is then to suppose 
(or presume) that at some typical time $t_{\mathrm{i}}$ an event horizon 
is formed with typical size $H_{\mathrm{i}}^{-1}$. Furthermore, since we are 
working in General Relativity, we shall also demand that $H_{\mathrm{i}} < 
M_{\mathrm{P}}$. Now {\em if} the Universe is sufficiently homogeneous 
inside the created event horizon, it will remain (approximately) homogeneous 
also later on, by definition of event horizon. In other words, if, inside 
the event horizon, $\delta\rho/\rho$ is sufficiently small, we can think 
of fitting inside a single event horizon at $t_{\mathrm{i}}$ the whole 
observable Universe. In practice, this condition translates into a typical size 
of $H_{\mathrm{i}}$ which should be such that $H_{\mathrm{i}} \leq 10^{-5} \,M_{\mathrm{P}}$ or, in  equivalent terms, an event horizon that is sufficiently 
large with respect to the Planck length, i.e. 
$H_{\mathrm{i}}^{-1} \gg  \ell_{\mathrm{P}}$. 

To fit the whole observable Universe inside the newly formed event horizon at the onset 
of inflation, the de Sitter (or quasi-de Sitter) phase must last for a sufficiently large amount 
of time. In equivalent terms it is mandatory that the scale factor grows of a sufficient amount. Since 
the growth of the scale factor is exponential (or quasi-exponential) it is common practice 
to quantify the growth of the scale factor in terms of the number of e-folds, denoted by $N$ and defined as  
\begin{equation}
e^{N} = \frac{a(t_{\mathrm{f}})}{a(t_{\mathrm{i}})} \equiv 
\frac{a_{\mathrm{f}}}{a_{\mathrm{i}}} , \qquad N = \ln{\biggl(\frac{a_{\mathrm{f}}}{a_{\mathrm{i}}} \biggr)}.
\end{equation}
To estimate the condition required on the number of e-folds $N$ 
we can demand that the whole (present) Hubble volume (blushifted at the epoch $t_{\mathrm{i}}$ when the event horizon is formed) is smaller than $H_{\mathrm{i}}^{-1}$.
In fully equivalent terms we can demand that $H_{\mathrm{i}}^{-1}$ redshifted at the 
present epoch is larger than (or comparable with) the present Hubble radius. 
By following this second path we are led to require that 
\begin{equation}
H_{\mathrm{i}}^{-1} \biggl(\frac{a_{\mathrm{i}}}{a_{\mathrm{f}}}\biggr)_{dS} 
\biggl(\frac{a_{\mathrm{r}}}{a_{\mathrm{f}}}\biggr)_{reh} \biggl(\frac{a_{\mathrm{r}}}{a_{\mathrm{eq}}}\biggr)_{rad} \biggl(\frac{a_{\mathrm{eq}}}{a_{0}}\biggr)_{mat} \geq H_{0}^{-1}.
\label{horcond1}
\end{equation}
In Eq. (\ref{horcond1}) the subscripts appearing in each round bracket indicate 
the specific phase during which the given amount of redshift is computed. 
Between the end of the de Sitter stage and the beginning 
of the radiation-dominated phase there should be an intermediate 
phase usually called reheating (or pre-heating) where the Universe 
makes a transition from the accelerated do the decelerated expansion. 
The rationale for the existence of this phase stems from the observation that, during 
the de Sitter phase, any radiation present at $t_{\mathrm{i}}$ is rapidly diluted 
and becomes soon negligible since, as we saw, $\rho_{\mathrm{R}}$ scales 
as $a^{-4}$. In equivalent terms we can easily appreciate that the temperature, as well 
as the entropy density (possibly present at $t_{\mathrm{i}}$) decay exponentially so that, 
as soon as the accelerated expansion proceeds, the Universe approaches 
 a configuration where the temperature and the entropy density are exponentially 
 vanishing. There is therefore the need of reheating the Universe at the end of inflation.  To estimate 
 the minimal number of e-folds $N$ we can rely on the sudden reheating approximation where, basically, $a_{\mathrm{f}} \simeq a_{\mathrm{r}}$. Consequently, under this approximation we can write Eq. (\ref{horcond1}) as 
 \begin{equation}
 e^{N} \geq \biggl(\frac{H_{\mathrm{i}}}{H_{0}} \biggr) \biggl(\frac{H_{\mathrm{eq}}}{H_{\mathrm{i}}}\biggr)^{1/2} (z_{\mathrm{eq}} +1)^{-1},
 \label{EF1}
 \end{equation}
 which can also be expressed, by taking the natural logarithm, as 
 \begin{equation}
 N \geq 62.2 + \frac{1}{2} \ln{\biggl(\frac{\xi}{10^{-5}}\biggr)} 
 - \ln{\biggl(\frac{h_{0}}{0.7}\biggr)}+ \frac{1}{4} \ln{\biggl(\frac{h_{0}^2 \Omega_{\mathrm{R}0}}{4.15\,\, 10^{-5}}\biggr)}.
 \label{EF2}
 \end{equation}
 In Eq. (\ref{EF2}) the quantity $\xi = H_{\mathrm{i}}/M_{\mathrm{P}}$ has been defined and 
it has also been used that
 \begin{equation}
 H_{\mathrm{eq}} =  \sqrt{ 2 \,\,\Omega_{\mathrm{M}0}} \, H_{0} \, \biggl(\frac{a_{0}}{a_{\mathrm{eq}}}\biggr)^{3/2},
 \label{Heq}
 \end{equation}
 which comes directly from Eq. (\ref{FL1}) by requiring 
 $\rho_{\mathrm{M}}(t_{\mathrm{eq}}) = \rho_{\mathrm{R}}(t_{\mathrm{eq}})$. During a quasi-de Sitter stage of expansion, quantum mechanical fluctuations of the inflaton will be amplified to observable values and their amplitude is exactly 
 controlled by $\xi$. To ensure that the amplified quantum-mechanical inhomogeneities will match the observed values of the angular power spectrum of temperature 
 inhomogeneities we have to require $\xi \simeq 10^{-5}$ which demands that 
 $N \geq 63$. 
 
\begin{figure}
\centering
\includegraphics[height=6cm]{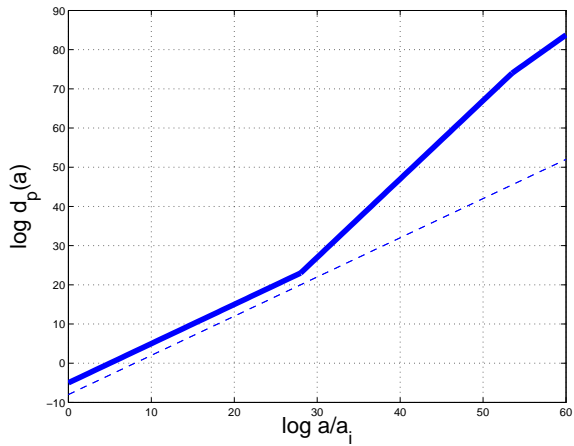}
\caption{The evolution of the particle horizon in the inflationary case (full line) is reported 
for a model Universe which passes from a de Sitter stage of expansion to a radiation-dominated 
stage which evolves, in turn, into a matter-dominated epoch. The evolution 
of a typical scale smaller than the event horizon is also reported. The first branch (where 
$d_{p}(a)$ evolves as the scale factor) illustrates the evolution of the typical size of causally 
connected regions during inflation. This quantity is formally divergent for $t\to -\infty$.}
\label{HOR3}      
\end{figure}

The same hierarchy of scales required to address the horizon problem, also 
relaxes the flatness problem. The flatness problem arises, in the SCM, from the 
observation that the contribution of the spatial curvature increases sharply, during the 
radiation and matter-dominated epochs. This observation
 entails that if $\Omega_{\mathrm{t}}\simeq {\mathcal O}(1)$ today, $\Omega_{\mathrm{t}}$ had to be fine-tuned to 1 also at the onset of the radiation-dominated 
evolution but with much greater precision. So, if today $\Omega_{\mathrm{t}} \simeq 1$ with an experimental error 
of, for instance, $0.1$, at the Planck scale $\Omega_{\mathrm{t}}$ had to be fine-tuned to $1$ with an accuracy of, roughly, ${\mathcal O}(10^{-60})$.

If the ordinary radiation-dominated evolution is preceded by a de Sitter (or quasi-de Sitter) phase of expansion the spatial curvature will be exponentially (or quasi-exponentially) suppressed with respect to the Hubble curvature $H_{\mathrm{i}}^2$ 
which is constant (or slightly decreasing). Thus, if the exponential growth of the 
Universe will last for a sufficient number of e-folds, the spatial curvature 
at the onset of the radiation dominated phase will be sufficiently suppressed to allow
for a subsequent growth of $k/(a H)^2$ during the radiation and matter-dominated epochs. 
The same number of e-folds required to address the horizon problem also guarantees 
that the spatial curvature will be sufficiently suppressed during the phase of exponential
expansion. In fact, while today 
\begin{equation}
\Omega_{\mathrm{t}}(t_{0})  - 1 = \frac{k}{a_{0}^2 \, H_{0}^2}, 
\label{OMt0}
\end{equation}
at the onset of the de Sitter phase 
\begin{equation}
\Omega_{\mathrm{t}}(t_{\mathrm{i}})  - 1
= \frac{k}{a_{\mathrm{i}}^2 \, H_{\mathrm{i}}^2}.
\label{OMti}
\end{equation}
Dividing Eq. (\ref{OMt0}) by Eq. (\ref{OMti}) we can also obtain, rather easily that 
\begin{equation}
\sqrt{|\Omega_{\mathrm{tot}}(t_{0}) -1|} = \frac{a_{\mathrm{i}} \, H_{\mathrm{i}}}{a_{0}
\, H_{0}}  \sqrt{|\Omega_{\mathrm{tot}}(t_{\mathrm{i}}) -1|}.
\label{ratio}
\end{equation}
Now, from Eq. (\ref{ratio}) it is clear that if $\Omega_{\mathrm{tot}}(t_{0})$ 
is tuned to $1$ with the precision of, say, $1$ \%, the pre-factor 
appearing at the right hand side of Eq. (\ref{ratio}) must be of the order $0.1$ 
if $|\Omega_{\mathrm{tot}} (t_{\mathrm{i}}) - 1| $ was of order $1$ 
at the onset of the de Sitter phase. Thus, more generally, we are led to require 
that 
\begin{equation}
\frac{a_{\mathrm{i}} H_{\mathrm{i}}}{a_{0} H_{0}} <1, 
\label{CURVcon}
\end{equation}
which becomes, after making explicit the redshift contribution, exactly 
Eq. (\ref{horcond1}). We then discover that if $N \geq 63$ the 
spatial curvature at the end of inflation will be small enough 
to guarantee that the successive growth (during radiation and matter) 
will not cause (today) $|\Omega_{\mathrm{tot}}(t_{0}) -1|$ to be of order 
$1$ (or even larger).
\begin{figure}
\centering
\includegraphics[height=6cm]{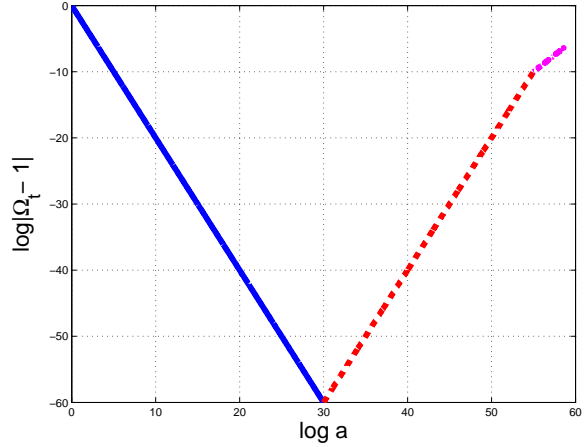}
\caption{The evolution of the logarithm (to base $10$) of $|\Omega_{\mathrm{t}} -1|$ as a 
function of the $\log{a}$ for an inflationary phase where $N \simeq 69$ and in the sudden reheating 
approximation. The full line denotes the evolution during inflation while the dashed 
and the dot-dashed lines denote the approximate evolution during radiation and matter.}
\label{SPCURV}      
\end{figure}
In Fig. \ref{SPCURV} the evolution of $|\Omega_{\mathrm{t}} -1|$ is reported as a 
function of the logarithm (to base $10$) of the scale factor for the situation 
where the Universe inflates during $69$ e-folds.

As it will be discussed in section \ref{sec5} 
that the inflationary dynamics can be modeled in terms of one (or more) 
minimally (or non-minimally) coupled light scalar degrees of freedom. 
Here the word light refers to the typical scale of the problem, i.e. 
$H_{\mathrm{i}}$ so that the mass of the scalar field should be small in units 
of $H_{\mathrm{i}}$. So suppose that, at $t_{\mathrm{i}}$, there is a scalar field
which has some typical inhomogeneities over different wavelengths. It is 
clear that the most generic evolution of such a system represents 
a tough numerical task under general circumstances. By this we mean 
that it is not said that the most generic thing a scalar field does will be to inflate.
However one can also guess that if the scalar field $\varphi$ is sufficiently homogeneous 
over a region $H_{\mathrm{i}}^{-1}$ one of the possibilities will be inflation provided 
the kinetic energy of the scalar field is sufficiently small in comparison with 
its potential energy \footnote{This condition can be, indeed, relaxed by noticing that, in the absence 
of potential, the (homogeneous) evolution of the inflaton is given by $\ddot{\varphi} + 3 H \dot{\varphi} \simeq 0$.
This relation (see section \ref{sec5}, Eq. (\ref{FLS3})) implies that $\dot{\phi}^2$ scales as $a^{-6}$ and may become, eventually, subleading in comparison with the potential energy.}. These observations lead to the following requirements:
\begin{equation}
\frac{|\nabla^2 \varphi| }{a_{\mathrm{i}}^2} \ll \frac{\partial V}{\partial\varphi},\qquad 
\dot{\varphi}^2 \ll V,\qquad \frac{(\nabla\varphi)^2}{a_{\mathrm{i}}^2} \ll {\dot{\varphi}}^2,
\end{equation}
at the time $t_{\mathrm{i}}$ and over a typical region $H_{\mathrm{i}}^{-1}$.
If the duration of inflation lasts just for $63$ (or $65$) e-folds 
it can happen that some initial spatial gradients (i.e. some initial spatial curvature) 
will still cause inhomogeneities inside the present Hubble volume. 

\subsection{Classical and quantum fluctuations}

Classical and quantum fluctuations, in inflationary cosmology, 
have similarities but also crucial differences. While classical fluctuations are given, once forever,
on a given space-like hypersurface, quantum 
fluctuations keep on reappearing all the time thanks to the zero-point energy 
of various quantum fields that are potentially present in de Sitter space-time.
If the the accelerated phase lasts just the minimal amount of e-folds 
required to solve the problem of the standard cosmological model classical 
fluctuations can definitely have an observational and physical relevance.
Suppose, indeed, that classical fluctuations are present prior to the onset of inflation
and suppose that their typical wavelength was of the order of $H_{\mathrm{i}}^{-1}$.
Then we can say that their wavelength today is 
\begin{equation}
\lambda(t_{0}) = H_{\mathrm{i}}^{-1} \frac{a_{0}}{a_{\mathrm{i}}}.
\label{lambda1}
\end{equation}
But $a_{\mathrm{0}}/a_{\mathrm{i}}$ is just the redshift factor required to 
fit the present Hubble patch inside the event horizon of our de Sitter phase.
From Eq. (\ref{lambda1}) it is clear that if $H_{\mathrm{i}}^{-1} = 10^{5} \ell_{\mathrm{P}}
\simeq 10^{-28} \,\mathrm{cm}$, then $\lambda(t_0) \simeq H_{0}^{-1}$ potentially relevant today.

If the inflationary phase lasts much more than the minimal amount of e-folds the classical fluctuations (possibly present at the onset of inflation) will be, in the 
future, redshifted to larger and larger length-scales (even much larger than the present 
Hubble pacth). In the future these wavelengths will be, in some sense, accessible 
since the Hubble patch increases as time goes by. Therefore, if the 
inflationary phase lasts much more than the minimal amount of e-folds,
the only fluctuations potentially accessible through satellite and terrestrial 
observations will be quantum-mechanically generated fluctuations that can be, under some conditions to be discussed later, parametrically amplified.

\begin{figure}
\centering
\includegraphics[height=6cm,width=10cm]{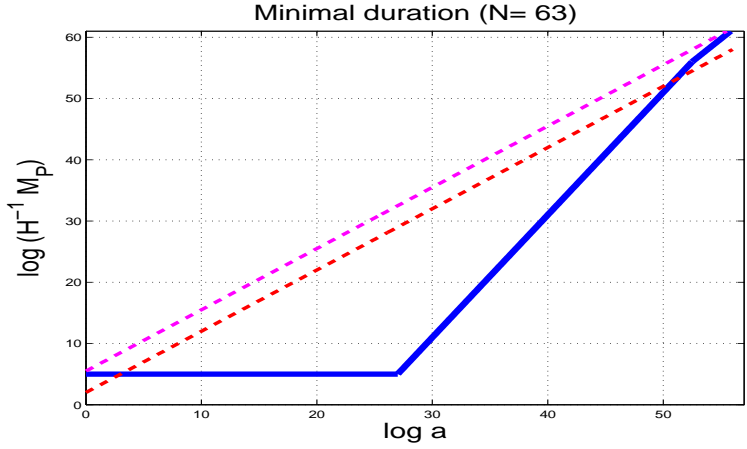}
\caption{The evolution of the Hubble radius in the case when the duration 
of inflation is minimal. With the dashed lines we also illustrate the evolution 
of different typical wavelengths.}
\label{mindur}      
\end{figure}

\begin{figure}
\centering
\includegraphics[height=6cm,width=10cm]{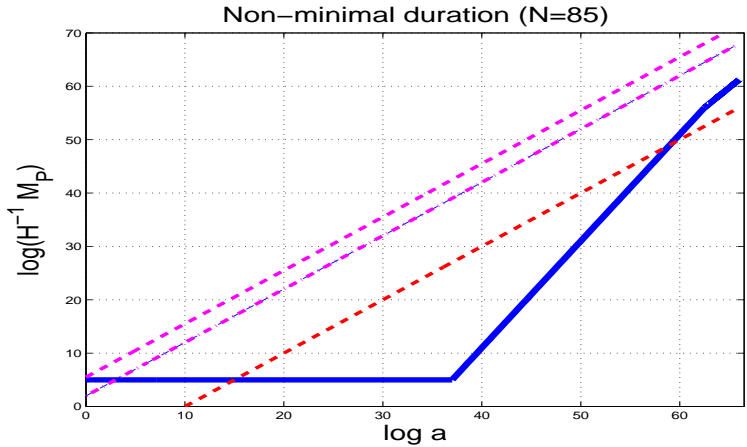}
\caption{The same quantity illustrated in Fig. \ref{mindur} but for a case when the duration of inflation 
is non-minimal.}
\label{nonmindur}      
\end{figure}

In Fig. \ref{mindur} the evolution of the Hubble radius (in Planck units) is reported as a function of the logarithm 
of the scale factor. In Fig. \ref{mindur} inflation lasts for the minimal amount (i.e. $ N =63$) while in Fig. \ref{nonmindur}
the duration of inflation is non-minimal (i.e. $N =85 \gg 63$). 
From Figs. \ref{mindur} and \ref{nonmindur} the difference in the behaviour of classical and quantum fluctuations 
is evident.  The dashed lines represent the wavelength of a given perturbation (either classical or quantum mechanical).
If the duration of inflation is minimal, it is plausible (see Fig. \ref{mindur}) that a classical fluctuation 
crosses the Hubble radius the second time around the epoch of matter-radiation equality. This means 
that the classical fluctuation may have an observational impact.
If, on the contrary, inflation lasts for much more than $63$ e-folds (85, in Fig. \ref{nonmindur})  there are no chances 
that a classical fluctuation present at the onset of inflation will cross the Hubble radius around the epoch of matter-radiation equality.
In this second case, the only fluctuations that will be eventually relevant will be the quantum mechanical ones.
Summarizing the discussion conducted so far, we can say that there are two 
physically different situations:
\begin{itemize}
\item{} if the duration off the inflationary phase lasts for the minimal amount of e-folds 
(i.e. $N\simeq 63$) then it is plausible that some (observable?) relics of a pre-inflationary dynamics can be eventually detected in CMB observations;
\item{} if, on the contrary, the duration of inflation greatly exceeds the minimal 
duration we can expect that any memory of the pre-inflationary phase will be lost 
thanks to the efficiency of exponential expansion.
\end{itemize}

\subsection{The entropy problem}
The entropy of the CMB refers 
to the entropy of the {\em matter sources} of the geometry. 
There could be, in principle, also a truly gravitational entropy associated with the gravitational field itself. 
The way one can attribute an entropy to the gravitational field is subject of debate.
This entropy, for instance, could be connected to the possibility of activating 
new degrees of freedom and can be measured, for instance according to Penrose, by 
the Weyl tensor \cite{penrose} (see also \cite{tod}).
So, for the moment, let us focus our attention on the entropy of the sources and 
let us recall that, today, this entropy seats in photons and it is given, in natural 
units, by $S_{\gamma} \simeq 10^{88}$. Since the evolution of the sources 
is characterized, in the SCM, by the covariant conservation of the total 
energy-momentum, the total entropy of the sources will also be 
conserved.

We are therefore in the situation where the entropy at the end 
of the inflationary phase must be of the order of $10^{88}$.
During inflation, at the same time, there is no reason why 
the evolution of the sources must not follow from the covariant 
conservation of the energy-momentum tensor. If this is the case, the entropy 
of our observable Universe must be produced, somehow, at the end 
of inflation. This is, indeed, what we can call the standard lore 
for the problem of entropy generation within the inflationary proposal.
In the standard lore entropy (as well as radiation) is generated 
at the end of inflation during a phase called reheating.

During reheating the degree of freedom that drives inflation (the 
inflaton, in single field inflationary models) decays and this process 
is non-adiabatic.  What was the 
entropy at the beginning of inflation? The answer to this question 
clearly depends upon the specific inflationary dynamics and, in 
particular, upon the way inflation starts. Let us try, however, to get 
a rather general (intuitive) picture of the problem.
Suppose that, at some time $t_{\mathrm{i}}$, an event horizon forms with 
typical size $H_{\mathrm{i}}^{-1}$. The source of this dynamics could be, in principle, a 
cosmological constant or, more realistically, the (almost) constant potential 
energy of a scalar degree of freedom. In spite of the nature of the source, it can be always 
argued that its energy density is safely estimated by $H_{\mathrm{i}}^2 M_{\mathrm{P}}^2$. When the event horizon forms, massless particles can be around. Suppose, for 
a moment, that all the massless species are in thermal equilibrium at a common temperature $T_{\mathrm{i}}$. Thus, their energy and entropy densities will be estimated, respectively, by $T_{\mathrm{i}}^4$ and by $T_{\mathrm{i}}^3$. The 
total entropy at $t_{\mathrm{i}}$ contained inside the newly formed event horizon 
can then be quantified as 
\begin{equation}
S_{\mathrm{i}} \simeq \biggl(\frac{T_{\mathrm{i}}}{H_{\mathrm{i}}}\biggr)^3.
\label{entroin}
\end{equation}
 As not yet discussed (but already 
mentioned) the quantum fluctuations amplified in the course of the inflationary evolution
force us to a value of $H_{\mathrm{i}}/M_{\mathrm{P}} \simeq 10^{-5}$.
On the other hand we will have that, for inflation to start,
\begin{equation}
H_{\mathrm{i}}^2 M_{\mathrm{P}}^2 \gg g_{\rho} T_{\mathrm{i}}^4,
\label{entroin2}
\end{equation}
where $g_{\rho}$ is the effective number of relativistic degrees of 
freedom at $t_{\mathrm{i}}$ (see Appendix  \ref{APPB}, Eq. (\ref{grho})).
From Eq. (\ref{entroin2}) it is easy to deduce that 
\begin{equation}
S_{\mathrm{i}} \simeq \biggl(\frac{T_{\mathrm{i}}}{H_{\mathrm{i}}}\biggr)^3,\qquad 
\frac{T_{\mathrm{i}}}{H_{\mathrm{i}}}  \ll g_{\rho}^{1/4} \biggl(\frac{H_{\mathrm{i}}}{M_{\mathrm{P}}}\biggr)^{-1/2}.
\label{entroin3}
\end{equation}
Recalling that $H_{\mathrm{i}} \simeq 10^{-5}\, M_{\mathrm{P}}$ it is then 
plausible, under the assumptions mentioned above, that the entropy 
at the onset of inflation is of order one and, anyway, much smaller 
than the present entropy sitting in the CMB photons.

During the development of inflation, if there is no 
significant energy and momentum exchange between the inflaton field 
and the photons, the temperature, the concentration 
of photons and the entropy density will all redshift exponentially so that 
we end-up eventually in a very flat and cold state where, however, the total 
entropy is still of order one thanks to the adiabaticity of the evolution. 
At some point, however, the inflaton will start decaying and massless
particles will be produced. Let us now try to estimate the entropy 
produced in this process. It will be, in general terms, of the order 
of 
\begin{equation}
S_{\mathrm{rh}} = \frac{4}{3} \pi \biggl(\frac{T_{\mathrm{rh}}}{H_{\mathrm{rh}}}\biggl)^3,
\label{entroin4}
\end{equation}
where $T_{\mathrm{rh}}$ is the reheating temperature and where $H_{\mathrm{rh}}^{-1}$ is the Hubble radius at the reheating. Let us assume, to begin with, that the reheating 
is instantaneous and perfectly efficient. This amounts to suppose that {\em all} the energy density of the inflaton is efficiently transformed into radiation at $t_{\mathrm{rh}}$. Recalling now that $H_{\mathrm{rh}}^{-1}$ can be usefully connected with 
$H_{\mathrm{i}}^{-1}$ as 
\begin{equation}
H_{\mathrm{rh}}^{-1} \simeq H_{\mathrm{i}}^{-1}
 \biggl(\frac{a_{\mathrm{f}}}{a_{\mathrm{i}}}\biggr),
 \label{entroin5}
 \end{equation}
 we will have that the effective number of e-folds should be 
 \begin{equation}
 N \geq 65.9 + \frac{1}{3}\ln{\biggl(\frac{\xi}{10^{-5}}\biggr)} - \ln{\biggl(\frac{T_{\mathrm{rh}}}{10^{15}\,\mathrm{GeV}}\biggr)},
 \label{entroin6}
 \end{equation}
 where we assumed that $H_{\mathrm{rh}} \simeq T_{\mathrm{rh}}^2/M_{\mathrm{P}}$.
So, if the inflationary phase is sufficiently long, the Hubble radius at reheating 
will be large enough to match the observed value of the entropy of the sources.
There are, at least, three puzzling features, among others, with the argument we just presented:
\begin{itemize}
\item{} the amount of entropy crucially depends upon the temperature 
of the reheating which depends upon the coupling 
of the inflaton to the degrees of freedom of the particle physics model;
\item{} the entropy should not exceed the observed one and, consequently, the solution 
of the entropy problem seems to imply a lower bound on the number of the 
inflationary e-folds;
\item{} the reheating may not be instantaneous and this will entail the possibility 
that the number of inflationary e-folds may be smaller since, during reheating, the 
Hubble radius may grow.
\end{itemize}
In some models of reheating the decay of the inflaton occurs through a 
phase where the inflaton field oscillates around the minimum 
of its potential. In this phase of coherent oscillations the Universe becomes, effectively, 
dominated by radiation and $a(t) \simeq t^{2/3}$. Consequently, the radiation of 
$H_{\mathrm{rh}}^{-1}$ to $H_{\mathrm{i}}^{-1}$ will be given by a different 
equation and, more specifically, by 
\begin{equation}
H_{\mathrm{rh}}^{-1} = e^{N} \biggl(\frac{H_{\mathrm{i}}}{H_{\mathrm{rh}}}\biggr)^{2/3}
H_{\mathrm{i}}^{-1} 
\label{relation}
\end{equation}
where the power $2/3$ in the last bracket accounts for the evolution during 
the reheating phase.
In this case, requiring that the total entropy exceeds a bit the observed 
entropy we will obtain the following condition, on $N$:
\begin{equation}
N\geq 60.1 + \frac{1}{3} \ln{\biggl(\frac{\xi}{10^{-5}}\biggr)}   + \frac{4}{3} \ln{\biggl(\frac{T_{\mathrm{rh}}}{10^{15} \,{\mathrm{GeV}}}\biggl)}.
\label{entroin8}
\end{equation}
Equation (\ref{entroin6}) gives a lower estimate for the number of inflationary e-folds
simply because the Universe was redshifting also in the intermediated (reheating) phase by, roughly, $5$ effective e-folds.  

According to the presented solution of the entropy problem the initial state of the Universe prior to inflation 
must have been rather ordered. Let us assume, indeed, the validity of the second law 
of thermodynamics
\begin{equation}
\dot{S}_{\mathrm{m}} + \dot{S}_{\mathrm{Gr}} \geq 0,
\label{secondlaw}
\end{equation}
where $S_{\mathrm{Gr}}$ denotes, quite generically, the entropy of the gravitational field itself. 
Equation (\ref{secondlaw}) is telling us that as we go back in time the Universe 
had to be always less and less entropic. The conclusion that the pre-inflationary Universe 
was rather special seems to clash with the idea that the initial conditions of inflation were 
somehow chaotic \cite{linde1}. The idea here is that inflation is realized by means of a scalar 
degree of freedom (probably a condensate, see section \ref{sec5}) initially displaced from the minimum of its own potential.
In some regions of space the inflaton will be sufficiently displaced from its minimum and its spatial 
gradients will be large in comparison with the potential. In some other regions 
the spatial gradients will be sufficiently small. This picture, here only swiftly described, 
is really chaotic and it is conceptually difficult to imagine that this chaos could avoid also 
a large entropy of the pre-inflationary stage.  A possible way out of this apparent impasse 
may be to include consistently the entropy of the gravitational field. 

We conclude by recalling that there are proposals on the possible measures of the entropy of the gravitational field.
On top of the proposal of Penrose already quoted in this section \cite{penrose,tod}, Davies \cite{davies1,davies2} proposed 
to associate an entropy to the cosmological backgrounds endowed with an event horizon. In this 
case $S_{\mathrm{Gr}} \simeq d_{\mathrm{e}}^2 M_{\mathrm{P}}^{-2}$. One can easily imagine models 
where the entropy of the sources decreases but the total entropy (i.e. sources and gravitational field) does not 
decrease \cite{mgv1,mgv2,mgv3,mgv4}. There is also the possibility of associating 
an entropy to the process of production of relic gravitons \cite{mgentr1,mgentr2} but we shall swiftly 
get back on this point later on in section \ref{sec6}.

\subsection{The problem of geodesic completeness}
 
 Inflation does not solve the problem of the initial singularity. This statement 
 can be appreciated by noticing that the expanding de Sitter space-time is not past
  geodesically complete \cite{alex2,alex3,alex4,alex5}. 
 Such an occurrence is equivalent (both technically and physically) to a singularity.
 The geodesic incompleteness of a given space-time simply means that causal 
 geodesics cannot be extended indefinitely in the past as a function of the affine 
 parameter that shall be denoted with $\lambda$. The causal geodesics 
 are either the time-like or null geodesics. Let us therefore consider, for 
 simplicity, the case of null geodesics i.e. 
 \begin{equation}
 dt^2 - a^2(t) d\vec{x}^2=0.
 \label{NULL}
 \end{equation}
 From the geodesic equation:
 \begin{equation}
 \frac{d^2 x^{\mu}}{d\lambda^2} + \Gamma^{\mu}_{\alpha\beta} \frac{d x^{\alpha}}{d\lambda}  \frac{d x^{\beta}}{d\lambda} =0,
 \label{GG}
 \end{equation}
 it is immediate to obtain the following pair of conditions:
 \begin{eqnarray}
 && \frac{d^2 t}{d\lambda^2} + 2 \dot{a} a \biggl(\frac{d\vec{x}}{d\lambda}\biggr)^2 =0,
 \label{GG1}\\
 && \frac{ d^2 x^{i}}{d\lambda^2} + 2 \frac{\dot{a}}{a} \frac{d t}{d\lambda} 
 \frac{dx^{i}}{d\lambda} = 0.
 \label{GG2}
 \end{eqnarray}
 Inserting Eq. (\ref{NULL}) into Eq. (\ref{GG1}) to eliminate $(d\vec{x}/d\lambda)^2$ we obtain  
 \begin{equation}
 \frac{d f}{d\lambda} + \frac{\dot{a}}{a} f^2 =0,\qquad f(\lambda) = \frac{d t}{d\lambda}.
 \label{NULL1}
 \end{equation}
But if we now recall that 
\begin{equation}
\frac{d f}{d\lambda} = \frac{d f}{d t} \frac{dt}{d\lambda} = \frac{df}{dt} f,
\end{equation}
Eq. (\ref{NULL1}) allows to express $f(t)$ in terms of the scale factor:
\begin{equation}
f(t) =\frac{d t}{d\lambda}= \frac{1}{a(t)},\qquad d \lambda = a(t) dt.
\label{NULL3}
\end{equation}
In the case of expanding de Sitter space-time we will have 
\begin{equation}
a(t) \simeq e^{Ht},\qquad t = \frac{1}{H} \ln{(H \lambda)}
\end{equation}
implying that $t(\lambda)\to -\infty$ for $\lambda \to 0^{+}$.
This means that null geodesics are past-incomplete.

To appreciate what would be a geodesically complete 
space-time let us consider the following example:
\begin{equation}
a(t) \simeq \cosh{H t},\qquad t = \frac{1}{H} 
\ln{[ H \lambda + \sqrt{H^2 \lambda^2 +1}]}.
\label{nullcosh}
\end{equation}
 In the case $\lambda \to 0$, $t(\lambda)\to 0$ and the geodesics 
 are complete. The background discussed in Eq. (\ref{nullcosh}) is a solution 
 of Einstein equations but in the presence of positive spatial curvature 
 while, in the present example, we considered, implicitly, a spatially flat background 
 manifold. In the case of pre-big bang models the geometry is geodesically complete in the past 
 and the potentially dangerous (curvature) singularities may arise but not in the far past \cite{PBB2,PBB3} (see also 
 \cite{reg1,reg2} for some possible mechanism for the regularization of the background).
 
\newpage
\renewcommand{\theequation}{5.\arabic{equation}}
\setcounter{equation}{0}
\section{Essentials of inflationary dynamics}
\label{sec5}
In this section we will swiftly discuss how can inflation be realized. 
Diverse models have been proposed so far and the purpose of the present section 
is to outline some general aspects that will be useful in the discussion of the evolution 
of the inhomogeneities.

\subsection{Fully inhomogeneous Friedmann-Lema\^itre equations}
The usual presentation of inflationary dynamics deals, predominantly, with {\em homogeneous}
equations for scalar degrees of freedom in the early Universe. It is then argued that, when the scalar 
potential dominates over the spatial gradients and over the kinetic energy of the scalar 
degree of freedom the geometry is led to inflate. 
In a slightly more quantitative perspective we shall demand that the aforementioned 
conditions should be verified over a spatial region of typical size $H_{\mathrm{i}}^{-1} > 10^{5} \ell_{\mathrm{P}}$ 
where $H_{\mathrm{i}}^{-1}$, as explained in section \ref{sec4}, is a newly formed event horizon at the 
cosmic time $t_{\mathrm{i}}$.  Why should we neglect spatial gradients during a phase 
of inflationary expasion? The answer to this question can be neatly formulated in terms 
of the inhomogenous form of Friedmann-Lema\^itre equations. The 
{\em homogeneous}  Friedmann-Lema\^itre equations (see Eqs. (\ref{FL1}), (\ref{FL2}) and (\ref{FL3}))
have been written neglecting all the spatial gradients. A very useful strategy will now be to write 
the Friedmann-Lema\^itre equations in a fully inhomogeneous form, i.e. in a form where 
the spatial gradients are not neglected. From this set of equations it will be possible to expand the metric 
to a given order in the spatial gradients, i.e. we will have that the zeroth-order solution will not contain 
any gradient, the first-order iteration will contain two spatial gradients, the second-order solution will 
contain four spatial gradients and so on.  This kind of perturbative expansion has been pioneered, 
in the late sixties and in the seventies, by Lifshitz, Khalatnikov \cite{lif1,lif2,KL0}, and by
 Belinskii and Khalatnikov \cite{BK1,BK2,BLK}. 
 
There are various applications of this formalism to inflationary cosmology \cite{dr1,dr2,dr3,KS} as well as 
to dark energy models (see, for instance, \cite{md1,md2,md3}  and references therein). 
In the present framework, the fully inhomogeneous approach will 
be simply employed in order to justify the following statements:
\begin{itemize}
\item{} if the (total) barotropic index of the sources is such that $w > - 1/3$ the spatial gradients 
will be relevant for large values of the cosmic time coordinate (i.e., formally, $t \to \infty$) but they 
will be negligible in the opposite limit (i.e. $t \to 0^{+}$);
\item{} if the total barotropic index is smaller than $-1/3$ the situation is reversed: the spatial 
gradients will become more and more subleading as time goes by but they will be of upmost importance 
in the limit of small cosmic times;
\item{} if $w= -1/3$ the contribution of the spatial gradients remains constant.
\end{itemize}
The second point of the above list of items will justify why spatial gradients can be neglected as 
inflation proceeds. At the same time, it should be stressed that the announced analysis does 
not imply that the inflationary dynamics is {\em generic}. On the contrary  it 
implies that, once inflation takes place, the spatial gradients will be progressively subleading.
Similarly, the present analysis will also show that prior to the onset of inflation the spatial gradients cannot be neglected. For the present purposes, a very convenient form of the line element is represented by 
\begin{equation}
ds^2 = g_{\mu\nu} dx^{\mu} dx^{\nu}= dt^2 - \gamma_{ij}(t, \vec{x}) dx^{i} dx^{j},
\label{metricinh1}
\end{equation}
where 
\begin{equation}
g_{00} =1, \qquad g_{ij} = -\gamma_{ij}(t,\vec{x}),\qquad g_{0i} =0.
\label{metricinh2}
\end{equation}
Since the four-dimensional metric $g_{\mu\nu}$ has ten independent degrees of freedom and 
since there are four available conditions to fix completely the coordinate system, Eqs. 
(\ref{metricinh1}) and (\ref{metricinh2}) encode all the relevant functions
allowing a faithful description of the dynamics: the tensor $\gamma_{ij}(t,\vec{x})$ being 
symmetric, contains $6$ independent degrees of freedom. The idea is now, in short, the following. 
The Einstein equations can be written in a form where the spatial gradients and the temporal gradients 
are formally separated. In particular, using Eqs. (\ref{metricinh1}) and (\ref{metricinh2}) it can be easily shown that 
the Christoffel connections can be written as
\begin{eqnarray}
&&\Gamma_{ij}^{0} = \frac{1}{2} \frac{\partial}{\partial t} \gamma_{ij} = - K_{ij},\qquad 
\Gamma_{0i}^{j} = \frac{1}{2} \gamma^{jk} \frac{\partial}{\partial t} \gamma_{ki}  = - K_{i}^{j},
\label{CCin0}\\
&& \Gamma_{ij}^{k} = \frac{1}{2} \gamma^{k\ell} \biggl[ - \partial_{\ell} \gamma_{ij} + \partial_{j} \gamma_{\ell i} + \partial_{i} \gamma_{j\ell} ],
\label{CCin}
\end{eqnarray}
where $K_{ij}$ is the so-called extrinsic curvature which is the inhomogeneous generalization of the Hubble parameter.
Notice, in fact, that when $\gamma_{ij}= a^2(t) \delta_{ij}$ (as it happens in the homogeneous case) 
$K_{i}^{j} = - H \delta_{i}^{j}$ where $H$ is the well known Hubble parameter. 
Using Eqs. (\ref{CCin}) the components of the Ricci tensor can be written as 
\begin{eqnarray}
&& R_{0}^{0} = \dot{K} - {\mathrm Tr}K^2,
\label{ricci1in}\\
&& R_{i}^{0} = \nabla_{i} K - \nabla_{k} K^{k}_{i},
\label{ricci2in}\\
&& R_{i}^{j} = \frac{\partial}{\partial t} K_{j}^{i} - K K_{i}^{j} - r_{i}^{j}, 
\label{ricci3in}
\end{eqnarray}
wherethe overdot denotes a 
 partial derivation with respect to $t$; $\nabla_{i}$ denotes 
 the covariant derivative defined in terms of $\gamma_{ij}$ and of Eq. (\ref{CCin}). In Eqs. (\ref{ricci1in}), (\ref{ricci2in}) 
 and (\ref{ricci3in}) the explicit meaning of the various quantities  is given by
\begin{equation}
{\rm Tr} K^2 = K_{i}^{j} \, K_{j}^{i}, \qquad K = K_{i}^{i},\qquad r_{i}^{j} = \gamma^{j k} r_{ki}.
\label{convinh}
\end{equation}
The three-dimensional Ricci tensor is simply given in terms of the Christoffel connections 
with spatial indices:
\begin{equation}
r_{i j} = \partial_{m} \Gamma^{m}_{i j} - \partial_{j} \Gamma^{m}_{m i} + 
\Gamma_{i j}^{m} \Gamma_{m \ell}^{\ell} - \Gamma_{j m}^{\ell} \Gamma_{i \ell}^{m},
\label{intr} 
\end{equation}
Equations (\ref{ricci1in}), (\ref{ricci2in}) and (\ref{ricci3in}) allow to write the Einstein equations 
in a fully inhomogeneous form. More specifically, assuming 
that the energy-momentum tensor is a perfect 
relativistic fluid 
\begin{equation}
T_{\mu}^{\nu} = (p+ \rho) u_{\mu}\,u^{\nu} - p \delta_{\mu}^{\nu},
\end{equation}
the Hamiltonian and momentum constraints are, respectively,
\begin{eqnarray}
&& K^2 - {\rm Tr} K^2  + r  =  16\pi G [ ( p + \rho) u_{0} u^{0} - p],
\label{E100}\\
&& \nabla_{i} K - \nabla_{k} K^{k}_{i} = 8\pi G u_{i} u^{0} (p + \rho).
\label{E10i}
\end{eqnarray}
The $(ij)$ components of Einstein equations lead instead to
\begin{eqnarray}
&&( \dot{K}_{i}^{j} - K \, K_{i}^{j} - \dot{K} \delta_{i}^{j}) + \frac{1}{2} \delta_{i}^{j}( K^2 + {\rm Tr} K^2) - (r_{i}^{j} - \frac{1}{2} r \delta_{i}^{j})  
\nonumber\\
&&= - 8\pi G [ (p + \rho) u_{i} u^{j} + p \delta_{i}^{j} ],
\label{E1ij}
\end{eqnarray}
A trivial  remark is that, in Eqs. (\ref{E100}), (\ref{E10i}) 
and (\ref{E1ij}), the indices are raised and lowered using directly 
$\gamma_{ij}(t,\vec{x})$. 
By combining the previous set of equations the following relation  can be 
easily deduced
\begin{equation}
q\, {\rm Tr} K^2 = 8\pi G \biggl[(p + \rho) u_{0}u^{0} + \frac{p - \rho}{2} \biggr]
\label{00cont}
\end{equation}
 where 
\begin{equation}
q(\vec{x},t) =  - 1 + \frac{\dot{K}}{{\rm Tr} K^2},
\label{defq}
\end{equation}
is the inhomogeneous generalization of the deceleration parameter.
In fact, in the homogeneous and isotropic limit, $ \gamma_{ij} = a^2(t) 
\delta_{ij}$, $K_{i}^{j} = - H \delta_{i}^{j}$ and, as expected, $q(t) \to - \ddot{a} a/ \dot{a}^2$.
Recalling the definition of ${\rm Tr}K^2$ it is rather easy to show that
\begin{equation}
{\rm Tr} K^2 \geq \frac{K^2}{3} \geq 0,
\label{trin}
\end{equation}
where the sign of equality (in the first relation) is reached, again, in 
the isotropic limit.
Since $\gamma^{ij}$ is always positive semi-definite, it is
also clear that 
\begin{equation}
 u_{0}\,u^{0} = 1 + \gamma^{ij} u_{i} u_{j} \geq 1,
\label{u0in}
\end{equation}
that follows from the condition $g^{\mu\nu} u_{\mu} u_{\nu}=1$.
From Eq. (\ref{00cont}) it follows
that $q(t,\vec{x})$ is always positive semi-definite if $(\rho + 3p) \geq 0$. This result 
is physically very important and it shows that spatial gradients cannot make gravity 
repulsive. One way of making gravity repulsive is instead to change the sources of the 
geometry and to violate the string energy condition. Eqs. (\ref{00cont}) and (\ref{defq}) 
generalize the relations already obtained in section \ref{sec2} and, in particular,
Eqs. (\ref{acc}) and (\ref{decpar}). Also in section \ref{sec2} it has been observed that 
the acceleration is independent on the contribution of the spatial curvature. Furthermore, 
it is easy to show that when the (negative) spatial curvature dominates over all the other 
sources the scale factor expands, at most, linearly in cosmic time (i.e. $a(t) \sim t$) 
and the deceleration parameter vanishes.

Equations (\ref{E100}), (\ref{E1ij}) and (\ref{E10i}) must be supplemented 
by the explicit form of the covariant conservation equations:
\begin{eqnarray}
&& \frac{1}{\sqrt{\gamma}} \frac{\partial}{\partial t} [ \sqrt{\gamma} ( p + \rho)
u^{0} u^{i}] - \frac{1}{\sqrt{\gamma}} \partial_{k}\{ \sqrt{\gamma} [ (p+ \rho) 
u^{k} u^{i} + p\gamma^{ki}]\}  
 -2 K^{i}_{\ell} u^{0} u^{\ell} (p + \rho)
 \nonumber\\
&& - \Gamma_{k \ell}^{i} 
[ (p + \rho) u^{k} u^{\ell} + p \gamma^{k\ell}] =0,
\label{con1}\\
&& \frac{1}{\sqrt{\gamma}} \frac{\partial }{\partial t}\{\sqrt{\gamma} [ (p + \rho) u_{0}u^{0} - p] \} - \frac{1}{\sqrt{\gamma}} \partial_{i} \{ \sqrt{\gamma} (p + \rho) 
u_{0}u^{i} \} 
\nonumber\\
&&- K_{k}^{\ell} [ (p + \rho) u^{k} u_{\ell} + p\delta_{\ell}^{k}] =0,
\label{con2}
\end{eqnarray}
where $\gamma = {\rm det}(\gamma_{ij})$.
It is useful to recall, from the Bianchi identities, that the intrinsic 
curvature tensor and its trace satisfy the following identity
\begin{equation}
\nabla_{j} r^{j}_{i} = \frac{1}{2} \nabla_{i} r.
\label{fromb}
\end{equation}

Note, finally, that combining  Eq. (\ref{E100}) with the trace 
of Eq. (\ref{E1ij}) the following equation is obtained:
\begin{equation}
{\rm Tr}K^2 + K^2 + r - 2 \dot{K} = 8\pi G( \rho - 3p).
\label{trace}
\end{equation}
Equation (\ref{trace}) allows to re-write Eqs. (\ref{E100}), (\ref{E1ij}) and (\ref{E10i})  as 
\begin{eqnarray}
&& \dot{K} - {\rm Tr} K^2 = 8\pi G\biggl[ (p+ \rho) u_{0}\,u^{0} +
\frac{p -\rho}{2} \biggr], 
\label{E200}\\
&& \frac{1}{\sqrt{\gamma}} \frac{\partial}{\partial t}\biggl(\sqrt{\gamma} \,K_{i}^{j}\biggr) - r_{i}^{j} = 8\pi G \biggl[ - (p+ \rho) u_{i} u^{j} 
+ \frac{p -\rho}{2} \delta_{i}^{j} \biggr],
\label{E2ij}\\
&& \nabla_{i} K - \nabla_{k} K^{k}_{i} = 8\pi G (p + \rho) u_{i}\,u^{0},
\label{E20i}
\end{eqnarray}
where we used the relation $2 K = - \dot{\gamma}/\gamma$.

Let us now look for solutions of the previous system of equations in the form 
of a gradient expansion. We shall be considering $\gamma_{ij}$ written in the form 
\begin{equation}
\gamma_{ik} = a^2(t) [ \alpha_{ik}(\vec{x}) + \beta_{ik}(t,\vec{x})],\,\,\,\,\,\,\,\,\,\,
\gamma^{kj} = \frac{1}{a^2(t)} [ \alpha^{kj} - \beta^{kj} (t, \vec{x})],
\label{exp}
\end{equation}
where $\beta(\vec{x},t)$ is considered to be the first-order correction 
in the spatial gradient expansion. Note that from Eq. (\ref{exp}) 
$\gamma_{ik}\gamma^{kj} =  \delta_{i}^{j} + {\cal O}(\beta^2)$.
The logic is now very simple since Einstein equations will determine 
the specific form of $\beta_{ij}$ once the specific form of $\alpha_{ij}$ 
is known. 

Using Eq. (\ref{exp}) into Eqs. (\ref{CCin0}) we obtain
\begin{equation}
K_{i}^{j} = - \biggl( H \delta_{i}^{j} + \frac{\dot{\beta}_{i}^{j}}{2} \biggr),
\qquad K = - \biggl( 3 H + \frac{1}{2}\dot{\beta}\biggr),
\qquad {\rm Tr} K^2 = 3 H^2 + H \dot{\beta},
\end{equation}
where, as usual, $ H = \dot{a}/a$.

From Eq. (\ref{E10i}) it also follows  
\begin{equation}
\nabla_{k} \dot{\beta}_{i}^{k} - \nabla_{i} \dot{\beta} = 16\pi G u_{i} \,\,u^{0} (p+ \rho).
\label{momex}
\end{equation}
The explicit form of the momentum constraint suggests to 
look for the solution in a separable form, namely, 
 $\beta_{i}^{j}(t,\vec{x}) = g(t) {\cal B}_{i}^{j}(\vec{x})$. Thus  
Eq. (\ref{momex}) becomes 
\begin{equation}
\dot{g} (\nabla_{k} {\cal B}^{k}_{i} - \nabla_{i} {\cal B}) = 16 \pi G u_{i} u^{0} (p + \rho).
\label{momex1}
\end{equation}
Using this parametrization and solving the constraint for $u_{i}$, Einstein
equations to second order in the gradient expansion reduce then 
to the following equation:
\begin{equation}
( \ddot{g} + 3 H \dot{g}) {\cal B}_{i}^{j} + H \dot{g} {\cal B} \delta_{i}^{j} + 
\frac{2 {\cal P}_{i}^{j}}{a^2} = \frac{w -1}{3 w + 1} (\ddot{g} + 2 H \dot{g}) B 
\delta_{i}^{j}.
\label{dec}
\end{equation}
In Eq. (\ref{dec}) the spatial curvature tensor has been parametrized as 
\begin{equation}
r_{i}^{j} = \frac{{\cal P}_{i}^{j}}{a^2}.
\label{defr}
\end{equation}
Recalling that 
\begin{equation}
H= H_{0} a^{- \frac{3( w + 1)}{2}}, \qquad \dot{H} = - \frac{3 ( w + 1)}{2} 
H^2,
\end{equation}
the solution for Eq. (\ref{dec}) can be written as
\begin{eqnarray}
&& {\cal B}_{i}^{j} = - \frac{4}{H_{0}^2 ( 3 w + 1) ( 3 w + 5)} \biggl( {\cal P}_{i}^{j} - \frac{ 5 + 6 w - 3 w^2}{4 ( 9 w + 5)}{\cal P} \delta_{i}^{j} \biggr),
\nonumber\\
&& {\cal B} = - \frac{\cal P}{H_{0}^2 ( 9 w + 5)},
\label{calB}
\end{eqnarray}
with  $g(t)$ simply given by 
\begin{equation}
g(t) = a^{3 w +1}.
\end{equation}
Note that, in Eq. (\ref{calB}), $H_{0} = 2/[ 3 (w + 1)\, t_0]$. 
Equation (\ref{calB}) can be also inverted, i.e. ${\cal P}_{i}^{j}$ can
be easily expressed in terms of ${\cal B}_{i}^{j}$ and ${\cal B}$:
\begin{equation}
{\cal P}_{i}^{j} = - \frac{H_{0}^2}{4} [ {\cal B} \delta_{i}^{j} ( 6w + 5 - 3 w^2) 
+ {\cal B}_{i}^{j} ( 3 w + 5) ( 3 w + 1) ].
\end{equation}
 
Using Eq. (\ref{fromb}) 
the peculiar velocity field and the energy density can also be written as 
\begin{eqnarray}
&& u^{0} u_{i} = - \frac{3}{8\pi G \rho} \biggl(\frac{w}{3 w+ 5}\biggr)a^{3 w + 1} H \partial_{i} {\cal B}(\vec{x}),
\nonumber\\
&& \rho = \frac{3 H_{0}^2}{8\pi G}\biggl[ a^{-3( w + 1)} 
- \frac{w +1}{2} {\cal B}(\vec{x}) a^{-2} \biggr].
\end{eqnarray}
Let us therefore rewrite the solution in terms of $\gamma_{ij}$, i.e. 
\begin{equation}
\gamma_{i j} = a^2(t)[ \alpha_{i j} (\vec{x}) + \beta_{ij} (\vec{x}, t) ] = 
a^{2}(t)\biggl[\alpha_{ij}(\vec{x}) +  a^{ 3 w + 1} {\cal 
B}_{i j} (\vec{x}) \biggr].
\label{qisol}
\end{equation}
Concerning this solution a few comments are in order:
\begin{itemize}
\item{} if $w > -1/3$, $\beta_{ij}$ becomes large as 
$a\to \infty$ (note that if $w= - 1/3$, $a^{3 w + 1}$ is constant);
\item{} if $w < - 1/3$, $\beta_{ij}$  vanishes as  $a\to \infty$;
\item{} if $w < -1$, $\beta_{ij}$ not only the gradients become sub-leading 
but the energy density also increases as $a\to \infty$.
\item{} to the following order in the perturbative expansion the time-dependence is easy to show: $\gamma_{ij} \simeq  a^2(t)[\alpha_{ij} + a^{3w + 1} {\cal B}_{ij} +a^{2(3w + 1)}  {\cal E}_{ij}]$ and so 
on for even higher order terms;
clearly the calculation of the curvature tensors will now be just a bit 
more cumbersome.
\end{itemize}
Equation (\ref{qisol}) then proves the statements illustrated at the beginning of the present 
section and justifies the use  of homogeneous equations for the 
analysis of the inflationary dynamics. Again, it should be stressed that homogeneous 
equations can be used for the description of inflationary dynamics.
The debatable issue on how inflation starts should however be discussed within the 
inhomogeneous approach. 

It is finally relevant to mention that the present formalism also answer an important 
question on the nature of the singularity in the standard cosmological model. Suppose 
that the evolution of the Universe is always decelerated (i.e. $\ddot{a}<0$) but 
expanding (i.e. $\dot{a}>0$). What should we expect in the limit $a\to0$? As 
emphasized in the past by Belinskii, Lifshitz and Khalatnikov (see for instance \cite{BLK}) 
close to the singularity the spatial gradients become progressively less important as also 
implied by Eq. (\ref{qisol}). This conclusion is very important since it means that the 
standard big-bang may be highly anisotropic but rather homogeneous. In particular, close
to the singularity the solution may fall in one of the metrics of the Bianchi classification \cite{ryan}
(see also Eqs. (\ref{bianchiI}) and (\ref{otherbianchi})). In more general terms it can also 
happen that the geometry undergoes anisotropic oscillations that are customarily 
named BKL (for Belinskii, Khalatnikov and  Lifshitz) oscillations. 

\subsection{Homogeneous evolution of a scalar field}

The Friedmann-Lema\^itre equations  imply that the scale factor can accelerate 
provided $w<-1/3$, where $w$ is the barotropic index  of the generic fluid 
driving the expansion.
This condition can be met, for instance, if one (or more) scalar degrees of freedom 
have the property that their potential dominates over their kinetic energy.
Consider, therefore, the simplest case where a single scalar degree of freedom 
is present in the game. The action can be written as 
\begin{equation}
S = \int d^{4} x \sqrt{-g}\biggl[ - \frac{R}{16\pi G} + 
g^{\alpha\beta} \partial_{\alpha}\varphi \partial_{\beta} \varphi - V(\varphi)\biggr],
\label{actionS}
\end{equation}
where $\varphi$ is the scalar degree of freedom and $V(\varphi)$ its related potential. 
The scalar field appearing in the action (\ref{actionS}) is said to be {\em minimally 
coupled}. There are of course other possibilities. For instance the scalar field $\varphi$ 
can be {\em conformally coupled} or even {\em non-minimally coupled}. These 
couplings arise when the scalar field action is written in the form 
\begin{equation}
S = \int d^{4} x \sqrt{-g}\biggl[ - \frac{R}{16\pi G} + 
g^{\alpha\beta} \partial_{\alpha}\varphi \partial_{\beta} \varphi - V(\varphi) - \alpha R \varphi^2\biggr].
\label{actionS2}
\end{equation}
Clearly the difference between Eq. (\ref{actionS}) and Eq. (\ref{actionS2}) is 
the presence of an extra term, i.e. $-\alpha R \varphi^2$. If $\alpha=0$ 
we recover the case of minimal coupling. If $\alpha =1/6$ the field is conformally 
coupled and its evolution equations are invariant under the Weyl rescaling of the metric.
In all other cases the field is said to be simply non-minimally coupled.
In what follows, for pedagogical reasons, we will stick to the case of minimal coupling.

By taking the variation of (\ref{actionS}) with respect to $g_{\mu\nu}$ and $\varphi$ 
we get, respectively, 
\begin{eqnarray}
&& R_{\mu}^{\nu} - \frac{1}{2}\delta_{\mu}^{\nu} R = 8\pi G T_{\mu}^{\nu},
\label{einS}\\
&& g^{\alpha\beta} \nabla_{\alpha} \nabla_{\beta} \varphi + \frac{\partial V}{\partial\varphi} =0,
\label{kleinS}
\end{eqnarray}
where 
\begin{eqnarray}
&& \nabla_{\alpha}\nabla_{\beta} \varphi = 
\partial_{\alpha} \partial_{\beta}\varphi - \Gamma_{\alpha\beta}^{\sigma}
\partial_{\sigma}\varphi,
\label{nablanabla}\\
&& T_{\mu}^{\nu} = \partial_{\mu}\varphi \partial^{\nu}\varphi - \delta_{\mu}^{\nu} \biggl[ \frac{1}{2} g^{\alpha\beta}
\partial_{\alpha}\varphi\partial_{\beta} \varphi - V(\varphi)\biggr].
\label{ENMOMPHI}
\end{eqnarray}
The  components of Eq. (\ref{ENMOMPHI}) can be written, in a spatially flat FRW metric, as   
\begin{eqnarray}
&& T_{0}^{0} \equiv \rho_{\varphi}= \biggl(\frac{{\dot{\varphi}}^2}{2} + V \biggr) + \frac{1}{2a^2} (\partial_{k} \varphi)^2, 
\label{T00S}\\
&& T_{i}^{j} = -\frac{1}{a^2} \partial_{i}\varphi \partial^{j} \varphi - \biggl( \frac{\dot{\varphi}^2}{2}  -V\biggr) \delta_{i}^{j} +
\frac{1}{2a^2} (\partial_{k}\varphi)^2 \delta_{i}^{j} 
\label{TijS}\\
&& T_{i}^{0} = \dot{\varphi} \partial_{i}\varphi
\label{Ti0S}
\end{eqnarray}
where, for the moment, the spatial gradients have been kept. To correctly identify the pressure and energy density 
of the scalar field the components of $T_{\mu}^{\nu}$ can be written as 
\begin{equation}
T_{0}^{0} = \rho_{\varphi}, \qquad  T_{i}^{j} = - p_{\varphi} \delta_{i}^{j} + \Pi_{i}^{j}(\varphi).
\label{CONVTS}
\end{equation}
where $\Pi_{i}^{j}$ is a traceless quantity and it is called anisotropic stress \footnote{The anisotropic stress
is rather relevant for the correct discussion of the pre-decoupling physics and, as we shall see, is mainly due, 
after weak interactions have fallen out of thermal equilibrium, to the quadrupole moment of the neutrino
phase space distribution.}.
By comparing Eqs. (\ref{T00S}), (\ref{TijS}) and (\ref{Ti0S}) with Eq. (\ref{CONVTS}) we will have 
\begin{eqnarray}
&& \rho_{\varphi} =\biggl(\frac{{\dot{\varphi}}^2}{2} + V \biggr) + \frac{1}{2a^2} (\partial_{k} \varphi)^2,
\label{ROPHI}\\
&& p_{\varphi} = \biggl(\frac{\dot{\varphi}^2}{2} - V\biggr) - \frac{1}{6 a^2} (\partial_{k} \varphi)^2,
\label{PPHI}\\
&& \Pi_{i}^{j}(\varphi) = - \frac{1}{a^2} \biggl[\partial_{i}\varphi \partial^{j}\varphi - \frac{1}{3} (\partial_{k}\varphi)^2 \delta_{i}^{j} \biggr].
\label{PijPHI}
\end{eqnarray}
Equations (\ref{ROPHI}) and (\ref{PPHI}) imply that the effective barotropic index for the scalar system under discussion is simply given by 
\begin{equation}
w_{\varphi} = \frac{p_{\varphi}}{\rho_{\varphi}} = 
\frac{\biggl(\frac{\dot{\varphi}^2}{2} - V\biggr) - \frac{1}{6 a^2} (\partial_{k} \varphi)^2}{\biggl(\frac{{\dot{\varphi}}^2}{2} + V \biggr) + \frac{1}{2a^2} (\partial_{k} \varphi)^2}.
\label{wPHI}
\end{equation}
Concerning Eq. (\ref{wPHI}) three comments are in order:
\begin{itemize}
\item{} if $\dot{\varphi}^2 \gg V$ and $\dot{\varphi}^2 \gg (\partial_{k}\varphi)^2/a^2$, then $p_{\varphi} \simeq \rho_{\varphi}$: in this 
regime the scalar field behaves as a stiff fluid;
\item{} if $V \gg \dot{\varphi}^2 \gg (\partial_{k}\varphi)^2/a^2$, then $w_{\varphi} \simeq -1$: in this regime 
the scalar field is an inflaton candidate;
\item{} if $ (\partial_{k}\varphi)^2/a^2 \gg  \dot{\varphi}^2 $ and $ (\partial_{k}\varphi)^2/a^2 \gg  V$, then $w_{\varphi} \simeq -1/3$:
in this regime the system is gradient-dominated and, according to the previous results the inhomogeneous 
deceleration parameter $q(t,\vec{x}) \simeq 0$.
\end{itemize}
Of course also intermediate situations are possible (or plausible).  
If the scalar potential dominates both over the gradients and over the kinetic energy for a sufficiently 
large event horizon at a given time the subsequent evolution is therefore likely to be rather homogeneous 
and the relevant equations will simply be:
\begin{eqnarray}
&&\overline{M}_{\mathrm{P}}^2 H^2 = \frac{1}{3}\biggl[ \frac{{\dot{\varphi}}^2}{2} + V \biggr] - 
\frac{k\,\,\overline{M}_{\mathrm{P}}^2}{a^2},
\label{FLS1}\\
&& \overline{M}_{\mathrm{P}}^2 \dot{H} = - \frac{{\dot{\varphi}}^2}{2} +\frac{k\,\,\overline{M}_{{\mathrm{P}}}^2}{a^2},
\label{FLS2}\\
&& \ddot{\varphi} + 3 H \dot{\varphi} + \frac{\partial V}{\partial\varphi} =0,
\label{FLS3}
\end{eqnarray}
where the reduced Planck mass has been defined according to the following chain of equalities:
\begin{eqnarray}
\overline{M}_{\mathrm{P}}^2 = \frac{1}{8\pi G} = \frac{M_{\mathrm{P}}^2}{8\pi}.
\label{PLdef}
\end{eqnarray}
Even if it is not desirable to introduce different definitions of the 
Planck mass, the conventions adopted in Eq. (\ref{PLdef}) are widely 
used in the study of inflationary dynamics so we will stick to them. 
Because of the factor $\sqrt{8\pi}$ in the denominator, $\overline{M}_{\mathrm{P}}$ will 
be roughly $5$ times smaller than $M_{\mathrm P}$.

\subsection{Classification(s) of inflationary backgrounds}
Inflationary backgrounds can be classified either in {\em geometric} or in {\em dynamical} terms. 
The geometric classification is based on the evolution of the Hubble parameter 
(or of the extrinsic curvature). The conditions $\ddot{a}>0$ and $\dot{a}>0$ can be realized 
for different evolutions of the Hubble parameter. Three possible cases arise naturally:
\begin{itemize}
\item{} de Sitter inflation (realized when $\dot{H} =0$);
\item{} power-law inflation (realized when $\dot{H} <0$);
\item{} superinflation (realized when $\dot{H}>0$).
\end{itemize}
The case of {\em exact} de Sitter inflation is a  useful simplification 
but it is, in a sense, unrealistic. On one hand 
it is difficult, for instance by means of a (single) scalar field, to obtain a pure de Sitter dynamics.
On the other hand, if $\dot{H}=0$ only the tensor modes of the geometry 
are excited by the time evolution of the background geometry.  This observation would imply 
that the scalar modes (so important for the CMB anisotropies) will not be produced.

 The  closest situations to a pure de Sitter dynamics is realized by means of a
 {\em quasi-de Sitter} phase of expansion  where $\dot{H} \laq 0$. 
 Quasi-de Sitter inflation is closely related with power-law inflation where 
the scale factor exhibits a power-law behaviour and $\dot{H} <0$. If the power 
of the scale factor is much larger than $1$ (i.e. $a(t) \simeq t^{\beta} $ with $\beta \gg 1$)
the quasi-de Sitter phase is 
essentially a limit of the power-law models which may be realized, 
for instance, in the case of exponential potentials as we shall see in a moment.
Finally, an unconventional case is the one of super-inflation. In standard 
Einstein-Hilbert gravity superinflation can only be achieved (in the absence of spatial curvature) if the 
dominant energy condition is violated, i.e. if the effective enthalpy of the sources 
is negative definite. This simple observation (stemming directly form Eq. (\ref{FL2})) implies 
that, in Einstein-Hilbert gravity, scalar field sources with positive kinetic terms cannot 
give rise to superinflationary dynamics. 

This impasse can be overcome in two different (but complementary) ways. If internal dimensions 
are included in the game, the overall solutions differ substantially from the simple 
four-dimensional case contemplated along these lectures. This possibility arises 
naturally in string cosmology and has been investigated \cite{norma1,norma2} in the context 
of the evolution of fundamental strings in curved backgrounds \cite{norma3}. If the Einstein-Hilbert 
theory is generalized to include a fundamental scalar field (the dilaton) 
different frames arise naturally in the problem. In this context, superinflation 
arises as a solution in the string frame as a result of the dynamics of the dilaton (which is, in turn, 
connected with the dynamics of the gauge coupling). This is the path followed, for instance, in the 
context of pre-big bang models (see \cite{PBB2,reg1} and references therein).

\begin{figure}
\centering
\includegraphics[height=5cm]{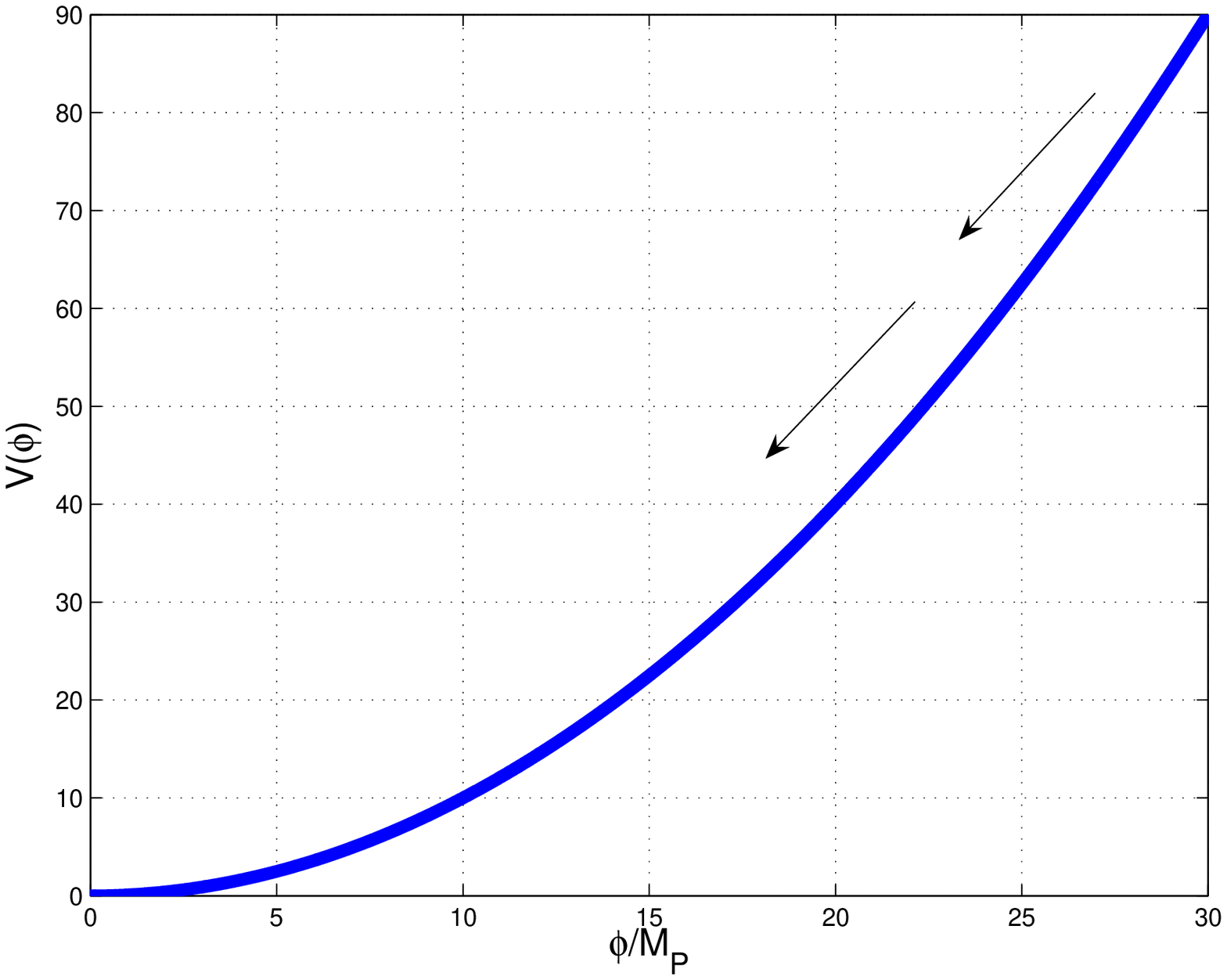}
\includegraphics[height=5cm]{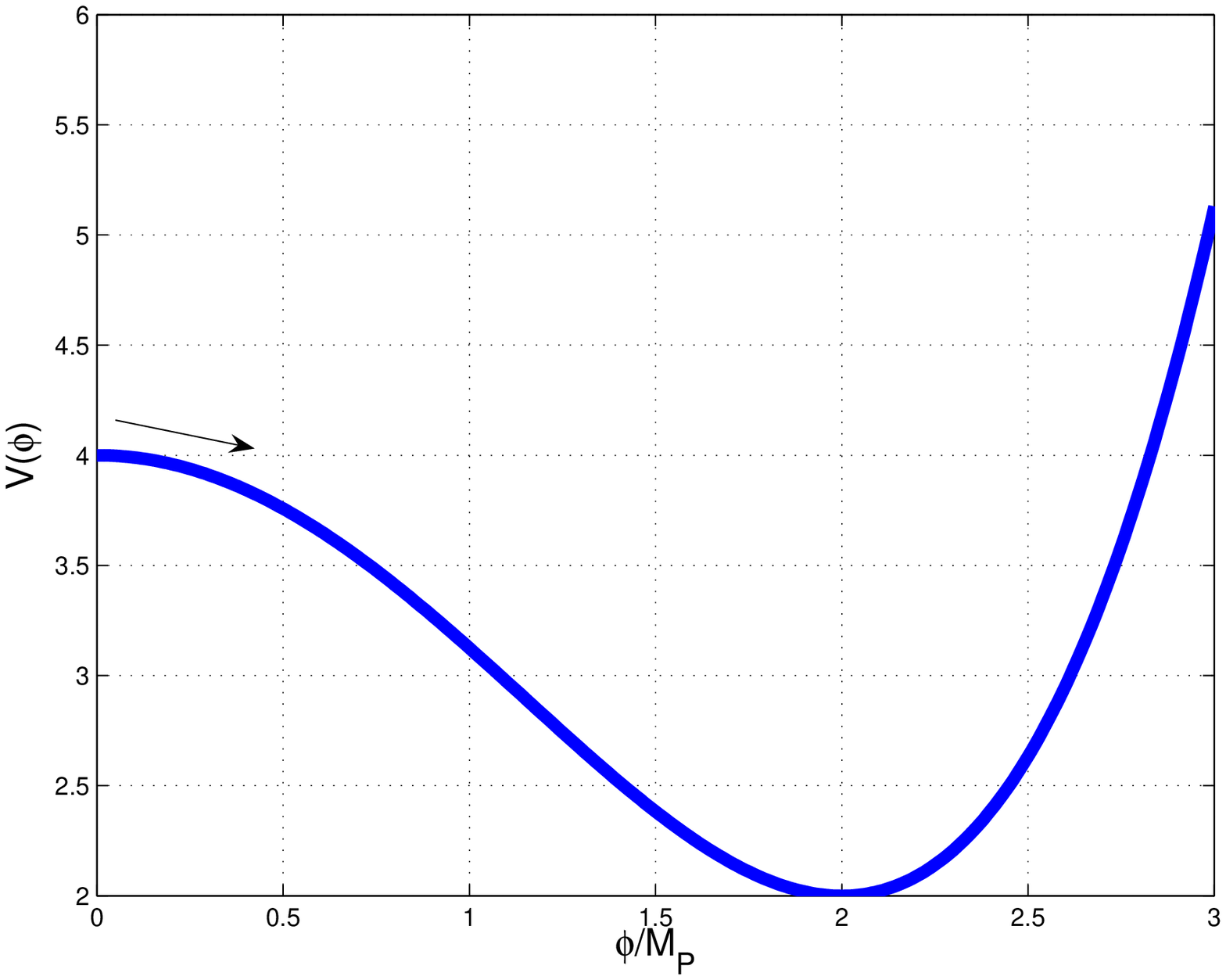}
\caption{Two schematic examples of inflationary potentials: small field models (at the right) and large field 
models (at the left). The directions of the arrows emphasize the evolution of the scalar degree of freedom}
\label{examples}      
\end{figure}

If inflation is realized by means of one (or many) scalar degrees of freedom, the classification 
of inflationary models is usually described in terms of the properties of the scalar potential.
This is a more dynamical classification that is, however, narrower than the geometric 
one introduced above. The rationale for this statement is that while the geometric classification 
is still valid in the presence of many (scalar) degrees of freedom driving inflation, 
the dynamical classification may slightly change depending upon the number and nature 
of the scalar sources introduced in the problem.
Depending on the way the slow-roll dynamics is realized, two cases 
are customarily distinguished 
\begin{itemize}
\item{} small field models (see Fig. \ref{examples}, right panel);
\item{} large field models (see Fig. \ref{examples} left panel).
\end{itemize}
In small field models the value of the scalar field in Planck units is usually smaller than one. In large field 
models the value of the scalar field at the onset of inflation is usually larger than one (ore even much larger than one) 
in Planck units. 
In Fig. \ref{examples} a schematic view of the large and small field models is provided.
In what follows examples of large and small field models will be given.

Let us finally comment on the relevance of spatial curvature. Inflation is safely described 
in the absence of spatial curvature since, as we saw, the inflationary dynamics washes out the spatial 
gradients quite efficiently. In spite of this statement, there can be situations (see second example in the following subsection)
where the presence of spatial curvature leads to {\em exact} inflationary backgrounds. Now, these 
solutions, to be phenomenologically relevant, should inflate for the {\em minimal} amount of e-folds. If, on the contrary, inflation lasts 
much more than the required $60$ or $65$ e-folds, the consequences of inflationary models endowed with spatial curvature 
will be indistinguishable, for practical purposes, from the consequences of those models where the spatial curvature 
is absent from the very beginning.

\subsection{Exact inflationary backgrounds}
As Eqs. (\ref{FL1}), (\ref{FL2}) and (\ref{FL3}), also Eqs. (\ref{FLS1}), (\ref{FLS2}) and (\ref{FLS3}) 
are not all independent. To illustrate inflationary dynamics, the following game can be, in some cases, successfully 
played : specify a given geometry, then obtain the scalar field profile by integrating (with respect to the cosmic time coordinate) Eq. (\ref{FLS2}). If the 
result of this manipulation is explicit and invertible, then Eq. (\ref{FLS1}) 
allows to determine the specific form of the potential. The drawback of this strategy concerns the range of applicability: few solvable examples are known and two of them will now be described in pedagogical 
terms. Consider the following power-law background:
\begin{eqnarray}
&& a(t) =a_{1} \biggl(\frac{t}{t_{1}}\biggr)^{\beta}, \qquad \beta > 1, \qquad t >0,
\label{SFPW}\\
&& H = \frac{\beta}{t}, \qquad \dot{H} = - \frac{\beta}{t^2},
\label{PLback}
\end{eqnarray}
where Eq. (\ref{PLback}) follows from Eq. (\ref{SFPW}) by using the definition of Hubble parameter.
Using now Eq. (\ref{FLS2}), $\dot{\varphi}$ can be swiftly determined as 
\begin{equation}
\varphi(t) = \varphi_{0} + \sqrt{2 \beta} \overline{M}_{\mathrm{P}} \ln{\biggl(\frac{t}{t_{1}}\biggr)}.
\label{PLphi}
\end{equation}
Inverting Eq. (\ref{PLphi}) we can easily obtain:
\begin{equation}
\biggl(\frac{t}{t_{1}}\biggr) = \exp[{\frac{\varphi - \varphi_0}{\sqrt{2\beta} \,\,\,\overline{M}_{\mathrm{P}}}}].
\label{INVphi}
\end{equation}
With the help of Eq. (\ref{INVphi}), Eq. (\ref{FLS1}) 
can be used to determine the specific form of the potential, i.e. 
\begin{eqnarray}
&& V(\varphi) = V_{0} e^{ - \sqrt{\frac{2}{\beta}} \frac{\varphi}{\overline{M}_{\mathrm{P}}}}
\label{POTphi}\\
&& V_{0} = \beta(3\beta-1) \overline{M}_{\mathrm{P}}^2 \, t_{1}^{-2} 
 e^{\sqrt{\frac{2}{\beta}} \frac{\varphi_{0}}{\overline{M}_{\mathrm{P}}}}.
 \label{constphi}
 \end{eqnarray}
With the geometry (\ref{PLback}), Eq. (\ref{kleinS}) 
is automatically satisfied provided $ \varphi$ is given by (\ref{PLphi}) and the potential is 
the one determined in Eqs. (\ref{POTphi})--(\ref{constphi}).  The example developed 
in this paragraph goes often under the name of power law infation \cite{PL1,PL2,PL3,PL4}.

Consider now 
a different example where the scale factor is given by 
\begin{eqnarray}
&& a(t) = a_{1} \cosh{(H_1 t)},\qquad  H_{1}>0,
\label{coshk}\\
&& H(t) = H_1 \tanh{(H_1 t)},\qquad \dot{H} = \frac{H_{1}^2}{\cosh^2{(H_1 t)}}.
\label{Hcosh}
\end{eqnarray}
This type of solution is not compatible with Eq. (\ref{FLS2}) 
if the spatial curvature vanish (or if it is negative). In these cases, in fact, $\dot{H}$ would be 
positive semi-definite and it should equal, by Eq. (\ref{FLS2}), $-\dot{\varphi}^2/2$
which is, overall negative definite.
Taking thus into account the necessary contribution of the spatial curvature,
Eq. (\ref{FLS2}) gives us $\dot{\varphi}$ and, after explicit integration, also 
$\varphi(t)$. The result of this procedure is that 
\begin{eqnarray}
&& \dot{\varphi} = \frac{A_{1}}{\cosh{H_1 t}},
\label{sol2dotphi}\\
&& \varphi(t) = \varphi_{0} + \frac{A_{1}}{H_{1}} \arctan[\sinh{(H_1 t)}],
\label{sol2phi}\\
&& A_{1} = \sqrt{2} \biggl(\frac{k}{a_1^2} - H_{1}^2\biggr)^{1/2}\,\overline{M}_{\mathrm{P}}.
\label{sol2A1}
\end{eqnarray}
Recalling now that, from Eq. (\ref{sol2phi}), 
\begin{equation}
\sinh{(H_1 t)} = \tan{\tilde{\varphi}}, \qquad \tilde{\varphi} = \frac{H_1}{A_{1}} ( \varphi - \varphi_0),
\label{sol2inv}
\end{equation}
Eq. (\ref{FLS1}) can be used to determine the potential which is 
\begin{equation}
V(\tilde{\varphi}) = \overline{M}^2_{\mathrm{P}} H_1^2 [ 3 - 2 \cos^2{\tilde{\varphi}}] + 2 \frac{k\,\,\overline{M}_{\mathrm{P}}^2}{a_{1}^2} \cos^2{\tilde{\varphi}}.
\label{sol2pot}
\end{equation}
The example developed in this paragraph goes also under the name of de Sitter bounce 
and has been studied in different contexts \cite{dsbounce1,dsbounce2}.
As already stressed, solvable examples are rather uncommon. 
It is therefore mandatory to devise general procedure 
allowing the discussion of the scalar field dynamics even in the situation when 
the exact solution is lacking. 

\subsection{Slow-roll  dynamics}
In the previous section it has been pointed out that the expanding de Sitter 
phase used for the first description of inflationary dynamics may not be exact 
and, therefore, we talked about quasi-de Sitter dynamics.
In inflationary dynamics a number of slow-roll parameters are customarily 
defined. They have the property of being small during the (quasi)-de Sitter stage 
of expansion. Thus they can be employed as plausible expansion parameters.
 As an example consider the following choice \footnote{Concerning the notation employed 
 for the second slow-roll parameter $\eta$ we remark that the same Greek letter has been also used 
 to denote the ratio between the concentration of baryons and photons (i.e. $\eta_{\mathrm{b}}$) introduced in Eq. (\ref{etab}). No confusion 
 is possible both because of the subscript and because the two variables never appear together 
 in this discussion. We warn the reader that, however, very often $\eta_{\mathrm{b}}$ is simply denoted by $\eta$ in the existing 
 literature and, therefore, it will only be the context to dictate the correct signification of the symbol.} :
 \begin{equation}
 \epsilon = - \frac{\dot{H}}{H^2},\qquad \eta = \frac{\ddot{\varphi}}{H\dot{\varphi}}.
 \label{SRpar}
 \end{equation}
 As we shall see, in the literature these parameters are often linearly combined.
The smallness of these two (dimensionless) parameters define 
the range of validity of a given inflationary solutions characterized 
by the dominance of the potential term in the field equations. In other 
words during the (slow-roll) inflationary phase $|\epsilon| \ll 1$ and 
$|\eta| \ll 1$. As soon as $\epsilon\simeq\eta \simeq 1$ inflation 
ends. 

During a slow-roll phase the (effective) 
evolution equations for the homogeneous part of the inflaton background can be 
written as 
\begin{eqnarray}
&& H^2 \overline{M}_{\mathrm{P}}^2 \simeq \frac{V}{3}, 
\label{SR1}\\
&& 3 H \dot{\varphi} + \frac{\partial V}{\partial \varphi} =0.
\label{SR2}
\end{eqnarray}
A naive example of slow-roll dynamics characterized 
by the following single-field potential:
\begin{equation}
V(\varphi) = V_1 - \frac{m^2}{2} \varphi^2 + \frac{\lambda}{4} \varphi^4+...,
\label{SRpot1}
\end{equation}
where $V_1$ is a constant. In the jargon, this is a rather simplistic 
example of what is called a small field model. The solution of Eqs. (\ref{SR1}) and 
(\ref{SR2}) implies, respectively, that 
\begin{eqnarray}
&& a(t) \simeq e^{H_1 t}, \qquad H_{1} \simeq \frac{\sqrt{V_1}}{\overline{M}_{\mathrm{P}}\sqrt{3} },
\label{aSRpot1}\\
&& \varphi \simeq \varphi_{1} e^{\frac{m^2}{3 H_1} t}, \qquad \frac{m^2}{3 H_1}t <1.
\label{aSRpot2}
\end{eqnarray}
The slow-roll phase lasts until the scalar field is approximately constant, i.e. until the cosmic time 
$t_{\mathrm{f}}$ that can be read-off from Eq. (\ref{aSRpot2}): 
\begin{equation}
t_{\mathrm{f}} \simeq  \frac{3 H_1}{m^2}, \qquad H_1 t_{\mathrm{f}} \simeq \frac{3 H_{1}^2}{m^2}.
\label{finpot2}
\end{equation}
From Eq. (\ref{finpot2}) the number of e-folds of this toy model 
can be computed and it is given by 
\begin{equation}
N \simeq \frac{3 H_{1}^2}{m^2} > 65, \qquad m^2 \leq \frac{3 H_1^2}{65}
\end{equation}
which shows that $m$ should be sufficiently small in units of $H_1$ to get 
a long enough inflationary phase.
In the case of the exact inflationary background discussed 
in Eq. (\ref{PLback}) the definitions of the slow-roll parameters given in 
Eq. (\ref{SRpar}) lead quite simply to the following expressions:
\begin{equation}
\epsilon = \frac{1}{\beta}, \qquad \eta = - \frac{1}{\beta},
\end{equation}
which can be smaller than $1$ provided $\beta > 1$ as already required in the 
process of deriving the solution.
\subsection{Slow-roll parameters}
The slow-roll parameters of Eq. (\ref{SRpar}) can be directly expressed in terms of the potential and of its derivatives by using Eqs. (\ref{SR1}) and (\ref{SR2}). 
The result of this calculation  is that 
\begin{eqnarray}
&& \epsilon = - \frac{\dot{H}}{H^2} = \frac{\overline{M}^2_{\mathrm{P}}}{2} \biggl(\frac{V_{,\varphi}}{V}\biggr)^2, 
\label{epsilon}\\
&& \eta = \frac{\ddot{\varphi}}{H \dot{\varphi}} = \epsilon - \overline{\eta}, \qquad 
\overline{\eta} =   \overline{M}^2_{\mathrm{P}} \biggl(\frac{V_{\varphi\varphi}}{V}\biggr),
\label{eta}
\end{eqnarray}
where $V_{,\varphi}$ and $V_{,\varphi\varphi }$ denote, respectively, the first and second derivatives of the 
potential with respect to $\varphi$.
Equations (\ref{epsilon}) and (\ref{eta}) follow from 
Eqs. (\ref{SRpar}) by using Eqs. (\ref{SR1}) and (\ref{SR2}).
From the definition of $\epsilon$ (i.e. Eqs. (\ref{SRpar})) we can write 
\begin{equation}
\epsilon = - \frac{1}{H^2} \frac{\partial H}{\partial\varphi} \dot{\varphi} = 
\frac{1}{3 H^3} \biggl(\frac{\partial H}{\partial\varphi} \biggr) 
\biggl(\frac{\partial V}{\partial\varphi} \biggr).
\label{epsex}
\end{equation}
But from Eq. (\ref{SR1}) it also follows that 
\begin{equation}
H \frac{\partial H}{\partial \varphi} = \frac{1}{6 \overline{M}^2_{\mathrm{P}}} \frac{\partial V}{\partial \varphi}
\label{epsex1}
\end{equation}
Inserting now Eq. (\ref{epsex1}) into Eq. (\ref{epsex}) and recalling Eq. (\ref{SR1}),
Eq. (\ref{epsilon}) is swiftly obtained. 

Consider next the definition of $\eta$ as it appears in Eq. (\ref{SRpar}) and 
let us write it as 
\begin{equation}
\eta = - \frac{1}{H\dot{\varphi}} \frac{\partial}{\partial t} \biggl[ \frac{1}{3H} \frac{\partial V}{\partial \varphi}\biggr] =  - \frac{1}{H\dot{\varphi}} \biggl[ - \frac{\dot{H}}{3 H^2} \frac{\partial V}{\partial\varphi} + \frac{\dot{\varphi}}{3 H} \frac{\partial^2 V}{\partial \varphi^2}
\biggr],
\label{etaex1}
\end{equation}
where the second time derivative has been made explicit.
Recalling now the definition of $\epsilon$ as well as Eq. (\ref{SR2}), 
Eq. (\ref{etaex1}) can be written as 
\begin{equation}
\eta = \epsilon - \overline{\eta},\qquad 
\overline{\eta} = \overline{M}_{\mathrm{P}}^2 \frac{V_{,\varphi\varphi}}{V}.
\label{etabar}
\end{equation}

It is now possible to illustrate the use of the slow-roll parameters by 
studying, in rather general terms, the total number of e-folds and by 
trying to express it directly in terms of the slow-roll parameters.
Consider, first of all, the following way of writing the total number 
of e-folds:
\begin{equation}
N = \int_{a_{\mathrm{i}}}^{a_{\mathrm{f}}} \frac{d a}{a} = \int_{t_{\mathrm{i}}}^{t_{\mathrm{f}}} H \, dt = \int_{\varphi_{\mathrm{i}}}^{\varphi_{\mathrm{f}}} \frac{H}{\dot{\varphi}}
 d \varphi.
\label{TOTN}
\end{equation}    
 Using now Eq. (\ref{SR1}) and, then,  Eq. (\ref{SR2}) inside 
 Eq. (\ref{TOTN}) we do get the following chain of equivalent 
 expressions:
 \begin{equation}
 N = - \int_{\varphi_{\mathrm{i}}}^{\varphi_{\mathrm{f}}} \frac{3 H^2}{V_{,\varphi}} d\varphi
 = \int_{\varphi_{\mathrm{f}}}^{\varphi_{\mathrm{i}}} 
 \frac{d\varphi}{\overline{M}_{\mathrm{P}} \sqrt{2 \epsilon}}
 \label{TOTN2}
 \end{equation}
 Let us now give a simple and well known example, i.e. the case 
 of a monomial potential \footnote{It is clear that monomial potentials are not so realistic for various reasons. 
 A more general approach to the study of generic polynomial potentials has been recently developed 
 \cite{NH1,NH2,NH3,NH4,NH4a,NH5}. In this framework the inflaton field is not viewed as a fundamental field 
 but rather as a condensate. Such a description bears many analogy with the Landau-Ginzburg
 description of superconducting phases. }
 Recently of the form $V(\varphi) \propto \varphi^{n}$.
 In this case Eqs. (\ref{epsilon}) and (\ref{TOTN2}) imply, respectively,  
 \begin{equation}
 \epsilon = \frac{\overline{M}^2_{\mathrm{P}}}{2} \frac{n^2}{\varphi^2},\qquad 
 N = \frac{ \varphi_{\mathrm{i}}^2 - \varphi_{\mathrm{f}}^2}{
 2 \, n\, \overline{M}^2_{\mathrm{P}}}.
 \label{MONpot}
 \end{equation}
Let us now ask the following pair of questions:
\begin{itemize}
\item{} what was the value of $\varphi$ say $60$ e-folds before the end 
of inflation?
\item{} what was the value of $\epsilon$ 60 e-folds before the end of inflation?
\end{itemize}
To answer the first question let us recall that inflation ends when 
$\epsilon(\varphi_{\mathrm{f}}) \simeq 1$. Thus from Eq. (\ref{MONpot}) 
we will have, quite simply, that 
\begin{equation}
\varphi_{60}^2 = \frac{n(n + 240)}{2} \, \overline{M}_{\mathrm{P}}^2.
\label{phi60}
\end{equation}
Consequently, the value of $\epsilon$ corresponding to $60$ e-folds 
before the end of inflation is given by
\begin{equation}
\epsilon(\varphi_{60}) = \frac{n}{n + 240},
\label{epsilon60}
\end{equation}
which is, as it should, smaller than one.

This last example, together with the definition 
of slow-roll parameters suggests a second class of inflationary models which has been 
illustrated in Fig. \ref{examples} (panel at the left).
The slow-roll 
dynamics is also realized if, in the case of monomial potential
$\varphi$ is sufficiently large in Planck units. These are the so-called 
large field models. Notice that to have a field $\varphi > \overline{M}_{\mathrm{P}}$ 
does not imply that the energy density of the field is larger than the Planck 
energy density.

The slow-roll algebra introduced in this section allows to express the spectral 
indices of  the scalar and tensor modes 
of the geometry in terms of $\epsilon$ and $\overline{\eta}$.
The technical tools 
appropriate for such a discussion are collected in section \ref{sec10}.
The logic is, in short, the following. The slow-roll parameters can be expressed in terms 
of the derivatives of the potential. Now, the spectra of the scalar and tensor fluctuations of the geometry 
(allowing the comparison of the model with the data of the CMB anisotropies) can be expressed, again, in terms 
of $\epsilon$ and $\overline{\eta}$. 
Consider, as an example, the case of single-field inflationary models. In this case the scalar and tensor 
power spectra (i.e. the Fourier transforms of the two-point functions of the corresponding 
fluctuations) are computed in section \ref{sec10} (see, in particular, the final formulas reported in Eqs. (\ref{EXscal11}), 
(\ref{EXscal12}) and (\ref{EXscal13})). Therefore, according to the results derived in section \ref{sec10} we 
will have, in summary, that the power spectra of the scalar and tensor modes can be parametrized as 
\begin{equation}
{\mathcal P}_{\mathrm{T}} \simeq k^{n_{\mathrm{T}}},\qquad {\mathcal P}_{{\mathcal R}}  \simeq k^{n_{\mathrm{s}} -1},
\label{PS}
\end{equation}
where $n_{\mathrm{T}}$ and $n_{\mathrm{s}}$ are, respectively,  the tensor and scalar spectral indices. 
Using the slow-roll algebra of this section and following the derivation of Appendix D, $n_{\mathrm{s}}$ and 
$n_{\mathrm{T}}$ can  be related to $\epsilon$ and $\overline{\eta}$ as 
\begin{equation}
n_{\mathrm{s}} = 1 - 6\epsilon + 2 \overline{\eta},\qquad n_{\mathrm{T}} = - 2 \epsilon.
\end{equation}
The ratio of the scalar and tensor power spectra is usually called $r$ and it is also a function 
of $\epsilon$, more precisely (see section \ref{sec10}), 
\begin{equation}
r = \frac{{\mathcal P}_{\mathrm{T}}(k_{\mathrm{p}})}{{\mathcal P}_{{\cal R}}(k_{\mathrm{p}})} = 16 \epsilon = - 8 
n_{\mathrm{T}}.
\label{cons}
\end{equation}
Equation (\ref{cons}) is often named consistency relation and the wave-number $k_{\mathrm{p}}$ is the 
pivot scale at which the scalar and tensor power spectra are normalized. A possible choice 
in order to parametrize  the inflationary predictions is to assign the amplitude of the scalar 
power spectrum, the scalar and tensor spectral indices and the the $r$-parameter.

The development of these lectures can now follow two complementary (but logically 
very different) approaches. In the first approach we may want to assume that 
the whole history of the Universe is known and it consists of an inflationary phase almost suddenly followed 
by a radiation-dominated stage of expansion which is replaced, after equality, by a the matter and by the dark energy epochs. In this first approach the initial conditions for the scalar and tensor fluctuations of the geometry 
will be set during inflation. There is a second approach where the initial conditions for CMB anisotropies are 
set after weak interactions have fallen out of thermal equilibrium. In this second approach 
the scalar and tensor power spectra are taken as free parameters 
and are assigned when the relevant wavelengths of the perturbations  are still larger 
than the Hubble radius after matter-radiation equality but prior to decoupling. In what follows 
the second approach will be developed. In section \ref{sec10} the inflationary power spectra 
will instead be computed within the first approach.
\newpage
\renewcommand{\theequation}{6.\arabic{equation}}
\setcounter{equation}{0}
\section{Inhomogeneities in FRW models}
\label{sec6}

All the discussion presented so far dealt with completely homogeneous 
quantities.  An essential tool for the discussion of CMB anisotropies is 
the theory of cosmological inhomogeneities of a FRW metric.  In the present section 
the following topics will be discussed:
\begin{itemize}
\item{} decomposition of inhomogeneities in FRW Universes;
\item{} gauge issues for the scalar modes;
\item{} evolution of the tensor modes;
\item{} quantum mechanical treatment of the tensor modes;
\item{} spectra of relic gravitons.
\end{itemize}
The first two topics are a (necessary)  technical interlude which will be of upmost importance 
for the remaining part of the present script. The evolution of the tensor modes and their 
quantum mechanical normalization will lead to the (simplified) calculation of the spectral properties 
of relic gravitons. For didactical reasons it is better to study first the evolution 
of the tensor modes. They have the property of not being coupled with the (scalar) matter sources. In the simplest 
case of FRW models they are only sensitive to the evolution of the curvature. Moreover,
 the amplification of quantum-mechanical (tensor) 
fluctuations is technically easier. 
The analog phenomenon (arising in the case of the scalar modes of the geometry) will be separately 
discussed for the simplest case of the fluctuations induced by a (single) scalar field (see, in particular,
section \ref{sec10} and Appendix \ref{APPA}). Sections \ref{sec7}, \ref{sec8} and \ref{sec9} 
will be devoted to the impact of the scalar modes on CMB anisotropies which is the theme 
mostly relevant for the present discussion.
For technical reasons, 
the conformal time parametrization is more convenient for the treatment of the fluctuations of FRW geometries
(recall Eqs. (\ref{FRW2}) and (\ref{FL1C})--(\ref{FL3C})).  According to the conventions previously adopted 
the derivation with respect to the conformal time coordinate will be denoted by a prime.

\subsection{Decomposition of inhomogeneities in FRW Universes}
Given a conformally flat background metric of FRW type  
\begin{equation}
\overline{g}_{\mu\nu}(\tau) = a^2(\tau) \eta_{\mu\nu}, 
\label{MBack}
\end{equation}
its first-order  fluctuations can be written as 
\begin{equation}
\delta g_{\mu\nu}(\tau,\vec{x}) = \delta_{\rm s} g_{\mu\nu}(\tau,\vec{x})
+  \delta_{\rm v} g_{\mu\nu}(\tau,\vec{x}) +
\delta_{\rm t} g_{\mu\nu}(\tau,\vec{x}),
\end{equation}
where the subscripts define, respectively, the scalar, vector and  tensor
perturbations classified according to rotations in the three-dimensional
Euclidean sub-manifold. Being a symmetric rank-two tensor in four-dimensions,  
the perturbed metric $\delta g_{\mu\nu}$  has, overall, 10 independent 
components whose explicit form will be parametrized as\footnote{Notice that the partial derivations 
with respect to the spatial indices arise as a result of the explicit choice of  dealing with a 
spatially flat manifold. In the case of a spherical or hyperbolic spatial manifold they will be replaced 
by the appropriate covariant derivative defined on the appropriate spatial section.}
\begin{eqnarray}
&& \delta g_{00} = 2 a^2 \phi,
\label{g001}\\
&& \delta g_{ij} = 2 a^2 ( \psi \delta_{ij} - \partial_{i} \partial_{j}E) 
- a^2 h_{ij} + a^2 ( \partial_{i} W_{j} + \partial_{j} W_{i}),
\label{gij1}\\
&& \delta g_{0i} = - a^2 \partial_{i} B - a^2 Q_{i},
\label{g0i1}
\end{eqnarray}
together with the conditions
\begin{equation}
\partial_{i} Q^{i} = \partial_{i} W^{i} =0 ,\,\,\,\,\,\,\,\,
h_{i}^{i} = \partial_{i}h^{i}_{j} =0.
\label{div1}
\end{equation}
The decomposition expressed by Eqs. (\ref{g001})--(\ref{g0i1}) and (\ref{div1}) 
 is the one normally employed in  the Bardeen formalism \cite{bardeen,PV1,PV2,MB,bardeen2} (see 
 also \cite{press,lyth})
 and it is the one adopted in \cite{THTH} to derive, consistently, the 
 results relevant for the theory of CMB anisotropies.
Concerning Eqs. (\ref{g001})--(\ref{g0i1}) few comments are in order:
\begin{itemize}
\item{} the scalar fluctuations of the geometry are parametrized by  4 scalar functions, i. e. 
$\phi$, $\psi$, $B$ and $E$;
\item{} the vector fluctuations 
are described by the two (divergenceless) vectors in three (spatial) dimensions
 $W_{i}$ and $Q_{i}$, i.e. 
by $4$ independent degrees of freedom;
\item{} the tensor modes 
are described by the three-dimensional rank-two tensor $h_{ij}$, leading, overall, to 2 independent components 
because of the last two conditions of Eq. (\ref{div1}).
\end{itemize}

The strategy will then be to obtain the evolution equations 
for the (separate) scalar, vector and tensor contributions. 
To achieve this goal we can either perturb the most appropriate form of the Einstein equation 
to first-order in the amplitude of the fluctuations, or we may perturb the action 
to second-order in the amplitude of the same fluctuations. Schematically, within the first 
approach we are led to compute 
\begin{equation}
\delta^{(1)} R_{\mu}^{\nu} - \frac{1}{2} \delta_{\mu}^{\nu} \delta^{(1)} R = 8\pi G 
\delta^{(1)} T_{\mu}^{\nu},
\label{scheme1}
\end{equation}
where $\delta^{(1)}$ denotes the first-order variation with respect either to the scalar, vector and tensor modes. 
Of course it will be very convenient to perturb also the covariant conservation of the sources. This will lead to 
a supplementary set of equations that are not independent from Eqs. (\ref{scheme1}).

The form of the energy-momentum tensor depends on the specific physical application.
For instance in the pre-decoupling physics, the matter sources are represented by the total energy-momentum tensor of the fluid (i.e. baryons 
photons, neutrinos and dark matter). 
During inflation, the matter content will be 
given by the scalar degrees of freedom whose dynamics produces the 
inflationary evolution. In the simplest case of a {\em single} scalar degree of freedom this 
analysis will be discussed in section \ref{sec10} (relevant complements are also
derived in Appendix \ref{APPC}).

As already mentioned, instead of perturbing the equations of motion 
to first-order in the amplitude of the fluctuations, it is possible  
to perturb the gravitational and matter parts of the action to second-order 
in the amplitude of the fluctuations, i.e. formally
\begin{equation}
\delta^{(2)} S = \delta^{(2)} S_{\mathrm{g}} + \delta^{(2)} S_{\mathrm{m}}.
\label{secondord}
\end{equation}
How to quantize a system of fluctuations evolving on a classical 
background? The standard procedure for this problem is to find 
the canonical normal modes of the system and to promote them to the 
status of quantum mechanical operators. For the success of such an approach
it is essential to perturb the action to second order in the amplitude 
of the fluctuations.  This step will give us the Hamiltonian of the fluctuations leading, ultimately, to the 
evolution of the field operators in the Heisenberg representation.

\subsection{Gauge issues for the scalar modes}

The discussion of the perturbations on a given background geometry is complicated 
by the fact that, for infinitesimal coordinate transformations, of the type
\begin{equation}
x^{\mu} \to \tilde{x}^{\mu} = x^{\mu} + \epsilon^{\mu},
\label{inf}
\end{equation}
 the fluctuation of a rank-two (four-dimensional) tensor changes according to the Lie derivative in the direction $\epsilon^{\mu}$. 
 It can be easily shown that the fluctuations of a tensor $T_{\mu\nu}$ change, under the transformation (\ref{inf}) as:
\begin{equation}
\delta T_{\mu\nu} \to \tilde{\delta T}_{\mu\nu} = \delta T_{\mu\nu} - 
T^{\lambda}_{\mu} \nabla_{\nu} \epsilon_{\lambda} - T^{\lambda}_{\nu} \nabla_{\lambda} \epsilon_{\mu} - \epsilon^{\lambda} \nabla_{\lambda} T_{\mu\nu},
\label{gaugetrgen}
\end{equation}
where the covariant derivatives are performed by using the background metric 
which is given, in our case, by Eq. (\ref{MBack}). Thus, for 
instance, we will have that 
\begin{equation}
\nabla_{\mu} \epsilon_{\nu} = \partial_{\mu}\epsilon_{\nu} - \overline{\Gamma}_{\mu\nu}^{\sigma} \epsilon_{\sigma},
\end{equation}
where $\overline{\Gamma}_{\mu\nu}^{\sigma}$ are Christoffel connections 
computed using the background metric (\ref{MBack}) and they are 
\begin{equation}
\overline{\Gamma}_{ij}^{0} = {\mathcal H} \delta_{ij},\qquad 
\overline{\Gamma}_{00}^{0} = {\mathcal H},\qquad \overline{\Gamma}_{i0}^{j} = {\mathcal H} \delta_{i}^{j}, \qquad 
{\mathcal H} = \frac{a'}{a}.
\label{backCR}
\end{equation}
If $T_{\mu\nu}$ coincides with the metric tensor, then the metricity condition 
allows to simplify (\ref{gaugetrgen}) which then becomes: 
\begin{equation}
\delta {g}_{\mu\nu} \to \tilde{\delta{g}}_{\mu\nu} = \delta g_{\mu\nu} - \nabla_{\mu} \epsilon_{\nu} - \nabla_{\nu} \epsilon_{\mu},
\label{T2}
\end{equation}
where 
\begin{equation}
\epsilon_{\mu} = a^2(\tau)(\epsilon_{0}, - \epsilon_{i}),
\label{gaugepar}
\end{equation}
is the shift vector that induces the explicit transformation of the coordinate system, 
namely:
\begin{equation}
\tau \to \tilde{\tau} = \tau + \epsilon_{0},\,\,\,\,\,\,\,\,\,\,\, {x}^{i} \to \tilde{x}^{i} = x^{i} + \epsilon^{i}.
\label{T1}
\end{equation}
Equation (\ref{T2}) can be also written as 
\begin{equation}
\delta {g}_{\mu\nu} \to \tilde{\delta{g}}_{\mu\nu} = \delta g_{\mu\nu}  - \Delta_{\epsilon},
\end{equation}
where, $\Delta_{\epsilon}$ is the Lie derivative in the direction $\epsilon_{\mu}$.
The functions $\epsilon_{0}$ and 
$\epsilon_{i}$ are often called gauge parameters since the infinitesimal 
coordinate transformations of the type (\ref{T1}) form a group which is 
in fact the gauge group of gravitation. 
The gauge-fixing procedure, amounts, in four space-time dimensions, to fix the four independent functions 
$\epsilon_{0}$ and $\epsilon_{i}$. As they are, the three gauge parameters $\epsilon_{i}$ (one for each axis) 
will affect both scalars and three-dimensional vectors. To avoid this possible confusion, 
the gauge parameters  $\epsilon_{i}$ can be separated 
into their divergenceless and divergencefull parts, i.e.
\begin{equation}
\epsilon_{i} = \partial_{i} \epsilon + \zeta_{i},
\end{equation}
where $ \partial_{i} \zeta^{i} =0$. The gauge transformations 
involving $\epsilon_{0}$ and $ \epsilon$ preserve the scalar nature of 
the fluctuations while the gauge transformations parametrized by $\zeta_{i}$ 
preserve the vector nature of the fluctuation.
The fluctuations in the tilded coordinate system, defined  by the 
transformation of  Eq. (\ref{T1}), can then be written as 
\begin{eqnarray}
&& \phi \to \tilde{\phi} = \phi - {\cal H} \epsilon_0 - \epsilon_{0}' ,
\label{phi}\\
&& \psi \to \tilde{\psi} = \psi + {\cal H} \epsilon_{0},
\label{psi}\\
&& B \to \tilde{B} = B +\epsilon_{0} - \epsilon',
\label{B}\\
&& E \to \tilde{E} = E - \epsilon,
\label{E}
\end{eqnarray}
in the case of the scalar modes of the geometry. As anticipated the gauge 
transformations of the scalar modes involve $\epsilon_{0}$ and $\epsilon$.
 Under a coordinate transformation preserving the 
vector nature of the fluctuation, i.e. $x^{i} \to \tilde{x}^{i} = x^{i} 
+ \zeta^{i}$ (with  
$\partial_{i} \zeta^{i} =0$), the rotational  
modes of the geometry transform as 
\begin{eqnarray}
&& Q_{i} \to \tilde{Q}_{i} = Q_{i} - \zeta_{i}',
\label{Q}\\
&& W_{i} \to \tilde{W}_{i}= W_{i} + \zeta_{i}.
\label{W}
\end{eqnarray}
The tensor fluctuations, in the parametrization of Eq. (\ref{gij1})
are automatically invariant under infinitesimal diffeomorphisms, i.e. 
$\tilde{h}_{ij} = h_{ij}$. It is possible to select appropriate 
combinations of the fluctuations of given spin that are invariant under infinitesimal coordinate 
transformations. This possibility is particularly clear in the case of the vector modes. If we define 
the quantity $V_{i} = Q_{i} + W_{i}'$, we will have, according to Eqs. (\ref{Q}) and (\ref{W}), that 
for $x^{i} \to \tilde{x}^{i} = x^{i} + \zeta^{i}$, $\tilde{V}_{i} = V_{i}$, i.e. $V_{i}$ is invariant 
for infinitesimal coordinate transformations or, for short, gauge-invariant. The same trick 
can be used in the scalar case. In the scalar case the most appropriate  gauge-invariant 
fluctuations depend upon the specific problem at hand. An example of fully gauge-invariant 
fluctuations arising, rather frequently, in the treatment of scalar fluctuations is given in section
\ref{sec10} (see in particular Eqs. (\ref{giphi}), (\ref{gipsi}) and (\ref{gichi})).
The perturbed components of the energy-momentum tensor can be written, for a single species $\lambda$, as:
\begin{equation}
\delta T_{0}^{0} = \delta \rho_{\lambda}, \,\,\,\,\,  \delta T_{i}^{j} = - \delta p_{\lambda} \delta_{i}^{j},\,\,\,\,\,\,\,\,
\delta T_{0}^{i} =  ( p_{\lambda} + \rho_{\lambda})\partial^{i} v^{(\lambda)},
\label{enmomf}
\end{equation}
where we defined $\delta u_{i}^{(\lambda)} = \partial_{i} v^{(\lambda)}$ and where the index $\lambda$ 
denotes the specific component of the fluid characterized by a given barotropic index and by a given sound 
speed.  It is also appropriate, for applications, to work directly with the divergence 
of the peculiar velocity field by defining a variable $\theta_{\lambda} = \nabla^2 v^{(\lambda)}$.
Under the infinitesimal 
coordinate transformations of Eq. (\ref{T1}) the fluctuations given in Eq. (\ref{enmomf}) 
 transform according to Eq. (\ref{gaugetrgen}) and the explicit results are 
\begin{eqnarray}
&& \delta \rho_{\lambda} \to \delta \tilde{\rho}_{\lambda} = \delta \rho_{\lambda} - \rho'_{\lambda} \epsilon_{0},
\label{drho}\\
&&  \delta p_{\lambda} \to \delta \tilde{p}_{\lambda} = \delta p_{\lambda} - w_{\lambda}\rho_{\lambda}' \epsilon_{0},
\label{dp}\\
&& \theta_{\lambda}\to \tilde{\theta}_{\lambda} = \theta_{\lambda} + \nabla^2\epsilon'.
\label{vL}
\end{eqnarray}
Using the covariant conservation equation for the background fluid density, 
the gauge transformation for the density contrast, 
i.e. $\delta_{(\lambda)} = \delta \rho_{(\lambda)}/\rho_{(\lambda)}$, follows easily from Eq. (\ref{drho}): 
\begin{equation}
\tilde{\delta}_{(\lambda)} = \delta_{(\lambda)} + 
3 {\cal H} ( 1 + w_{(\lambda)}) \epsilon_{0}.
\label{denscontr}
\end{equation}
There are now, schematically,  three possible strategies
\begin{itemize}
\item{} a specific gauge can be selected  by fixing (completely or partially) 
the coordinate system; this will amount to fix, in the scalar case, the two independent 
functions $\epsilon_{0}$ and $\epsilon$;
\item{}  gauge-invariant fluctuations of 
the sources and of the geometry can be separately defined;
\item{} gauge-invariant fluctuations mixing the perturbations  
of the sources and of the geometry can be employed.
\end{itemize}

The vector modes are not so relevant in the conventional scenarios and will not be specifically discussed. 
If the Universe is expanding, the vector modes will always 
be damped depending upon the barotropic index of the sources 
of the geometry. This result has been obtained long ago \cite{rot0}.
However, if the geometry contracts or if internal dimensions are present in the game \cite{rot1,rot2}, such a statement 
is no longer true. These topics involve unconventional completions of the standard cosmological model and, therefore,
will not be discussed here.

\subsection{Evolution of the tensor modes and superadiabatic amplification}
The evolution of the tensor modes of the geometry can be obtained, as 
stressed before, either from the Einstein equations perturbed to first-order 
or from the action perturbed to second order. 
Consider now the case of the tensor modes of the geometry, i.e., according to 
Eq. (\ref{div1}), the two polarization of the graviton:
\begin{equation}
\delta_{\rm t} g_{i j} = - a^2 h_{ij},\,\,\,\,\,\,\,\,\,\,\,
\delta_{\rm t}  g^{i j} = \frac{h^{ij}}{a^2}.
\end{equation}
The tensor contribution to the 
fluctuation of the connections can then be expressed as 
\begin{eqnarray}
&& \delta_{\rm t} \Gamma_{ij}^{0} =  \frac{1}{2} ( h_{ij}' + 2 {\cal H}
h_{i j}),
\nonumber\\
&&  \delta_{\rm t} \Gamma_{i0}^{j} = \frac{1}{2} {h_{i}^{j}}',
\nonumber\\
&& \delta_{\rm t} \Gamma_{ij}^{k} = \frac{1}{2} [ \partial_{j} h_{i}^{k} + 
\partial_{i} h_{j}^{k} - \partial^{k} h_{ij}].
\label{TCHR}
\end{eqnarray}
Inserting these results into the perturbed expressions 
of the Ricci tensors it is easy to obtain:
\begin{eqnarray}
&& \delta_{\rm t} R_{i j} = \frac{1}{2}[ h_{i j}'' + 2 {\cal H} 
h_{ij}' +  2 ( {\cal H}' + 2 {\cal H}^2) h_{i j} - \nabla^2 h_{ij} ],
\label{driccit1}\\
&& \delta_{\rm t} R_{i}^{j} = - \frac{1}{2 a^2}[ {h_{i}^{i}}'' + 2 {\cal H} 
{h_{i}^{j}}' - \nabla^2 h_{i}^{j}],
\label{driccit2}
\end{eqnarray}
where $\nabla^2 = \partial_{i} \partial^{i}$ is the usual four-dimensional Laplacian.
In order to pass from Eq. (\ref{driccit1}) to Eq. (\ref{driccit2}) we may recall that 
\begin{equation}
\delta_{\mathrm{t}} R_{i}^{j} = \delta_{\mathrm{t}}(g^{jk} R_{ki}) = \delta_{\mathrm{t}} g^{jk}  \overline{R}_{ki} + 
\overline{g}^{jk} \delta_{\mathrm{t}} R_{ij},
\label{intricci1}
\end{equation}
where the relevant Ricci tensor, i.e. $\overline{R}_{ij}$ is simply given (see also Eqs. (\ref{Ricciconformal})):
\begin{equation}
\overline{R}_{ij} = ({\mathcal H}' + 2 {\mathcal H}^2) \delta_{ij}.
\end{equation}
Since both the fluid sources and the scalar fields do not contribute 
to the tensor modes of the geometry the evolution equation for $h_{i}^{j}$ is simply 
given, in Fourier space, by 
\begin{equation}
{h_{i}^{i}}'' + 2 {\cal H} {h_{i}^{j}}' +k^2 h_{i}^{j} =0. 
\label{T1a}
\end{equation}
Thanks to the conditions $\partial_{i} h^{i}_{j} = h_{k}^{k} =0$ (see Eq. (\ref{div1})), 
the direction of propagation can be chosen to lie along the 
third axis and, in this case the two physical polarizations
of the graviton will be
\begin{equation}
h_{1}^{1} = - h_{2}^{2} = h_{\oplus},\,\,\,\,\,\,\,\,\,\,\,\, h_{1}^{2} = 
h_{2}^{1} = h_{\otimes},
\end{equation}
where $h_{\oplus}$ and $h_{\otimes}$ 
obey the same evolution equation (\ref{T1a}) and will be denoted, in the 
remaining part of this section, by $h$. 
Equation (\ref{T1a}) can also be written in one of the following two  equivalent forms:
\begin{eqnarray}
&&h_{k}'' + 2 {\mathcal H} h_{k}' + k^2 h_{k} =0,
\label{firstform}\\
&& \mu_{k}'' + \biggl[ k^2 - \frac{a''}{a} \biggr] \mu_{k}=0,
\label{secondform}
\end{eqnarray}
where $\mu_{k} = a h_{k}$. Equation (\ref{secondform}) simply follows from Eq. (\ref{firstform}) 
by eliminating the first time derivative. Concerning Eq. (\ref{secondform}) two comments are in order:
\begin{itemize} 
\item{} in the limit where $k^2$ dominates over $|a''/a|$ the solution 
of the Eq. (\ref{secondform})  are simple plane waves;
\item{} in the opposite limit, i.e. 
$ |a''/a|\gg k^2$ the solution may exhibit, under certain conditions a growing mode.
\end{itemize}
In more quantitative terms, the solutions of Eq. (\ref{secondform}) in the two aforementioned limits are 
\begin{eqnarray}
&& k^2 \gg |a''/a|,\qquad \mu_{k}(\tau) \simeq e^{\pm i k\tau}
\label{plane}\\
&& k^2 \ll |a''/a|,\qquad \mu_{k}(\tau) \simeq A_{k} a(\tau) + B_{k} a(\tau) \int^{\tau} \frac{dx}{a^2(x)}.
\label{grow}
\end{eqnarray}
The oscillatory regime is sometimes called  {\em adiabatic} since, in this regime, 
$h_{k} \simeq a^{-1}$. If the initial fluctuations are normalized 
to quantum mechanics (see the following part of the present section) $\mu_{k} \simeq 
1/\sqrt{k}$ initially and, therefore 
\begin{equation}
\delta_{h} \simeq k^{3/2}|h_{k}(\tau)| \simeq \frac{k}{a} \simeq \omega,
\end{equation}
where $\omega(\tau) = k/a(\tau)$ denotes, in the present context, the physical wave-number 
while $k$ denotes the comoving wave-number.
Recalling that $\mu_{k} = a h_{k}$, Eq. (\ref{grow}) implies that 
\begin{equation}
h_{k}(\tau) \simeq A_{k} + B_{k} \int^{\tau} \frac{dx}{a^2(x)}.
\label{hgrow}
\end{equation}
This solution describes what is often named  {\em super-adiabatic amplification}. In particular 
cases (some of which of practical interest) Eq. (\ref{hgrow}) implies 
the presence of a decaying mode and of a constant mode. Since in the adiabatic 
regime $h_{k} \simeq 1/a$, the presence of a constant mode would imply a 
growth with respect to the adiabatic solution hence the name super-adiabatic \cite{gr1,gr2}.
We pause here for a moment to say that the adjective {\em adiabatic} is sometimes
used not in direct relation with thermodynamic notions as we shall also see 
in the case of the so-called adiabatic perturbations. It should be recalled that the first
author to notice that the tensor modes of the geometry can be amplified in FRW backgrounds
was L. P. Grishchuk \cite{gr1,gr2} (see also \cite{ALLEN,SAHNI,SOL,GGGW}).

Equation (\ref{secondform}) suggests an interesting analogy for 
the evolution of the tensor modes of the geometry since it 
can be viewed, for practical purposes, as a Schr\"odinger-like equation 
where, the analog of the wave-function does not depend on a spatial 
coordinate (like in the case of one-dimensional potential barriers) but 
on a time coordinate (the conformal time in the case of Eq. (\ref{secondform})).
The counterpart of the potential barrier is represented by the term $a''/a$ sometimes 
also called {\em pump field}. The physics of the process is therefore rather simple:
energy is transferred from the background geometry to the corresponding fluctuations.
This does not always happen since the properties of the background enter crucially.
For instance, the pump field $a''/a$ vanishes in the case of a radiation-dominated 
Universe. In this case the evolution equations of the tensor modes are said to be 
conformally invariant (or, more correctly, Weyl invariant) since with an appropriate 
rescaling the evolution equations have the same form they would have in the 
Minkowskian space-time. On the contrary, in the case of de Sitter 
expansion\footnote{In the cosmic time parametrization the (expanding) de Sitter metric is parametrized as $a(t) = e^{H_{1}t}$. Recalling that $\tau = \int dt/a(t)$ it follows, after integration, that $a(\tau)\simeq \tau^{-1}$.}
 $a(\tau) \simeq (-\tau_{1}/\tau)$ and $a''/a= 2/\tau^2$. 
It should be appreciated that expanding (exact) de Sitter space-time 
supports the evolution of the tensor modes of the geometry while 
the scalar modes, in the pure de Sitter case, are not amplified.
To get amplification of scalar modes during inflation it will be mandatory 
to have a phase of quasi-de Sitter expansion. 

Therefore, a more realistic model of the evolution of the background geometry 
can be achieved by a de Sitter phase that evolves into a radiation-dominated epoch 
which is replaced, in turn, by a matter-dominated stage of expansion.
In the mentioned case the evolution of the scale factor can be 
parametrized as:
\begin{eqnarray}
&& a_{\rm i} (\tau) = \biggl(-\frac{\tau}{\tau_1}\biggr)^{-\beta},\,\,\,\,\,\,\,\,\,\,\, \tau\leq - \tau_{1},
\label{ai}\\
&& a_{\rm r}(\tau) = \frac{\beta \tau + (\beta + 1) \tau_1}{\tau_{1}}, \,\,\,\,\,\,\,\,\,\, - \tau_{1} \leq \tau \leq \tau_{2},
\label{ar}\\
&& a_{\rm m} (\tau) = 
\frac{[ \beta (\tau + \tau_{2}) + 2 \tau_{1} (\beta + 1)]^2}{4 \tau_1 [ \beta \tau_{2} + (\beta + 1) \tau_1]},\,\,\,\,\,\,\,\,\
\tau > \tau_{2},
\label{am}
\end{eqnarray}
where the subscripts in the scale factors refer, respectively, to the inflationary, radiation and matter-dominated stages.
As already discussed, a generic power-law inflationary phase is characterized by a 
power $\beta$. In the case $\beta = 1$ we have the case of the expanding branch of de Sitter
space. During the radiation-dominated epoch the scale factor expands linearly in conformal time 
while during matter it expands quadratically. Notice that the form of the scale factors 
given in Eqs. (\ref{ai})--(\ref{am}) is continuous and differentiable at the transition points, i.e. 
\begin{eqnarray}
&& a_{\mathrm{i}}(-\tau_{1}) = a_{\mathrm{r}}(-\tau_{1}),\qquad   a_{\mathrm{i}}'(-\tau_{1}) = a_{\mathrm{r}}'(-\tau_{1}),
\nonumber\\
&&  a_{\mathrm{r}}(-\tau_{2}) = a_{\mathrm{m}}(-\tau_{2}),\qquad   a_{\mathrm{r}}'(-\tau_{2}) = a_{\mathrm{m}}'(-\tau_{2}).
\end{eqnarray}
The continuity of the scale factor and its derivative
 prevents the presence of divergences in the pump field, given by $a''/a$. 
\begin{figure}
\centering
\includegraphics[height=6cm,width=10cm]{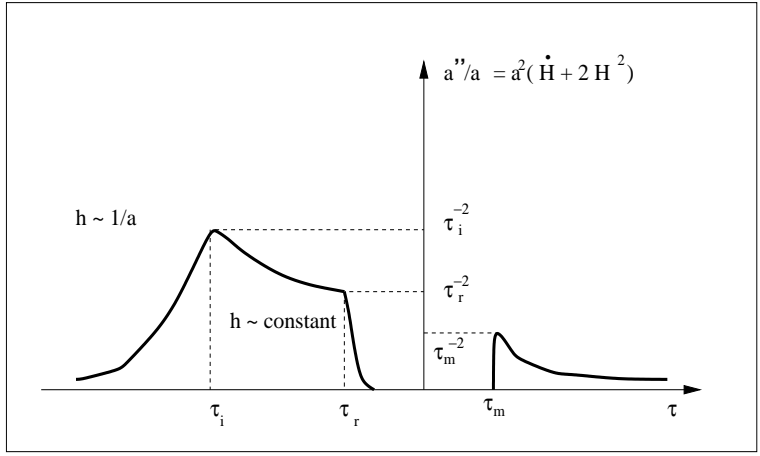}
\caption{The effective ``potential" appearing in Eq. (\ref{secondform}) is illustrated as a function 
of the conformal time coordinate $\tau$ in the case when the background passes through different stages 
of expansion. Being conformally invariant in the case of radiation, $a''/a=0$ in the central part of the plot.}
\label{POT}      
\end{figure}
In Fig. \ref{POT} the structure of the potential barrier is reproduced. Notice 
that, by using known identities together with the definition 
of the conformal time $\tau$ in terms of the cosmic time $t$, it is possible to express 
$a''/a$ in terms of the Hubble parameter and its first (cosmic) time derivative 
\begin{equation}
\frac{a''}{a} = {\mathcal H}^2 + {\mathcal H}' = a^2[ \dot{H} + 2 H^2],
\label{adoublpr}
\end{equation}
where, as usual, the prime denotes a derivation with respect to $\tau$ and a dot 
denotes a derivation with respect to $t$.

\subsection{Quantum mechanical description of the tensor modes}
The tensor 
modes of the geometry will now be discussed along a quantum-mechanical perspective.  
Such a treatment is essential in order 
to normalize properly the fluctuations, for instance during an initial inflationary phase or during any initial 
stage of the Universe when the only relevant fluctuations are the ones of quantum mechanical origin.
The calculation proceeds, in short, along the following steps:
\begin{itemize}
\item{} obtain the action perturbed to second order in the amplitude 
of the tensor modes of the geometry;
\item{} define the appropriate normal modes and promote them to the 
status of (quantum) field operators in the Heisenberg representation;
\item{} solve the evolution of the system and compute the number of produced particles.
\end{itemize}
To comply with the first step, let us observe that the second-order action
can be written, up to (non-covariant) total derivatives, as 
\begin{equation}
\delta_{\mathrm{t}}^{(2)} S = \frac{1}{64\pi G} \int d^{4} x a^2(\tau) \eta^{\alpha\beta} 
\partial_{\alpha} h_{i}^{j} \partial_{\beta} h_{j}^{i}.
\label{secondaction}
\end{equation}
Recalling that the polarization can be chosen as 
\begin{equation}
h_{1}^{1} = - h_{2}^{2}= h_{\oplus},\qquad h_{1}^{2} = h_{2}^{1} = h_{\otimes}.
\label{defoplot}
\end{equation}
and recalling the definition of reduced Planck mass (see Eq. (\ref{PLdef}))
\begin{equation}
\ell_{\mathrm{P}} = \sqrt{8\pi G} \equiv \frac{1}{\overline{M}_{\mathrm{P}} }\equiv \frac{\sqrt{8\pi}}{M_{\mathrm{P}}},
\label{ellP}
\end{equation}
Eq. (\ref{secondaction}) can be written as 
\begin{equation}
\delta_{\mathrm{t}}^{(2)} S = \frac{1}{4\ell_{\mathrm{P}^2} }\int d^{3}x\, d\tau\,a^{2}(\tau)\bigl[ {h_{\oplus}'}^2 + 
{h_{\otimes}'}^2 - (\partial_{i} h_{\oplus})^2 - (\partial_{i} h_{\otimes})^2 \bigr],
\label{secon1}
\end{equation}
becoming, for a single tensor polarization,  
\begin{equation}
S_{\mathrm{gw}} = \frac{1}{2} \int d^{3}x\, d\tau\,a^{2}(\tau)\bigl[ {h'}^2 - (\partial_{i} h)^2 \bigr],
\label{secon2}
\end{equation}
where 
\begin{equation}
h = \frac{h_{\oplus}}{\sqrt{2} \ell_{\mathrm{P}}} =  \frac{h_{\otimes}}{\sqrt{2} \ell_{\mathrm{P}}},
\end{equation}
denotes, indifferently, each of the two polarization of the graviton.
Defining now the appropriate canonical normal mode of the action (\ref{secon2}), i.e. $\mu = a h$ we get to the action 
\begin{equation}
\tilde{S}_{\mathrm{gw}} = \frac{1}{2} \int d^{3}x\, d\tau \bigl[ {\mu'}^2 + {\mathcal H}^2 \mu^2 - 2 {\mathcal H} \mu\mu' - 
(\partial_{i} \mu)^2 \bigr].
\label{secon3}
\end{equation}
From Eq. (\ref{secon2}) non-covariant total derivatives can be dropped. With this method it is clear that 
the term going as $- 2 {\mathcal H} \mu\mu'$ can be traded for $({\mathcal H} \mu^2)'$ by paying the prize 
of a new term proportional to ${\mathcal H}' \mu^2$. Hence, up to total derivatives Eq. (\ref{secon3}) gives:
\begin{equation}
S_{\mathrm{gw}} = \frac{1}{2} \int d^{3}x\, d\tau \bigl[ {\mu'}^2 + ({\mathcal H}^2 + {\mathcal H}')\mu^2 - 
(\partial_{i} \mu)^2 \bigr].
\label{secon4}
\end{equation}
From Eq. (\ref{secon4}) it follows that the Lagrangian and the Hamiltonian of the tensor modes can be 
expressed in terms of the appropriate Lagrangian and Hamiltonian densities, namely, 
\begin{equation}
L_{\mathrm{gw}}(\tau) = \int d^{3} x \, {\mathcal L}_{\mathrm{gw}}(\tau, \vec{x}), \qquad 
H_{\mathrm{gw}}(\tau) = \int d^{3}x \bigl[\pi \mu' - {\mathcal L}_{\mathrm{gw}}\bigr].
\label{LH}
\end{equation}
The quantity $\pi$ appearing in Eq. (\ref{LH}) is the conjugate momentum. From the actions (\ref{secon3}) and (\ref{secon4}) 
the derived conjugate momenta are different and, in particular, they are, respectively:
\begin{equation}
\tilde{\pi} = \mu' - {\mathcal H} \mu,\qquad \pi = \mu'.
\label{LH2}
\end{equation}
Equation (\ref{LH2}) implies that the form of the Hamiltonian changes depending on the specific form of the action. 
This is a simple reflection of the fact that, in the Lagrangian formalism, the inclusion (or exclusion) of a total derivative 
does not affect the Euler-Lagrange equations. Correspondingly, in the Hamiltonian formalism, the total Hamiltonian 
will necessarily change by a  time derivative of the generating functional of the canonical transformation. This 
difference will have, however, no effect on the Hamilton equations. Therefore, the Hamiltonian derived 
from Eq. (\ref{secon4}) can be simply expressed as:
\begin{equation}
H_{\mathrm{gw}}(\tau) = \frac{1}{2} \int d^{3} x \, \biggl[ \pi^2 -\frac{a''}{a} \mu^2 + (\partial_{i} \mu)^2\biggr].
\label{HAM1}
\end{equation}
The Hamiltonian derived from Eq. (\ref{secon3}) can be instead written as: 
\begin{equation}
\tilde{H}_{\mathrm{gw}}(\tau) =  \frac{1}{2} \int d^{3} x \, \biggl[ \tilde{\pi}^2 + 2 {\mathcal H} \mu \tilde{\pi} +(\partial_{i} \mu)^2\biggr].
\label{HAM2}
\end{equation}
Suppose than to start with the Hamiltonian of Eq. (\ref{HAM2}) and define the appropriate generating functional 
of the (time-dependent) canonical transformation, i.e. 
\begin{equation}
{\mathcal F}(\mu,\pi,\tau) = \int d^{3} x \biggl[ \mu\pi - \frac{{\mathcal H}}{2} \mu^2 \biggr],
\label{CAN}
\end{equation}
which is, by definition, a functional of the new momenta (i.e. $\pi$). Thus, we will have that 
\begin{eqnarray}
&& \tilde{\pi} = \frac{\delta {\mathcal F}}{\delta \mu} = \pi  - {\mathcal H} \mu,
 \label{newold}\\
 && H_{\mathrm{gw}}(\tau) = \tilde{H}_{\mathrm{gw}}(\tau) + \frac{\partial {\mathcal F}}{\partial \tau}.
 \label{HfromtildeH}
 \end{eqnarray}
 Equation (\ref{newold}) gives the new momenta as a function of the old ones so that, if we start with Eq. (\ref{HAM2}) 
 we will need to bear in mind that $\pi = \tilde{\pi} + {\mathcal H}\mu$ and substitute into $\tilde{H}_{\mathrm{gw}}(\tau)$. 
 Equation (\ref{HfromtildeH}) will then allow to get the  $H_{\mathrm{gw}}(\tau)$ reported in Eq. 
 (\ref{HAM1}), as it can be directly verified. 
 
This digression on the canonical properties of time-dependent Hamiltonians is useful not so much at 
the classical level (since, by definition of canonical transformation, the Hamilton equations are invariant) 
but rather at the quantum level \cite{MGTP2}. Indeed, the vacuum will be the state minimizing a given Hamiltonian.
It happens that some 
non-carefully selected Hamiltonians may lead to initial vacua that, indeed, would lead to an energy density 
of the initial state which is (possibly) larger than the one of the background geometry \cite{MGTP1}. 
The Hamiltonian of Eq. (\ref{HAM1}) is valuable in this respect since 
the initial vacuum (i.e. the state minimizing (\ref{HAM1}) possesses an energy density which is usually much smaller than the background, as it should be
to have a consistent picture (see also Ref. \cite{THTH} for the discussion of the so-called 
transplankian ambiguities). It should be finally remarked that all the imaginable Hamiltonians (connected by time-dependent 
canonical transformations) lead always to the same quantum evolution either in the Heisenberg or in the Schr\"odinger 
description. Even there, however, there are (practical) differences. For instance Eq. (\ref{HAM2}) 
seems more convenient in the Schr\"odinger description. Indeed at the quantum level the time evolution operator 
would contain, in the exponential, operator producs as $\tilde{\pi}\mu$ which are directly related to the so-called squeezing 
operator in the theory of optical coherence (see the last part of the present section). 
In a complementary perspective and always at a practical level, the 
Hamiltonian defined in Eq. (\ref{HAM1}) is more suitable for the Heisenberg description.

The quantization of the canonical Hamiltonian of Eq. (\ref{HAM1})  is
performed  by promoting the normal modes of the action  to field operators in the Heisenberg 
description and by imposing (canonical) equal-time commutation relations:
\begin{equation}
[ \hat{\mu}(\vec{x},\tau), \hat{\pi}(\vec{y},\tau)] = i \delta^{(3)} (\vec{x} - \vec{y}).
\label{cancomma}
\end{equation}
the operator corresponding to the  Hamiltonian (\ref{HAM1}) becomes: 
\begin{equation}
\hat{H}(\tau) = \frac{1}{2} \int d^3 x 
\biggl[ \hat{\pi}^2 - \frac{a''}{a} \hat{\mu}^2 
+  (\partial_{i} \hat{\mu})^2\biggr].
\label{ham2a}
\end{equation}
In Fourier space the quantum fields $\hat{\mu}$ and $\hat{\pi}$ can be expanded as
\begin{eqnarray}
&&\hat{\mu}(\vec{x},\tau) = \frac{1}{2 (2\pi)^{3/2} } \int d^3 k \biggl[ \hat{\mu}_{\vk} e^{- i \vec{k} \cdot \vec{x} }
+ \hat{\mu}_{\vk}^{\dagger}  e^{ i \vec{k} \cdot \vec{x} }\biggr],
\nonumber\\
&& \hat{\pi}(\vec{y},\tau) = \frac{1}{2 (2\pi)^{3/2} } \int d^3 p \biggl[ \hat{\pi}_{\vk} e^{- i \vec{p} \cdot \vec{y} }
+ \hat{\pi}_{\vk}^{\dagger}  e^{ i \vec{p} \cdot \vec{y} }\biggr].
\label{expansiona}
\end{eqnarray}
Demanding the validity of the canonical commutation relations of 
Eq. (\ref{cancomma})  the Fourier components must obey:
\begin{eqnarray}
&& [ \hat{\mu}_{\vk}(\tau), \hat{\pi}_{\vp}^{\dagger}(\tau) ] = i \delta^{(3)}(\vec{k} - \vec{p}),
\qquad
[ \hat{\mu}_{\vk}^{\dagger}(\tau), \hat{\pi}_{\vp}(\tau) ]= i \delta^{(3)}(\vec{k} - \vec{p}),
\nonumber\\
&& [ \hat{\mu}_{\vk}(\tau), \hat{\pi}_{\vp}(\tau) ]= i \delta^{(3)}(\vec{k} + \vec{p}),
\qquad [ \hat{\mu}_{\vk}^{\dagger}(\tau), \hat{\pi}_{\vp}^{\dagger}(\tau) ]= i \delta^{(3)}(\vec{k} + \vec{p}).
\label{fcomma}
\end{eqnarray}
Inserting now Eq. (\ref{expansiona}) into Eq. (\ref{HAM1}) the Fourier space representation
of the quantum Hamiltonian \footnote{ Notice that in order to derive the following equation, 
the relations $\hat{\mu}_{-\vk}^{\dagger} \equiv  \hat{\mu}_{\vk}$ and 
$\hat{\pi}_{-\vk}^{\dagger} \equiv  \hat{\pi}_{\vk}$ should be used .} can be obtained:
\begin{equation}
\hat{H}(\tau) = \frac{1}{4} \int d^3 k \biggl[(\hat{\pi}_{\vk} \hat{\pi}^{\dagger}_{\vk} + 
\hat{\pi}_{\vk}^{\dagger} \hat{\pi}_{\vk}) + \biggl( k^2 - \frac{a''}{a}\biggr) (\hat{\mu}_{\vk} \hat{\mu}^{\dagger}_{\vk} + 
\hat{\mu}_{\vk}^{\dagger} \hat{\mu}_{\vk}) \biggr].
\label{ham3a}
\end{equation} 
The evolution of $\hat{\mu}$ and $\hat{pi}$  is therefore dictated, in the Heisenberg representation,  by:
\begin{eqnarray}
i \hat{\mu}' = [\hat{\mu},\hat{H}],\qquad
i \hat{\pi}' = [\hat{\pi},\hat{H}].
\label{pieq1}
\end{eqnarray}
Using now the mode expansion (\ref{expansiona}) and the Hamiltonian in the form (\ref{ham3a})
the evolution for the Fourier components of the operators is 
\begin{equation}
\hat{\mu}_{\vk}'= \hat{\pi}_{\vk},\qquad
 \hat{\pi}_{\vk}' = - \biggl( k^2  - \frac{a''}{a}\biggr) \hat{\mu}_{\vk},
\label{pieq2a}
\end{equation}
implying 
\begin{equation}
\hat{\mu}_{\vk}'' +\biggl[k^2 - \frac{a''}{a}\biggr] \mu_{\vk}=0.
\label{thirdform}
\end{equation}
It is not a surprise that the evolution equations of the field operators, in the Heisenberg 
description, reproduces, for $\hat{\mu}_{\vk}$ the classical evolution equation derived 
before in Eq. (\ref{secondform}).

The general solution of the system is then 
\begin{eqnarray}
&& \hat{\mu}_{\vk}(\tau) = \hat{a}_{\vk}(\tau_0) f_{\mathrm{i}}(k,\tau) + \hat{a}_{-\vk}^{\dagger}(\tau_0) f^{\ast}_{\mathrm{i}}(k,\tau),
\label{solmu}\\
&& \hat{\pi}_{k}(\tau) = \hat{a}_{\vk}(\tau_0) g_{\mathrm{i}}(k,\tau) + \hat{a}_{-\vk}^{\dagger}(\tau_0)g^{\ast}_{\mathrm{i}}(k,\tau),
\label{solpi}
\end{eqnarray}
where the mode function $f_{\mathrm{i}}$ obeys 
\begin{equation}
f_{\mathrm{i}}'' + \biggl[ k^2 - \frac{a''}{a}\biggr] f_{\mathrm{i}} =0,
\label{fkeq}
\end{equation}
and\footnote{Of course if the form of the Hamiltonian is different by a time-dependent canonical transformation, also the canonical momenta will differ and, consequently, the relation of $g_{\mathrm{i}}$ to $f_{\mathrm{i}}$ may be different.} $g_{\mathrm{i}} = f_{\mathrm{i}}'$.  In the case when the scale factor has a power dependence, in cosmic 
time, the scale factor will be, in conformal time 
$a(\tau) = (-\tau/\tau_{1})^{-\beta}$ (with $\beta = p/(p -1)$ and $a(t)\simeq t^{p}$). The solution of Eq. (\ref{fkeq})  is then 
\begin{eqnarray}
f_{\mathrm{i}}(k,\tau) &=& \frac{{\cal N}}{\sqrt{2 k}} \sqrt{-x} H^{(1)}_{\mu}(- x),
\label{fk}\\
g_{\mathrm{i}}(k,\tau) &=& = f_{k}'=  - {\cal N}\sqrt{\frac{k}{2}} \sqrt{-x}\biggl[ H^{(1)}_{\mu -1} (-x) +
\frac{(1 -2 \mu)}{2(-x)} H^{(1)}_{\mu} (-x)\biggr] ,
\label{gk}
\end{eqnarray}
where $ x =  k\tau$ and 
\begin{equation}
{\cal N} = \sqrt{\frac{\pi}{2}} e^{\frac{i}{2}(\mu + 1/2)\pi},~~~~~\mu = \beta+\frac{1}{2}.
\end{equation}
The functions $H^{(1)}_{\mu}(-x) = J_{\mu}(-x)+ iY_{\mu}(-x)$ is the Hankel function 
of first kind \cite{abr,tric} and the other linearly independent solution will be $H^{(2)}_{\mu}(z) =  {H^{(1)}}^{\ast}_{\mu}(z)$.
Notice that the phases appearing in Eqs. (\ref{fk}) and (\ref{gk}) are carefully selected in such a 
way that for $\tau \to -\infty$, $ f_{\mathrm{i}} \to e^{- i\,k\,\tau}/\sqrt{2 k} $.

A possible application of the formalism developed so far is the calculation of 
 the energy density of the gravitons produced, for instance, in the transition from 
a de Sitter stage of inflation and a radiation-dominated stage of (decelerated) expansion.
This corresponds to a scale factor that, for $\tau< - \tau_{1}$ goes as in Eq. (\ref{ai}) with  $\beta=1$.
For $\tau > - \tau_{1}$ the scale factor is, instead, exactly the one reported in Eq. (\ref{ar}). 
Consequently, from Eq. (\ref{fk}) and (\ref{gk}), the mode functions 
\begin{eqnarray}
&&f_{\mathrm{i}}(k,\tau) = \frac{1}{\sqrt{2k}}\biggl(1 - \frac{i}{k\tau} \biggr) e^{- i k\tau},\qquad \tau \leq -\tau_{1},
\label{fi}\\
&& g_{\mathrm{i}}(k,\tau) = \sqrt{\frac{k}{2}} \biggl(\frac{i}{k^2 \tau^2} - \frac{1}{k\tau} - i\biggr)e^{-ik\tau},\qquad 
\tau \leq - \tau_{1}.
\label{gi}
\end{eqnarray}
For $\tau > - \tau_{1}$ the field operators can be expanded in terms of a new set of creation and 
annihilation operators, i.e. 
\begin{eqnarray}
&& \hat{\mu}_{\vk}(\tau) = \hat{b}_{\vk}(\tau_1) \tilde{f}_{\rm r}(k,\tau) + \hat{b}_{-\vk}^{\dagger}(\tau_{1})
\tilde{f}_{\rm r}^{\ast}(k,\tau),\qquad \tau > - \tau_{1}
\nonumber\\
&&\hat{\pi}_{\vk}(\tau) = \hat{b}_{\vk}(\tau_1) \tilde{g}_{\rm r}(k,\tau) + \hat{b}_{-\vk}^{\dagger}(\tau_{1})
\tilde{g}_{\rm r}^{\ast}(k,\tau),\qquad \tau > - \tau_{1},
\label{secsol}
\end{eqnarray}
where, now, $f_{\mathrm{r}}(k,\tau)$ are simply appropriately normalized plane waves since, in this phase, $a'' =0$:
\begin{equation}
\tilde{f}_{\rm r}(k,\tau) = \frac{1}{\sqrt{2 k} } e^{ - i y},\qquad \tilde{g}_{\rm i}(k,\tau) = -i  \sqrt{\frac{k}{2}}  e^{- i y},\qquad 
\tau > - \tau_{1},
\label{radmode}
\end{equation}
where $ y = k [ \tau + 2 \tau_{1}]$. 
Since the creation and annihilation operators must always be canonical, $\hat{b}_{\vk}$ and $\hat{b}_{\vec{k}}^{\dagger}$ can
be expressed as a linear combination of  $\hat{a}_{\vk}$ and $\hat{a}_{\vec{k}}^{\dagger}$, i.e. 
\begin{eqnarray}
&&\hat{b}_{\vk} =  B_{+}(k) \hat{a}_{\vk}+ B_{-}(k)^{\ast}\hat{a}_{-\vk}^{\dagger},
\nonumber\\
&& \hat{b}_{\vk}^{\dagger} = B_{+}(k)^{\ast} \hat{a}_{\vk}^{\dagger}+ B_{-}(k)\hat{a}_{-\vk}.
\label{BOG1}
\end{eqnarray}
Equation (\ref{BOG1}) is a special case of a Bogoliubov-Valatin transformation. But because 
\begin{equation}
[\hat{a}_{\vk}, \hat{a}_{\vec{p}}^{\dagger}] = \delta^{(3)}(\vk -\vp), \qquad 
[\hat{b}_{\vk}, \hat{b}_{\vec{p}}^{\dagger}] = \delta^{(3)}(\vk -\vp)
\label{BOG1a}
\end{equation}
we must also have:
\begin{equation}
|B_{+}(k)|^2 - |B_{-}(k)|^2 =1.
\label{UN}
\end{equation}
 Equation (\ref{BOG1}) can be inserted into 
Eq. (\ref{secsol}) and the following expressions can be easily obtained:
\begin{eqnarray}
&& \hat{\mu}_{\vk}(\tau) = \hat{a}_{\vk} [   B_{+}(k) f_{\rm r} + B_{-}(k) 
f_{\rm r}^{\ast}]+ \hat{a}_{-\vk}^{\dagger}[   B_{+}(k)^{\ast} 
f_{\rm r}^{\ast} + B_{-}(k) ^{\ast}f_{\rm r}],
\label{solmurad}\\
&& \hat{\pi}_{k}(\tau) = \hat{a}_{\vk} [  B_{+}(k) g_{\rm r} + B_{-}(k) g_{\rm r}^{\ast}  ]  
+ \hat{a}_{-\vk}^{\dagger}  [  B_{+}(k)^{\ast} g_{\rm r}^{\ast} + B_{-}^{\ast}(k) g_{\rm r}  ] .
\label{solpirad}
\end{eqnarray}
Since the evolution of the canonical fields must be continuous, Eqs. (\ref{solmu})--(\ref{solpi}) together 
with  Eqs. (\ref{solmurad})--(\ref{solpirad}) imply 
\begin{eqnarray}
&& f_{\rm i}(-\tau_1) = B_{+}(k) f_{\rm r}(-\tau_1) + B_{-}(k)f_{\rm r}^{\ast}(-\tau_1),
\nonumber\\
&& g_{\rm i}(- \tau_1)  =  B_{+}(k) g_{\rm r}(-\tau_1) + B_{-}(k) g_{\rm r}^{\ast}(-\tau_1),
\label{match1}
\end{eqnarray}
which allows to determine the coefficients of the Bogoliubov transformation $B_{\pm}(k)$, i.e. 
\begin{eqnarray}
&& B_{+}(k) = e^{2 i x_{1}} \biggl[ 1 - \frac{i}{x_{1}} - \frac{1}{2 x_{1}^2}\biggr],
\nonumber\\
&& B_{-}(k) = \frac{1}{2 x_{1}^2},
\label{BPMex}
\end{eqnarray}
where $x_{1} = k \tau_{1}$ and where Eq. (\ref{UN}) is trivially satisfied.  Between 
$B_{+}(k)$ and $B_{+}(k)$ the most important quantity is clearly $B_{-}(k)$ since 
it defines the amount of ``mixing" between positive and negative frequencies.
In the case when the gravitational interaction is switched off, the positive/negative
frequencies will not mix and $B_{-}(k)$ would vanish. The presence of a time-dependent 
gravitational field, however, implies that outgoing waves will mix, in a semiclassical language,
 with ingoing waves. This mixing simply signals that energy has been transferred from the background geometry 
 to the quantum fluctuations (of the tensor modes, in this specific example). This aspect can be 
 appreciated by computing the mean number of produced pairs of gravitons. Indeed, if a graviton 
 with momentum $\vec{k}$ is produced, also a graviton with momentum $-\vec{k}$ is produced 
 so that the total momentum of the vacuum (which is zero) is conserved:
 \begin{equation}
 \overline{n}_{k}^{\mathrm{gw}} = \frac{1}{2} \langle 0| \hat{N} |0\rangle = \frac{1}{2}\langle 0|[\hat{b}_{\vk}^{\dagger} \hat{b}_{\vk} +  
 \hat{b}_{-\vk}^{\dagger} \hat{b}_{-\vk}]|0\rangle = |B_{-}(k)|^2.
\label{nk}
\end{equation}
Now the total energy of the produced gravitons can be computed recalling that 
\begin{equation}
d\rho_{\mathrm{gw}} = 2\times \frac{d^{3}k}{(2\pi)^{3}} \overline{n}_{k}^{\mathrm{gw}},
\label{rhogw}
\end{equation}
where the factor $2$ counts the two helicities.
Using now the result of Eq. (\ref{BPMex}) into Eqs. (\ref{nk}) and (\ref{rhogw}) we do obtain the following interesting 
result, i.e. 
\begin{equation}
\frac{ d\rho_{\mathrm{gw}}}{d\ln{k}} = \frac{k^4}{\pi^2} \overline{n}_{k} = \frac{H_{1}^4}{\,\pi^2}
\label{endensgrav}
\end{equation}
where $|H_{1}/a_{1}|= \tau_{1}^{-1} = H_{1}$ [since in our parametrization of the scale factor $a(-\tau_{1}) =1$).
The result expressed by Eq. (\ref{endensgrav}]  implies that, in conventional inflationary models, 
the spectrum of relic gravitons is, in the best case, flat. In more realistic cases, in fact, it is quasi-flat (i.e. slightly decreasing) since 
the de Sitter phase, most likely, is not exact. As it is evident from 
the pictorial illustration of the effective potential the modes inheriting flat spectrum are the ones leaving 
the potential barrier during the de Sitter stage and re-entering during the radiation-dominated phase, i.e. 
comoving wave-numbers $ k_{2} < k< k_{1}$ where $k_{1} = \tau_{1}^{-1}$ and $k_{2} =\tau_{2}^{-1}$.
It is also clear that for sufficiently infra-red modes, we must also take into account the second relevant transition 
of the background from radiation to matter-dominated phase. This second transition will lead, for $k< k_{2}$
a slope $k^{-2}$ in terms of the quantity defined in Eq. (\ref{endensgrav}).
In analogy with what done in the case of black-body emission is practical to
parametrize the energy density of the relic gravitons in terms of the parameter 
\begin{equation}
\Omega_{\mathrm{GW}}(\nu,\tau) = \frac{1}{\rho_{\mathrm{crit}}} \frac{ d\rho_{\mathrm{gw}}}{d\ln{\nu}} 
\label{OMGW}
\end{equation}
where $\nu = k/[2\pi\,a(\tau)]$ is the physical frequency which is conventionally 
evaluated at the present time since our detectors of gravitational radiation are at the present epoch.
\begin{figure}
\centering
\includegraphics[height=6cm, width=10cm]{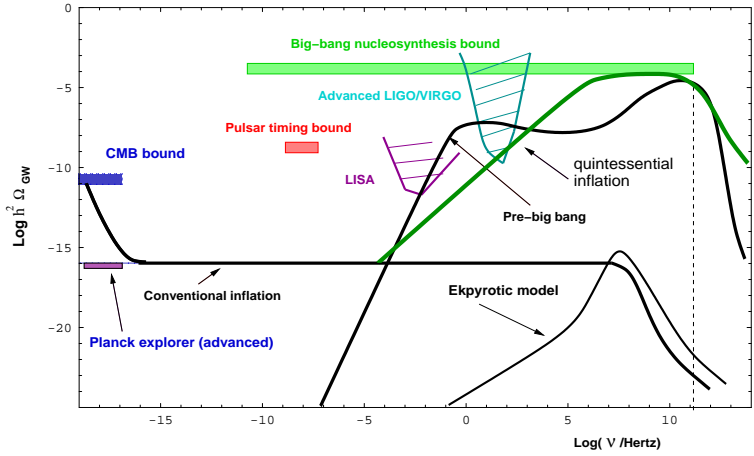}
\caption{The logarithmic energy spectrum of relic gravitons is illustrated in different models 
of the early Universe as a function of the present frequency, $\nu$.}
\label{GRAV}      
\end{figure}
In Fig. \ref{GRAV} different models are illustrated in terms of their energy spectrum. The calculation 
performed above estimates the flat plateau labeled by ``conventional inflation". To compare directly the plot with the result 
of the calculation one must, however, also take into account the redshift of the energy density 
during the matter-dominated phase. 
\subsection{Spectra of relic gravitons}
The spectrum reported in Fig. \ref{GRAV} consists of two branches 
a soft branch ranging  between 
$\nu_0 \simeq 10^{-18}~h_{0}$ Hz and 
$\nu_{{\rm dec}}\simeq  10^{-16}$ Hz. 
For 
$\nu >\nu_{{\rm dec}}$ we have instead the  hard
 branch consisting of high frequency gravitons mainly produced 
thanks to the transition from the inflationary regime to radiation. In the 
soft branch $\Omega_{{\rm GW}}(\nu,\tau_0) \sim \nu^{-2}$. 
In the hard branch $\Omega_{{\rm GW}}(\nu,\tau_0) $
 is  constant in frequency (or slightly decreasing in the quasi-de Sitter case). 
The large-scale observation of the first multipole moments 
of the temperature anisotropy imply a bound for the relic graviton background. 
The rationale for this statement is very simple since relic gravitons contribute 
to the integrated Sachs-Wolfe effect as discussed in section \ref{sec7}.  
The gravitational wave contribution to the Sachs-Wolfe 
integral 
cannot be larger than the (measured) amount of anisotropy directly
detected. The soft branch 
of the  spectrum is then constrained and  the bound reads \cite{THTH,allen2,SCHUTZ}
\begin{equation}
h_{0}^2\Omega_{{\rm GW}}(\nu,\tau_0) \laq~ 6.9~ \times 10^{-11},
\label{cobe}
\end{equation}
for  $\nu\sim \nu_{0} \sim 10^{-18} {\rm Hz}$. The very small size of the fractional timing error in the 
arrivals of the millisecond plusar's pulses imply that also the hard
branch is bounded according to \cite{taylor}
\begin{equation}
\Omega_{{\rm GW}}(\nu, \tau_0)~ \laq~ 10^{-8},
\label{puls}
\end{equation}
for $\nu\sim 10^{-8}$ Hz corresponding, roughly, to the inverse of the 
observation time during which the various millisecond pulsars have been 
monitored.

The two constraints of Eqs. (\ref{cobe}) and (\ref{puls}) are reported 
in Fig. \ref{GRAV}, at the two relevant frequencies.
The Sachs-Wolfe and millisecond pulsar constraints are  differential
since they limit, locally, the logarithmic derivative of the 
gravitons energy density. There exists also an  integral 
bound coming from standard BBN analysis  
and constraining the integrated graviton energy spectrum \cite{schw}: 
\begin{equation}
h_{0}^2\int_{\nu_{{\rm n}}}^{\nu_{{\rm max}}} \Omega_{{\rm GW}}(\nu,\tau_0) d\ln{\nu}
\laq ~ 0.2\,\,\,h_{0}^2 \Omega_{\gamma} \simeq 10^{-6}.
\label{NS}
\end{equation}
In Eq. (\ref{NS})  $\nu_{{\rm max}}$ 
corresponds to the (model dependent) ultra-violet cut-off 
of the spectrum and $\nu_{{\rm n}}$ is the frequency corresponding 
to the horizon scale at nucleosynthesis. 
 Notice 
that the BBN 
constraint of Eq. (\ref{NS}) has been derived in the 
context of the simplest BBN model, namely, 
assuming that no inhomogeneities and/or matter anti--matter domains 
are present at the onset of nucleosynthesis. In the presence of  
matter--antimatter domains for scales comparable with the 
neutron diffusion scale  this bound is relaxed \cite{inBBN3}. 

From Fig. \ref{GRAV} we see that also  the global bound of 
 Eq. (\ref{NS})  is satisfied and the typical amplitude of the 
logarithmic energy spectrum in critical units 
for frequencies $\nu_{I} \sim 100 $ Hz (and larger)
cannot exceed $10^{-14}$. This amplitude 
has to be compared with the LIGO sensitivity to a flat
$\Omega_{{\rm GW}}(\nu_{I}, \tau_0)$ 
which could be {\em at most} of the order of 
$h_{0}^2 \Omega_{{\rm GW}}(\nu_{I},\tau_0)
= 5\times 10^{-11}$ after four months of
 observation with $90\% $ confidence. At the moment there is 
 no direct detection of relic gravitons (and more generally of GW)
 from any detectors. For an introduction to various detectors of gravitational waves
 see, for instance, \cite{ground} (see also \cite{THTH}).
 
 Even if gravitational waves of high frequency are not central for the present discussion,
 it should be borne in mind that there exist cosmological scenarios where, for frequencies 
 larger than $10^{-3}$ Hz,  $\Omega_{\mathrm{GW}}(\nu)$ can deviate from 
 the inflationary (nearly scale-invariant) spectrum. In  particular, in Fig. \ref{GRAV} 
 the expected signals from quintessential inflationary models \cite{quint1,quint2} (see also \cite{quint3}) and from 
 pre-big bang models \cite{pbb1c,pbb2c,pbb3c} are reported. In quintessential inflationary models 
 the rise in the spectrum occurs since the inflaton and the quintessence field are unified in a single 
 scalar degree of freedom. Consequently, during a rather long phase (after inflation and before 
 the radiation epoch) the Universe is dominated by the kinetic energy of the inflaton. This 
 dynamics enhances the graviton spectrum at high frequencies. 
 In fact, pre big-bang models are formulated in the framework of the low-energy 
 string effective action where the Einstein-Hilbert term is naturally coupled to the dilaton. 
 The evolution of the tensor modes will then be slightly different from the one 
 derived in the present section and will be directly sensitive to the evolution of the dilaton. Both in quintessential inflation and in pre-big bang models
 the spectrum of relic gravitons is larger at high frequencies suggesting
 that superconducting cavities are a promising tool for the experimental 
 investigation in this range of frequencies (see \cite{microw1,microw2,microw3} and references therein).  Another very interesting (complementary) approach along this 
 direction has been reported in \cite{microw4,microw5} where a prototype 
 detector working in the $100$ MHz region has been described.

\subsection{More on the quantum state of cosmological perturbations}

The evolution of the cosmological inhomogeneities 
has been described, so far, in the Heisenberg representation. 
To investigate the correlation properties of the fluctuations 
and their semiclassical limit it is often useful 
to work within the Schr\"odinger representation where   
the evolution can be pictured as the spreading 
of a quantum mechanical wave-functional. The initial wave-functional 
will be constructed as the direct product of states minimizing the 
indetermination relations for the different harmonic oscillators forming the quantum 
field. The quantum mechanical states of the fluctuations 
will then be a generalization of the concept of coherent state firstly 
introduced in \cite{luciano,stoler}. These states are essentially 
coherent states associated to Lie algebras of
 non-compact groups (such as $SU(1,1)$ which is isomorphic to the algebra 
 of $SO(2,1)$ and $SL(2,R)$).  Since their discovery it has been 
 understood that their typical quantum mechanical property was to minimize the 
 indetermination relations \cite{stoler}. It was then understood they 
 can be obtained, as the coherent states, by the action of a unitary operator 
 acting on the vacuum. Following the pioneering work of Yuen \cite{yuen}
 the squeezed states  
 have been experimentally investigated in quantum optics with the hope of obtaining 
 "squeezed light".  This light could be of upmost importance 
 for various devices since it would allow to have one of the conjugate 
 (quantum) variables fluctuating above the quantum limit while the 
 the other variable fluctuates below the quantum limit preserving, overall
 the minimal uncertainty. To have the flavor of the manifold 
 applications of squeezed states to quantum optics the reader 
 can consult two classical textbooks \cite{sq1,sq2} 
 and also two (not so recent) review articles \cite{sq3,sq4}.
 While the experimental evidence is that squeezed light is rather
 hard to produce for large values of the squeezing parameter\footnote{In the language of this 
 discussion $r \gg 1$, see below.} 
 (which would be the interesting range for applications),
 the squeezed states formalism has been applied 
 with success in the analysis of the correlation properties 
 of quantum fluctuations produced in the early Universe (see, for instance,  \cite{sq5,sq6,sq9,sq10,sq11} and references  therein). In particular 
 a natural definition of coarse grained entropy arises in the squeezed 
 state formalism \cite{mgentr1,mgentr2}.

In what 
follows, instead of giving the full discussion of the problem in the Schr\"odinger 
representation, the squeezed states will be analyzed not in the case of a quantum 
field but in the case of a single (quantum) harmonic oscillator. This will allow to 
get an interesting physical interpretation of Eq. (\ref{BOG1}). To be more precise, the
analog of Eq. (\ref{BOG1}) can be realized with a two-mode squeezed 
state. However, to be even simpler, only one-mode squeezed states will be discussed.
Let us therefore rewrite Eq. (\ref{BOG1}) in its simplest form, namely 
\begin{equation}
\hat{b} = c(r) \hat{a} + s(r) \hat{a}^{\dagger},\qquad 
\hat{b}^{\dagger} =  s(r) \hat{a} + c(r) \hat{a}^{\dagger},
\label{UN1}
\end{equation}
where $c(r) = \cosh{r}$ and $s(r) = \sinh{r}$; $r$ is the so-called 
squeezing parameter \footnote{There should be no ambiguity in this notation. It is true that, in the present script, the variable $r$ may also denote the ratio between the tensor and the scale power spectrum arising in the consistency condition (see Eqs. (\ref{cons}) and (\ref{EXscal13})). However, since the squeezing parameter and the tensor to scalar ratio are never used in the same context, there is 
no possible confusion.}. Since $[\hat{a},\hat{a}^{\dagger}]=1$ 
and $[\hat{b},\hat{b}^{\dagger}]=1$, the transformation that allows to pass 
from the $\hat{a}$ and $\hat{a}^{\dagger}$ to the $\hat{b}$ and $\hat{b}^{\dagger}$
is clearly unitary (extra phases may appear in Eq. (\ref{UN1}) whose 
coefficients may be complex; in the present exercise we will stick to the case 
of real coefficients). Notice that, in Eq. (\ref{UN1}) the index referring to the momentum has been suppressed since we are dealing here with a single harmonic oscillator with
Hamiltonian 
\begin{equation}
\hat{H}_{a} = \frac{\hat{p}^2}{2} + \frac{\hat{x}^2}{2} = 
\hat{a}^{\dagger} \hat{a} + \frac{1}{2},
\label{UN2}
\end{equation}
where 
\begin{equation}
\hat{a} = \frac{1}{\sqrt{2}} ( \hat{x} + i \hat{p}),\qquad 
\hat{a}^{\dagger} = \frac{1}{\sqrt{2}} ( \hat{x} - i \hat{p}).
\label{UN3}
\end{equation}
Equation (\ref{UN1}) can also be written as 
\begin{equation}
\hat{b} = S^{\dagger}(z)\, \hat{a}\,S(z),\qquad 
\hat{b}^{\dagger} =  S^{\dagger}(z)\, \hat{a}^{\dagger}\,S(z),
\label{UN4}
\end{equation}
where the (unitary) operator $S(z)$ is the so-called squeezing 
operator defined as 
\begin{equation}
S(z) = e^{\frac{1}{2}(z {\hat{a}^{\dagger\,2}} - z^{*} \hat{a}^2)}.
\label{UN5}
\end{equation}
In Eq. (\ref{UN5}), in general, $ z= r e^{i\vartheta}$. In the case of Eq. (\ref{UN1}) 
$\vartheta =0$ and $z=r$.  A squeezed state is, for instance, the state 
$|z \rangle = S(z) |0\rangle$. The same kind of operator 
arises in the field theoretical description of the process of production 
of gravitons (or phonons, as we shall see, in the case of scalar 
fluctuations).  The state $|0\rangle$ is the state annihilated by $\hat{a}$ 
and minimizing the Hamiltonian (\ref{UN2}). In the coordinate representation, therefore,
the wave-function of the vacuum will be, in the coordinate representation
\begin{equation}
\psi_{0}(x)= \langle x|0\rangle = N_0 e^{- \frac{x^2}{2}},\qquad \hat{a}|0\rangle =0,
\label{UN6}
\end{equation} 
where $N_0$ is a constant fixed by normalizing to 1 the integral over $x$ of 
$|\psi_{0}(x)|^2$.  Obviously the wave-function in the $p$-representation 
will also be Gaussian. Notice that Eq. (\ref{UN6}) is simply obtained from the 
condition $\hat{a} |0\rangle =0$ by recalling Eq. (\ref{UN3}) 
(where $ \hat{p} = - i \frac{\partial}{\partial x}$). By applying the same trick, we can obtain 
the wave-function (in the coordinate representation) for the state $|z\rangle$. By 
requiring $\hat{b}|0\rangle=0$, using Eq. (\ref{UN1}) together with Eq. (\ref{UN3}) 
we will have the following simple differential equation
\begin{equation}
[c(r) + s(r)] x \psi_{z}+  [c(r) - s(r)]\frac{\partial \psi_{z}}{\partial x} =0,\qquad \psi_{z}(x)
=\langle x| z\rangle = N_{z} e^{- \frac{ x^2}{2\sigma^2 }},
\label{UN7}
\end{equation}
implying that the wave-function will still be Gaussian but with a different variance
(since $\sigma = e^{- r}$) and with a different normalization (since $N_{z} \neq N_{0}$). In this case the wave-function gets squeezed in the $x$-representation while it gets broadened in the $p$-representation in such a 
way that the indetermination relations $\Delta \hat{x} \Delta\hat{p} = 1/2$.
It should be borne in mind that the broadening (or squeezing) 
of the Gaussian wave-function(al) corresponds to a process of particle production 
when we pass, by means of a unitary transformation, from one vacuum to the other.
In fact, using Eq. (\ref{UN1}) we also have 
\begin{equation}
\langle 0| \hat{a}^{\dagger}\hat{a} |0\rangle = 0,\qquad 
\langle 0|\hat{b}^{\dagger}\hat{b} |0\rangle  = \sinh^2{r}.
\label{UN8}
\end{equation}
So, while the initial vacuum has no particles, the "new" vacuum is, really and truly, a many-particle system.  In the case of the amplification of fluctuations 
driven by the gravitational field in the early Universe the squeezing parameter is always 
much larger than one and the typical mean number of particles per Fourier mode can be 
as large as $10^{4}$--$10^{5}$.

Squeezed states, unitarily connected to the vacuum, minimize 
the indetermination relations as the well known coherent states 
introduced by Glauber (see, for instance, \cite{sq1}). The 
usual coherent states can be obtained in many different ways, but the simplest way 
to introduce them is to define the so-called Glauber displacement operator:
\begin{equation}
{\mathcal D}(\alpha) = e^{\alpha \hat{a}^{\dagger} - \alpha^{*} \hat{a}}, \qquad 
|\alpha\rangle = {\mathcal D}(\alpha) |0\rangle,
\label{UN9}
\end{equation}
where $\alpha$ is a complex number and $|\alpha\rangle$ is a 
coherent state such that $\hat{a} |\alpha\rangle =  \alpha |\alpha\rangle$.
It is clear that while the squeezing operator of Eq. (\ref{UN5}) is quadratic 
in the creation and annihilation operators, the Glauber operator is linear 
in $\hat{a}$ and $\hat{a}^{\dagger}$.
By using the Baker-Campbell-Hausdorff (BCH) formula it is possible to get, from Eq. 
(\ref{UN9}) the usual expression of a coherent state in whatever basis (such as
the Fock basis or the coordinate basis). In the coordinate basis, coherent states 
are also Gaussians but rather than squeezed they are simply not centered 
around the origin.  From the BCH formula it is possible to understand 
that the Glauber operator is simply given by the product of the generators 
of the Heisenberg algebra, i.e. $\hat{a}$, $\hat{a}^{\dagger}$ and the identity.
So the Glauber coherent states are related with the Heisenberg algebra. The squeezed 
states, arise, instead in the context of non-compact groups. 

To see this intuitively, consider the Hamiltonian 
\begin{equation}
\hat{H}_{b} = \hat{b}^{\dagger}\hat{b} + \frac{1}{2},
\label{UN10}
\end{equation}
and express it in terms of $\hat{a}$ and $\hat{a}^{\dagger}$ according to Eq. (\ref{UN1}).
The result of this simple manipulation is 
\begin{equation}
\hat{H}_{b} = \cosh{2r} \biggl(\hat{a}^{\dagger}\hat{a} + \frac{1}{2} \biggr) +
\sinh{2 r} \biggl( \frac{\hat{a}^2}{2} +  \frac{\hat{a}^{\dagger\,\,2}}{2} \biggr).
\label{UN11}
\end{equation}
But, in Eq. (\ref{UN11}), the operators 
\begin{eqnarray}
&&\biggl(\hat{a}^{\dagger}\hat{a} + \frac{1}{2} \biggr) = L_{0},
\nonumber\\
&&\frac{\hat{a}^2}{2} = L_{-},\qquad \frac{\hat{a}^{\dagger\,\,2}}{2} = L_{+},
\label{U12}
\end{eqnarray}
form a realization of the $SU(1,1)$ Lie algebra since, as it can be 
explicitly verified:
\begin{equation}
[L_{+}, L_{-}] = - 2 L_{0}, \qquad [L_{0}, L_{\pm}] = \pm L_{\pm}.
\label{U13}
\end{equation}
Note that the squeezing operator of Eq. (\ref{UN5}) 
can be written as the exponential of $L_{\pm}$ and $L_{0}$, i.e. 
more precisely
\begin{equation}
S(z) = e^{\frac{1}{2}(z {\hat{a}^{\dagger\,2}} - z^{*} \hat{a}^2)}= e^{\tanh{r}
L_{+}} \,\,e^{-\ln[\cosh{r}]L_{0}} \,\, e^{-\tanh{r}
L_{-}},
\label{U14}
\end{equation}
where the second equality follows from the BCH relation \cite{sq12,sq13}.
This is the rationale for the statement that the squeezed states are 
the coherent states associated with SU(1,1). 
Of course more complicated states can be obtained from the squeezed 
vacuum states. These have various applications in cosmology. 
For instance we can have the squeezed coherent states, i.e.
\begin{equation}
|\alpha,z\rangle = S(z) \,\, {\cal D}(\alpha)|0\rangle.
\end{equation}
Similarly one can define the squeezed number states (squeezing operator applied
to a state with $n$ particles) or the squeezed thermal states (squeezed states 
of a thermal state \cite{sq9}.
The squeezed states are also important to assess precisely the semiclassical 
limit of quantum mechanical fluctuations. On a purely formal ground 
the semiclassical limit arises in the limit $\hbar\to0$. However, on a more 
operational level, the classical limit can be addressed more physically 
by looking at the number of particles produced via the pumping action of 
the gravitational field. In this second approach the squeezed states 
represent an important tool. The squeezed state formalism therefore 
suggests that what we call {\em classical} fluctuations are the limit of 
quantum mechanical states in the same sense a laser beam is 
formed by coherent photons. Also the laser beam has a definite 
classical meaning, however, we do know that coherent light 
is different from thermal (white) light. This kind of distinction 
follows, in particular, by looking at the effects of second-order 
interference such as the Hanbury-Brown-Twiss effect \cite{sq1,sq2}.

In closing this section it is amusing to get back to the problem of entropy \cite{mgentr1,mgentr2}.
In fact, in connection with squeezed states of the tensor modes of the geometry it is 
possible to define a coarse-grained entropy in which the loss of information associated to the 
reduced density matrix is represented by an increased dispersion in a superfluctuant field operator 
which is the field-teoretical analog of the quantum mechanical momentum $\hat{p}$. 
The estimated entropy goes in this case as $r(\nu)$ (now a function of the present frequency)
i.e. as $\ln{\overline{n}_{\nu}}$ where $\overline{n}_{\nu}$ is the number of produced gravitons. 
Consequently the entropy can be estimated by integrating over all the frequencies 
of the graviton spectrum presented, for instance, in Fig. \ref{GRAV}. The result 
will be that 
\begin{equation}
S_{\mathrm{gr}} = V \int_{\nu_{0}}^{\nu_{1}} r(\nu) \,\nu^2\, d\nu \simeq (10^{29})^3 \biggl(\frac{H_{1}}{M_{\mathrm{P}}}\biggr)^{3/2}.
\label{ENTRSQ}
\end{equation}
The factor $10^{29}$ arises from the hierarchy between the  lower frequency of the spectrum (i.e. $\nu_{0} \simeq 10^{-18}\,\,
{\mathrm Hz}$) and the higher frequency (i.e. $\nu_{\mathrm{1}} \simeq 10^{11} \,\,(H_{1}/M_{\mathrm{P}})^{1/2}\,\, \mathrm{Hz}$).
From Eq. (\ref{ENTRSQ}) it follows that this gravitational entropy is of the same order of the thermodynamic entropy 
provided the curvature scale at the inflation-radiation transition is sufficiently close to the Planck scale. 

\newpage
\renewcommand{\theequation}{7.\arabic{equation}}
\setcounter{equation}{0}
\section{A primer in CMB anisotropies}
\label{sec7}
The description of CMB anisotropies will be presented by successive 
approximations. In the present section a simplistic (but still reasonable) 
account of the physics of CMB anisotropies will be derived in a two-fluid model.
In section \ref{sec8} a more realistic account of pre-decoupling physics 
will be outlined within an (improved) fluid description. In section \ref{sec9} 
the Einstein-Boltzmann hierarchy will be introduced. The following derivations will be specifically discussed. 
\begin{itemize}
\item{} the Sachs-Wolfe effect for the tensor modes of the geometry;
\item{} the Sachs-Wolfe effect for the scalar modes of the geometry;
\item{} the evolution equations of the scalar fluctuations in the 
pre-decoupling phase;
\item{} a simplified solution of the system allowing the estimate 
of the (scalar) Sachs-Wolfe contribution;
\item{} the concept of adiabatic and non-adiabatic modes.
\end{itemize}

\subsection{Tensor Sachs-Wolfe effect} 
After decoupling, the photon mean free path 
becomes comparable with the actual size of the present Hubble patch (see, for instance, 
Eq. (\ref{phex}) and derivations therein).
Consequently, the photons will travel to our detectors without suffering 
any scattering. In this situation what happens is that the photon 
geodesics are slightly perturbed by the presence of inhomogeneities 
and what we ought to compute is therefore the temperature 
fluctuation induced by scalar and tensor fluctuations. Also vector 
fluctuations may induce important sources of anisotropy but will 
be neglected in the first approximation mainly for the reason that, in the 
conventional scenario, the vector modes are always decaying 
both during radiation and, a fortiori, during the initial inflationary phase (see also 
\cite{magn1,magn2} for the case when the pre-decoupling sources support vector modes).

Since the Coulomb rate is much larger than Thompson rate of interactions around equality (see 
Eqs. (\ref{COULRAD})--(\ref{THOMRAD}) and (\ref{COULMAT})--(\ref{THOMMAT})), 
 baryons and electrons are more tightly coupled than photons 
and baryons. Still, prior to decoupling, it is rather plausible to treat the whole 
baryon-lepton-photon fluid as a unique physical entity.

Let us therefore start by studying the null geodesics in a conformally 
flat metric of FRW type where $g_{\alpha\beta} = a^{2}(\tau) \tilde{g}_{\alpha\beta}$
where $\tilde{g}_{\alpha\beta}$ coincides, in the absence of metric fluctuations, with 
the Minkowski metric. If metric fluctuations are present $\tilde{g}_{\alpha\beta}$ will have the form of a (slightly inhomogeneous) Minkowski metric. The latter observation implies that
\begin{eqnarray}
&&\tilde{g}_{00}= 1 + \delta_{\mathrm{s}} \tilde{g}_{00}, \qquad \delta_{\mathrm{s}} \tilde{g}_{00} = 2 \phi,
\label{00pert}\\
&& \tilde{g}_{ij} = -\delta_{ij} + \delta_{\mathrm{s}} \tilde{g}_{ij} +  \delta_{\mathrm{t}} \tilde{g}_{ij}, \qquad \delta_{\mathrm{s}} \tilde{g}_{ij} = 2 \psi \delta_{ij}, \qquad
\delta_{\mathrm{t}} \tilde{g}_{ij} = - h_{ij},
\label{ijpert}
\end{eqnarray}
with $\partial_{i}h^{i}_{j} =0 = h_{i}^{i} $ (see Eq. (\ref{div1})) and where the scalar fluctuations 
of the geometry have been introduced in the longitudinal (or conformally
Newtonian) gauge characterized by the two non-vanishing degrees of freedom $\phi$ and 
$\psi$ (see Eqs. (\ref{g001}) and (\ref{gij1})).

Neglecting the inhomogeneous contribution, the lowest-order geodesics of the photon in the background 
$g_{\alpha\beta} = a^2(\tau) \eta_{\alpha\beta}$ are  
\begin{eqnarray}
&&\frac{d^2 x^{\mu}}{d\lambda^2} + \Gamma_{\alpha\beta}^{\mu} 
\frac{dx^{\alpha}}{d\lambda} \frac{dx^{\beta}}{d\lambda} =0,
\label{geod1}\\
&& g_{\alpha\beta}\frac{d x^{\alpha}}{d\lambda} \frac{d x^{\beta}}{d\lambda} =0,
\label{geod2}
\end{eqnarray}
where $\lambda$ denotes the affine parameter. 
Recalling Eq. (\ref{backCR})
the $(0)$ component of Eq. (\ref{geod1})  and Eq. (\ref{geod2}) can be written, respectively,  as
\begin{eqnarray}
&& \frac{d^2 \tau}{d\lambda^2} + {\mathcal H} \biggl(\frac{d\vec{x}}{d\lambda}\biggr)^2
+ {\mathcal H}  \biggl(\frac{d \tau}{d\lambda}\biggr)^2=0.
\label{geod3}\\
&& \biggl(\frac{d\vec{x}}{d\lambda}\biggr)^2 = \biggl(\frac{d \tau}{d\lambda}\biggr)^2.
\label{null} 
\end{eqnarray}
Using Eq. (\ref{null}),  Eq. (\ref{geod3}) can be rewritten as 
\begin{equation}
\frac{d^2 F}{d\lambda^2} + 2 {\mathcal H} F^2 =0,\qquad F = \frac{d\tau}{d\lambda}.
\label{geod4}
\end{equation}
With a simple manipulation Eq. (\ref{geod4}) can be solved 
and the result will then be 
\begin{equation}
F = \frac{d\tau}{d\lambda} = \frac{1}{a^2(\tau)}.
\label{geod5}
\end{equation}
Equation (\ref{geod4}) implies that if  the affine parameter and the metric are changed  as 
\begin{equation}
d\lambda \to a^2(\tau) d\tau, \qquad g_{\alpha\beta} \to a^2(\tau) \tilde{g}_{\alpha\beta}
\label{change}
\end{equation}
the new geodesics will be exactly the same as before. In particular, as a function of $\tau$ we will have that the unperturbed geodesics will be 
\begin{equation}
x^{\mu} = n^{\mu}\tau,\qquad n^{\mu}= (n^{0},\,n^{i}),
\label{geod6}
\end{equation}
where ${n^{0}}^2 = n_{i}n^{i} =1 $. 
Consider now  the energy of the photon as measured in the reference 
frame of the baryonic fluid, i.e. \footnote{The internal energy of a thermodynamic system 
has been denoted by ${\mathcal E}$ in Appendix \ref{APPB}. Now ${\mathcal E}$ will denote 
the energy of a photon in the reference frame of the baryon fluid. These possible ambiguity 
is harmless since the two concepts will never interfere in the present treatment.}
\begin{equation}
{\mathcal E} = g_{\mu\nu} u^{\mu} P^{\nu},
\label{energyphot}
\end{equation}
where $u^{\mu}$ is the four-velocity of the fluid and $P^{\nu}$ is the photon 
momentum defined as 
\begin{equation}
P^{\mu} = \frac{d x^{\mu}}{d\lambda} = \frac{E}{a^2} \frac{dx^{\mu}}{d\tau},
\label{Pmu}
\end{equation}
where $E$ is a parameter (not to be confused with one of the off-diagonal 
entries of the perturbed metric) defining the photon energy.  
If the geodesic is perturbed by a tensor fluctuation we will have that 
Eq. (\ref{Pmu}) becomes 
\begin{equation}
P^{\mu} = \frac{E}{a^2} \biggl[ n^{\mu} + \frac{d \delta_{\mathrm{t}} x^{\mu}}{d\tau} \biggr].
\label{Pmu2}
\end{equation}
Since the condition $u^{\mu} u^{\nu} g_{\mu\nu} =1$ implies that $u^{0} =1/a$,  Eqs. (\ref{energyphot}) 
and (\ref{Pmu2}) lead, respectively, to the following two more explicit expressions:
\begin{equation}
{\mathcal E} = \frac{E}{a} \biggl[ 1 +\frac{d \delta_{\mathrm{t}} x^{0}}{d\tau} \biggr],\qquad\frac{d^{2} \delta_{\mathrm{t}} x^{0}}{d\tau^2} = - \delta \tilde{\Gamma}_{ij}^{0} n^{i} n^{j}.
\label{enphottens}
\end{equation}
The quantity $\delta_{\mathrm{t}} \tilde{\Gamma}_{ij}^{0}$ can be computed from $\delta_{\mathrm{t}} \Gamma_{ij}^{0}$
(see, in particular, the first expression in Eq. (\ref{TCHR})) by setting ${\mathcal H} =0$. The result is: 
\begin{equation}
\delta_{\mathrm{t}} \tilde{\Gamma}_{ij}^{0} = \frac{1}{2} h_{ij}'.
\end{equation}
Consequently Eq. (\ref{enphottens}) can be be rearranged as  
\begin{eqnarray}
&&\frac{d \delta_{\mathrm{t}} x^{0}}{d\tau} = -\frac{1}{2} \int_{\tau_{\mathrm{i}}}^{\tau_{\mathrm{f}}} h_{ij}' \, n^{i} \, n^{j} d\tau,
\label{tens1}\\
&&{\mathcal E} = \frac{E}{a} \biggl[ 1 - \frac{1}{2}  \int_{\tau_{\mathrm{i}}}^{\tau_{\mathrm{f}}} h_{ij}' \, n^{i} \, n^{j} d\tau\biggr],
\label{tens2}
\end{eqnarray}
where $\tau_{\mathrm{i}} = \tau_{0}$ and $\tau_{\mathrm{f}} = \tau_{\mathrm{dec}}$.
The temperature fluctuation  due to the tensor modes of the geometry 
can then be computed as 
\begin{equation}
\biggl( \frac{\Delta T}{T} \biggr) = \frac{a_{\mathrm{f}} {\mathcal E}_{\mathrm{f}} - a_{\mathrm{i}} {\mathcal E}_{\mathrm{i}}}{a_{\mathrm{i}} {\mathcal E}_{\mathrm{i}}}.
\label{tens3}
\end{equation}
By making use of Eq. (\ref{tens2}), Eq. (\ref{tens3}) simply becomes:
\begin{equation}
\biggl(\frac{\Delta T}{T}\biggr)_{\rm t} = - \frac{1}{2} \int_{\tau_i}^{\tau_{f}} h_{ij}' n^{i} n^{j} d\tau. 
\end{equation}
The only contribution to the tensor  Sachs-Wolfe effect is given by the Sachs-Wolfe integral. It is clear that since during the matter dominated stage the evolution of the 
tensor modes is not conformally invariant there will be a tensor contribution to the SW 
effect. The absence of positive detection places bounds on the possible 
existence of a stochastic background of gravitational radiation 
for present frequencies of the order of $10^{-18}$ Hz (see Fig. \ref{GRAV}).

\subsection{The scalar Sachs-Wolfe effect}

According to the observations discussed in Eqs. (\ref{ijpert}),
in the case of the scalar modes of the geometry the perturbed geodesics can be written as 
\begin{equation}
\frac{d^{2}\delta  x^{\mu}}{d\tau^2 } + \delta_{\rm s} 
\tilde{\Gamma}^{\mu}_{\alpha\beta} \frac{d x^{\alpha}}{d\tau} \frac{d x^{\beta}}{d\tau} = 0,
\label{geods0}
\end{equation}
where now $\delta_{\mathrm{s}}$ denotes the scalar fluctuation of the Christoffel 
connection computed with respect to $\tilde{g}_{\alpha\beta}$ which 
is the inhomogeneous Minkowski metric. In the conformally Newtonian 
gauge the fluctuations of the Christoffel connections of a perturbed Minkowski metric 
are \footnote{These expressions can be obtained from Eqs. (\ref{christ}) and (\ref{christ2}) of Appendix \ref{APPC}) by setting 
${\mathcal H} =0$.}:
\begin{equation}
\delta_{\mathrm{s}} \tilde{\Gamma}_{00}^{0} = \phi',\qquad 
\delta_{\mathrm{s}} \tilde{\Gamma}_{0i}^{0} = \partial_{i} \phi,\qquad
\delta_{\mathrm{s}} \tilde{\Gamma}_{ij}^{0} = -\psi'.
\label{chconn}
\end{equation}
Thus, using Eq. (\ref{chconn}), Eq. (\ref{geods0}) becomes
\begin{equation}
\frac{d}{d\tau} \biggl[ \frac{d\delta_{\mathrm{s}} x^{0}}{d\tau}\biggr] = - \delta_{\rm s} \tilde{\Gamma}_{00}^{0} n^{0} n^{0} - \delta_{\rm s} \tilde{\Gamma}_{ij}^{0} n^{i} n^{j}- 
2 \delta_{\rm s} \tilde{\Gamma}_{i0}^{0} n^{i} n^{0} \equiv \psi' - \phi' - 2 \partial_{i}\phi n^{i}.
\label{geods1}
\end{equation}
Since
\begin{equation}
\frac{d \phi}{d\tau} = \phi' + n^{i} \partial_{i} \phi,
\label{int1geods}
\end{equation}
we will also have that 
\begin{equation}
 \frac{d\delta x^{0}}{d\tau} = \int_{\tau_{i}}^{\tau_{f}} (\psi' + \phi') d\eta - 2 \phi.
\label{deltax0}
\end{equation}

The quantity to be computed, as previously anticipated, 
 is the photon energy as measured in the frame of reference 
of the fluid. 
Defining $u^{\mu}$ as the four-velocity 
of the fluid and $P^{\nu}$ as the photon four-momentum, 
 the photon energy will exactly be the one given in Eq. (\ref{energyphot})  
 but with a different physical content for $P^{\nu}$ and $u^{\mu}$. The rationale 
 for this statement is that while the tensor modes 
 do not contribute to $u^{\mu}$, the scalar modes affect the $0$-component
 of $u^{\mu}$.
The logic will now be to determine $u^{\mu}$, $g_{\mu\nu}$ and $P^{\mu}$ to first 
order in the scalar fluctuations of the geometry. This will allow to compute the 
right hand side of Eq. (\ref{energyphot})  as a function of the inhomogeneities of the metric.
The first-order variation of $g^{\mu\nu} u_{\mu} u_{\nu} =1$  leads to
\begin{equation}
\delta_{\rm s}  g_{00} u^{0} = -2 \delta_{\rm s} u^{0} g_{00},
\label{umu2}
\end{equation}
so that in the longitudinal coordinate system 
Eq. (\ref{umu2}) gives, to first-order in the metric fluctuations, 
\begin{equation}
u^{0} = \frac{1}{a}( 1 - \phi).
\label{u0}
\end{equation}
The divergencefull peculiar  velocity field is given by
\begin{equation}
\delta_{\rm s} u^{i} = \frac{v^{i}}{a} \equiv \frac{1}{a} \partial^{i} v.
\label{ui}
\end{equation}
The relevant peculiar velocity field will be, in this derivation, the baryonic 
peculiar velocity since this is the component emitting and observing (i.e. absorbing)
the radiation. In the following this identification will be undesrtood and, hence,
$v^{i} = v^{i}_{\rm b}$. 

The energy of the photon in the frame of reference of the fluid becomes, then 
\begin{equation}
{\cal E} =  g_{\mu\nu} u^{\mu} P^{\nu} = g_{00} u^{0} P^{0} + g_{i j} u^{i} P^{j}.
\label{photen}
\end{equation}
Recalling 
the explicit forms of $g_{00}$ and $g_{ij}$ to first order in the 
metric fluctuations we have 
\begin{equation}
{\cal E} = \frac{E}{a}\biggl[ 1 + \phi - n_{i}v^{i}_{\rm b} + \frac{d \delta x^{0}}{d\tau} \biggr].
\label{SW1}
\end{equation}
Assuming, as previously stated,  that the observer, located at 
the end of a photon geodesic, is  at $\vec{x}=0$, Eq. (\ref{SW1}) can be expressed as 
\begin{equation}
{\cal E} = \frac{E}{a} \biggl\{ 1 - \phi - n_{i} v^{i}_{\rm b}+ \int_{\tau_{i}}^{\tau_{f}} (\psi' + \phi') d\tau\biggr\}.
\label{SW2}
\end{equation}
The temperature fluctuation can be expressed by taking the difference between the final 
and initial energies, i.e. 
\begin{equation}
\frac{\delta T}{T} =
 \frac{ a_{\mathrm{f}} {\mathcal E}(\tau_{\mathrm{f}}) - 
a_{\mathrm{i}}{\mathcal E}_{\mathrm{i}}}{a_{\mathrm{i}}{\mathcal E}_{\mathrm{i}}}.
\label{SW3}
\end{equation}
The final and initial photon energy are also affected by an intrinsic contribution, i.e.
\begin{equation}
\frac{a_{\mathrm{f}}E_{\mathrm{f}}}{a_{\mathrm{i}} E_{\mathrm{i}}} = \frac{T_{0} - \delta T_{\mathrm{f}}}{T_{i} - \delta T_{\mathrm{i}}}
\label{INTSWS1}
\end{equation}
where we wrote that 
\begin{equation}
T_{0} = T_{\mathrm{f}} + \delta_{\mathrm{s}} T_{\mathrm{f}},\qquad T_{\mathrm{dec}} = T_{\mathrm{i}} + \delta_{\mathrm{s}} T_{\mathrm{i}}.
\end{equation}
Now the temperature variation at the present epoch can be neglected while the 
intrinsic temperature variation at the initial time (i.e. the last scattering surface) cannot be 
neglected and it is given by 
\begin{equation}
\frac{\delta_{\mathrm{s}} T_{\mathrm{i}}}{T_{\mathrm{i}}} = \frac{\delta_{\gamma}}{4}(\tau_{\mathrm{i}}),
\end{equation}
since $\rho_{\gamma}(\tau_{\mathrm{i}}) \simeq T_{\mathrm{i}}^4$; $\delta_{\gamma}$ 
is the fractional variation of photon energy density.

The final expression for the SW effect induced by scalar fluctuations can be written as 
\begin{equation}
\biggl(\frac{\Delta T}{T}\biggr)_{\rm s} = \frac{\delta_{\rm r}(\tau_{i})}{4} - [\phi]_{\tau_{i}}^{\tau_{f}} -
 [n_{i} v^{i}_{\rm b}]_{\tau_{i}}^{\tau_{f}} + 
\int_{\tau_{i}}^{\tau_{f}} ( \psi' + \phi') d\tau.
\label{SW0}
\end{equation}
Sometimes, for simplified esitmates, the temperature fluctuation can then be written, in explicit terms,  as
\begin{equation}
\biggl(\frac{\Delta T}{T}\biggr)_{\rm s} = \biggr[\frac{\delta_{\rm r}}{4} + \phi + n_{i} v^{i}_{\rm b}\biggr]_{\tau_{i}} + 
\int_{\tau_{i}}^{\tau_{f}}( \psi' +\phi') d\tau. 
\label{scalSW}
\end{equation}

Equation  (\ref{scalSW}) has three contribution
\begin{itemize}
\item{} the ordinary SW effect given by the first two terms at the righ hand side of Eq. (\ref{scalSW}) 
i.e. $\delta_{\rm r}/4$ and $\phi$;
\item{} the Doppler term (third term in Eq. (\ref{scalSW}));
\item{} the integrated SW effect (last term in Eq. (\ref{scalSW})).
\end{itemize}
The ordinary SW effect is due both to the intrinsic temperature  inhomogeneities on the last 
scattering surface and to the inhomogeneities of the metric.  On large angular scales 
the ordinary SW contribution dominates. The Doppler term arises thanks to the relative 
velocity of the emitter and of the receiver. At large angular scales its contribution is subleading 
but it becomes important at smaller scales, i.e. in multipole space, for $\ell \sim 200$
corresponding to the first peak in Fig. \ref{F4}.
The SW integral contributes to the temperature anisotropy if $\psi$ and $\phi$ depend on time.

\subsection{Scalar modes in the pre-decoupling phase}
To estimate the various scalar Sachs-Wolfe contributions, the evolution of the 
metric fluctuations and of the perturbations of the sources must be discussed when the background 
is already dominated by matter, i.e. after equality but before decoupling. 
In this regime, for typical wavelengths larger than the Hubble radius at the corresponding epoch, 
the primeval fluctuations produced, for instance, during inflation, will serve as initial conditions 
for the fluctuations of the various plasma variables such as the density contrasts and the peculiar velocities (see 
section \ref{sec10} for a discussion of the amplification of scalar modes during inflation).
As already mentioned in different circumstances the wavelengths of the fluctuations are 
larger than the Hubble radius provided the corresponding wave-numbers satisfy the condition $k\tau <1$ where 
$\tau$ is the conformal time coordinate that has been consistently employed throughout the whole 
discussion of inhomogeneities in FRW models.
The first step along this direction is to write down the evolution equations 
of the metric perturbations which will now be treated in the longitudinal gauge.
For the moment the explicit expression for the fluctuations of the matter 
sources will be left unspecified.

The perturbed components the Christoffel connections 
are obtained  in Appendix \ref{APPC} (see Eqs. (\ref{christ}) and (\ref{christ2})). 
The perturbed form of the components of the Ricci tensor  can be readily 
obtained (see Eqs. (\ref{riccisf}) and (\ref{mixedR}). Finally, from the first-order 
form of the components of the Einstein tensor (see Eqs. (\ref{dg00}), (\ref{dgij}) and 
(\ref{dg0i})), the perturbed Einstein equations can then  be formally written as:
\begin{eqnarray}
&& \delta_{\rm s} {\cal G}_{0}^{0} = 8 \pi G \delta_{\rm s} T_{0}^{0},
\label{l00a}\\
&& \delta_{\rm s} {\cal G}_{i}^{j} = 8 \pi G \delta_{\rm s} T_{i}^{j},
\label{lija}\\
&& \delta_{\rm s} {\cal G}_{0}^{i} = 8\pi G \delta_{\rm s} T_{0}^{i}.
\label{l0ia}
\end{eqnarray}
The fluctuations of total the energy-momentum tensor are 
written as the sum of the fluctuations over the various 
species composing the plasma, i.e. according to Eqs. (\ref{dT00})--(\ref{dT0i2}),
\begin{eqnarray}
&& \delta_{\rm s} T_{0}^{0} = \delta \rho = \sum_{\lambda} \rho_{\lambda} \delta_{\lambda},
\label{dT00la}\\
&& \delta_{\rm s} T_{i}^{j} = -\delta_{i}^{j} \delta p + \Pi_{i}^{j}= 
- \delta_{i}^{j} \sum_{\lambda} w_{\lambda} \rho_{\lambda} \delta_{\lambda} + \Pi_{i}^{j},
\label{dTija}\\
&& \delta_{\rm s} T_{0}^{i} = (p + \rho) v^{i} = \sum_{\lambda} ( 1 + w_{\lambda}) \rho_{\lambda} v^{i}_{\lambda}.
\label{dToia}
\end{eqnarray}
Concerning Eqs. (\ref{dT00la})--(\ref{dToia}) few comments are in order:
\begin{itemize}
\item{}the sum over $\lambda$ runs, in general, over the different 
species of the plasma (in particular, photon, baryons, neutrinos and CDM particles);
\item{} $\delta_{\lambda}$ denotes the density contrast  for each single species of the plasma 
(see Eq. (\ref{denscontr}) for the properties of $\delta_{\lambda}$ under infinitesimal gauge transformations);
\item{} the term $\Pi_{i}^{j}$ denotes the contribution of the anisotropic stress to the spatial components 
of the (perturbed) energy-momentum tensor of the fluid mixture.
\end{itemize}
Concerning the last point of this list, it should be borne in mind that the relevant 
physical situation is the one where Universe evolves for temperatures 
smaller than the MeV. In this case neutrinos 
have already decoupled and form a quasi-perfect (collisionless) fluid.  For 
this reason, neutrinos will be the dominant source of anisotropic stress 
and such a contribution will be directly proportional to the quadrupole 
moment of the neutrino phase-space distribution. 

In what follows the full content of the plasma will be drastically reduced with the 
purpose of obtaining (simplified) analytical estimates. In sections \ref{sec8} and 
(\ref{sec9}) the simplifying assumptions adopted 
in the present section will be relaxed. While the dynamics of the whole  system
will emerge as more complex, it is the hope of the author that its simple 
properties will still be evident.

Equations (\ref{l00a})--(\ref{l0ia}) lead then to the following system 
of equations:
\begin{eqnarray}
&&\nabla^2 \psi- 3 {\cal H} ( {\cal H} \phi + \psi') = 4\pi G a^2 \delta \rho,
\label{p00l}\\
&& \nabla^2( {\cal H} \phi + \psi') =- 4 \pi G a^2 (p + \rho) \theta,
\label{p0il}\\
&& \biggl[ \psi'' + {\cal H} ( 2\psi' +\phi') + ( 2 {\cal H}' + {\cal H}^2) \phi + 
\frac{1}{2} \nabla^2(\phi - \psi) \biggr] \delta_{i}^{j}  
\nonumber\\
&& - \frac{1}{2} \partial_{i}\partial^{j} ( \phi - \psi) = 
 4\pi G a^2 (\delta p \delta_{i}^{j} - \Pi_{i}^{j}).
\label{pijl}
\end{eqnarray}
where the divergence of the total velocity field has been defined as:
\begin{equation}
( p + \rho) \theta = \sum_{\lambda} ( p_{\lambda} + \rho_{\lambda})\theta_{\lambda},
\label{thetadefinition}
\end{equation}
with $\theta = \partial_{i} v^{i}$ and 
$\theta_{\lambda}= \partial_{i} v^{i}_{\lambda}$. Equations  (\ref{p00l}) and 
 (\ref{p0il}) are, respectively, the Hamiltonian and the 
momentum constraint. The enforcement of these two constraints is
crucial for the regularity  of the initial conditions. 
Taking the trace of Eq. (\ref{pijl}) and recalling that 
the anisotropic stress is, by definition,  traceless (i.e. $\Pi_{i}^{i}=0$) 
it is simple to obtain 
\begin{equation}
\psi'' + 2 {\cal H} \psi' + {\cal H} \phi' + ( 2 {\cal H}' + {\cal H}^2)\phi + 
\frac{1}{3} \nabla^2 (\phi - \psi) = 4 \pi G a^2 \delta p.
\label{B1}
\end{equation}
The difference of Eqs. (\ref{pijl}) and (\ref{B1}) leads to 
\begin{equation}
\frac{1}{6} \nabla^2( \phi - \psi)\delta_{i}^{j}  - \frac{1}{2} \partial_{i} 
\partial^{j} ( \phi - \psi) = - 4 \pi a^2 G \, \Pi_{i}^{j}.
\label{B2}
\end{equation}
By now applying to both sides of Eq. (\ref{B2}) the differential 
operator $\partial_{j}\partial^{i}$ we are led to the following 
expression 
\begin{equation}
\nabla^4(\phi- \phi) = 12\pi G a^2 \partial_{j}\partial^{i} \Pi_{i}^{j}.
\label{B2a}
\end{equation}
The right hand side of Eq. (\ref{B2a}) can be usefully parametrized as 
\begin{equation}
\partial_{j}\partial^{i} \Pi_{i}^{j} = \sum_{\lambda} (p_{\lambda} + \rho_{\lambda}) \nabla^2 \sigma_{\lambda}.
\label{B2b}
\end{equation}
This parametrization may now appear baroque but it is helpful since $\sigma_{\lambda}$, in the case of neutrinos, is easily 
related to the quadrupole moment of the (perturbed) neutrino phase space distribution (see section \ref{sec9}).
Equations (\ref{p00l})--(\ref{pijl}) may be 
supplemented with the perturbation of the 
covariant conservation of the energy-momentum tensor \footnote{See Appendix \ref{APPC} (and, in particular, 
Eqs. (\ref{pertconsgen1}), (\ref{T0a}) and (\ref{T0b})) for further details 
on the derivation.}:
\begin{eqnarray}
\delta_{\lambda}' &=& ( 1 + w_{\lambda}) ( 3 \psi' - \theta_{\lambda}) + 
3 {\cal H} \biggl[ w_{\lambda} - 
\frac{\delta p_{\lambda}}{\delta \rho_{\lambda}}\biggr] \delta_{\lambda}, 
\label{delta}\\
\theta_{\lambda}' &=& ( 3 w_{\lambda} -1) {\cal H} \theta_{\lambda} - 
\frac{w_{\lambda}'}{w_{\lambda} + 1} 
\theta_{\lambda} - \frac{1}{w_{\lambda}+1} \frac{\delta 
p_{\lambda}}{\delta \rho_{\lambda}} \nabla^2 \delta_{\lambda} 
\nonumber\\
&+& \nabla^2 
\sigma_{\lambda} - \nabla^2 \phi.
\label{theta}
\end{eqnarray}
In Eq. (\ref{theta}) $\sigma_{\lambda}$ appears and it stems directly 
from the correct fluctuation of the spatial components of energy-momentum 
tensor of the fluid mixture.
In Eqs. (\ref{delta}) and (\ref{theta}) the energy and momentum 
exchange has been assumed to be negligible between the various components of the plasma. This approximation 
is not so realistic as far as the baryon-photon system is concerned as it will be explained in section \ref{sec8}.

\subsection{CDM-radiation system}
The content of the plasma 
is formed by four different species, namely dark matter particles, 
photons, neutrinos and baryons. The following simplifying 
assumptions will now be proposed:
\begin{itemize}
\item{}  the pre-decoupling plasma is only formed by a radiation 
component (denoted by a subscript r)  and by a CDM component (denoted by a subscript c);
\item{} neutrinos will be assumed to be a component of the radiation fluid but their anisotropic stress 
will be neglected: hence, according to Eqs. (\ref{B2}) and (\ref{B2a}), the two 
longitudinal fluctuations of the metric will then be equal, i.e. $\phi =\psi$;
\item{} the energy-momentum exchange between photons and baryons will be neglected.
\end{itemize}
These three assumptions will allow two interesting exercises:
\begin{itemize}
\item{} a simplified estimate of the large-scale Sachs-Wolfe contribution;
\item{} a simplified introduction to the distinction between adiabatic and non-adiabatic modes.
\end{itemize}
Since the neutrinos are absent, then we can set $\psi =\phi$ in 
Eqs. (\ref{p00l}), (\ref{p0il}) and (\ref{pijl}) whose explicit form 
will become, in Fourier space
\begin{eqnarray}
&&-k^2 \psi - 3 {\mathcal H}({\mathcal H} \psi + \psi') = 4\pi G a^2 \delta \rho,
\label{PS1}\\
&& \psi'' + 3 {\mathcal H} \psi' + ({\mathcal H}^2  + 2 {\mathcal H}') \psi = 4\pi G a^2 \delta p,
\label{PSI2}\\
&& k^2 ( {\mathcal H}\psi + \psi') = 4\pi G a^2 (p + \rho) \theta.
\label{PSI3}
\end{eqnarray}
Since the only two species of the plasma are, in the present discussion, radiation and 
CDM particles we will have, according to Eqs. (\ref{dT00la}) and (\ref{thetadefinition})
\begin{equation}
(p + \rho ) \theta = \frac{4}{3} \rho_{\mathrm{r}} \theta_{\mathrm{r}} + \rho_{\mathrm{c}} \theta_{\mathrm{c}},
\qquad \delta\rho = \rho_{\mathrm{r}} \delta_{\mathrm{r}} + \rho_{\mathrm{c}} \delta_{\mathrm{c}}.
\end{equation}
There are two different regimes where this system can be studied, i.e. either 
before equality or after equality. Typically, as we shall see, initial conditions 
for CMB anisotropies are set deep in the radiation-dominated regime. However, in the 
present example we will solve the system separately during 
the radiation and the matter-dominated epochs.
During the radiation-dominated epoch, i.e. prior to $\tau_{\mathrm{eq}}$ the evolution 
equation for $\psi$ can be solved exactly by noticing that, in this regime, 
$3\delta p = \delta\rho$. Thus, by linearly combining Eqs. (\ref{PS1}) and (\ref{PSI2}) in such a way 
to eliminate the contribution of the fluctuations of the energy density and of the pressure,
 the (decoupled) evolution equation for $\psi$ can be written as 
\begin{equation}
  \psi'' + 4 {\cal H} \psi' + \frac{k^2}{3} \psi = 0.
  \label{psirad}
  \end{equation}
  Since during radiation $a(\tau)\sim \tau$, Eq. (\ref{psirad}) can be solved as a combination 
  of Bessel functions of order $3/2$ which can be expressed, in turn, as a combination 
  of trigonometric functions weighted by inverse powers of their argument \cite{abr,tric}
  \begin{equation}
  \psi(k,\tau) = A_{1}(k) \frac{ y \cos{y} - \sin{y}}{y^3} + B_{1}(k) \frac{y \sin{y}  + \cos{y}}{y^3},
  \label{psiradsol}
  \end{equation}
  where $ y = k\tau/\sqrt{3}$.
 If the solution parametrized by the arbitrary constant $A_{1}(k)$, $\psi\to \psi_{\rm r}$ for $k\tau \ll 1$
 ( where $\psi_{\rm r}$ is a constant). This is the case of purely adiabatic initial conditions. If, on the 
 contrary, $A_{1}(k)$ is set to zero, then $\psi$ will not go to a constant. This second solution is important 
 in the case of the non-adiabatic modes. At the end of this section the physical distinction 
 between adiabatic and non-adiabatic modes will be specifically discussed.

 In the case of adiabatic fluctuations the constant mode $\psi_{\rm r}$ matches to a constant mode 
 during the subsequent matter dominated epoch. In fact, during the matter dominated epoch 
 and under the same assumptions of absence of anisotropic stresses the equation 
 for $\psi$ is
 \begin{equation}
 \psi'' + 3 {\cal H} \psi' =0.
 \label{psimat}
 \end{equation}
 Since, after equality, $a(\tau)\sim \tau^2$, the solution of Eq. (\ref{psimat}) is then 
 \begin{equation}
 \psi(k,\tau) = \psi_{\rm m} + D_{1}(k) \biggl(\frac{\tau_{\rm eq}}{\tau} \biggr)^{5},
 \label{psimatsol}
 \end{equation}
 where $\psi_{\rm m}$ is a constant. The values 
 of $\psi_{\rm r}$ and $\psi_{\rm m}$ are different but can be easily connected. In fact 
 we are interested in wave-numbers $k\tau < 1$ after equality and, in this regime
  \begin{equation}
 \psi_{\rm m} = \frac{9}{10} \psi_{\rm r}.
 \label{psimpsir}
 \end{equation}
 Disregarding the complication 
 of an anisotropic stress (i.e., from Eqs. (\ref{B2})--(\ref{B2a}), $\phi =\psi$)
from Eqs. (\ref{delta})--(\ref{theta}) the covariant conservation equations become
 \begin{eqnarray}
 && \delta_{\rm c}' = 3 \psi' - \theta_{\rm c},
 \label{Cdelc}\\
 && \theta_{\rm c}' = - {\cal H} \theta_{\rm c} + k^2 \psi,
 \label{Cthc}\\
 && \delta_{\rm r}' = 4 \psi' - \frac{4}{3} \theta_{\rm r},
 \label{Cdelr}\\
 && \theta_{\rm r}' = \frac{k^2}{4} \delta_{\rm r} + k^2 \psi.
 \label{Cthr}
 \end{eqnarray}
Combining Eqs. (\ref{Cdelr}) and (\ref{Cthr})  
in the presence of the constant adiabatic mode $\psi_{\rm m}$ and during the 
 matter-dominated phase  
 \begin{equation}
 \delta_{\rm r}'' + k^2 c_{\rm s}^2 \delta_{\rm r} = - 4 c_{\rm s}^2 k^2 \psi_{\rm m},
 \label{harmonic}
 \end{equation}
 where  $ c_{\rm s} = 1/\sqrt{3}$.
 The solution of Eq. (\ref{harmonic}) can be obtained with elementary methods. In 
 particular it will be 
 \begin{eqnarray}
 \delta_{\mathrm{r}} &=& c_1 \cos{k c_{\mathrm{s}}\tau} +c_{2} \sin{k c_{\mathrm{s}}\tau}
 \nonumber\\
 &-& 4 c_{\rm s}^2 k^2 \psi_{\rm m} \int_{0}^{\tau} d\xi [ \cos{k c_{\mathrm{s}}\xi}
 \sin{k c_{\mathrm{s}}\tau} - \sin{k c_{\mathrm{s}}\xi}\cos{k c_{\mathrm{s}}\tau}].
 \end{eqnarray}
  The full solution of Eqs. (\ref{Cthc})--(\ref{Cthr})  will then be:
 \begin{eqnarray}
 && \delta_{\rm c} = - 2 \psi_{\rm m} - \frac{\psi_{\rm m}}{6} k^2 \tau^2,
 \label{deltam}\\
 && \theta_{\rm c} = \frac{k^2 \tau}{3} \psi_{\rm m},
 \label{thetam}\\
 && \delta_{\rm r} = \frac{4}{3} \psi_{\rm m} [ \cos{(k c_{\rm s}\tau)} - 3],
 \label{deltar}\\
 && \theta_{\rm r} = \frac{k \psi_{\rm m}}{\sqrt{3}} \sin{(k c_{\rm s} \tau)}.
 \label{thetar}
 \end{eqnarray}
Notice that:
\begin{itemize}
\item{} for $k\tau \ll 1$, $\theta_{\rm c} \simeq \theta_{\rm r}$;
\item{}  for $k\tau \ll 1$, $\delta_{\mathrm{r}}= 4 \delta_{\mathrm{c}}/3$.
\end{itemize}
These relations have a rather interesting physical interpretation that will be scrutinized at 
the end of the present section.
  The ordinary SW effect can now be roughly estimated.  
 Consider Eq. (\ref{SW0}) in the case 
 of the pure adiabatic mode. Since the longitudinal degrees of freedom of the metric are roughly constant
 and equal, inserting the solution of Eqs. (\ref{deltam})--(\ref{thetar}) into  
 Eq. (\ref{SW0}) the following result can be obtained 
  \begin{equation}
 \biggl( \frac{\Delta T}{T}\biggr)^{\rm ad}_{k, {\rm s}} = 
 \biggl(\frac{\delta_{\rm r}}{4} + \psi\biggl)_{\tau\simeq \tau_{\rm dec}}
  \equiv  \frac{\psi_{\rm m}}{3} \cos{(k \,c_{\rm s}\,\tau_{\rm dec})} = \frac{3}{10} \psi_{\rm r}
   \cos{(k\, c_{\rm s} \, \tau_{\rm dec})},
  \label{SWscalad}
 \end{equation}
where the third equality follows from the relation between the constant modes during radiation 
and matter, i.e. Eq. (\ref{psimpsir}).  Concerning Eq. (\ref{SWscalad}) few comments 
are in order:
\begin{itemize}
\item{} for superhorizon modes the baryon peculiar velocity does not contribute to the leading 
result of the SW effect;
\item{} for $k\,c_{\rm s}\,\tau_{\rm dec} \ll 1$ the temperature fluctuations induced by the adiabatic mode are simply $\psi_{\rm m}/3$ ;
\item{} even if more accurate results on the temperature fluctuations on small angular scales 
can be obtained from a systematic expansion in the inverse of the differential optical depth (tight 
coupling expansion, to be discussed in section 8), Eq. (\ref{SWscalad}) suggests that 
the first true peak  in the temperature fluctuations is located at $k c_{\rm s} \eta_{\rm dec} \simeq \pi$.
\end{itemize}
In this 
discussion, the r\^ole of the baryons has been completely neglected. In section 
\ref{sec8} a more refined 
picture of the acoustic oscillations will be developed and it will be shown that the inclusion of baryons induces a shift of the first Doppler peak.

\subsection{Adiabatic and non-adiabatic modes}

The solution obtained in Eqs. (\ref{deltam}), (\ref{thetam}), (\ref{deltar}) and (\ref{thetar}) obeys, in the limit $k\tau \ll 1$ the 
following interesting  condition
\begin{equation}
\delta_{\mathrm{r}} = \frac{4}{3} \delta_{\mathrm{c}}.
\label{adcon}
\end{equation}
A solution obeying Eq. (\ref{adcon}) for the radiation 
and matter density contrasts is said to be {\em adiabatic}. 
A distinction playing a key r\^ole in the theory of the CMB 
anisotropies is the one between {\em adiabatic} and {\em isocurvature} \footnote{To avoid misunderstandings  it would be more appropriate to use the 
terminology non-adiabatic since the term isocurvature may be interpreted 
as denoting a fluctuation giving rise to a uniform curvature. In the 
following the common terminology will be however used.}
modes. Consider, again, the idealized 
case of a plasma where the only  fluid variables are the ones 
associated with CDM particles and radiation.
The entropy per dark matter particle will then be given by 
$ \varsigma = T^3/n_{\rm c}$ where $n_{\rm c}$ is the number density 
of CDM particles and $\rho_{\rm c} = m_{\rm c} n_{\rm c}$ is the associated
 energy density.
Recalling that $\delta_{\rm r} = \delta \rho_{\rm r}/\rho_{\rm r}$ and $\delta_{\rm c} = \delta \rho_{\rm c}/\rho_{\rm c}$ 
are, respectively the density contrast in radiation and in CDM, the fluctuations of the specific entropy will then be 
\begin{equation}
{\cal S} = \frac{\delta{\varsigma}}{{\varsigma}} = 3 \frac{\delta T}{T} - \delta_{\rm c}=  \frac{3}{4} \delta_{\rm r} 
- \delta_{\rm c},
\label{SE}
\end{equation}
where the second equality follows recalling that $\rho_{\rm r} \propto T^4$. 
If the fluctuations in the specific entropy vanish, at large-scales, then a 
chacteristic relation 
between the density contrasts of the various plasma quantities appears, i.e. 
for a baryon-photon-lepton fluid with CMD particles,
\begin{equation}
\delta_{\gamma }\simeq \delta_{\nu} \simeq \frac{4}{3} \delta_{\rm c} 
\simeq \frac{4}{3} \delta_{\rm b}.
\label{ad}
\end{equation}
Eq. (\ref{SE}) can be generalized to the case of a mixture of different fluids with
arbitrary equation of state. For 
instance, in the case of two fluids ${\rm a}$ and ${\rm b}$ with barotropic indices 
$w_{\rm a}$ and $w_{\rm b}$ the fluctuations in the specific entropy are
\begin{equation}
{\cal S}_{{\rm a\,b}} = \frac{\delta_{\rm a}}{1 + w_{{\rm a}}}  - \frac{\delta_{\rm b}}{1 +  w_{{\rm b}}},
\label{SEgen}
\end{equation}
where $\delta_{\rm a}$ and $\delta_{\rm b}$ are the density contrasts of the two species.
It is appropriate to stress that, according to Eq. (\ref{denscontr}), giving the gauge variation 
of the density contrast of a given species,  ${\cal S}_{{\rm a\,b}}$ is gauge-invariant (see, in fact, 
Eq. (\ref{denscontr})).
As a consequence of the mentioned  distinction the total pressure density can be connected to the total fluctuation of the energy density as 
\begin{equation}
\delta p= c_{\rm s}^2 \delta \rho + \delta p_{\rm nad},
\label{dpnad}
\end{equation}
where 
\begin{equation}
c_{\rm s}^2 = \biggl(\frac{\delta p}{\delta\rho}\biggr)_{\varsigma} = 
\biggl(\frac{p'}{\rho'}\biggr)_{\varsigma},
\label{cs2}
\end{equation}
is the speed of sound {\em computed from the variation 
of the total pressure and energy density at constant specific entropy}, i.e. 
$\delta \varsigma =0$. The second term appearing 
in Eq. (\ref{dpnad}) is the pressure density variation 
produced by the fluctuation in the specific entropy 
at  constant energy density, i.e.
\begin{equation}
\delta p_{\rm nad} = 
\biggl( \frac{\delta p}{\delta \varsigma }\biggr)_{\rho} \delta \varsigma,
\label{deltapnaddef1}
\end{equation}
accounting for the 
non-adiabatic contribution to the total pressure perturbation.
 
If only one species is present with equation of state 
$p = w \rho$, then it follows from the 
definition that $ c_{s}^2 =w$ and the 
non-adiabatic contribution vanishes. 
As previously anticipated around Eq. (\ref{SE}), a 
 sufficient condition in order to have $\delta p_{\rm nad} \neq 0$
is that the fluctuation in the specific entropy $\delta \varsigma$ 
is not vanishing. Consider, for simplicity, the case of a plasma 
made of radiation and CDM particles. In this case the speed of 
sound and the non-adiabatic contribution can be easily computed 
and they are:
\begin{eqnarray}
&& c_{\rm s}^2 = \frac{p'}{\rho'} =
 \frac{p_{\rm r}' + p_{\rm c}'}{\rho_{\rm r}' + \rho_{\rm c}'} \equiv 
\frac{4}{3} \biggl( \frac{ \rho_{\rm r}}{3 \rho_{\rm c} + 4 \rho_{\rm r}} \biggr),
\label{cs2ex}\\
&&  \varsigma\biggl( \frac{\delta p}{\delta \varsigma}\biggr)_{\rho} = 
\frac{4}{3} \frac{\delta \rho_{\rm r}}{3 \frac{\delta\rho_
{\rm r}}{\rho_{\rm r}} - 4 \frac{\delta\rho_{\rm c}}{\rho_{\rm c}}} \equiv 
\frac{4}{3} \biggl(\frac{ \rho_{\rm c}\rho_{\rm r}}{ 3 
\rho_{\rm c} + 4 \rho_{\rm r}} \biggr) \equiv \rho_{\rm c} c_{\rm s}^2.
\label{entrex1} 
\end{eqnarray}
To obtain the final expression appearing at the right-hand-side 
of Eq. (\ref{cs2ex}) the conservation equations for the two species 
(i.e. $\rho_{\rm r}' = - 4 {\cal H} \rho_{\rm r}$ and $\rho_{\rm c}'  = 
- 3 {\cal H} \rho_{\rm c}$) have been 
used.  Concerning  Eq. (\ref{entrex1}) the following remarks are in order:
\begin{itemize}
\item{} the first equality follows from the fluctuation 
of the specific entropy computed in Eq. (\ref{SE}); 
\item{} the second 
equality appearing in Eq. (\ref{entrex1}) follows from the observation that 
the  increment of the pressure should be computed for constant (total) 
energy density, i.e. $\delta \rho = \delta \rho_{\rm r} + 
\delta \rho_{\rm c} =0$, implying 
$\delta\rho_{\rm c} = - \delta \rho_{\rm r}$;
\item{} the third equality (always 
in Eq. (\ref{entrex1})) is a mere consequence of the explicit 
expression of $c_{\rm s}^2$ obtained in Eq. (\ref{cs2ex}). 
\end{itemize}
As in the case of Eq. (\ref{SEgen}), the analysis presented up to now 
can be easily generalized to a mixture of fluids $``{\rm a}"$ and $``{\rm b}"$ with barotropic 
indices $w_{\rm a}$ and $w_{\rm b}$.
The generalized speed of sound  is then  given by 
\begin{equation}
c_{\rm s}^2 = \frac{w_{\rm a} (w_{\rm a} + 1) \rho_{\rm a} + w_{\rm b} ( w_{\rm b} + 1) \rho_{\rm b}}{ 
(w_{\rm a } + 1 )\rho_{\rm a} + (w_{\rm b} + 1) \rho_{\rm b}}.
\end{equation}
From Eq. (\ref{cs2ex}) and from the definition of the (total) barotropic index it  
follows that, for a CDM-radiation fluid
\begin{equation}
c_{\rm s}^2 = \frac{4}{3} \frac{1}{3 a + 4},\qquad w = \frac{1}{3} \frac{1}{1 + a}.
\label{wint}
\end{equation}
Using Eqs. (\ref{deltapnaddef1})--(\ref{entrex1}) the non-adiabatic contribution to the 
total pressure fluctuation becomes
\begin{equation}
\delta p_{\rm nad} = \frac{4}{3} \rho_{\rm c}  \frac{{\cal S}}{3 a + 4},
\label{nadcdm2}
\end{equation}
where the definition given in Eq. (\ref{SE}), i.e. ${\cal S} = (\delta \varsigma)/\varsigma$, has been used.
Using the splitting of the total pressure density fluctuation into a adiabatic 
and a non-adiabatic parts, Eq. (\ref{p00l}) can be multiplied by a 
factor $c_{\rm s}^2$ and subtracted from Eq. (\ref{B1}). The result 
of this operation leads to a  formally simple expression
for  the evolution of curvature fluctuations in the longitudinal 
gauge, namely:
\begin{eqnarray}
&&\psi'' + {\cal H}[ \phi' + ( 2 + 3 c_{\rm s}^2 ) \psi'] + 
[ {\cal H}^2 ( 1 + 2 c_{\rm s}^2) + 2 {\cal H}'] \phi 
\nonumber\\
&&- c_{\rm s}^2 \nabla^2 \psi + \frac{1}{3}\nabla^2( \phi - \psi) =
4\pi G a^2 \delta p_{\rm nad},
\label{B5}
\end{eqnarray}
which  is indipendent of the specific form of $\delta p_{\rm nad}$. 
The left hand side of Eq. (\ref{B5}) can be written as 
the (conformal) time derivative of a single scalar function whose specific form is,   
\begin{equation}
{\cal R} = - \biggl( \psi+ 
\frac{{\cal H} ( \psi' + {\cal H} \phi)}{{\cal H}^2 
- {\cal H}'} \biggr).
\label{defR2}
\end{equation}
Taking now  the first (conformal) 
time  derivative of ${\cal R}$ as expressed by 
Eq. (\ref{defR2}) and using the definition of $c_{\rm s}^2$ we arrive at the 
following expression 
\begin{equation}
{\cal R}' = -\frac{ {\cal H}}{ 4 \pi G a^2 ( \rho + p)}\{ \psi'' 
+ {\cal H}[ ( 2 + 3 c_{\rm s}^2) \psi' + \phi'] + 
[ 2 {\cal H}' + ( 3 c_{\rm s}^2 + 1 ) {\cal H}^2] \phi \}.
\label{derR}
\end{equation}
Comparing now Eqs. (\ref{derR}) and (\ref{B5}), it is clear that Eq. 
(\ref{derR}) reproduces Eq. (\ref{B5}) but only up to the spatial 
gradients. Hence, using Eq. (\ref{derR}) into Eq. (\ref{B5}) the following 
final expression can be obtained:
\begin{equation}
{\cal R}' = - \frac{{\cal H}}{p + \rho} \delta p_{\rm nad} -
\frac{k^2 {\cal H}}{ 12 \pi G a^2 ( p + \rho)} (\phi -\psi) 
+\frac{ c^2_{\mathrm{s}} {\mathcal H}}{4\pi G a^2 ( p +\rho)} k^2\psi.
\label{evolR}
\end{equation}
Equation (\ref{evolR})  is very useful in different situations. 
Suppose, as a simple exercise, to consider the evolution of modes with wavelengths 
larger than the Hubble radius at the transition between matter and radiation.
Suppose also that $\delta p_{\mathrm{nad}}=0$. In this case Eq. (\ref{evolR}) 
implies, quite simply, that across the radiation-matter transition ${\mathcal R}$ is 
constant up to corrections of order of $k^2 \tau^2$ which are small when the 
given wavelngths are larger than the Hubble radius. Now it happens so that 
the relevant modes for the estimate of the ordinary Sachs-Wolfe effect are exactly 
the ones that are still larger than the Hubble radius at the transition between matter and radiation.
This observation allows to derive Eq. (\ref{psimpsir}). In fact, using the definition of ${\mathcal R}$ and 
recalling that during radiation and matter the longitudinal fluctuations of the geometry 
are constants we will have 
\begin{equation}
{\mathcal R}_{\mathrm{m}} = - \frac{5}{3} \psi_{\mathrm{m}}, \qquad 
{\mathcal R}_{\mathrm{r}} = - \frac{3}{2} \psi_{\mathrm{r}}.
\label{rmrr}
\end{equation}
But since ${\mathcal R}_{\mathrm{m}} = {\mathcal R}_{\mathrm{r}}$, Eq. (\ref{psimpsir}) easily 
follows. The result expressed by Eqs. (\ref{psimpsir}) and (\ref{rmrr}) holds in the case when 
neutrinos are not taken into account. This result can be however generalized to the case 
where neutrinos are present in the system, as it will be discussed in section \ref{sec8}.
Equation (\ref{evolR}) can be also used in order to obtain the evolution of $\psi$. Consider, again, the case 
of adiabatic initial conditions. In this case, as already mentioned, deep in the radiation-dominated 
regime ${\mathcal R}_{\mathrm{r}} = - 3 \psi_{\mathrm{r}}/2$. From the definition of ${\mathcal R}$ we can 
write the evolution for $\psi$ using, as integration variable, the scale factor. In fact, recalling that across
the radiation transition
\begin{equation}
p = w \rho,\qquad w(a) = \frac{1}{3 ( 1 + a)}, \qquad \frac{\rho + p}{\rho} = \frac{3 a + 4}{3 (a + 1)},
\label{defw}
\end{equation}
from Eq. (\ref{defR2}) it is easy to obtain the following (first order) differential equation:
\begin{equation}
\frac{d \psi}{da} + \frac{5 a + 6}{2 a ( a + 1)} \psi = \frac{3}{4} \biggl(\frac{3 a + 4}{a + 1}\biggr) \psi_{\mathrm{r}},
\label{psia1}
\end{equation}
which can be also written as
\begin{equation}
\frac{\sqrt{a + 1}}{a^3} \frac{d}{d a} \biggl(\frac{a^3}{\sqrt{a +1}} \psi \biggr) = 
\frac{3}{4} \psi_{\mathrm{r}}\biggl(\frac{3 a + 4}{a(a+ 1)}\biggr).
\label{psia2}
\end{equation}
By integrating once the result for $\psi(a)$ is 
\begin{equation}
 \psi(a) = \frac{\psi_{\rm r}}{10 a^3 }\{16 ( \sqrt{ a + 1} -1)  + a [ a ( 9 a + 2) -8]\};
 \label{solPSI1}
 \end{equation}
the  limit for $a\to \infty$ (matter-dominated phase)  of the right hand side of Eq. (\ref{solPSI1}) 
leads to $(9/10)\psi_{\rm r}$.
In the simplistic case of CDM-radiation plasma a rather instructive derivation of the 
gross features of the non-adiabatic mode can also be obtained. 
If $\delta p_{\rm nad}$ is given by Eq.  (\ref{nadcdm2}), Eq. (\ref{evolR}) can be simply written as 
\begin{equation}
\frac{d{\cal R}}{d a} = -  \frac{4 {\cal S}}{ ( 3 a + 4)^2} + {\mathcal O}(k^2 \tau^2).
\label{evolRCDM}
\end{equation}
Eq. (\ref{evolRCDM}) can be easily obtained inserting Eq. (\ref{nadcdm2}) into Eq. (\ref{evolR}) and recalling 
that, in the physical system under consideration, $( p + \rho) = \rho_{\rm c} + (4/3) \rho_{\rm r}$.
In the case of the CDM-radiation isocurvature mode, the non-adiabatic contribution 
is non-vanishing and proportional to ${\cal S}$. Furthermore, it can be easily shown that   
the fluctuations of the entropy density, ${\cal S}$ are roughly constant (up to logarithmic 
corrections) for $k \tau \ll 1$, i.e. for the modes which are relevant for the SW effect after equality.
This conclusion can be easily derived by subtracting Eq. (\ref{Cdelc}) from  $3/4$ of Eq. (\ref{Cdelr}). 
Recalling the definition of ${\cal S}$ the result is 
\begin{equation}
{\cal S}' = - (\theta_{\rm r} -\theta_{\rm c}). 
 \label{entropyevol}
 \end{equation}
Since $\theta_{\rm r}$ and $\theta_{\rm c}$ vanish in the limit $k \tau \ll 1$,  ${\cal S}$ is 
indeed constant. 
Eq. (\ref{evolRCDM}) can then be integrated in explicit terms,  across the radiation--matter 
transition
\begin{equation}
{\cal R} = - 4 {\cal S} \int_{0}^{a} \frac{db}{(3 b + 4)^2} \equiv - {\cal S} \frac{a}{3 a + 4},
\label{RCDMsol1}
\end{equation}
implying that ${\cal R} \to 0 $ for $a \to 0$ (radiation-dominated phase)  and that 
${\cal R} \to - {\cal S}/3$ for $a\to \infty$ (matter-dominated phase). 
 
Recalling again the explicit form of ${\cal R}$ in terms of $\psi$, i.e. Eq. (\ref{defR2}),  Eq. (\ref{RCDMsol1}) 
leads to a simple equation giving the evolution of $\psi$ for modes $k \tau \ll 1$, i.e. 
\begin{equation}
\frac{d \psi}{d a } + \frac{ 5 a + 6}{ 2 a ( a + 1)} \psi = \frac{{\cal S}}{2 ( a + 1)}.
\label{PSICDMeq1}
\end{equation}
The solution of Eq. (\ref{PSICDMeq1}) can be simply obtained imposing the isocurvature 
boundary condition, i.e. $\psi(0) \to 0$:
\begin{equation}
\psi(a ) = \frac{{\cal S}}{5 a^3} \{ 16 ( 1 - \sqrt{ a + 1}) + a [ 8 + a ( a - 2)]\}.
\label{PSICDMsol1}
\end{equation}
Eq. (\ref{PSICDMsol1})  is similar to Eq. (\ref{solPSI1}) but with 
few crucial differences. According to Eq. (\ref{PSICDMsol1}) 
(and unlike Eq. (\ref{solPSI1})), $\psi(a)$ vanishes, 
for  $a\to 0$,  as $ {\cal S} a /8$. In the limit $ a \to \infty$ 
$\psi(a) \to {\cal S}/5$. This is the growth of the adiabatic 
mode triggered, during the transition from radiation to matter, by the 
presence of the non-adiabatic pressure density 
fluctuation.  

Having obtained the evolution of $\psi$, the evolution of the total density contrasts
and of the total peculiar velocity field can be immediately obtained by solving 
the Hamiltonian and momentum  constraints of Eqs. (\ref{p00l}) and (\ref{p0il}) 
 with respect to $\delta\rho$ and $\theta$ 
\begin{eqnarray}
&& \delta = \frac{\delta \rho}{\rho} \equiv \frac{\delta_{\rm r} }{ a + 1} + \delta_{\rm c} 
\frac{a}{a + 1} = -2 \biggl(\psi + \frac{d \psi}{ d\ln{a}}\biggr), 
\label{drhonad}\\
&&\theta = \frac{ 2 k^2 ( a + 1)}{( 3 a + 4)} \biggl(\psi +\frac{ d \psi}{d\ln{a}}\biggr),
\label{dtheta}
\end{eqnarray} 
which also implies, for $k \tau \ll 1$, 
\begin{equation}
\theta =- \frac{k^2 (a + 1) }{( 3 a + 4)} \delta.
\label{THsub}
\end{equation}
Equation (\ref{THsub})  is indeed consistent with the result that the total velocity field is negligible 
for modes outside the horizon. Inserting Eq. (\ref{PSICDMsol1}) into Eqs. 
(\ref{drho})--(\ref{theta})
it can be easily argued that the total density contrast goes to zero for $a\to 0$, while, for 
$a \to \infty$ we have  the following relations 
\begin{equation}
\delta_{\rm c} \simeq - \frac{2}{5} {\cal S} \simeq - 2\psi \simeq - \frac{1}{2} \delta_{\rm r}
\label{psitoentr}
\end{equation}
The first two equalities in Eq. (\ref{psitoentr})  follow from the asymptotics of Eq. (\ref{drhonad}), the last equality 
follows from the conservation law (valid for isocurvature modes) which can be 
derived from Eq. (\ref{Cdelr}), i.e. 
\begin{equation}
\delta_{\rm r} \simeq 4 \psi.
\label{drpsi}
\end{equation}
Thanks to the above results, the contribution to the scalar Sachs-Wolfe effect 
can be obtained in the case of the CDM-radiation non-adiabatic mode. From Eq. 
(\ref{SW0}) we have 
 \begin{equation}
 \biggl( \frac{\Delta T}{T}\biggr)^{\rm nad}_{k, {\rm s}} = 
 \biggl(\frac{\delta_{\rm r}}{4} + \psi\biggl)_{\tau\simeq \tau_{\rm dec}} \equiv 2 \psi_{\rm nad}
 \equiv  \frac{2}{5} {\cal S},
 \label{SWscalnad}
 \end{equation}
where the second equality follows from Eq. (\ref{drpsi}) and the third equality 
follows from Eq. (\ref{PSICDMsol1}) in the limit $a\to \infty$ (i.e. $a \gg a_{\rm eq}$).

The following comments are in order:
\begin{itemize}
\item{} as in the case of the adiabatic mode, also in the case of  non-adiabatic mode in the CDM-radiation
system, the peculiar velocity does not contribute to the SW effect; 
\item{} for $k\,c_{\rm s}\,\tau_{\rm dec} \ll 1$ the temperature fluctuations induced by the adiabatic mode are simply $ 2\psi_{\rm nad}$ (unlike the adiabatic case) ;
\item{}  Equation (\ref{SWscalnad}) suggests that 
the first true peak  in the temperature fluctuations is located at $k c_{\rm s} \eta_{\rm dec} \simeq \pi/2$.
\end{itemize}
The last conclusion comes from an analysis similar to the one conducted in the case 
of the adiabatic mode but with the crucial difference that, in the case of the isocurvature 
mode, $\psi$ vanishes as $\tau$ at early times. This occurrence implies the 
presence of sinusoidal (rather than cosinusoidal) oscillations. This point will 
be further discussed in section \ref{sec8}.
\begin{figure}
\centering
\includegraphics[height=6cm]{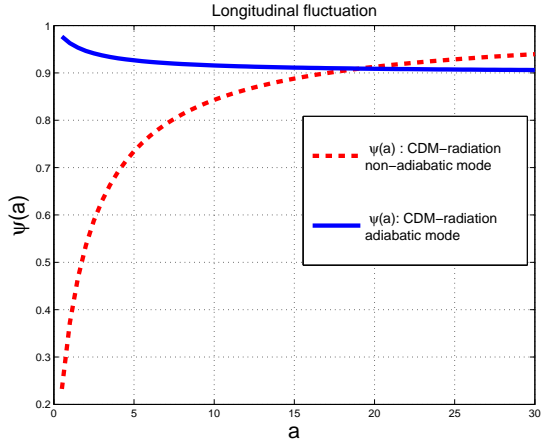}
\caption{The longitudinal fluctuation of the metric is plotted as a function of the scale factor 
in the case of the adiabatic mode (see Eq. (\ref{solPSI1})) and in the case of the non-adiabatic mode
(see Eq. (\ref{PSICDMsol1})).}
\label{LOG}      
\end{figure}
\begin{figure}
\centering
\includegraphics[height=6cm]{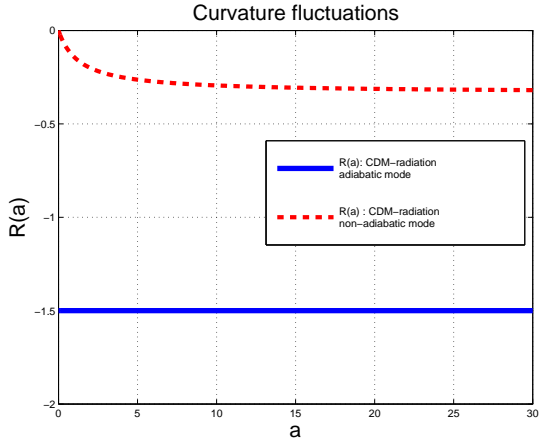}
\caption{The curvature fluctuation ${\mathcal R}$ is plotted as a function of the scale factor 
in the case of the adiabatic (full line) and in the case of the non-adiabatic (dashed line) modes. In both 
cases the curves are reported for the CDM-radiation system.}
\label{CURVFL}      
\end{figure}
In Fig. \ref{LOG} the evolution of $\psi(a)$ is illustrated for the adiabatic and for the non-adiabatic mode.
In the case of the adiabatic mode, deep in the radiation epoch (i.e. $a\to0$) $\psi\to \psi_{\mathrm{r}}$ (conventionally chosen to be $1$ 
in Fig. \ref{LOG}). Always in the case of the adiabatic mode, 
for $a\to \infty$ (i.e. during the matter epoch) $\psi_{\mathrm{m}} = 9\psi_{\mathrm{r}}/10$.
In Fig. \ref{CURVFL} the evolution of ${\mathcal R}(a)$ is reported. For the adiabatic mode 
${\mathcal R}$ is constant. In fact, according to Eq. (\ref{evolR}) the non-adiabtic 
pressure variation vanishes.
In the case of the non-adiabatic mode $\delta p_{\mathrm{nad}}\neq 0$ and ${\mathcal R}$ goes to zero 
deep in the radiation epoch (i.e. $a\to 0$) and it goes to a constant in the matter epoch. Correspondingly, in the case 
of the non-adiabatic mode
$\psi(a)$ goes to zero in the limit $a\to0$ and it goes to a constant for $a\to \infty$.

\newpage
\renewcommand{\theequation}{8.\arabic{equation}}
\setcounter{equation}{0}
\section{Improved fluid description of pre-decoupling physics}
\label{sec8}
 In spite 
of the drastic simplifications adopted in the mathematical analysis of the problem, the 
results obtained in the previous section are reasonable but they are only meaningful for large enough scales or, 
equivalently, for sufficiently small harmonics $\ell$. For larger harmonics (i.e. for smaller length-scales) 
the approach introduced in section \ref{sec7} fails. Let us recall that the angular separation 
appearing in the expression of the angular power spectrum is related to the harmonics as 
\begin{equation}
\theta \simeq \frac{\pi}{\ell},\qquad k \simeq\ell\, h_{0}\, 10^{-4} \,\, \mathrm{Mpc}^{-1}
\label{relationsell}
\end{equation}
where the second relation gives the comoving wave-number in terms of $\ell$ for 
a Universe with $\Omega_{\mathrm{M}0} \simeq 0.3$ and $\Omega_{\Lambda0} \simeq 0.7$.
In section \ref{sec7} the general system of fluctuations has been (artificially) reduced to the case when only 
radiation and CDM particles were present. In the present section this assumption will be dropped and 
the following topics will be treated:
\begin{itemize}
\item{} the general four-components plasma;
\item{} tight coupling between photons and baryons;
\item{} the general solution for the adiabatic modes;
\item{}numerical solutions in the tightly coupled regime.
\end{itemize}

\subsection{The general four components plasma}
In the general case, the plasma contains essentially four components, namely, photons, baryons, 
neutrinos and CDM particles. Under the assumption that the dark-energy component 
is parametrized by a cosmological constant, there are no extra sources of inhomogeneity to be 
considered on top of the metric fluctuations which will be treated, as in section \ref{sec7}, within the 
longitudinal coordinate system.
Since the neutrinos are present in the game with their anisotropic stress, it will not be possible any longer 
to consider the case $\phi=\psi$.  When all the four species are simultaneously 
present in the plasma, Eq. (\ref{p00l}) can be written, in Fourier space, as 
\begin{equation}
- k^2 \psi - 3 {\mathcal H}({\mathcal H} \phi + \psi') = 4\pi G\,a^2[ \delta \rho_{\gamma} + 
\delta \rho_{\nu} + \delta\rho_{\mathrm{c}} + \delta \rho_{\mathrm{b}}].
\label{00k}
\end{equation}
Defining as $\delta_{\nu}$, $\delta_{\gamma}$, $\delta_{\rm b}$ and 
$\delta_{\rm c}$ the neutrino, photon, baryon and CDM 
density contrasts, the Hamiltonian constraint of Eq. (\ref{00k}) 
can also be written as 
\begin{equation}
- 3 {\mathcal H} ( {\mathcal H} \phi + \psi') - k^2 \psi = \frac{3}{2} {\mathcal H}^2 
[ ( R_{\nu} \delta_{\nu} + ( 1 - R_{\nu}) \delta_{\gamma})  + 
\Omega_{\rm b} \delta_{\rm b} + \Omega_{\rm c} \delta_{\rm c}],
\label{p00lex}
\end{equation}
where, for $N_{\nu}$ species of massless neutrinos,   
\begin{equation}
R= \frac{7}{8} N_{\nu} \biggl( \frac{4}{11}\biggr)^{4/3},\,\,\,\,\,\,\, 
R_{\nu}  = \frac{R}{1 + R},\,\,\,\,\,\,\,\, R_{\gamma} = 1 - R_{\nu},
\label{DEFRNU}
\end{equation}
so that $R_{\nu}$ and $R_{\gamma}$ represent the fractional 
contributions of photons and neutrinos to the total density at early times 
deep within the radiation-dominated epoch.
Eq. (\ref{FL1C}) has been used (in the case of vanishing spatial curvature) into Eq. (\ref{00k}) in order to 
eliminate the explicit dependence upon the total energy density 
of the background.  Notice that Eq. (\ref{p00lex}) has been written under the assumption that the total 
energy density of the sources is dominated by radiation since this is the regime where the 
initial conditions for the numerical integration are customarily set.
From the momentum constraint of Eq. (\ref{p0il}), 
and from Eq. (\ref{B1}) the following pair of equations can be derived:
\begin{eqnarray}
&& k^2 ( {\mathcal H} \phi + \psi') = \frac{3}{2} {\cal H}^2
 \biggl[ \frac{4}{3} (R_{\nu} \theta_{\nu} + R_{\gamma} 
\theta_{\gamma}) + 
\theta_{\rm b} \Omega_{\rm b} + \theta_{\rm c} \Omega_{\rm c} \biggr],
\label{p0ilex}\\
&& \psi'' + ( 2 \psi' + \phi') {\mathcal H} + (2 {\cal H}' + {\mathcal H}^2) \phi
 - \frac{k^2}{3} ( \phi - \psi) = \frac{{\mathcal H}^2}{2} 
( R_{\nu} \delta_{\nu} +
\delta_{\gamma}  R_{\gamma}),
\label{pijlex}
\end{eqnarray}
where, following Eq. (\ref{thetadefinition}),  the divergence of the (total) peculiar velocity field 
has been separated for the different species, i.e. 
\begin{equation}
(p + \rho) \theta = \frac{4}{3} \rho_{\nu} \theta_{\nu} + \frac{4}{3} \rho_{\gamma} \theta_{\gamma}  + \rho_{\mathrm{c}} \theta_{\mathrm{c}} + \rho_{\mathrm{b}} \theta_{\mathrm{b}}.
\label{thetasum}
\end{equation}
Furthermore, 
in Eqs (\ref{p0ilex})--(\ref{pijlex}),  Eq. (\ref{FL1C}) has 
been used in order to eliminate the explicit dependence upon the 
(total) energy and pressure densities. 
Finally, according to Eqs. (\ref{B2})--(\ref{B2a}) the neutrino anisotropic stress fixes 
the difference between the two longitudinal fluctuations of the 
geometry. Recalling Eqs. (\ref{B2}) and (\ref{B2a}) we will have 
\begin{equation}
\nabla^4 (\phi - \psi) = 12 \pi G a^2 \partial_{i}\partial^{j} \Pi_{j}^{i}.
\label{ANIS2}
\end{equation}
For temperatures smaller than the MeV the only collisionless 
species of the plasma are neutrinos (which will be assumed to be massless, for simplicity). Thus 
neutrinos will provide, in the absence of large-scale magnetic fields, the dominant 
contribution to the neutrinos anisotropic stress. The term at the right hand side of Eq. (\ref{ANIS2}) 
can then be parametrized, for future convenience, as 
\begin{equation}
\partial_{i}\partial^{j} \Pi_{j}^{i} = (p_{\nu} + \rho_{\nu}) \nabla^2 \sigma_{\nu}.
\end{equation}
Hence, using Eqs. (\ref{FL1C}) and (\ref{FL2C}),  Eq. (\ref{ANIS2}) can then be written as 
\begin{equation}
k^2 (\phi - \psi) = -6 {\mathcal H}^2 R_{\nu} \sigma_{\nu},
\label{pineqjex}
\end{equation}
The evolution equations for the various species will now be discussed.
The CDM and neutrino components are only coupled to the fluctuations of the 
geometry. Their evolution equations are then give, respectively by
\begin{eqnarray}
&&\theta_{\rm c}' + {\cal H} \theta_{c} = k^2 \phi,
\label{CDM1}\\
&& \delta_{\rm c}' = 3 \psi' - \theta_{\rm c}.
\label{CDM2} 
\end{eqnarray}
and by 
\begin{eqnarray}
&& \delta_{\nu}' = - \frac{4}{3} \theta_{\nu} + 4\psi',
\label{nu1}\\
&& \theta_{\nu}' = \frac{k^2}{4} \delta_{\nu} - k^2 \sigma_{\nu} + k^2 \phi,
\label{nu2}\\
&& \sigma_{\nu}' = \frac{4}{15} \theta_{\nu} - \frac{3}{10} k {\cal F}_{\nu 3}.
\label{nu3}
\end{eqnarray}
Equations (\ref{nu1}) and (\ref{nu2}) are directly obtained from Eqs. 
(\ref{delta}) and (\ref{theta}) 
in the case $ w_{\nu} = 1/3$ and $\sigma_{\nu} \neq 0$. Eq. (\ref{nu3}) 
is not obtainable in the fluid approximation and the full Boltzmann hierarchy has to be introduced.
The quantity  ${\cal F}_{\nu 3}$   introduced in 
 Eq. (\ref{nu3}),  is the octupole term of the neutrino phase space distribution. For a derivation of Eqs. (\ref{nu1}), 
 (\ref{nu2}) and (\ref{nu3}) see Eqs. (\ref{mom0}), (\ref{mom1}) and  (\ref{mom2}) in section \ref{sec10}.
 
\subsection{Tight coupling between photons and baryons}

Prior to decoupling, protons and electrons are tightly coupled. 
The strength of the Coulomb coupling justifies the consideration of a unique 
proton-electron component which will the the so-called baryon fluid. The baryon fluid 
is however also coupled, through Thompson scattering, to the photons.
Since the photon-electron cross section is larger than the photon-protron cross section, the 
momentum exchange between the two components will be dominated by electrons. 
The evolution equation of the photon component can be written as:
\begin{eqnarray}
&& \delta_{\gamma}' = - \frac{4}{3} \theta_{\gamma} + 4 \psi',
\label{phot1}\\
&& \theta_{\gamma}' = \frac{k^2}{4} \delta_{\gamma} + k^2 \phi  
+ a x_{\mathrm{e}} \,n_{\rm e} \sigma_{\rm T} ( \theta_{\rm b} - \theta_{\gamma} ).
\label{phot2}
\end{eqnarray}
For the baryon-lepton fluid, the two relevant equations are instead:
\begin{eqnarray}
&& \delta_{\rm b}' = 3 \psi' - \theta_{\rm b},
\label{baryon1}\\
&& \theta_{\rm b}' = - {\cal H} \theta_{\rm b} +  k^2  \phi +  
\frac{4}{3} \frac{\rho_{\gamma}}{\rho_{\rm b}} a n_{\rm e} 
x_{\rm e} \sigma_{\rm T} ( \theta_{\gamma} - \theta_{\rm b}).
\label{baryon2} 
\end{eqnarray}
From Eqs. (\ref{phot2}) and (\ref{baryon2}) it can be argued that the Thompson scattering terms 
drag the system to the final configuration where $\theta_{\gamma}\simeq \theta_{\mathrm{b}}$.
In short the argument goes as follows. By taking the difference of Eqs. (\ref{phot2}) and (\ref{baryon2}) 
the following equation can be easily obtained:
\begin{equation}
(\theta_{\gamma} -\theta_{\mathrm{b}})' + \Gamma_{\mathrm{T}} (\theta_{\gamma} - \theta_{\mathrm{b}}) = 
J(\tau,\vec{x}),
\label{diffthgthb}
\end{equation}
where 
\begin{eqnarray}
&&\Gamma_{\mathrm{T}} = a n_{\mathrm{e}} x_{\mathrm{e}}  \sigma_{\mathrm{T}}
\biggl( 1 + \frac{4}{3} \frac{\rho_{\gamma}}{\rho_{\mathrm{b}}} \biggr) \equiv  a n_{\mathrm{e}} x_{\mathrm{e}}  \sigma_{\mathrm{T}}\biggl(\frac{R_{\mathrm{b}} +1}{R_{\mathrm{b}}}\biggr),
\label{GT}\\
&& J(\tau,\vec{x}) = \frac{k^2}{4} \delta_{\gamma} + {\mathcal H} \theta_{\mathrm{b}}.
\label{Jdef}
\end{eqnarray}
For future convenience, in Eq. (\ref{Jdef}) the baryon-to-photon ratio $R_{\mathrm{b}}$ has been introduced, i.e. 
\begin{equation}
R_{\mathrm{b}}(z) =  \frac{3}{4} \frac{\rho_{\mathrm{b}}}{\rho_{\gamma}} = \biggl(\frac{698}{z +1}\biggr) \biggl(\frac{h_{0}^2 \Omega_{\mathrm{b}}}{0.023}\biggr).
\label{Rb}
\end{equation}
From Eq. (\ref{diffthgthb}) it can be easily appreciated that any deviation of $(\theta_{\gamma}-\theta_{\mathrm{b}})$ swiftly 
decays away in spite of the strength of the source term $J(\tau,\vec{x})$. In fact, from Eq. (\ref{diffthgthb}),
the characteristic time for the synchronization of the baryon and photon velocities is of the order of $(x_{\mathrm{e}}
n_{\mathrm{e}}\sigma_{\mathrm{T}} )^{-1}$ which is small in comparison with the expansion time. In the limit 
$\sigma_{\mathrm{T}}\to \infty$  the tight coupling is exact and the photon-baryon velocity field is a unique physical entity which will be denoted by $\theta_{\gamma\mathrm{b}}$.  The evolution equation for $\theta_{\gamma\mathrm{b}}$ can be easily obtained by summing up Eq. (\ref{phot2}) and Eq. (\ref{baryon2}) with a relative weight 
(given by $R_{\mathrm{b}}$) allowing the mutual cancellation of the scattering terms. The result of this procedure 
implies that the whole baryon-photon system can be written as 
\begin{eqnarray}
&& \delta_{\gamma}' = 4\psi' - \frac{4}{3} \theta_{\gamma{\mathrm b}},
\label{pb0}\\
&& \delta_{\mathrm b}' = 3 \psi' - \theta_{\gamma{\mathrm b}},
 \label{pb1}\\
&& \theta_{\gamma{\mathrm b}}' + \frac{{\cal H} R_{\rm b}}{(1 + R_{\mathrm b})} \theta_{\gamma{\mathrm b}} = \frac{k^2 \delta_{\gamma}}{4 ( 1 + R_{\rm b})} + 
k^2 \phi.
\label{pb2}
\end{eqnarray}
The introduction of the baryon-photon velocity field also slightly modifies the form of the 
momentum constraint of Eq. (\ref{p0ilex}) which now assumes the form:
\begin{equation}
k^2 ( {\mathcal H} \phi + \psi') = \frac{3}{2} {\cal H}^2
 \biggl[ \frac{4}{3} R_{\nu} \theta_{\nu} +\frac{4}{3} R_{\gamma} (1 + R_{\mathrm{b}}) \theta_{\gamma\mathrm{b}}
+ \theta_{\rm c} \Omega_{\rm c} \biggr].
\label{p0iexsec}
\end{equation}
By performing a (conformal) time derivation of both sides of Eq. (\ref{pb0}) and by using repeatedly Eq. (\ref{pb2}) 
to eliminate $\theta_{\gamma\mathrm{b}}$ and $\theta_{\gamma\mathrm{b}}'$ the evolution equation 
for $\delta_{\gamma}$ becomes 
\begin{equation}
\delta_{\gamma}'' + \frac{{\mathcal H} R_{\mathrm{b}}}{R_{\mathrm{b}} +1} \delta_{\gamma}' + 
c_{\mathrm{s}\mathrm{b}}^2 \delta_{\gamma} = 4 \biggl[ \psi'' + 
 \frac{{\mathcal H} R_{\mathrm{b}}}{R_{\mathrm{b}} +1} \psi' - \frac{k^2}{3} \phi \biggr],
 \label{deltagamma}
 \end{equation}
 where 
\begin{equation}
c_{\mathrm{s}\mathrm{b}}= \frac{1}{\sqrt{3 (R_{\mathrm{b}} +1)}},
\label{bphss}
\end{equation}
is the speed of sound in the baryon-photon system. 
\subsection{A particular example: the adiabatic solution}
We are now in the position of giving an explicit example of solution of the whole 
generalized system of fluctuations in the case of the adiabatic mode.
Consider first the Hamiltonian constraint of Eq. (\ref{p00lex}) deep in the radiation 
dominated epoch, i.e. for temperatures smaller than the temperature of the neutrino decoupling 
and temperatures larger than the equality temperature. In this case 
a solution can be found where the longitudinal fluctuations of the geometry are both constant in time, i.e. 
\begin{equation}
\phi(k,\tau) = \phi_{\mathrm{i}}(k),\qquad \psi(k,\tau) = \psi_{\mathrm{i}}(k).
\label{constantmode}
\end{equation}
Equation (\ref{p00lex}) implies then, to lowest order in $k\tau$ that the radiation 
density contrasts are also constant and given by 
\begin{equation}
\delta_{\gamma}(k,\tau) \simeq \delta_{\nu}(k,\tau) \simeq - 2 \phi_{\mathrm{i}}(k).
\end{equation}
Imposing now that the entropy fluctuations vanish we will also have 
that:
\begin{equation}
\delta_{\mathrm{c}}(k,\tau) \simeq \delta_{\mathrm{b}}(k,\tau) \simeq - \frac{3}{2}\phi_{\mathrm{i}}(k) + 
{\mathcal O}(k^2 \tau^2).
\end{equation}
Direct integration of Eqs. (\ref{CDM1}), (\ref{nu2}) and (\ref{pb2}) implies, always to lowest 
order in $k\tau$ that 
\begin{equation}
\theta_{\mathrm{c}}(k,\tau)\simeq \theta_{\nu}(k,\tau) \simeq \theta_{\gamma\mathrm{b}}(k,\tau) \simeq 
\phi_{\mathrm{i}}(k)\frac{k^2 \tau}{2}.
\label{velad}
\end{equation}
It can be checked that the momentum constraint is also satisfied in the radiation epoch. The relations 
expressed by Eq. (\ref{velad}) express a general property of the adiabatic mode: to lowest order the peculiar velocities 
of the various species are equal and much smaller than the density contrasts. 
Equations (\ref{pineqjex}) and (\ref{nu3}) imply that the following two important relations:
\begin{eqnarray}
&& \sigma_{\nu}(k,\tau) \simeq \frac{k^2\tau^2}{15} \phi_{\mathrm{i}}(k),
\label{sigmanuad}\\
&& \psi_{\mathrm{i}}(k) = \biggl(1 + \frac{2}{5} R_{\nu}\biggr) \phi_{\mathrm{i}}(k).
\label{psiphi}
\end{eqnarray}
Recalling the definition of ${\mathcal R}$, it is also possible to relate the longitudinal fluctuations to the 
curvature fluctuations ${\mathcal R}$, i.e. 
\begin{equation}
\psi_{\mathrm{i}}(k) = - \frac{2 ( 5 + 2 R_{\nu})}{15 + 4 R_{\nu}} {\mathcal R}_{\mathrm{i}}(k),\qquad 
\phi_{\mathrm{i}}(k) = - \frac{10}{15 + 4 R_{\nu}} {\mathcal R}_{\mathrm{i}}(k).
\label{psitoR}
\end{equation}
Clearly, in the case $R_{\nu}=0$ the relations of Eq. (\ref{psitoR}) reproduces the one already obtained 
and discussed in section \ref{sec7}. 
The "initial conditions" expressed through the Fourier components $\phi(k)$, $\psi(k)$ or ${\mathcal R}_{\mathrm{i}}(k)$ 
are computed by using the desired model of amplification. In particular, in the case of conventional 
inflationary models, the spectra of scalar fluctuations will be computed in section \ref{sec10}. Notice already that, in the case 
of the adiabatic mode, only one spectrum is necessary to set consistently the pre-decoupling initial conditions. 
Such a spectrum can be chosen to be either the one of $\phi(k)$ or the one of ${\mathcal R}_{\mathrm{i}}(k)$ .
As a final remark, it is useful to point out that the solution obtained for the adiabatic mode 
holds, in the present case, well before decoupling. In Eqs. (\ref{deltam}), (\ref{thetam}), (\ref{deltar}) and (\ref{thetar}) 
the solution has been derived, instead, during the matter epoch. These two solutions are physically 
different also because neutrinos (as well as baryons) have not been taken into account in section \ref{sec7}.

\subsection{Numerical solutions in the tight coupling approximation}
For the purposes of this presentation it is useful to discuss some simplified example of numerical integration 
in the tight coupling approximation. 
The idea will be to integrate numerically a set of equations where:
\begin{itemize}
\item{} baryons and photons are tightly coupled and the only relevant velocity fields are $\theta_{\mathrm{c}}$ and 
$\theta_{\gamma\mathrm{b}}$;
\item{} neutrinos will be assumed to be absent;
\item{} the evolution of the scalar modes will be implemented by means of ${\mathcal R}$ and $\psi$.
\end{itemize}
This 
approximation is appropriate for presentation but it is not necessary since neutrinos can be easily 
introduced. 
Since neutrinos are absent there 
is no source of anisotropic stress and the two longitudinal fluctuations of the metric are 
equal, i.e. $\phi = \psi$. Consequently, the system of equations to be solved becomes 
\begin{eqnarray}
&& {\cal R}' = \frac{k^2 c_{s}^2 {\cal H}}{{\cal H}^2 - {\cal H}'} \psi - \frac{{\cal H}}{p_{\rm t} + \rho_{\rm t}} \delta p_{\rm nad},
\label{simple1}\\
&& \psi' = - \biggl( 2 {\cal H} -\frac{ {\cal H}'}{{\cal H}} \biggr)\psi - \biggl({\cal H} - \frac{{\cal H}'}{{\cal H}}\biggl) {\cal R},
\label{simple2}\\
&& \delta_{\gamma}' = 4 \psi' - \frac{4}{3} \theta_{\gamma{\rm b}},
\label{simple3}\\
&& \theta_{\gamma{\rm b}}' = - \frac{{\cal H} R_{\rm b}}{R_{\rm b} + 1} \theta_{\gamma{\rm b} }+ 
\frac{k^2 }{4 ( 1 + R_{\rm b})} \delta_{\gamma} + k^2 \psi,
\label{simple4}\\
&& \delta_{\mathrm  c}' = 3 \psi' - \theta_{\mathrm c},
\label{simple5}\\
&& \theta_{\mathrm c}'  = - {\mathcal H} \theta_{\mathrm c} + k^2 \psi.
\label{simple6}
\end{eqnarray}

We can now use the explicit form of the scale factor discussed in Eq. (\ref{aint}) which implies:
\begin{eqnarray}
&&{\cal H} = \frac{1}{\tau_1} \frac{2 ( x +1)}{x ( x + 2)}, 
\nonumber\\
&& {\cal H}'  = - \frac{2}{\tau_1^2} \frac{ x^2 + 2 x + 4}{x^2 ( x + 2)^2},
\nonumber\\
&& {\cal H}^2 - {\cal H}' = \frac{1}{\tau_{1}^2} \frac{ 2 ( 3 x^2 + 6 x + 4)}{x^2 ( x + 2)^2},
\label{hubbles}
\end{eqnarray}
where $x = \tau/\tau_{1}$. It is also relevant 
to recall, from the results of section \ref{sec7}, that 
\begin{equation}
c_{\mathrm{s}}^2 = \frac{4}{3} \frac{1}{3 a + 4}, \qquad \delta p_{\mathrm{nad}} = \rho_{\mathrm{c}} c_{\mathrm{s}}^2 
{\mathcal S}.
\end{equation}
With these specifications the evolution equations given in (\ref{simple1})--(\ref{simple6}) 
become
\begin{eqnarray}
&&\frac{d {\mathcal R}}{d x} = \frac{4}{3} \frac{ x ( x+ 1) ( x + 2)}{( 3 x^2 + 6 x + 4)^2} \kappa^2 \psi 
- \frac{8 ( x + 1) {\mathcal S}}{(3 x^2 + 6 x + 4)^2},
\label{simpleex1}\\
&& \frac{d \psi }{d x} = - \frac{3x^2 + 6 x + 4}{x ( x + 1 )( x + 2)} {\mathcal R} - 
\frac{ 5 x^2 + 10 x + 6}{x (x + 1) ( x + 2)} \psi,
\label{simpleex2}\\
&& \frac{ d \delta_{\gamma}}{d x} = - \frac{ 4 ( 3 x^2 + 6 x + 4)}{x ( x + 1) ( x +2)} {\mathcal R} - \frac{4( 5 x^2 + 10 x + 6)}{x (x + 1) ( x + 2)} \psi - \frac{4}{3} \tilde{\theta}_{\gamma{\rm b}},
\label{simpleex3}\\
&& \frac{ d \tilde{\theta}_{\gamma{\rm b}}}{d x} = - \frac{ 2 R_{\rm b}}{R_{\rm b} + 1} \frac{ ( x+ 1)}{x ( x + 2)} + \frac{\kappa^2 }{4 ( 1 + R_{\rm b})} \delta_{\gamma} + \kappa^2 \psi,
\label{simpleex4}\\
&& \frac{d\delta_{\mathrm c}}{dx} =  - \frac{3(3x^2 + 6 x + 4)}{x ( x + 1 )( x + 2)} {\mathcal R} - 
\frac{3( 5 x^2 + 10 x + 6)}{x (x + 1) ( x + 2)} \psi - \tilde{\theta}_{\mathrm c},
\label{simpleex5}\\
&& \frac{d \tilde{\theta}_{\mathrm c}}{d x} =  -\frac{2 ( x +1)}{x ( x + 2)}\tilde{\theta}_{\mathrm c} + \kappa^2 \psi.
\label{simpleex6}
\end{eqnarray}
In Eqs. (\ref{simpleex1})--(\ref{simpleex6}) the following rescalings have been used (recall the role 
of $\tau_{1}$ arising in Eq. (\ref{aint})):
\begin{equation}
\kappa = k \tau_{1}, \qquad \tilde{\theta}_{\gamma{\mathrm b}} = \tau_{1} \theta_{\gamma{\mathrm b}},\qquad \tilde{\theta}_{{\mathrm c}} = \tau_{1} \theta_{{\mathrm c}}.
\end{equation}
The system of equations (\ref{simpleex1})--(\ref{simpleex6}) 
can be readily integrated by giving initial conditions for at $x_{\mathrm i} \ll 1$. In 
the case of the adiabatic mode (which is the one contemplated by Eqs. (\ref{simpleex1})--(\ref{simpleex6}) since we set $\delta p_{\mathrm nad}=0$) the initial conditions are as 
follows:
\begin{eqnarray}
&&{\mathcal R}(x_{\mathrm i}) = {\mathcal R}_{*},\qquad \psi(x_{\mathrm i}) = 
- \frac{2}{3} {\mathcal R}_{*} ,
\nonumber\\
&&\delta_{\gamma}(x_{\mathrm i}) = -2 \psi_{*}, \qquad \tilde{\theta}_{\gamma{\rm b}}(x_{\mathrm i}) =0,
\nonumber\\
&& \delta_{\mathrm c}(x_{\mathrm i}) =  \delta_{\mathrm b}(x_{\mathrm i})= - \frac{3}{2}\psi_{*},\qquad \tilde{\theta}_{{\rm c}}(x_{\mathrm i})=0.
\label{adinsimpl}
\end{eqnarray}
It can be shown by direct numerical integration that the system (\ref{simpleex1})--(\ref{simpleex6}) 
gives a reasonable semi-quantitative description of the acoustic oscillations. To simplify initial 
conditions even further we can indeed assume a flat Harrison-Zeldovich spectrum and set ${\mathcal R}_{*} = 1$.

\begin{figure}
\centering
\includegraphics[height=5.5cm]{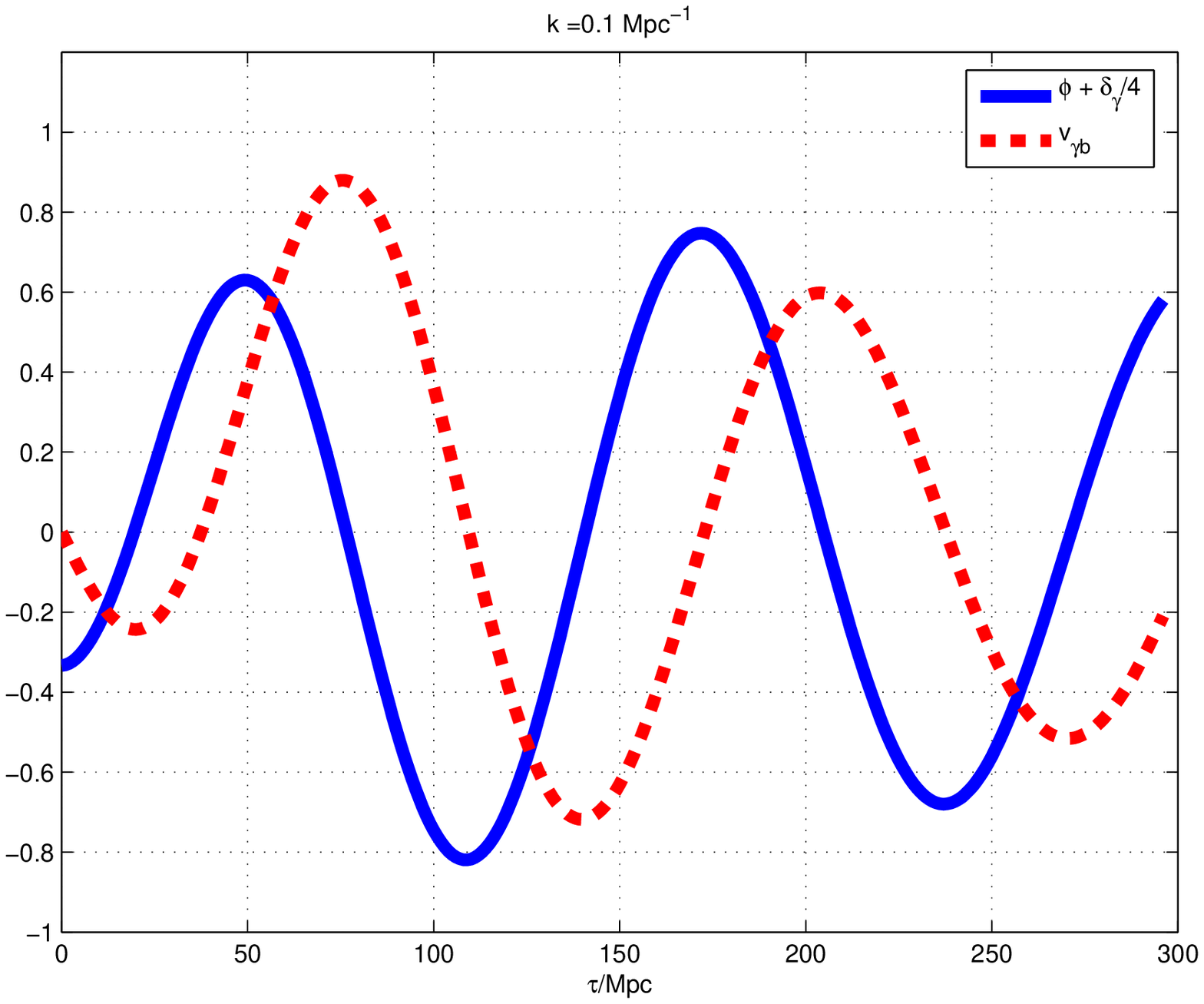}
\includegraphics[height=5.5cm]{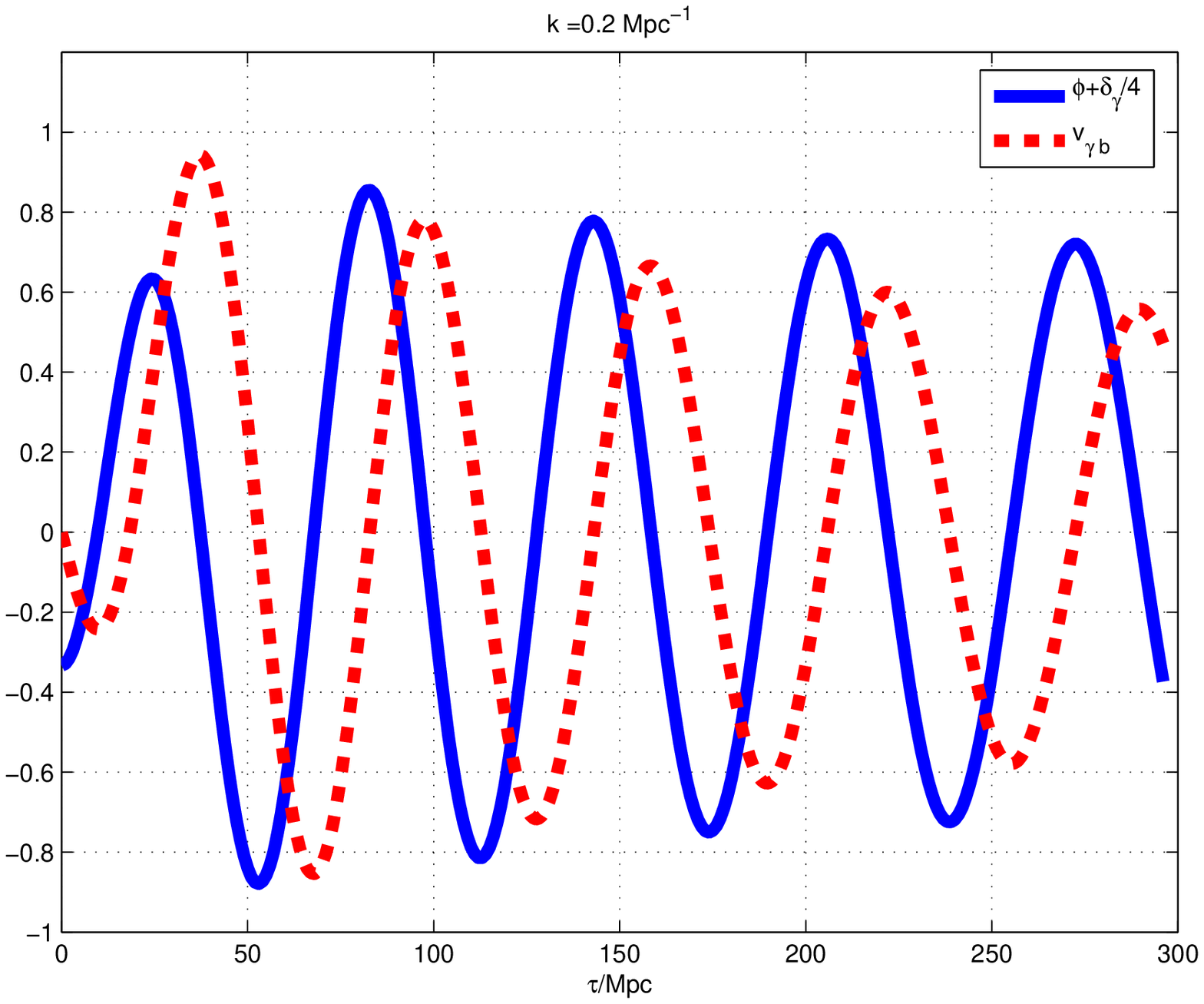}
\caption{The (ordinary) SW and Doppler contributions are illustrated as a function 
of the conformal time at fixed comoving wave-number. The initial conditions are adiabatic.}
\label{ADOSC}      
\end{figure}

The same philosophy used to get to this simplified form can be used to integrate the full system.
In this case, however, we would miss the important contribution of polarization 
since, to zeroth order in the tight-coupling expansion, the CMB is not polarized. 
In Fig. \ref{ADOSC} the so-called Doppler (or Sakharov) oscillations are reported 
for fixed comoving momentum $k$ and as a function of the cosmic time 
coordinate. The two plots illustrate to different values of $k$ in units of $\mathrm{Mpc}^{-1}$ and in the case 
of adiabatic initial conditions (see Eq. (\ref{adinsimpl})).
In each plot the ordinary Sachs-Wolfe contribution and the Doppler contributions 
are illustrated, respectively, with full and dashed lines. To make the plot more clear 
we just plotted the Fourier mode and not the Fourier amplitude (which differs from the 
Fourier mode by a factor $k^{3/2}$). The quantity $v_{\gamma b}$ is simply $\theta_{\gamma b}/(\sqrt{3} k)$.
From Fig. \ref{ADOSC} two general features emerge:
\begin{itemize}
\item{}in the adiabatic case  the ordinary SW contribution oscillates as a cosine;
\item{} in the adiabatic case the Doppler contribution (proportional to the 
peculiar velocity of the baryon-photon fluid) oscillates as a sine.
\end{itemize}
This rather naive observation has rather non-trivial consequences. In particular, the present 
discussion does not include polarization. However, the tight-coupling approximation can be 
made more accurate by going to higher orders. This will allow to treat polarization (see section \ref{sec9}). 
Now, the $Q$ stokes parameter evaluated to first-order in the tight-coupling expansion 
will be proportional to the zeroth-order dipole (see section \ref{sec9}). 
It is also useful to observe that in the units used in Fig. \ref{ADOSC} the decoupling occurs, as 
discussed in connection with Eq. (\ref{aint}), for $\tau_{\mathrm{dec}} \simeq 284$ Mpc. The 
equality time is instead for $\tau_{\mathrm{eq}} \simeq 120$ Mpc.
\begin{figure}
\centering
\includegraphics[height=5.5cm]{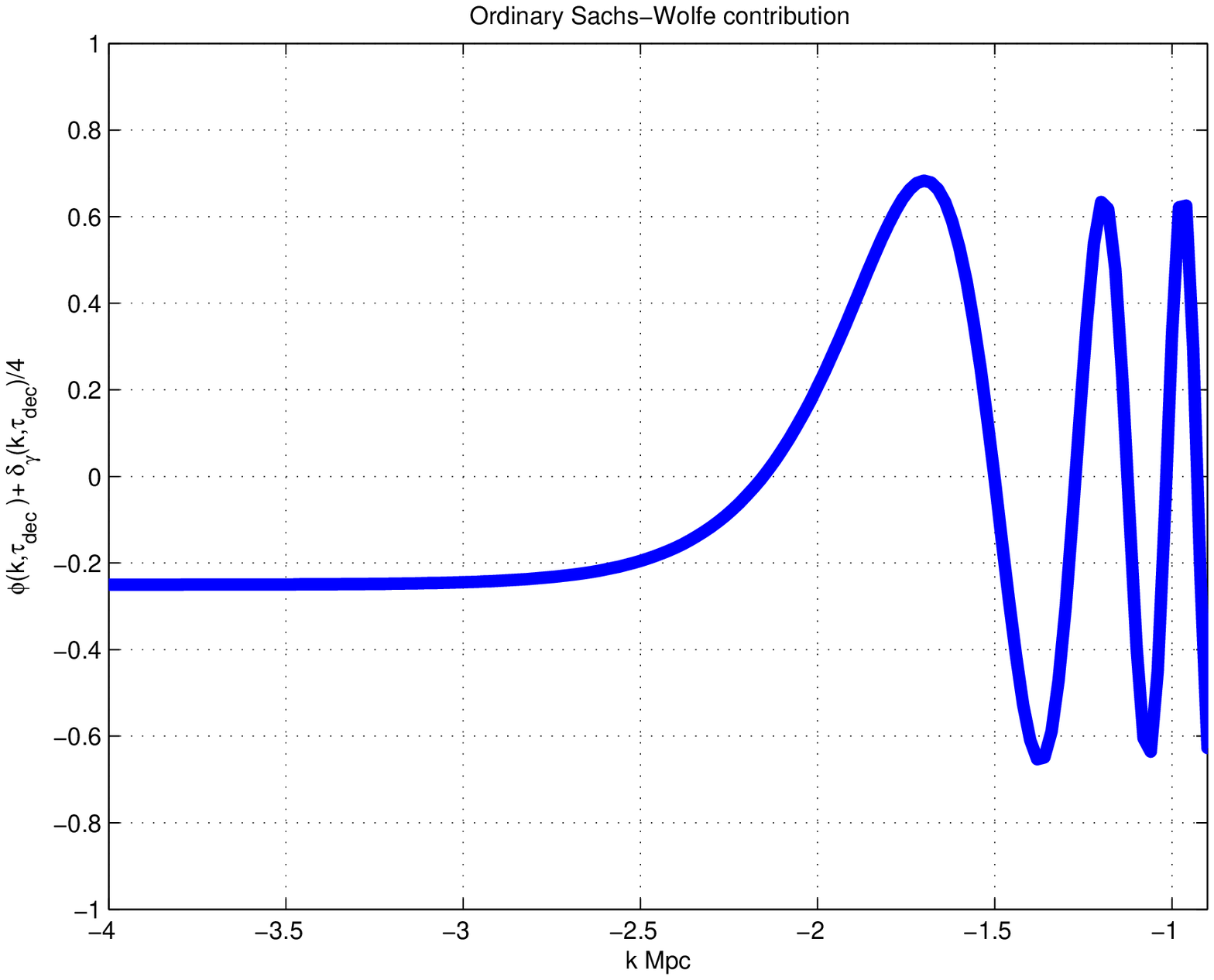}
\includegraphics[height=5.5cm]{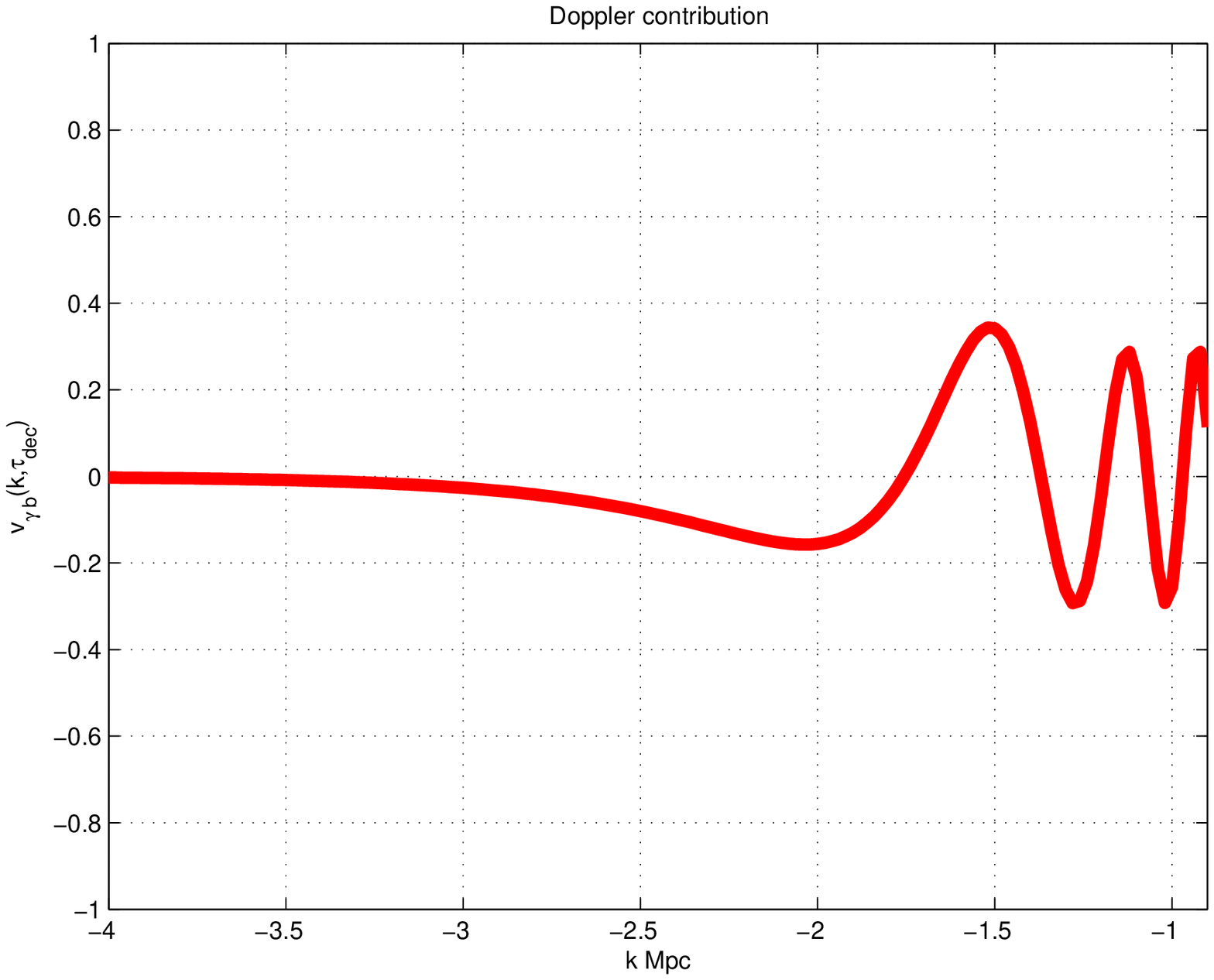}
\caption{The (ordinary) SW and the Doppler contributions are illustrated at fixed conformal time 
and as a function of $k$. The initial conditions are adiabatic.}
\label{ADK}      
\end{figure}
In Fig. \ref{ADK} the ordinary SW contribution and the Doppler contribution are 
illustrated at fixed time (coinciding with $\tau_{\mathrm{dec}}$) and for different comoving 
wave-numbers. It is clear that the ordinary SW contribution gives a peak (the so-called Doppler 
peak) that corresponds to a mode of the order  of the sound horizon at decoupling. Note 
that because of the phase properties of the SW contribution there is a region where the ordinary SW contribution 
is quasi-flat. This is the so-called Sachs-Wolfe plateau. Recall that, for sake of simplicity, the 
curvature fluctuation has been normalized to $1$ and the spectrum has been assumed scale-invariant.
This is a rather crude approximation that has been adopted only for the purpose of illustration.
Finally it should be remarked  that diffusive effects, associated with Silk damping, have been 
completely neglected. This is a bad approximation for scales that are shorter than the scale 
of the first peak in the temperature autocorrelation. It should be however mentioned that 
there are semi-analytical ways of taking into account the Silk damping also in the 
framework of the tight coupling expansion. In the tight coupling expansion the Silk damping arises 
naturally when going to second-order in the small parameter that is used in the expansion and that 
corresponds, roughly, to the inverse of the photon mean free path.

\begin{figure}
\centering
\includegraphics[height=5.5cm]{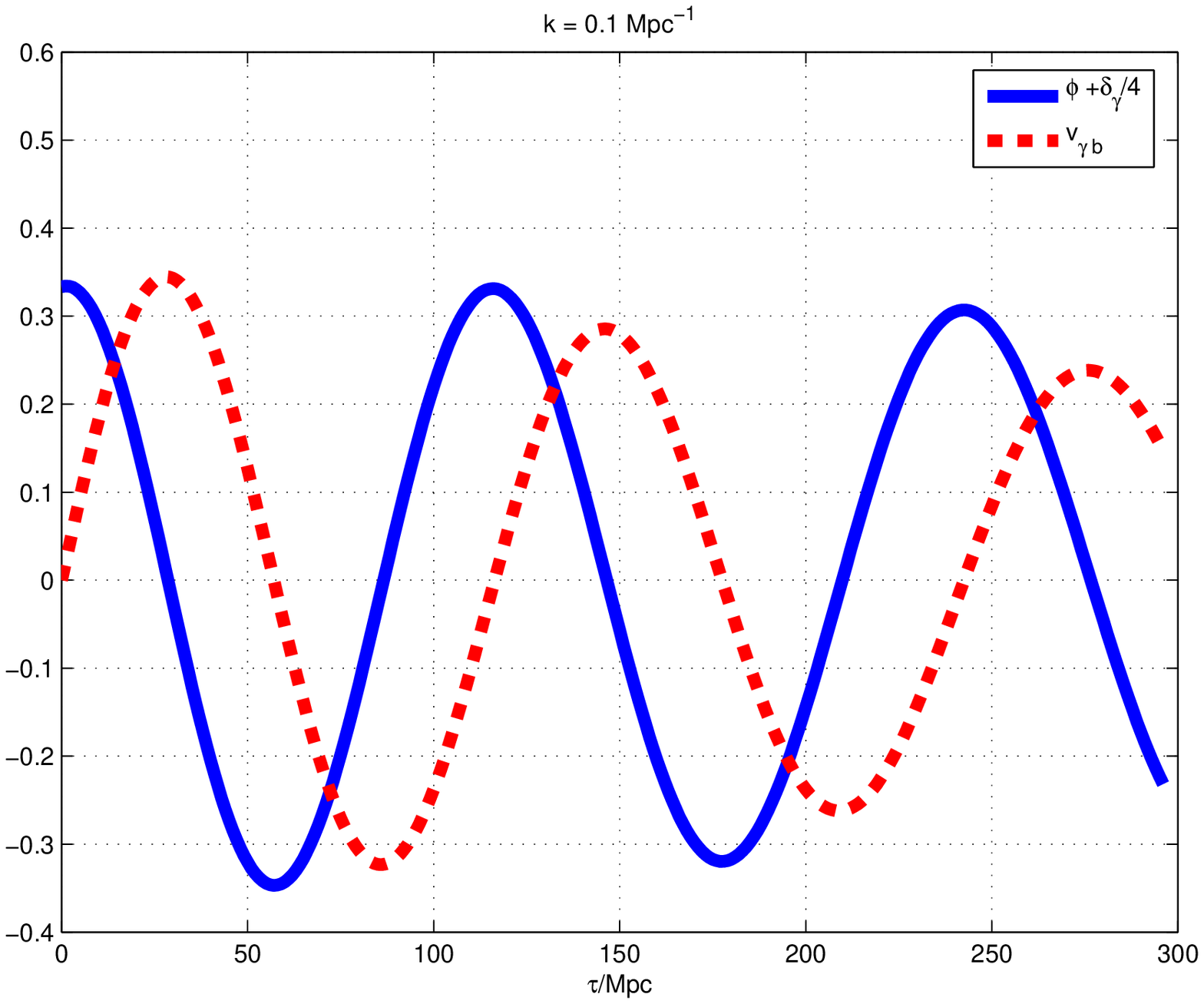}
\includegraphics[height=5.5cm]{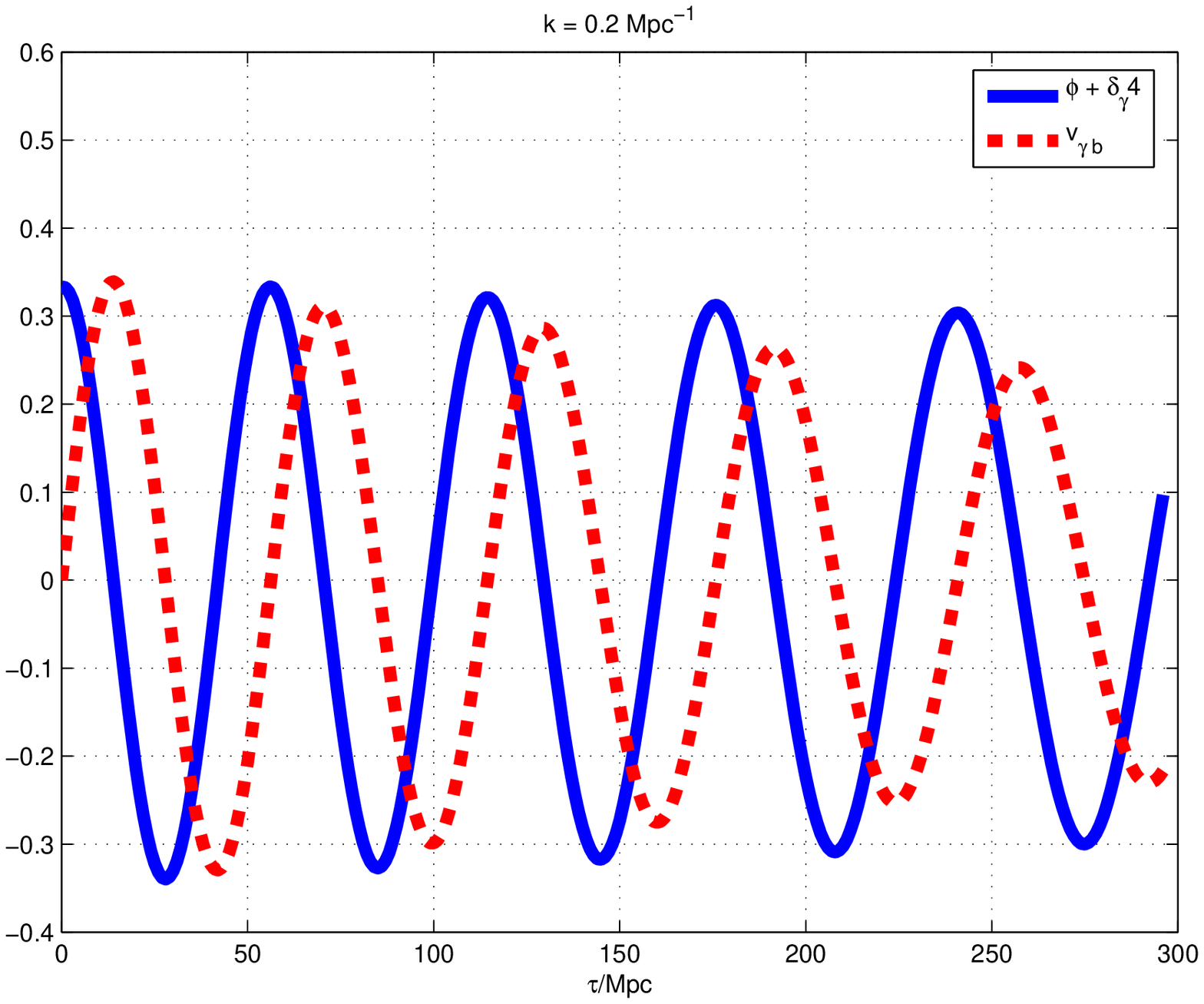}
\caption{The (ordinary) SW and Doppler contributions are illustrated as a function 
of the conformal time at fixed comoving wave-number. The initial conditions are non-adiabatic.}
\label{ISOOSC}      
\end{figure}

\begin{figure}
\centering
\includegraphics[height=4cm]{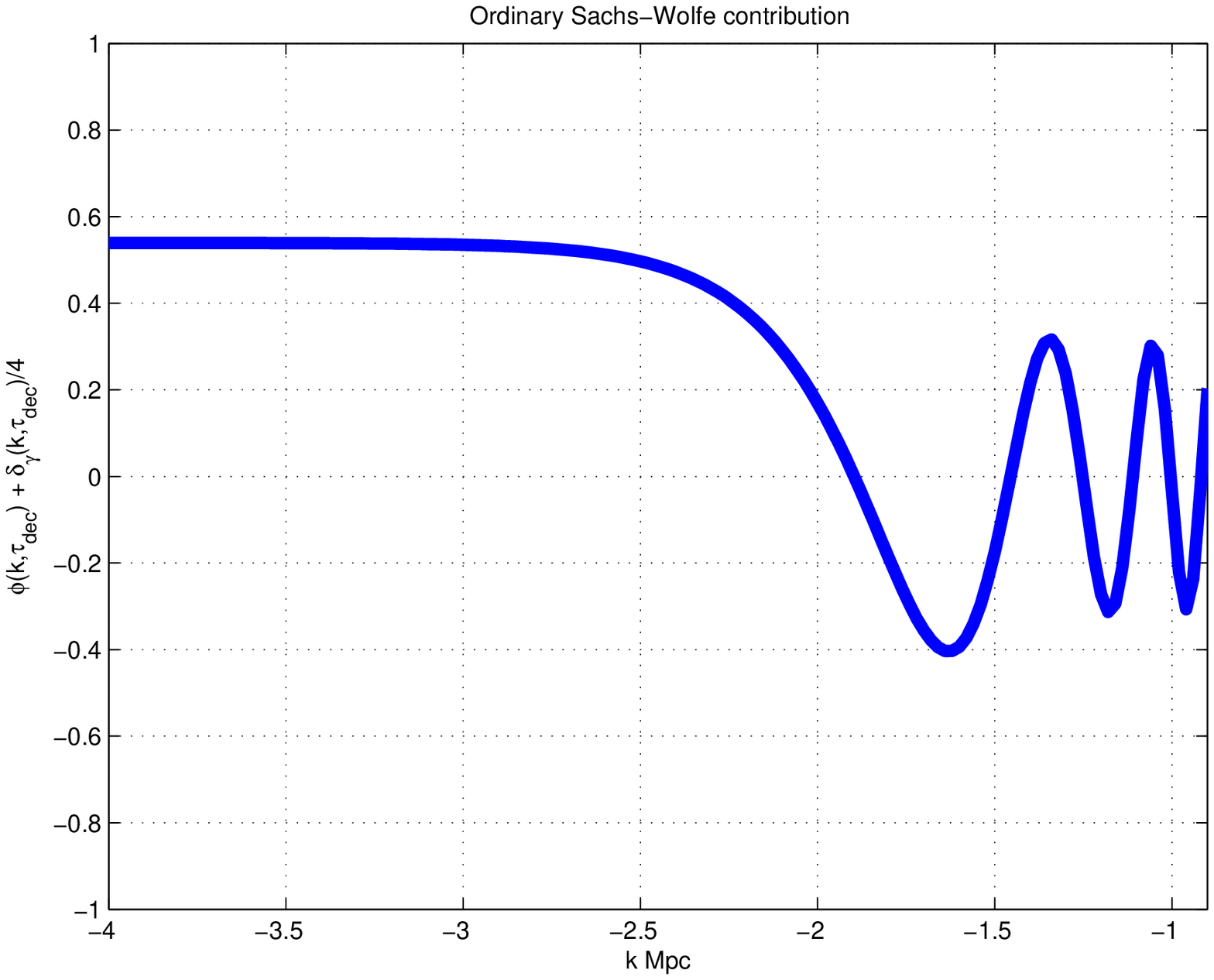}
\includegraphics[height=4cm]{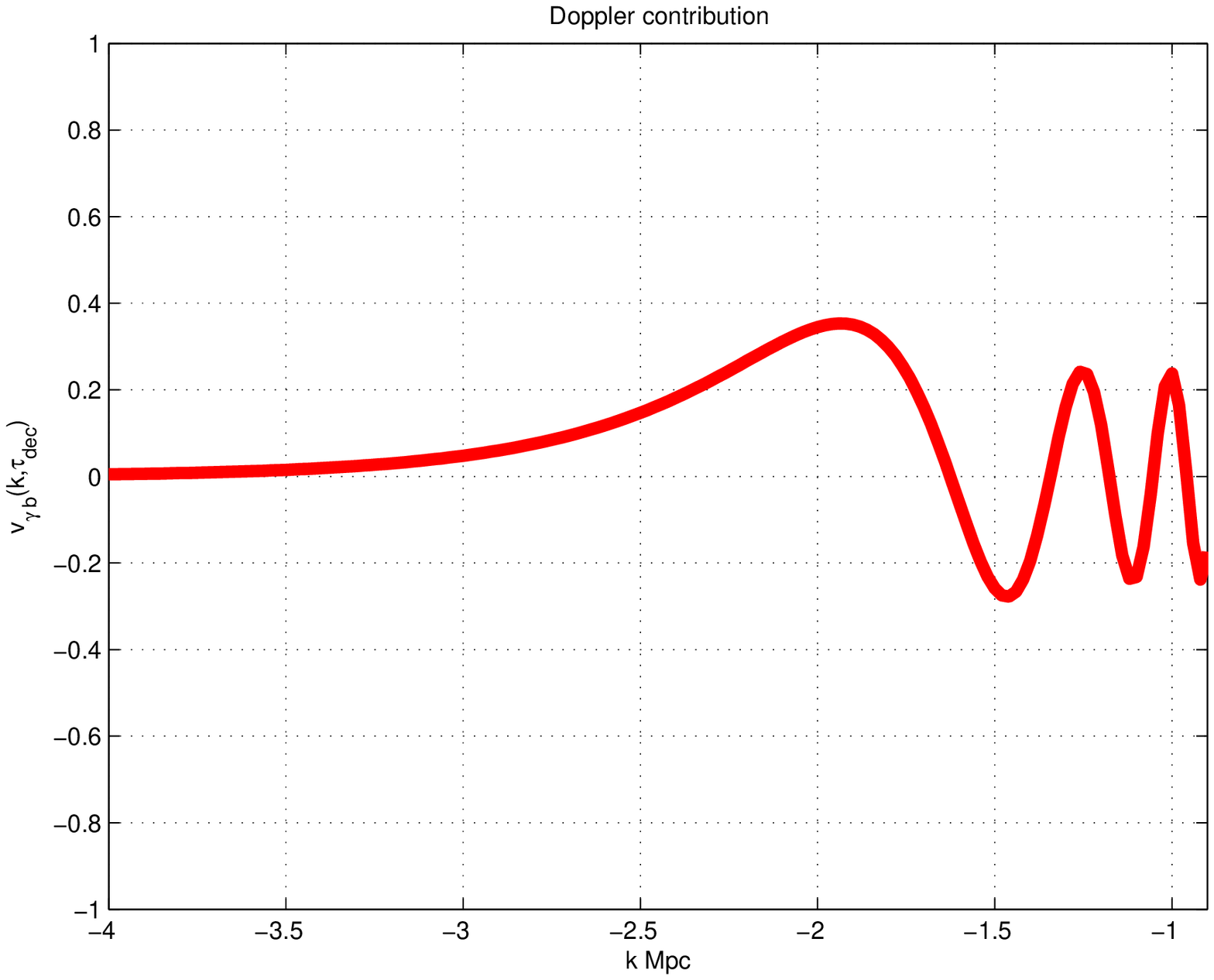}
\caption{The (ordinary) SW and the Doppler contributions are illustrated at fixed conformal time 
and as a function of $k$. The initial conditions are non-adiabatic.}
\label{ISOK}      
\end{figure}

Let us now move to the case of the non-adiabatic initial conditions. In this case, as already discussed, 
the curvature fluctuations vanish in the limit $x \to 0$ and, in particular, the CDM-radiation 
non-adiabatic mode implies that 
\begin{eqnarray}
&&{\mathcal R}(x_{\mathrm i}) = - \frac{{\mathcal S}_{*}}{3} x_{\mathrm{i}},\qquad \psi(x_{\mathrm i}) =
 \frac{{\mathcal S}_{*}}{4} x_{\mathrm{i}} ,
\nonumber\\
&&\delta_{\gamma}(x_{\mathrm i}) =  {\mathcal S}_{*} x_{\mathrm{i}} + \frac{4}{3} {\mathcal S}_{*}, \qquad \tilde{\theta}_{\gamma{\rm b}}(x_{\mathrm i}) =0,
\nonumber\\
&& \delta_{\mathrm c}(x_{\mathrm i}) = \delta_{\mathrm b}(x_{\mathrm i})=  \frac{3}{4}{\mathcal S}_{*}x_{\mathrm{i}} ,\qquad \tilde{\theta}_{{\rm c}}(x_{\mathrm i})=0.
\label{isoinsimpl}
\end{eqnarray}
In Fig. \ref{ISOOSC} the (ordinary) SW and Doppler contributions are 
reported for fixed comoving wave-numbers and as a function of the conformal time. 
Fig. \ref{ISOOSC} is the non-adiabatic counterpart of Fig. \ref{ADOSC}. It is 
clear that, in this case, the situation is reversed. While the adiabatic mode oscillates as cosine 
in the ordinary SW contribution, the non-adiabatic mode oscillates as a sine. Similarly, while the 
(adiabatic) Doppler contribution oscillates as a sine, the non-adiabatic Doppler term oscillates as cosine.
The different features of adiabatic and non-adiabatic contributions are even more evident in Fig. \ref{ISOK} 
which is the non-adiabatic counterpart of Fig. \ref{ADK}. 
Purely non-adiabatic initial conditions are excluded by current experimental data. However, a mixture 
of non-adiabatic and adiabatic initial conditions may be allowed as in the case of 
isocurvature modes induced by large-scale magnetic fields (see \cite{magn3,magn4,magn5}).

\newpage
\renewcommand{\theequation}{9.\arabic{equation}}
\setcounter{equation}{0}
\section{Kinetic hierarchies of multipole moments}
\label{sec9}
The effect of metric inhomogeneities on the properties  
of the radiation field will now be analyzed using the radiative transfer (or 
radiative transport) equations.
A classical preliminary reference  
 is the textbook of Chandrasekar
\cite{CHAN1} (see in particular chapter 1 in light of the calculation 
of the collision term of  Thompson scattering that is quite relevant for the present ends).
Another recent reference is \cite{PER1}. In broad terms the radiative transfer 
equations describe the evolution of the Stokes parameters of the radiation field through 
some layer of matter which could be, for instance, the stellar atmosphere or, in the present case, 
the primeval plasma prior to and beyond decoupling.

Radiative transfer equations have a further complication with 
respect to the flat space case:  the 
collisionless part of the Boltzmann equation  is modified by the inhomogeneities 
of the geometry. These 
inhomogeneities induce a direct coupling of the Boltzmann equation 
to the perturbed Einstein equations.  An interesting system 
of equations naturally emerges: the Einstein-Boltzmann
system of equations which is, in some approximation,  exactly 
what has been described in sections \ref{sec7} and \ref{sec8}.
 In that case the perturbed Einstein equations 
were coupled to a set of fluid equations for the density contrasts 
and for the peculiar velocities. These are, indeed, the first two terms (i.e. the 
monopole and the dipole)
in the Boltzmann hierarchy.  Truncated 
Boltzmann hierarchies are a useful tool for the analysis of
initial conditions, but  their limitations have been already 
emphasized in connection  with the description of collisionless 
particles.

While the general  conventions established  in the previous sections 
will be consistently enforced, further conventions related to the specific way  the brightness 
perturbations  are defined \footnote{See Eqs. (\ref{deBRf1})--(\ref{deBRf4}) below for the 
definition of brightness perturbation.}.   Denoting by $\Delta$ a  brightness 
perturbation (related generically to one of the four Stokes parameters 
of the radiation field), the expansion of $\Delta$ in terms of Legendre 
polynomials will be written, in this paper, as 
\begin{equation}
\Delta(\vec{k}, \hat{n}, \tau) = \sum_{\ell} (-i)^{\ell}\,\, (2 \ell + 1)\,\, \Delta_{\ell} (\vec{k},\tau)\,\,P_{\ell}(\hat{k}\cdot\hat{n}).
\label{CONV}
\end{equation}
where $\vec{k}$  is the momentum of the Fourier expansion, $\hat{k}$ its direction; $ \hat{n}$ 
is the direction of the photon momentum; $P_{\ell}(\hat{k}\cdot\hat{n})$ are the Legendre polynomials. The same 
expansion will be consistently employed for the momentum averaged phase-space 
density perturbation (see below Eq. (\ref{reddist})). This quantity will be also 
called, for short, reduced phase-space density and it is related to the brightness 
perturbation by a numerical factor. The conventions 
of \cite{HS1,HS2,HS3} (see also \cite{HS4})  are such that the factor $(2\ell +1)$  {\em  is not} 
included in the expansion.  Furthermore, in \cite{HS1,HS2,HS2a} the metric fluctuations 
are parametrized in terms of the Bardeen potential while in \cite{HS3} 
the treatment follows the conformally Newtonian gauge. 
Finally, in \cite{MB}  the conventions are the same as the ones of Eq. (\ref{CONV})
but  the metric convention is mostly plus (i.e. 
$-,+, +, +$) and the definition of the longitudinal 
degrees of freedom is inverted (i. e.  Ref.  \cite{MB} calls $\psi$ what we call $\phi$ and viceversa).
In \cite{HZ1,HZ2,HZ3} (see also \cite{kos2,KL})  the expansion of the brightness perturbation is different with respect 
to Eq. (\ref{CONV}) since the authors {\em do not} include the factor $(-i)^{\ell}$ in the expansion. In the latter case
the collision terms are modified by a sign difference in the 
dipole terms (involving a mismatch of $(-i)^2$ with respect to the conventions fixed by Eq. (\ref{CONV})). 

\subsection{Collisionless Boltzmann equation}
If the space-time would be 
homogeneous the position variables $x^{i}$ and  the conjugate momenta $P_{j}$ could 
constitute a practical set of pivot variables for the analysis of Boltzmann equation in curved 
backgrounds. However, since, in the 
present case, the space-time 
is not fully homogeneous, metric perturbations 
do affect the definition of conjugate momenta.
Hence,  for practical reasons, the approach usually followed is to write
the Boltzmann equations in terms of the proper moementa, i.e. 
the momentum measured by an observer at a fixed value of the spatial coordinate. 
Consider, for simplicity, the case of massless particles (like photons or massless neutrinos).
Their mass-shell condition can be written, in a curved background, as 
\begin{equation}
g_{\alpha\beta}P^{\alpha} P^{\beta}=0,
\label{MS}
\end{equation}
where  $g_{\alpha\beta}$  is now  the full
metric tensor (i.e. background plus inhomogeneities).
Equation (\ref{MS}) implies, quite trivially, 
\begin{equation}
g_{00} P^{0} P^{0} = - g_{ij} P^{i} P^{j} \equiv \delta_{ij} p^{i} p^{j},
\end{equation}
where the second equality is the definition of the physical
three momentum $p_{i}$.  Recalling that, to first order and in the 
longitudinal gauge, 
$g_{00}= a^2( 1 + 2\phi)$ and   $g_{i j} = - a^2 ( 1 - 2 \psi) \delta_{ij}$, then 
the relation between the conjugate momenta and the physical three-momenta 
can be easily obtained by expanding the obtained expressions 
for small $\phi$ and $\psi$. The result is simply 
\begin{eqnarray}
&& P^{0} = \frac{p}{a} ( 1 - \phi) = \frac{q}{a^2} ( 1 - \phi),\,\,\,\,\,\,\,\,\,\,\, P_{0} = a p ( 1 + \phi) = q ( 1 + \phi),
\nonumber\\
&& P^{i} =\frac{p^{i}}{a} ( 1 + \psi) = \frac{q^i}{a^2} ( 1 + \psi),\,\,\,\,\,\,\,\,\,\,\, P_{i} = - a p_{i} ( 1 - \psi) =- q_{i} ( 1 - \psi).
\label{CtoP}
\end{eqnarray}
The quantity $q_{i}$ defined in Eq. (\ref{CtoP}) are nothing but the 
comoving three-momenta, i.e.  $ p_{i} a = q_i$, while 
 $q =p a $ is the modulus of the comoving three-momentum. 
 Generalization of Eq. (\ref{CtoP}) is trivial since, in the 
 massive case, the mass-shell condition implies that 
 $g_{\alpha\beta}P^{\alpha} P^{\beta} = m^2$ and, for instance 
 $P^{0} = \sqrt{q^2 + m^2 a^2} ( 1 - \phi)$.
 In terms of the  modulus and direction of the comoving three-momentum \cite{bond1}, i.e. 
\begin{equation}
q_{i} = q n_{i}, \,\,\,\,\,\, n_{i}n^{i} = n_{i} n_{j} \delta^{ij} =1,
\label{C3m}
\end{equation}
the Boltzmann equation can be written as  
\begin{equation}
\frac{D f}{D \tau} = \frac{\partial f}{\partial \tau} + \frac{\partial x^{i}}{\partial\tau} \frac{\partial f}{\partial x^{i}} + 
\frac{\partial f}{\partial q} \frac{\partial q}{\partial \tau} + \frac{\partial f}{\partial n_{i}} \frac{\partial n^{i}}{\partial \tau} = 
{\cal C}_{\rm coll},
\label{BZ1}
\end{equation}
where a generic collision term, ${\cal C}_{\rm coll}$ has  been included for future convenience. 
Eq. (\ref{BZ1}) can now be perturbed around a configuration of local 
thermodynamic equilibrium by writing
\begin{equation}
f( x^i, q, n_{j}, \tau)= f_{0}(q) [ 1 + f^{(1)}( x^i, q, n_{j}, \tau)],
\label{expBZ}
\end{equation}
where $f_0(q)$ is the Bose-Einstein (or Fermi-Dirac in the case of fermionic 
degrees of freedom) distribution. Notice that $f_{0}(q)$  does not depend on
$n^{i}$ but only on $q$.
Inserting Eq. (\ref{expBZ}) into Eq. (\ref{BZ1}) the first-order form of the perturbed 
Boltzmann equation  can be readily obtained 
\begin{equation}
f_{0}(q)\,\, \frac{\partial f^{(1)}}{\partial \tau} + f_0(q)\,\,\frac{\partial f^{(1)}}{\partial x^{i}} \,\,n^{i}+ 
\frac{\partial f_0}{\partial q} \frac{\partial q}{\partial \tau}= 
{\cal C}_{\rm coll},
\label{BZ2}
\end{equation}
by appreciating  that a pair of terms 
\begin{equation}
\frac{\partial f^{(1)}}{\partial q} \frac{\partial q}{\partial \tau}, \,\,\,\,\,\,\,\,\,\,\,\,\,\,\,\,\,
 \frac{\partial f^{(1)}}{\partial n_{i}} \frac{\partial n^{i}}{\partial \tau},
\end{equation}
are of higher order (i.e. ${\cal O}(\psi^2)$) and have been neglected to first-order.
Dividing by $f_{0}$ Eq. (\ref{BZ2}) can also be written as 
\begin{equation}
\frac{\partial f^{(1)}}{\partial \tau} +   \frac{\partial f^{(1)}}{\partial x^{i}}  \,n^{i}+ 
\frac{\partial \ln{f_0}}{\partial q} \frac{\partial q}{\partial \tau}= 
\frac{1}{f_{0}}{\cal C}_{\rm coll}.
\label{BZ3}
\end{equation}
Notice that in Eq. (\ref{BZ2})--(\ref{BZ3})  the generalization  of known 
special relativistic expressions 
\begin{equation}
\frac{ d x^{i}}{d\tau} = \frac{P^{i}}{P^{0}} = \frac{q^{i}}{q} = n^{i},
\end{equation}
has been used.
To complete the derivation, $dq/d\tau$ must be written 
in explicit terms.  The geodesic 
equation gives essentially the first time derivative of the 
conjugate momentum, i.e. 
\begin{equation}
\frac{d P^{\mu}}{d s } = P^{0} \frac{d P^{\mu}}{d \tau} = - \Gamma^{\mu}_{\alpha\beta} P^{\alpha} P^{\beta},
\label{geod1a}
\end{equation}
where $s$ is the affine parameter.
As before in this section, $\Gamma^{\mu}_{\alpha\beta}$ denotes the full Christoffel connection (background 
plus fulctuations).  Using the values of the perturbed connections in the longitudinal gauge 
 Eq. (\ref{geod1a}) becomes  
\begin{equation}
\frac{d P^{i}}{d\tau} = - \partial^{i} \phi P^{0} + 2 \psi' P^{i} - 2 {\cal H} P^{i} - 
\frac{P^{j} P^{k}}{P^{0}} [ \partial^{i} \psi \delta_{jk} - \partial_{k} \psi \delta_{j}^{i} - \partial_{j} \psi \delta^{i}_{k}].
\label{geod2a}
\end{equation}
Recalling now that $q = q_{i} n^{i}$, the explicit form of $d q/d\tau$ will be 
\begin{equation}
\frac{d q}{d\tau} = \biggl[ \frac{\partial P^{i}}{\partial \tau} a^2 ( 1 - \psi) + 2 {\cal H} a^2 (1 - \psi ) P^{i} - a^2 \psi' P^{i} \biggr] n_{i}  - P^{i} a^2 \partial_{j}  \psi n^{j} n_{i}. 
\label{qtau}
\end{equation}
Inserting now Eq. (\ref{geod2a}) into Eq. (\ref{qtau}) 
the explicit form of $d q/d\tau$ becomes\footnote{To derive Eq. (\ref{qtau2}) from Eq. (\ref{qtau}),
the factors $P^{i}$ and $P^{0}$ appearing at the right hand side 
 Eq. (\ref{geod2a}) have to be replaced with their first-order 
expression in terms of the comoving-three momentum $q^{i}$ (and $q$) 
 as previously discussed in Eqs. (\ref{CtoP}).}
\begin{equation}
\frac{d q }{ d\tau} = q \psi' - q n_{i} \partial^{i} \phi.
\label{qtau2}
\end{equation}
Finally, using Eq. (\ref{qtau2}) into Eq.  (\ref{BZ3}) to eliminate $dq/d\tau$ the final form of the 
Boltzmann equation for massless particles becomes:
\begin{equation}
\frac{\partial f^{(1)}}{\partial \tau} +  n^{i} \frac{\partial f^{(1)}}{\partial x^{i}} + 
\frac{\partial \ln{f_0}}{\partial\ln q} [\psi' -  n_{i} \partial^{i} \phi] = 
\frac{1}{f_{0}}{\cal C}_{\rm coll},
\end{equation}
which can be also written, going to Fourier space, as 
\begin{equation}
\frac{\partial f^{(1)}}{\partial \tau} +  i k\mu f^{(1)} + 
\frac{\partial \ln{f_0}}{\partial\ln q} [\psi' -  i k\mu \phi] = 
\frac{1}{f_{0}}{\cal C}_{\rm coll},
\label{BZ4}
\end{equation}
where we have denoted, according to the standard notation, $k$ as the Fourier mode and 
$\mu = \hat{k}\cdot \hat{n} $ as the projection of the Fourier mode along 
the direction of the photon momentum \footnote{Notice that here there may be, in principle, a 
clash of notations since, in section \ref{sec6}
we denoted with $\mu$ the 
normal modes for the tensor action; in the present section $q$ and $\mu$ 
denote, on the contrary the comoving three-momentum and the cosine between the Fourier 
mode and the photon direction. The two sets of variables never appear together and there should not
be confusion.}. Clearly,  given the axial symmetry of the problem 
it will be natural to identify  the direction of $\vec{k}$ with the $\hat{z}$ direction in which 
case $\mu =\cos{\theta}$.
The result obtained so far can be easily generalized to the case of massive particles 
\begin{equation}
\frac{\partial f^{(1)}}{\partial \tau} +  i \alpha(q,m) k\mu f^{(1)} + 
\frac{\partial \ln{f_0}}{\partial\ln q} [\psi' -  i  \alpha(q,m) k\mu \phi] = 
\frac{1}{f_{0}}{\cal C}_{\rm coll},
\label{BZ5}
\end{equation}
where $\alpha(q,m) = q/\sqrt{q^2 + m^2 a^2}$ and where, now, the appropriate mass dependence 
has to appear in the equilibrium distribution $f_{0}(q)$. 

\subsection{Boltzmann hierarchy for massless neutrinos}
The Boltzmann equations derived in Eqs. (\ref{BZ4}) and (\ref{BZ5}) are general.  
In the following, two relevant cases will be discussed, namely the case of massless 
neutrinos and the case of photons. 
In order to proceed further with the case of massless neutrinos 
let us define the reduced phase-space distribution as 
\begin{equation}
{\cal F}_{\nu}( \vec{k}, \hat{n}, \tau) = \frac{\int q^{3} d q f_{0} f^{(1)}}{\int q^{3} d q f_{0}}.
\label{reddist}
\end{equation}
Eq. (\ref{BZ4}) becomes, in the absence of collision term,
\begin{equation}
\frac{\partial {\cal F}_{\nu}}{\partial \tau} + i k\mu {\cal F}_{\nu} = 4 (\psi' - i k \mu \phi).
\label{bz3} 
\end{equation}
 The factor $4$ appearing in Eq. (\ref{bz3}) follows 
from the explicit expression of the equilibrium Fermi-Dirac distribution and observing that integration by parts 
implies 
\begin{equation}
\int_{0}^{\infty} q^{3} d q \frac{\partial f_{0}}{ \partial \ln{q}} = - 4  \int_{0}^{\infty} q^{3} d q f_{0}.
\label{4fact}
\end{equation}
The reduced phase-space distribution of Eq. (\ref{reddist})  can be expanded in series of Legendre 
polynomials as defined in Eq. (\ref{CONV})
\begin{equation}
{\cal F}_{\nu}( \vec{k}, \hat{n}, \tau) = \sum_{\ell} (-i)^{\ell} ( 2 \ell + 1) {\cal  F}_{\nu\ell}(\vec{k},\tau) P_{\ell}(\mu).
\label{expF}
\end{equation}
Equation (\ref{expF}) will now be inserted  into Eq. (\ref{bz3}). 
The orthonormality relation for Legendre polynomials \cite{abr,tric},
\begin{equation}
\int_{-1} ^{1}  P_{\ell}(\mu) P_{\ell'}( \mu) d\mu  = \frac{2}{2 \ell + 1} \delta_{\ell\ell'},
\label{norm}
\end{equation}
together with the well-known recurrence relation 
\begin{equation}
(\ell + 1) P_{\ell + 1}(\mu) = (2 \ell +1) \mu P_{\ell}(\mu) - \ell P_{\ell -1}(\mu),
\label{rec1}
\end{equation}
allows to get a hierarchy of differential equations coupling together 
the various multipoles.  After having multiplied
each of the terms of Eq. (\ref{bz3}) by $\mu$, integration of the obtained 
quantity will be performed over $\mu$ (varying between $-1$ and $1$); in formulae:
\begin{eqnarray}
&& \int_{-1}^{1} P_{\ell'}(\mu) {\cal F}_{\nu} d\mu = 2 (-i)^{\ell'} {\cal F}_{\nu\ell'},
\label{r1a}\\
&& ik \int_{-1}^{1} \mu P_{\ell'}(\mu) {\cal F}_{\nu} 
d\mu = 2 i k \biggl[ (-i)^{\ell' + 1} \frac{\ell' +1}{ 2 \ell' + 1}  {\cal F}_{\nu(\ell' +1)}  
\nonumber\\
&&+ (-i)^{\ell' -1}\frac{\ell'}{2\ell' + 1} {\cal F}_{\nu (\ell' -1)}\biggr],  
\label{r2}\\
&& 4\int_{-1}^{1}  \psi' P_{\ell'}(\mu) d\mu = 8\psi' \delta_{\ell'0} ,\,\,\,\,\,\,\,\,\,\,\,,\,- 4 i \phi \int_{-1}^{1} \mu P_{\ell'}(\mu) d\mu  = - \frac{8}{3} i k \phi \delta_{\ell' 1}.
\label{r4}
\end{eqnarray}
Equation (\ref{r2}) follows from the relation
\begin{equation}
\int_{-1}^{1} \mu P_{\ell}(\mu) P_{\ell'}(\mu) d\mu = \frac{2}{2\ell + 1} \biggl[ \frac{\ell' +1 }{2\ell' + 1} \delta_{\ell,\ell'+1} 
+ \frac{\ell'}{2\ell' + 1} \delta_{\ell,\ell' -1} \biggr]
\end{equation}
that can be easily derived using Eqs. (\ref{rec1}) and (\ref{norm}).
Inserting Eqs. (\ref{r1a})--(\ref{r4}) into Eq. (\ref{bz3}) the first example of Boltzmann hierarchy can be derived:
\begin{eqnarray}
&& {\cal F}_{\nu 0}' = - k {\cal F}_{\nu 1} + 4 \psi',
\label{mom0}\\
&& {\cal F}_{\nu 1}' = \frac{k}{3} [ {\cal F}_{\nu 0} - 2 {\cal F}_{\nu 2}] + \frac{4}{3} k \phi,
\label{mom1}\\
&& {\cal F}_{\nu\ell}' = \frac{k}{2\ell +1} [ \ell {\cal F}_{\nu,(\ell-1)}  - (\ell+1) {\cal F}_{\nu (\ell+1)}].
\label{mom2}
\end{eqnarray}
Equation (\ref{mom2}) holds for $\ell \geq 2$. Eqs. (\ref{mom0}) and (\ref{mom1}) 
are nothing but the evolution equations for the density contrast and for the neutrino 
velocity field. This aspect can be easily appreciated by computing, in explicit terms, 
the components of the energy-momentum tensor as a function 
of the reduced neutrino phase-space density. In general terms, the energy-momentum 
tensor can be written, in the kinetic approach, as 
\begin{equation}
T_{\mu}^{\nu} = - \int \frac{d^{3} P}{\sqrt{-g}} \frac{P_{\mu}P^{\nu}}{P^{0}} f(x^{i}, P_{j}, \tau).
\label{defFTmunu}
\end{equation}
According to Eq. (\ref{CtoP}), establishing the connection between conjugate momenta 
and comoving three-momenta, the $(00)$ component of Eq. (\ref{defFTmunu}) becomes, for a 
completely homogeneous distribution,
\begin{equation}
\rho_{\nu} = \frac{1}{a^4} \int d^{3} q q  f_{0}(q),
\end{equation}
i.e. the homogeneous energy density.  Using instead the first-order phase space density, 
the density contrast, the peculiar velocity field and the neutrino anisotropic stress 
are connected, respectively, to the monopole, dipole and quadrupole moments of the 
reduced phase-space distribution:
\begin{eqnarray}
&& \delta_{\nu} = \frac{1}{4\pi} \int d \Omega {\cal F}_{\nu}(\vec{k},\hat{n},\tau) = {\cal F}_{\nu 0},
\label{def1a}\\
&& \theta_{\nu} = \frac{3i}{16\pi} \int d\Omega (\vec{k}\cdot \hat{n}) {\cal F}_{\nu}(\vec{k}, \hat{n},\tau) 
= \frac{3}{4} k {\cal F}_{\nu 1},
\label{def0a}\\
&& \sigma_{\nu} = -\frac{3}{16\pi }\int d\Omega \biggl[ (\vec{k}\cdot \hat{n})^2 - \frac{1}{3}\biggr] {\cal F}_{\nu}(\vec{k},\hat{n},\tau) =  \frac{{\cal F}_{\nu 2}}{2}.
\label{def2a}
\end{eqnarray}
Inserting Eqs. (\ref{def1a}) and (\ref{def2a}) into Eqs. (\ref{mom0})--(\ref{mom2}), the system following from 
the perturbation of the covariant conservation equations can be partially recovered 
\begin{eqnarray}
&& \delta_{\nu}' = - \frac{4}{3} \theta_{\nu} + 4\psi',
\label{mom4a}\\
&& \theta_{\nu}' = \frac{k^2}{4} \delta_{\nu} - k^2 \sigma_{\nu} + k^2 \phi,
\label{mom4b}\\
&& \sigma_{\nu}' = \frac{4}{15} \theta_{\nu} - \frac{3}{10} k {\cal F}_{\nu 3},
\label{mom4}
\end{eqnarray}
with the important addition of the quadrupole (appearing in Eq. (\ref{mom4a})) and of the whole 
Eq. (\ref{mom4}), which couples the quadrupole, the peculiar velocity field, and the octupole ${\cal F}_{\nu3}$.
For the adiabatic mode, after neutrino decoupling, ${\cal F}_{\nu 3}=0$. 
The problem of dealing with neutrinos 
while setting initial conditions for the evolution of the CMB anisotropies can be now fully
 understood. The fluid approximation 
implies that the dynamics of neutrinos can be initially described, after neutrino decoupling, by 
the evolution of the monopole and dipole of the neutrino phase space distribution. However, in order 
to have an accurate description of the initial conditions one should solve an infinite 
hierarchy of equations for the time derivatives of higher order moments 
of the photon distribution function. 

Eqs. (\ref{mom0})--(\ref{mom2}) hold for massless neutrinos but a similar hierarchy 
can be derived also in the case of the photons or, more classically, in the case of the 
brightness perturbations of the radiation field to be discussed below. The spatial 
gradients of the longitudinal fluctuations of the metric are sources of the equations 
for the lowest multipoles, i.e. Eqs. (\ref{mom0}) and (\ref{mom1}). For $\ell >2$, each multipole 
is coupled to the preceding (i.e. $(\ell -1)$) and to the following (i.e. $(\ell +1)$) multipoles. To 
solve numerically the hierarchy one could truncate the system at a certain $\ell_{\rm max}$.
This is, however, not the best way of dealing with the problem since \cite{MB} the effect 
of the truncation could be an unphysical reflection of power down through the lower 
(i.e. $\ell <\ell_{\rm max}$) multipole moments. This problem can be efficiently 
addressed with the method of line-of-sight integration (to be discussed later in this section) that 
is also a rather effective in the derivation of approximate expressions, for instance, of the polarization 
power spectrum. The method of line-of-sight integration is the one used, for instance, in CMBFAST 
\cite{CMBF,zaldsel}.

\subsection{Brightness perturbations of the radiation field}
Unlike  neutrinos, photons are a collisional species, so the generic collision term appearing in 
Eq. (\ref{BZ5}) has to be  introduced.  With this warning in mind, all the results derived so
far can be simply translated to the case of photons (collisionless part of Boltzmann equation, 
relations between the moments of the reduced phase-space and the components of the energy-momentum tensor...)
provided the Fermi-Dirac equilibrium distribution is replaced by the Bose-Einstein distribution.
 
Thompson scattering leads to a collision term  that depends both on the baryon velocity 
field \footnote{Since the electron-ion collisions are sufficiently rapid, it is 
normally assumed, in analytical estimates of CMB effects, that electrons 
and ions are in kinetic equilibrium at a common temperature $T_{\rm eb}$}
and on the direction cosine $\mu$ \cite{CHAN1}.  
The collision term is different for the brightness function describing 
the fluctuations of the total intensity of the radiation field (related to the 
Stokes parameter $I$) and for the brightness functions describing the degree 
of polarization of the scattered radiation (related to the Stokes parameters 
$U$ and $V$).

The conventions 
for the Stokes parameters and their well known properties will now be summarized: they 
can be found in standard electrodynamics textbooks
\cite{jackson} ( see also \cite{kos1,zalexp,huwhpol}  for phenomenological 
introduction to the problem of CMB polarization and \cite{kos2} for a more theoretical
perspective). Consider, for simplicity, 
a monochromatic  radiation field decomposed according to its linear polarizations and 
travelling along the $z$ axis:
\begin{equation}
\vec{E} = [ E_{1} \hat{e}_{x} + E_{2} \hat{e}_{y}] e^{i ( k z - \omega t)}.
\end{equation}
The decomposition according to circular polarizations can be written as 
\begin{equation}
\vec{E}= [ \hat{\epsilon}_{+} E_{+} + \hat{\epsilon}_{-} E_{-}] e^{i ( k z - \omega t)},
\end{equation}
where 
\begin{eqnarray}
 &&\hat{\epsilon}_{+} = \frac{1}{\sqrt{2}}( \hat{e}_{x} + i \hat{e}_{y}),\,\,\,\,\,\,\,\,\,
 \label{posh}\\
 && \hat{\epsilon}_{-} = \frac{1}{\sqrt{2}}( \hat{e}_{x} - i \hat{e}_{y}).
\label{neghel}
\end{eqnarray}
Eq. (\ref{posh}) is defined to be, conventionally, a {\em positive} helicity, while 
Eq. (\ref{neghel}) is the {\em negative} helicity. Recalling that  $E_{1}$ and $E_{2}$ can be written as 
\begin{equation}
E_{1} = E_{x} e^{i \delta_{x}},\,\,\,\,\,\,\,\,\,\,\,\,\,\,E_{2} = E_{y} e^{i \delta_{y}},
\end{equation}
the polarization properties of the radiation field can be described  in terms of $4$ real numbers
given by the projections of the radiation field over the linear and circular polarization unit vectors, i.e.
\begin{equation}
(\hat{e}_{x} \cdot \vec{E}),\,\,\,\,\,\,(\hat{e}_{y} \cdot \vec{E}),\,\,\,\,\,\,
(\hat{\epsilon}_{+} \cdot \vec{E}),\,\,\,\,\,\,(\hat{\epsilon}_{-} \cdot \vec{E}).
\end{equation}
The four  Stokes parameters are then, in the linear polarization basis
\begin{eqnarray}
&& I = |\hat{e}_{x} \cdot \vec{E}|^2 +  |\hat{e}_{y} \cdot \vec{E}|^2= E_{x}^2 + E_{y}^2,
\label{Idef}\\
&& Q = |\hat{e}_{x} \cdot \vec{E}|^2  - |\hat{e}_{y} \cdot \vec{E}|^2= E_{x}^2 - E_{y}^2 ,
\label{Qdef}\\
&& U = 2 {\rm Re}[(\hat{e}_{x} \cdot \vec{E})^{\ast} (\hat{e}_{y} \cdot \vec{E})]= 2 E_{x} E_{y} \cos{(\delta_y - \delta_{x})},
\label{Udef}\\
&& V= 2 {\rm Im}[(\hat{e}_{x} \cdot \vec{E})^{\ast} (\hat{e}_{y} \cdot \vec{E})]= 2 E_{x} E_{y} \sin{(\delta_y - \delta_{x})}.
\label{Vdef}
\end{eqnarray}
Stokes parameters are not all invariant under rotations. Consider a two-dimensional (clock-wise)
rotation of the coordinate system, namely
\begin{eqnarray}
&& \hat{e}_{x}' = \cos{\varphi} \hat{e}_{x}  + \sin{\varphi} \hat{e}_{y},
\nonumber\\
&&  \hat{e}_{y}' = -\sin{\varphi} \hat{e}_{x}  + \cos{\varphi} \hat{e}_{y}.
\label{rotation}
\end{eqnarray}
Inserting Eq. (\ref{rotation}) into Eqs. (\ref{Idef})--(\ref{Vdef}) it can be easily shown that 
$I'= I$ and $V'= V$ where the prime denotes the expression of the Stokes parameter 
in the rotated coordinate system. However, the remaining two parameters 
mix, i.e.
\begin{eqnarray}
&& Q' = \cos{2\varphi} Q + \sin{2\varphi} U,
\nonumber\\
&& U' = -\sin{2 \varphi} Q + \cos{2\varphi} U.
\label{clockwise}
\end{eqnarray}
From the last expression it can be easily shown that the polarization degree $P$ is invariant
\begin{equation}
P= \sqrt{Q^2 + U^2}= \sqrt{{Q'}^2 + {U'}^2}, 
\label{DP}
\end{equation}
while $U/Q= \tan{2\alpha}$ transform as $U'/Q' = \tan{2(\alpha -  \varphi)}$.

Stokes parameters are not independent (i.e. it holds that 
$I^2 = Q^2 + U^2 + V^2$ ), they only depend on the difference of the phases (i.e. $(\delta_{x} - \delta_{y})$) 
but not on their sum (see Eqs. (\ref{Idef})--(\ref{Vdef}) ).
 Hence the polarization tensor of the electromagnetic field can be written in matrix notation as 
\begin{equation}
\rho = \left(\matrix{I + Q 
& U - i V &\cr
U + i V  & I - Q &\cr}\right) \equiv \left(\matrix{E_{x}^2 
& E_{x} E_{y} e^{ - i \Delta} &\cr
 E_{x} E_{y} e^{ i \Delta} & E_{y}^2 &\cr}\right),
\label{matrixpol}
\end{equation}
where $\Delta = (\delta_{y} - \delta_{x})$. If the radiation field would be treated in a second 
quantization approach, Eq. (\ref{matrixpol})  can be promoted to the status of density matrix 
of the radiation field \cite{kos2}.

The evolution equations for the brightness functions will now be derived. 
Consider, again Eq. (\ref{BZ4}) written, this time, in the case of photons. As in the case 
of neutrinos we can define a reduced phase space distribution ${\cal F}_{\gamma}$, just changing $\nu$ 
with $\gamma$ in Eq. (\ref{reddist}) and using the Bose-Einstein 
instead of the Fermi-Dirac equlibrium distribution. 
The reduced photon phase-space density describes 
the fluctuations of the intensity of the radiation field (related to the Stokes parameter 
$I$); a second reduced  phase-space distribution, be it  ${\cal G}_{\gamma}$, can be 
defined  for the difference of the two intensities
(related to the stokes parameter Q). The equations for ${\cal F}_{\gamma}$ and ${\cal G}_{\gamma}$ 
can be written as 
\begin{eqnarray}
&&\frac{\partial F_{\gamma}}{\partial \tau} + i k\mu F_{\gamma} - 4 (\psi' - i k \mu \phi) = {\cal C}_{I},
\nonumber\\
&&\frac{\partial G_{\gamma}}{\partial \tau} + i k\mu G_{\gamma}  = {\cal C}_{Q},
\label{FGC}
\end{eqnarray}
The collision terms for these two equations are different \cite{bes1,bes2} and can be 
obtained following the derivation reported in the chapter 1 of Ref. \cite{CHAN1} or by 
following the derivation of Bond (with different notations) in the appendix C of Ref. \cite{bond} (see from p. 638).
Another way of deriving the collision 
terms for the evolution equations of the brightness perturbations is by employing 
the total angular momentum method \cite{tot1} that will be swiftly discussed in connection 
with CMB polarization.

Before writing the explicit form of the equations, including the collision terms,  it is useful to pass 
directly to the brightness perturbations. 
For the fluctuations of the total intensity  of the radiation field
the brightness perturbations is simply given by 
\begin{equation}
f( x^{i}, q, n_{j}, \tau) = f_{0}\biggl( \frac{q}{1 + \Delta_{\rm I}}\biggr).
\label{deBRf1}
\end{equation}
Recalling  now that, by definition,
\begin{equation}
f_{0}\biggl( \frac{q}{1 + \Delta_{\rm I}}\biggr) = f_{0}(q) + \frac{\partial f_{0}}{\partial q} [ q( 1 - \Delta_{\rm I}) - q],
\label{deBRf2}
\end{equation}
the perturbed phase-space distribution and the brightness perturbation must satisfy:
\begin{equation}
f_0(q) [ 1 + f^{(1)}(x^{i}, q,n_{j}, \tau) ] = f_{0}(q) \biggl[1 - \Delta_{{\rm I}}(x^{i}, q,n_{j}, \tau)\frac{\partial \ln{f_{0}}}{\partial\ln{ q}}\biggr],
\label{deBRf3}
\end{equation}
that also implies 
\begin{equation}
\Delta_{\rm I} = - f^{(1)} \biggl(\frac{\partial \ln{f_0}}{\partial \ln{q}} \biggr)^{-1},\,\,\,\,\,\,\,\,\,\,\,\,
F_{\gamma} = - \Delta_{\rm I} \frac{\int q^{3} d q  
f_{0}\frac{\partial f_{0}}{\partial \ln{q}}}{\int q^{3} d q f_{0}} = 
4 \Delta_{\rm I},
\label{deBRf4}
\end{equation}
where the second equality follows from integration by parts as in Eq. (\ref{4fact}).

The Boltzmann equations for the perturbation of the brightness are then
\begin{eqnarray}
&& \Delta_{\rm I}' + i k\mu ( \Delta_{\rm I} + \phi) = \psi' + 
\epsilon' \biggl[ - \Delta_{\rm I} + \Delta_{{\rm I}0 } +  \mu v_{b} - 
\frac{1}{2} P_{2}(\mu) S_{\rm Q}\biggr],
\label{BRI}\\
&& \Delta_{\rm Q}' + i k\mu  \Delta_{\rm Q}  =  \epsilon' \biggl\{ - \Delta_{\rm Q}  + 
\frac{1}{2} [1- P_{2}(\mu)] S_{\rm Q}\biggr\},
\label{BRQ}\\
&& \Delta_{\rm U}' + i k\mu  \Delta_{\rm U}  = - \epsilon'  \Delta_{\rm U} ,
\label{BRU}\\
&& \Delta_{\rm V}' + i k \mu \Delta_{\rm V} = - \epsilon' \biggl[ \Delta_{\rm V} +
\frac{3}{2} \,\,i\mu\,\, \Delta_{{\rm V}1}\biggr],
\label{BRV}
\end{eqnarray}
where we defined, for notational convenience and for homogeneity with the notations of other authors \cite{HZ1}
\begin{equation}
v_{\rm b} = \frac{\theta_{\rm b}}{i k}
\label{defvb}
\end{equation}
and 
\begin{equation}
S_{\rm Q}=  \Delta_{{\rm I}2} + \Delta_{{\rm Q}0} + \Delta_{{\rm Q}2}.
\end{equation}
In Eqs. (\ref{BRQ})--(\ref{BRU}), $P_{2}(\mu) = (3 \mu^2 -1)/2$ is the Legendre 
polynomial of second order, which appears in the collision operator of the Boltzmann 
equation for the photons 
due to the directional nature of Thompson scattering. Eq. (\ref{BRV}) is somehow decoupled from the system. So 
if, initially, $\Delta_{\rm V}=0$ it will also vanish at later times.
In Eqs. (\ref{BRI})--(\ref{BRU}) the function $\epsilon'$ denotes the differential 
optical depth for Thompson scattering\footnote{Notice that, in comparison with Eq. (\ref{Txs}), the ionization 
fraction has been taken out from the definition of electron density. This notation is often used in this context 
even if the notation used in section \ref{sec2} can be also employed. Notice also that, conventionally, the differential 
optical depth is denoted by $\tau'$. This notation would be highly ambiguous in the present case 
since $\tau$ denotes, in our lectures, the conformal time  coordinate. This is the reason why the differential optical 
depth will be denoted by $\epsilon'$.}
\begin{equation}
\epsilon' = x_{\rm e} n_{\rm e} \sigma_{\rm T} \frac{a}{a_0} =\frac{x_{\rm e} n_{\rm e} \sigma_{\rm T} }{z + 1},
\label{OPD1}
\end{equation}
having denoted with $x_{\rm e}$ the ionization fraction and $z = a_{0}/a -1$ the redshift. 
Defining with $\tau_{0}$ the time at which the 
signal is received, the optical depth will then be 
\begin{equation}
\epsilon(\tau,\tau_0) = \int_{\tau}^{\tau_0} x_{\rm e} n_{\rm e} \sigma_{\rm T} \frac{a(\tau)}{a_0} d\tau.
\label{OPD2}
\end{equation}
There are two important limiting cases. In the optically thin limit $\epsilon \ll 1$ absorption along the 
ray path is negligible so that the emergent radiation is simply the sum of the contributions 
along the ray path.  In the opposite case $\epsilon \gg 1$ the plasma is said to be 
optically thick.  

To close the system the evolution of the baryon velocity field can be 
rewritten as 
\begin{equation}
 v_{\rm b}' + {\cal H} v_{\rm b} + i k\phi  + \frac{\epsilon'}{R_{\mathrm{b}}}
 \biggl( 3 i \Delta_{{\rm I}1} + v_{b} \biggr) =  0,
\label{vb}
\end{equation}
where $R_{\mathrm{b}}(z)$ has been already defined in Eq. (\ref{Rb}).
At the decoupling epoch occurring for $z_{\mathrm{dec}} \simeq 1100$, $R_{\mathrm{b}}(z_{\mathrm{dec}})\sim 7/11$ for a typical
baryonic content of $h_{0}^2 \Omega_{\rm b0} \sim 0.023$.
Notice that the photon velocity field has been eliminated, in Eq. (\ref{vb}) with the 
corresponding expression involving the monopole of the brightness function.

As pointed out  in Eq. (\ref{DP}), while $Q$ and $U$ change under rotations, the 
degree of linear polarization is invariant. Thus, it is sometimes useful to combine Eqs. (\ref{BRI}) 
and (\ref{BRQ}). The result of this combination is 
\begin{eqnarray}
&& \Delta_{\rm P}' + ( i k\mu + \epsilon')  \Delta_{\rm P}  =  \frac{3}{4} \epsilon' ( 1 - \mu^2) S_{\rm P},
\nonumber\\
&& S_{\rm P}=   \Delta_{{\rm I}2} + \Delta_{{\rm P}0} + \Delta_{{\rm P}2}.
\label{BRP}
\end{eqnarray}
With the same notations Eq. (\ref{BRI}) can be written as 
\begin{equation}
\Delta_{\rm I}' + (i k \mu + \epsilon')\Delta_{\rm I} = 
\psi' - i k\mu \phi  + \epsilon'[ \Delta_{{\rm I}\,0}   + \mu v_{\rm b}- \frac{1}{2} P_{2}( \mu) S_{\rm P}].
\label{BRIPA}
\end{equation}
By adding a $\phi'$ and $\epsilon'\phi$ both at the left and right hand sides of Eq. (\ref{BRIPA}), 
the equation for the temperature fluctuations can also be written as:
\begin{equation}
(\Delta_{\rm I} + \phi)' + (i k \mu + \epsilon') ( \Delta_{\rm I} + \phi) = 
(\psi' + \phi') + \epsilon'[ (\Delta_{{\rm I}\,0} + \phi)  + \mu v_{\rm b}- \frac{1}{2} P_{2}( \mu) S_{\rm P}].
\label{BRIP}
\end{equation}
 This form of the equation is relevant in order to 
find formal solutions of the evolution of the brightness equation (see below the discussion 
of the line of sight integrals). 

\subsubsection{Visibility function}

An important function appearing naturally in various subsequent expressions is the 
so-called {\em visibility function}, ${\cal K}(\tau)$, giving the probability that a CMB photon was last 
scattered between $\tau$ and $\tau + d\tau$; the definition of ${\cal K}(\tau)$ is 
\begin{equation}
{\cal K}(\tau) = \epsilon' e^{- \epsilon(\tau,\tau_{0})},
\label{visibilityf}
\end{equation}
usually denoted by $ g(\tau)$ in the literature. The function ${\cal K}(\tau) $ is a rather important 
quantity since it is sensitive to the whole ionization history of the Universe.
The visibility function is strongly peaked around the decoupling time $\tau_{\rm dec}$ and 
can be approximated, for analytical purposes, by a Gaussian with variance of the order of few
$\tau_{\rm dec}$ \cite{wyse}.  In ${\rm Mpc}$  the width of the visibility function is about $70$.
A relevant limit is the so-called sudden decoupling 
limit  where the visibility function 
can be approximated by a Dirac delta function and its integral, i.e. the optical depth, can be approximated 
by a step function; in formulae:
\begin{equation}
{\cal K}(\tau) \simeq \delta(\tau -\tau_{\rm dec}),\,\,\,\,\,\,\,\,\,\,\,\, e^{- \epsilon(\tau,\tau_{0})} \simeq 
\theta( \tau - \tau_{\rm dec}).
\label{suddendec}
\end{equation}
This approximation will be used, below, for different applications and it is justified since the free electron density 
diminishes suddenly at decoupling.  In spite of this occurrence  there are 
 convincing indications that, at some epoch after decoupling, the 
Universe was reionized. 

\subsubsection{Line of sight integrals}

 Equations  (\ref{BRP}) and (\ref{BRIPA})--(\ref{BRIP})  can be formally written as 
 \begin{equation}
 {\cal M}(\vec{k}, \tau)' + ( i k \mu + \epsilon') {\cal M}(\vec{k},\tau) = {\cal N}(\vec{k},\tau),
\label{FORM1}
 \end{equation}
 where ${\cal M}(\vec{k},\tau)$ are appropriate functions changing from case to case and 
 ${\cal N}(\vec{k},\tau)$ is a source term which also depends on the specific equation 
 to be integrated.

The formal solution of the class of equations parametrized in the form (\ref{FORM1}) can be written as  
\begin{equation}
{\cal M}(\vec{k},\tau_0) = e^{- A(\vec{k},\tau_0)} \int_{0}^{\tau_{0}} e^{A(\vec{k},\tau)} {\cal N}(\vec{k},\tau) d \tau,
\label{FORM2}
\end{equation}
where the boundary term for $\tau\to 0$ can be dropped since it is unobservable \cite{HS1,HZ2}. The 
function $A(\vec{k},\tau)$ determines the solution of the homogeneous equations and it is:
\begin{equation}
A(\vec{k},\tau) 
= \int_{0}^{\tau} (i k \mu + \epsilon') d\tau  = i k\mu \tau + \int_{0}^{\tau} x_{\rm e} n_{\rm e} \sigma_{\rm T} 
\frac{a}{a_0} d\tau.
\label{FORM3}
\end{equation}
Using the results of  Eqs. (\ref{FORM1})--(\ref{FORM3}),  the solution of Eqs. (\ref{BRP}) and (\ref{BRIP})
 can be formally  written  as 
\begin{eqnarray}
(\Delta_{\rm I} + \phi)(\vec{k},\tau_0) &=& \int_{0}^{\tau_{0}} \,\,d\tau \,\,e^{- i k\mu \Delta \tau - \epsilon(\tau,\tau_{0})} 
(\phi' + \psi') 
\nonumber\\
&+& \int_{0}^{\tau_{0}} \,\,d\tau\,\,\,
 {\cal K}(\tau)\biggl[ ( \Delta_{{\rm I}\,0} + \phi + \mu v_{\rm b}) - \frac{1}{2} P_{2 }(\mu)
S_{\rm P}(k,\tau) \biggr],
\label{LSI}
\end{eqnarray}
and as
\begin{equation}
\Delta_{\rm P}( \vec{k}, \tau_{0}) =\frac{3}{4} \int_{0}^{\tau_{0}} {\cal K}(\tau)  e^{- i k \mu \Delta\tau} (1-\mu^2)
S_{\rm P}(k, \tau) d\tau,
\label{LSP}
\end{equation}
where $\epsilon(\tau,\tau_{0})$ is the optical depth already introduced in Eq. (\ref{OPD2})  and $\Delta \tau = (\tau_{0} 
-\tau)$ is the (conformal time) increment between the reception of the signal (at $\tau_0$)
and the emission (taking place for $\tau\simeq \tau_{\rm dec}$). In Eqs.
(\ref{LSI}) and (\ref{LSP}) the visibility function ${\cal K}(\tau)$, already defined in Eq. (\ref{visibilityf}), 
has been explicitly introduced.

Equations (\ref{LSI}) and (\ref{LSP}) are called for short line of sight integral solutions. There are at least two important applications of Eqs. (\ref{LSI}) 
and (\ref{LSP}). The first one is numerical and will be only swiftly described. The second one is 
analytical and will be exploited both in the present section and in the following. 

The formal solution  of Eq. (\ref{BRP}) can be written in a different form 
if the term $\mu^2$ is integrated by parts (notice, in fact, that the $\mu$ enters also the exponential). 
The boundary terms arising as a result of the integration by parts can be dropped because they are 
vanishing in the limit $\tau\to 0$ and are irrelevant for $\tau=\tau_{0}$ (since only an unobservable monopole 
is induced). The result the integration by parts of the $\mu^2$ term in Eq. (\ref{LSP}) can be expressed as 
\begin{eqnarray}
&& \Delta_{\rm P}(\vec{k},\tau_0) = \int_{0}^{\tau_{0}} e^{- i k\mu \Delta\tau} {\cal N}_{\rm P}(k,\tau)\,\, d\tau,
\label{LSPb}\\
&& {\cal N}_{\rm P}(\vec{k},\tau) = \frac{3}{4 k^2} [ {\cal K}( S_{\rm P}'' + k^2 S_{\rm P}) + 2 
{\cal K}' S_{\rm P}' + S_{\rm P} {\cal K}''],
\label{sourcePb}
\end{eqnarray}
where, as usual $\Delta\tau = (\tau_{0} -\tau)$.
The same exercise can be performed in the case of  Eq. (\ref{BRIPA}).
Before giving the general result, let us just integrate by parts the term $-ik\mu \phi$ 
appearing at the right hand side of Eq. (\ref{BRIPA}). The result of this manipulation
is 
\begin{eqnarray}
&&\Delta_{\rm I}(\vec{k},\tau_{0}) = \int_{0}^{\tau_{0}}e^{i k \mu (\tau - \tau_0) - \epsilon(\tau,\tau_{0})} (\psi' + \phi') \,\,d\tau
\nonumber\\
&& + \int_{0}^{\tau_{0}} {\cal K}(\tau) 
e^{i k \mu (\tau- \tau_{0})}\,\,d\tau\,\biggl[ \Delta_{{\rm I}\,0} + \phi + \mu v_{\rm b} - 
\frac{1}{2} P_{2}(\mu) S_{\rm P} \biggr].
\label{SWBZ1}
\end{eqnarray}
Let us now exploit the sudden decay approximation illustrated around Eq. (\ref{suddendec}) 
and assume that the (Gaussian) visibility function ${\cal K}(\tau)$ is indeed a Dirac delta function 
centered around $\tau_{\rm dec}$ (consequently the optical depth $\epsilon(\tau,\tau_{0})$ will be 
a step function). Then Eq. (\ref{SWBZ1}) becomes 
\begin{equation}
\Delta_{{\rm I}}(\vec{k}, \tau_{0} ) = \int_{\tau_{\rm dec}}^{\tau_{0}} e^{i k \mu (\tau - \tau_0)}[ \psi' + \phi'] d\tau + 
e^{i k \mu(\tau_{\rm dec} - \tau_{0})} [ \Delta_{{\rm I}\,0} + \phi + \mu v_{\rm b}]_{\tau_{\rm dec}},
 \label{SWBZ2}
 \end{equation}
 where the term $S_{\rm P}$ has been neglected since it is subleading at large scales. Equation (\ref{SWBZ2}) 
 is exactly (the Fourier space version of) Eq. (\ref{scalSW}) already derived with a different chain of arguments and we can directly recognize the integrated SW term (first term at the right hand side), the ordinary SW effect (proportional 
 to \footnote{Recall, in fact, that 
because of the relation between brightness and perturbed energy-momentum tensor, i.e. Eqs. (\ref{def1a}) 
and (\ref{deBRf4}), $4 \Delta_{{\rm I}\,0} = \delta_{\gamma}$.} 
 $ (\Delta_{{\rm I}0} + \phi)$) and the Doppler term receiving contribution from the 
 peculiar velocity of the observer and of the emitter. 
 
 If all the $\mu$ dependent terms appearing in Eq. (\ref{BRIPA}) are integrated by parts the result will be 
\begin{eqnarray}
&& \Delta_{\rm I}(\vec{k},
\tau_0) = \int_{0}^{\tau_{0}} e^{- i k\mu \Delta\tau - \epsilon(\tau,\tau_0)} ( \psi' + \phi')\,\, d\tau + 
 \int_{0}^{\tau_{0}} e^{- i k\mu \Delta\tau} {\cal N}_{\rm I}(k,\tau) d \tau,
 \label{LSIb}\\
 && {\cal N}_{\rm I}(k,\tau) = \biggl\{{\cal K}(\tau)\biggl[ \Delta_{{\rm I}, 0} + \frac{S_{\rm P}}{4} + \phi + 
 \frac{i}{k} v_{\rm b}' + \frac{3}{4 k^2} S_{\rm P}''\biggr] 
 \nonumber\\
 &&+ {\cal K}' \biggl[ \frac{i}{k} v_{\rm b} + \frac{3}{2 k^2} S_{\rm P}' \biggr] + \frac{3}{4 k^2} {\cal K}'' S_{\rm P}\biggr\}.
 \label{sourceIb}
 \end{eqnarray}

\subsubsection{Angular power spectrum and observables}

Equations (\ref{SWBZ2}) together with the results summarized in section 4 for the initial conditions 
of the metric fluctuations after equality allow the estimate of the angular power spectra in the case 
of adiabatic and isocurvature initial conditions.
The $C_{\ell}$ spectrum will now be derived for few interesting examples. 
Consider, for instance, the adiabatic mode.
In this case Eq. (\ref{SWBZ2}) (or, Eq. (\ref{scalSW})) has vanishing integrated contribution and vanishing
Doppler contribution at large scales (as discussed in section 4). Using then the result of Eq. (\ref{SWscalad}), 
the adiabatic contribution to the temperature fluctuations can be written as \begin{equation}
\Delta^{\rm ad}_{{\rm I}}(\vec{k}, \tau_{0} ) = 
e^{- i k \mu \tau_{0}} [ \Delta_{{\rm I}\,0} + \phi ]_{\tau_{\rm dec}}
 \simeq e^{- i k \mu \tau_{0}} \frac{1}{3} \psi_{\rm m}^{\rm ad}(\vec{k}),
 \label{ADSW3}
\end{equation}
noticing that, in the argument of the plane wave $\tau_{\rm dec}$ can be dropped since 
$\tau_{\rm dec} \ll \tau_{0}$.
The plane wave appearing in Eq. (\ref{ADSW3}) can now be expanded in series of Legendre polynomials 
and, as a result,
\begin{equation}
\Delta^{\rm ad}_{{\rm I},\ell}(\vec{k},\tau_{0}) = \frac{j_{\ell}(k\tau_{0})}{3} \psi_{\rm m}^{\rm ad}(\vec{k}),
\label{deltaAD}
\end{equation}
where $j_{\ell}(k\tau_{0})$ are defined as 
 \begin{equation}
 j_{\ell}( k \tau_{0}) = \sqrt{ \frac{\pi}{2 k \tau_{0}}} J_{\ell + 1/2}( k\tau_{0}).
 \label{BESSEL}
 \end{equation}

Assuming now that $\psi_{\rm m}^{\rm ad}(\vec{k})$ are the Fourier components of a Gaussian and isotropic 
random field (as, for instance, implied by some classes of inflationary models) then 
\begin{equation}
\langle \psi_{\rm m}^{\rm ad}(\vec{k})  \psi_{\rm m}^{\rm ad}(\vec{k}') \rangle = 
\frac{2\pi^2}{k^3} {\cal P}^{\rm ad}_{\psi}(k) \delta^{(3)}(\vec{k} -\vec{k}'),\,\,\,\,\,\,\, {\cal P}^{\rm ad}_{\psi}(k) = \frac{k^3}{2\pi^2} |\psi^{\rm ad}_{\rm m}(k)|^2 ,
\label{CORRAN}
\end{equation}
where ${\cal P}^{{\rm ad}}_{\psi}(k)$ is the power spectrum of the longitudinal fluctuations of the metric after equality.
Then, Eq. (\ref{deltaAD}) can be inserted into Eq. (\ref{AVALM}) and from Eq. (\ref{CORRAN}) (together with the orthogonality 
of spherical harmonics)
In this case 
\begin{equation}
C_{\ell}^{({\rm ad})} = \frac{4\pi}{9}\int_{0}^{\infty} \frac{d k}{k} {\cal P}^{\rm ad}_{\psi}(k) j_{\ell}(k\tau_{0})^2.
\label{CL1}
\end{equation}
To perform the integral it is customarily assumed that the power spectrum of adiabatic fluctuations 
has a power-law dependence characterized by a single spectral index $n$ 
\begin{equation}
{\cal P}^{\rm ad}_{\psi}(k) = \frac{k^3}{2\pi^2} |\psi_{k}|^2 = A_{\rm ad} \biggl( \frac{k}{k_{\rm p}}\biggr)^{n -1}.
\label{ANSPPSI}
\end{equation}
Notice that $k_{\rm p}$ is a typical pivot scale which is conventional since the whole dependence 
on the parameters of the model is encoded in $A_{\rm ad}$ and $n$. For instance, the WMAP 
collaboration \cite{map1,VERDE}, chooses to normalize $A$ at 
\begin{equation}
k_{\rm p} = k_1 = 0.05 \,\,{\rm Mpc}^{-1},
\end{equation}
while the scalar-tensor ratio (defined in section 6) is evaluated at a scale 
\begin{equation}
k_{0} = 0.002\,\,\, {\rm Mpc}^{-1} \equiv 6.481 \times 10^{-28}\,\,{\rm cm}^{-1} = 1.943\times 
10^{-17}\,\,{\rm Hz},
\end{equation}
recalling that $1\,\,{\rm Mpc} = 3.085\times 10^{24} \,\,{\rm cm}$.

Inserting Eq. (\ref{ANSPPSI}) into Eq. (\ref{CL1}) and recalling the explicit form of the spherical Bessel 
functions in terms of ordinary Bessel functions 
\begin{equation}
C^{({\rm ad})}_{\ell} = \frac{2\pi^2}{9} (\tau_{0} \,\,k_{\rm p})^{1 - n}\,\, A_{\rm ad}\,\,\int_{0}^{\infty} dy  y^{n -3}
 J^2_{\ell + 1/2}(y),
\label{CLMAP}
\end{equation}
where $ y = k \tau_{0}$.
The integral appearing in Eq. (\ref{CLMAP}) can be performed for 
$-3 < n< 3$ with the result
\begin{equation}
\int_{0}^{\infty} dy  y^{n -3}
 J^2_{\ell + 1/2}(y)  = \frac{1}{ 2 \sqrt{\pi}} \frac{ 
\Gamma\biggl( \frac{3 -n}{2}\biggr)
 \Gamma\biggl(\ell + \frac{n}{2} - \frac{1}{2} \biggr)}{\Gamma\biggl(\frac{4 - n}{2}\biggr) \Gamma\biggl( \frac{5}{2} + \ell  - \frac{n}{2} \biggr)}.
\label{CL3}
\end{equation}
To get the standard form of the $C_{\ell}$ use now the 
duplication formula for the $\Gamma$ function, namely in our case 
\begin{equation}
\Gamma\biggl( \frac{3 - n}{2}\biggr) = \frac{\sqrt{2\pi} \Gamma( 3 - n) }{
2^{5/2 -n} \Gamma\biggl( \frac{4 - n}{2}\biggr) }.
\label{rel2a}
\end{equation}
Insert now Eq. (\ref{rel2a}) into Eq. (\ref{CL3}); inserting then 
Eq. (\ref{CL3}) into Eq. (\ref{CLMAP}) we do get 
\begin{eqnarray}
&& C^{({\rm ad})}_{\ell} = \frac{\pi^2 }{36}  A_{\rm ad} {\cal Z}(n, \ell)
\nonumber\\
&&{\cal  Z}(n,\ell) = (\tau_{0} \,\,k_{\rm p})^{1 - n}\,\,2^{n}\frac{\Gamma( 3 - n) 
\Gamma\biggl( \ell + \frac{n}{2} - 
\frac{1}{2}\biggr)}{\Gamma^2\biggl(\frac{4 - n}{2}\biggr)
 \Gamma\biggl( \frac{5}{2} + \ell - \frac{n}{2} \biggr) },
\label{CLUS}
\end{eqnarray}
where the function ${\cal Z}(n,\ell)$ has been introduced for future convenience. Notice, as 
a remark, that for the approximations made in the evaluation of the SW effects, Eq. (\ref{CLUS}) 
holds at large angular scales, i.e. $\ell < 30$.

The $C^{({\rm ad})}_{\ell}$ have been given in the case 
of the spectrum of $\psi$.  There is a specific relation between the spectrum 
of $\psi$ and the spectrum of curvature perturbations which implies, quite trivially, 
${\cal P}^{\rm ad}_{\cal R} = (25/9) {\cal P}^{\rm ad}_{\psi}$. Finally, the spectrum of the longitudinal 
fluctuations of the geometry may also be related to the spectrum of the same quantity but computed 
before equality: this entails the $(9/10)$ factor discussed in Eq. (\ref{psimpsir}). 

The same calculation performed in the case of the adiabatic mode can be repeated, with minor 
(but relevant) modifications for the CDM-radiation non-adiabatic mode. In the specific case 
of this non-adiabatic mode, Eq. (\ref{deltaNAD}) is modified as 
\begin{equation}
\Delta^{({\rm nad})}_{{\rm I},\ell}(\vec{k},\tau_{0}) = 2 j_{\ell}(k\tau_{0})\psi_{\rm m}^{\rm nad}(\vec{k}),
\label{deltaNAD}
\end{equation}
as it follows directly from Eq. (\ref{SWBZ2}) in the case of non-adiabatic initial conditions after equality 
(see also Eq. (\ref{SWscalnad})).  Performing the same computation 
Eq. (\ref{CLUS}) becomes 
\begin{equation}
C^{({\rm nad})}_{\ell} =\pi^2   A_{\rm nad} {\cal Z}(n_{\rm nad}, \ell),
\end{equation}
where the power spectrum of non-adiabatic fluctuations has been defined as 
\begin{equation}
{\cal P}_{\psi}^{\rm nad}= A_{\rm nad} \biggl(\frac{k}{k}_{\rm p}\biggr)^{n_{\rm nad} -1}. 
\end{equation}
 Again, 
following the considerations reported in Eqs. (\ref{psitoentr}) and (\ref{SWscalnad}), the 
spectrum of non-adiabatic fluctuations can be directly expressed in terms of the fluctuations 
of ${\cal S}$, i.e. the fluctuations of the specific entropy 
 (see Eqs. (\ref{SE}) and (\ref{SWscalnad})), with 
the result that ${\cal P}_{\psi} = (1/25){\cal P}_{{\cal S}}^{\rm nad}$. Of course 
the major difference between adiabatic and non-adiabatic fluctuations will be much more 
dramatic at smaller angular scales (i.e. say between $\ell \sim 200$ and $\ell \sim 350$) 
where the patterns of acoustic oscillations have a crucial phase difference (this aspect will be 
discussed in the context of the tight coupling expansion).

There could be physical situations where adiabatic and non-adiabatic modes are simultaneously present with some 
degree of correlation. In this case the derivations given above change qualitatively, but not crucially.
The contribution to the SW effect will then be the sum of the adiabatic and non adiabatic contributions 
(weighted by the appropriate coefficients) i.e. 
\begin{equation}
\Delta^{\rm tot}_{{\rm I}}(\vec{k}, \tau_{0} ) 
 \simeq e^{- i k \mu \tau_{0}} \biggl[ \frac{1}{3} \psi_{\rm m}^{\rm ad}(\vec{k}) + 2 \psi_{\rm m}^{\rm nad}(\vec{k})\biggr].
 \label{ADSW4}
\end{equation}
While taking expectation values, there will not only be the adiabatic and non-adiabatic 
power spectra, i.e. ${\cal P}_{\psi}^{\rm ad}(k)$ and ${\cal P}^{\rm nad}_{\psi}(k)$, but also 
the power spectrum of the correlation between the two modes arising from 
\begin{eqnarray}
&&\langle \psi_{\rm ad}(\vec{k}) \psi_{\rm nad}(\vec{k}') \rangle = \frac{2\pi^2}{k^3} {\cal P}_{\psi}^{\rm cor}(k) \delta^{(3)}(\vec{k} -\vec{k}'), 
\nonumber\\
&&{\cal P}_{\psi}^{\rm cor}(k)= \sqrt{A_{\rm ad} A_{\rm nad}} \biggl(\frac{k}{k_{\rm p}}\biggr)^{n_{\rm c} -1}
\cos{\alpha_{\rm c}},
\label{CORRPWS}
\end{eqnarray}
where the angle $\alpha_{\rm c}$ parametrizes the degree of correlation between the adiabatic and non-adiabatic mode. The total angular power spectrum will then be given not only 
by the adiabatic and non-adiabatic contributions, but also by their correlation, i.e.  
\begin{equation}
C_{\ell}^{({\rm cor})} = \frac{\pi^2}{3} \sqrt{ A_{\rm ad} A_{\rm nad} } \cos{\alpha_{\rm c} } {\cal Z}(n_{\rm c},\ell).
\end{equation}

Consider, finally, the specific case of adiabatic fluctuations with Harrison-Zeldovich, i.e. the case $n=1$ 
in eq. (\ref{CLUS}). In this case
\begin{equation}
\frac{\ell (\ell +1)}{2\pi} C^{({\rm ad})}_{\ell} = \frac{A_{\rm ad}}{9}.
\label{CLAD}
\end{equation}
If the fluctuations were of purely adiabatic nature, then large-scale anisotropy experiments 
(see Fig. \ref{F4}) imply\footnote{To understand fully the quantitative features 
of Fig. \ref{F4} it should be borne in mind that sometimes the $C_{\ell}$ are given not in absolute units 
(as implied in Eq. (\ref{CLAD}) but they are  multiplied by the CMB temperature. To facilitate 
the conversion recall that the CMB temperature is $T_{0} = 2.725 \times 10^{6}
\,\,\mu {\rm K}$. For instance the WMAP collaboration normalizes the power 
spectrum of the curvature fluctuations at the pivot scale $k_{\rm p}$ 
as ${\cal P}_{{\cal R}} =(25/9)\times (800\pi^2/T_{0}^2) \times \tilde{A}$
where $\tilde{A}$ is not the $A$ defined here but it can be easily related to it.}
$A\sim 9 \times 10^{-10}$.
Up to now the large angular scale anisotropies have been treated. In the following 
the analysis of the smaller angular scales will be introduced in the 
framework of the tight coupling approximation.

\subsection{Tight coupling expansion}

The tight coupling approximation has been already implicitly used in section 4 where 
it has been noticed that, prior to recombination, 
 for comoving scales shorter than the mean free path of CMB photons, the baryons 
 and the photons evolve as a single fluid.

If tight coupling is  exact,  photons and baryons 
are synchronized so well that the photon phase-space distribution 
is isotropic in the baryon rest frame. In other words since the typical time-scale 
between two collisions is set by $\tau_{\rm c} \sim 1/\epsilon'$,  the scattering 
rate is rapid enough to equilibrate the photon-baryon fluid.  Since the photon distribution is 
isotropic, the resulting radiation is not polarized. The idea is then to tailor a systematic expansion
in $\tau_{\rm c} \sim 1/\epsilon'$ or, more precisely, in 
$k \tau_{\rm c} \ll 1$ and $\tau_{\rm c}{\cal H} \ll 1$.  
 
Recall the expansion of the brightness perturbations:
\begin{eqnarray}
&& \Delta_{\rm I}(\vec{k}, \hat{n}, \tau) = \sum_{\ell} (- i)^{\ell}( 2 \ell + 1) \Delta_{{\rm I}\ell}(\vec{k},\tau) P_{\ell}(\mu), 
\nonumber\\
&& \Delta_{\rm Q}(\vec{k}, \hat{n}, \tau) = \sum_{\ell} (- i)^{\ell}( 2 \ell + 1)  \Delta_{{\rm Q}\ell}(\vec{k},\tau) P_{\ell}(\mu), 
\label{expD}
\end{eqnarray}
$\Delta_{{\rm I}\ell}$ and $  \Delta_{{\rm Q}\ell}$ being the  $\ell$-th multipole of the brightness 
function $\Delta_{{\rm I}}$ and $\Delta_{{\rm Q}}$.

The idea is now to expand Eqs. (\ref{BRI}) and (\ref{BRQ}) 
in powers of the small parameter $\tau_{\rm c}$.  Before doing the expansion, it is useful to derive the hierarchy for the brightness 
functions in full analogy with what is  discussed in the appendix for the case of the neutrino phase-space 
distribution. 
To this aim, each side of Eqs. (\ref{BRI})--(\ref{BRQ}) and (\ref{vb}) will be multiplied 
by the various Legendre polynomials  and the  integration  over $\mu$ will be performed.
Noticing that, from the orthonormality relation for Legendre polynomials (i. e. Eq. (\ref{norm})),
\begin{equation}
\int_{-1}^{1} P_{\ell}(\mu) \Delta_{\rm I} d\mu = 2 (-i)^{\ell} \Delta_{{\rm I} \ell},\,\,\,\,\,\,\,\,\,\,\,
\int_{-1}^{1} P_{\ell}(\mu) \Delta_{\rm Q} d\mu = 2 (-i)^{\ell} \Delta_{{\rm Q} \ell},
\label{INTPL}
\end{equation}
and recalling that
\begin{equation}
P_{0}(\mu) =1,\,\,\,\,\,P_{1}(\mu) = \mu,\,\,\,\,\,\,\,\,P_{2}(\mu)= \frac{1}{2}(3 \mu^2 -1),\,\,\,\,\,\, P_{3}(\mu) = \frac{1}{2}( 5\mu^3 -3 \mu),
\end{equation}
Eqs. (\ref{BRI})--(\ref{BRQ}) and (\ref{vb}) 
allow the determination of the first three sets of equations for  the hierarchy of the brightness.
More specifically, multiplying Eqs. (\ref{BRI})--(\ref{BRQ}) and (\ref{vb}) by $P_{0}(\mu)$ and integrating over $\mu$, the following relations can be obtained
\begin{eqnarray}
&& \Delta_{{\rm I}0}' + k \Delta_{{\rm I}1} =  \psi',
\label{L01}\\
&& \Delta_{{\rm Q}0} '+k \Delta_{{\rm Q}1} = \frac{\epsilon'}{2} [ \Delta_{{\rm Q}2} + \Delta_{{\rm I}2} - \Delta_{{\rm Q}0} ],
\label{L02}\\
&& v_{b}' + {\cal H} v_{\rm b} = - i k \phi - \frac{\epsilon'}{R_{\mathrm{b}}} ( 3 i \Delta_{{\rm I}1} + v_{b} ).
\label{L03}
\end{eqnarray}
If Eqs. (\ref{BRI})--(\ref{BRQ}) and (\ref{vb}) are multiplied by  $P_{1}(\mu)$, both at right and left-hand sides, 
the integration  over $\mu$ of the various terms implies, using Eq.  (\ref{INTPL}):
\begin{eqnarray}
&& - \Delta_{{\rm I} 1}' - \frac{2}{3}k \Delta_{{\rm I}2} + \frac{k}{3} \Delta_{{\rm I}0} = - \frac{k}{3}  \phi + \epsilon' \biggl[ \Delta_{{\rm I} 1} + 
\frac{1}{3 i} v_{\rm b}\biggr],
\label{L11}\\
&& - \Delta_{{\rm Q}1}' - \frac{2}{3} k \Delta_{{\rm Q}2} + \frac{k}{3} \Delta_{{\rm Q}0} = \epsilon' \Delta_{{\rm Q} 1},
\label{L12}\\
&& v_{b}' + {\cal H} v_{b} = - i k \phi - \frac{\epsilon'}{R_{\mathrm{b}}} ( 3 i \Delta_{{\rm I}1} + v_{b} ).
\label{L13}
\end{eqnarray}
The same  procedure, using $P_{2}(\mu)$, leads to
\begin{eqnarray}
&& - \Delta_{{\rm I} 2}' - \frac{3}{5} k \Delta_{{\rm I}3} + \frac{2}{5} k \Delta_{{\rm I} 1} = \epsilon'\biggl[ \frac{9}{10} \Delta_{{\rm I}2} - \frac{1}{10} (\Delta_{{\rm Q}0} + 
\Delta_{{\rm Q} 2} )\biggr],
\label{L21}\\
&&  - \Delta_{{\rm Q} 2}' - \frac{3}{5} k \Delta_{{\rm Q}3} + \frac{2}{5} k \Delta_{{\rm Q} 1} = \epsilon'\biggl[ \frac{9}{10} \Delta_{{\rm Q}2} - \frac{1}{10} (\Delta_{{\rm Q}0} + 
\Delta_{{\rm I} 2} )\biggr],
\label{L22}\\
&& v_{b}' + {\cal H} v_{b} = - i k \phi - \frac{\epsilon'}{R_{\mathrm{b}}} \biggl( 3 i \Delta_{{\rm I}1} + v_{b} \biggr).
\end{eqnarray}
For $\ell\geq 3$ the hierarchy of the brightness can be determined in general terms by using 
the recurrence relation for the 
Legendre polynomials reported in Eq. (\ref{rec1}):
\begin{eqnarray}
&&\Delta_{{\rm I}\ell}' + \epsilon' \Delta_{{\rm I}\ell} 
= \frac{k}{2 \ell + 1} [ \ell \Delta_{{\rm I}(\ell-1)} - (\ell + 1) \Delta_{{\rm I}(\ell + 1)}],
\nonumber\\
&& \Delta_{{\rm Q}\ell}' + \epsilon' \Delta_{{\rm Q}\ell} 
= \frac{k}{2 \ell + 1} [ \ell \Delta_{{\rm Q}(\ell-1)} - (\ell + 1) \Delta_{{\rm Q}(\ell + 1)}].
\end{eqnarray}

\subsubsection{Zeroth order in the tight coupling expansion: acoustic oscillations}
We are now ready to compute the evolution of the various terms to a given order in the tight-coupling expansion parameter $\tau_{\rm c} = |1/\epsilon'|$. 
After expanding the various moments of the brightness function and the velocity field in $\tau_{\rm c}$ 
\begin{eqnarray}
&&\Delta_{{\rm I}\ell} = \overline{\Delta}_{{\rm I}\ell} + \tau_{\rm c} \delta_{{\rm I}\ell},
\nonumber\\
&& \Delta_{{\rm Q}\ell} = \overline{\Delta}_{{\rm Q}\ell} + \tau_{\rm c} \delta_{{\rm Q}\ell},
\nonumber\\
&&v_{\rm b} = \overline{v}_{\rm b} + \tau_{\rm c} \delta_{v_{\rm b}},
\label{DEFEXP}
\end{eqnarray}
the obtained expressions can be inserted  into Eqs. (\ref{L01})--(\ref{L13}) and the evolution of the various moments of the brightness 
function can be found order by order.

To zeroth order in the tight-coupling approximation, the evolution equation for the baryon velocity field, i.e. Eq. (\ref{L03}), leads to:
\begin{equation}
\overline{v}_{b} = - 3 i  \overline{\Delta}_{{\rm I}1},
\label{vb1}
\end{equation} 
while Eqs. (\ref{L02}) and (\ref{L12}) lead, respectively, to
\begin{equation}
\overline{\Delta}_{{\rm Q}0} = \overline{\Delta}_{{\rm I}2} + \overline{\Delta}_{{\rm Q}2},\,\,\,\,\,\,\,\,\,\,\, \overline{\Delta}_{{\rm Q}1} =0.
\label{int1Q}
\end{equation}
Finally Eqs. (\ref{L21}) and (\ref{L22}) imply
\begin{equation}
9\overline{\Delta}_{{\rm I}2}  = \overline{\Delta}_{{\rm Q}0} + \overline{\Delta}_{{\rm Q}2},
\,\,\,\,\,\,\,\,\,\,\,\,\, 9\overline{\Delta}_{{\rm Q}2}  = \overline{\Delta}_{{\rm Q}0} + \overline{\Delta}_{{\rm I}2}.
\label{int2Q}
\end{equation}
Taking together the four conditions expressed by Eqs. (\ref{int1Q}) and (\ref{int2Q}) we have, to zeroth order in the 
tight-coupling approximation:
\begin{equation}
\overline{\Delta}_{{\rm Q}\ell} =0,\,\,\,\,\,\,\,\,\,\,\ell\geq 0,\,\,\,\,\,\,\,\,\overline{\Delta}_{{\rm I}\ell} =0,\,\,\,\,\,\, \ell \geq 2.
\label{int3}
\end{equation}
Hence, to zeroth order in the tight coupling, the relevant equations are 
\begin{eqnarray}
&& \overline{v}_{b} = - 3i \overline{\Delta}_{{\rm I}1},
\label{zerothorder1}\\
&& \overline{\Delta}_{{\rm I}0}' + k \overline{\Delta}_{{\rm I}1} =  \psi'.
\label{zerothorder2}
\end{eqnarray}
This means, as anticipated, that to zeroth order in the tight-coupling expansion the CMB is not polarized 
since $\Delta_{\rm Q}$ is vanishing.

A decoupled evolution equation for the monopole can be derived. Summing up
Eq. (\ref{L11}) (multiplied by $ 3 i$) and  Eq. (\ref{L13})  (multiplied by $R_{\mathrm{b}}$) we get, to zeroth 
order in the tight coupling expansion:
\begin{equation}
R_{\mathrm{b}} \overline{v}_{\rm b}' - 3 i \overline{\Delta}_{{\rm I}1}' + i k \phi (R_{\mathrm{b}} + 1) -
2i k \overline{\Delta}_{{\rm I}2} + i k \overline{\Delta}_{{\rm I}0} + R_{\mathrm{b}} {\cal H} \overline{v}_{\rm b} =0.
\label{MON1}
\end{equation}
Recalling now
 Eq. (\ref{zerothorder1})  to eliminate $\overline{v}_{\rm b}$  from Eq. (\ref{MON1}),
 the following equation can be obtained
\begin{equation}
(R_{\mathrm{b}}+ 1) \overline{\Delta}_{{\rm I}1}' + {\cal H} R_{\mathrm{b}} 
\overline{\Delta}_{{\rm I} 1} - \frac{k}{3} \overline{\Delta}_{{\rm I}0} 
= 0.
\label{int4}
\end{equation}
Finally, the dipole term can be eliminated from Eq. (\ref{int4}) using Eq. (\ref{zerothorder2}). By doing so, Eq. (\ref{int4}) leads to  the wanted 
decoupled equation for the monopole:
\begin{equation}
\overline{\Delta}_{{\rm I}0}''  + \frac{R_{\mathrm{b}}'}{R_{\mathrm{b}} + 1} \overline{\Delta}_{{\rm I}0}' + 
k^2 c_{{\rm sb}}^2 \overline{\Delta}_{{\rm I}0} = 
 \biggl[ \psi'' + \frac{R_{\mathrm{b}}'}{R_{\mathrm{b}} +1} \psi' - \frac{k^2}{3} \phi \biggr],
 \label{monopole}
\end{equation}
where $c_{{\rm sb}}$  has been already defined in Eq. (\ref{bphss}) and it is the speed 
of sound of the baryon-photon system. 
The term  $k^2 c_{\rm sb}^2 \overline{\Delta}_{{\rm I}0}$ is the photon pressure.
Defining, from Eq. (\ref{bphss}), the sound horizon as 
\begin{equation}
r_{\rm s}(\tau) = \int_{0}^{\tau} c_{\rm sb}(\tau') d \tau',
\label{Shor}
\end{equation}
the photon pressure cannot be neglected for modes $k r_{\rm s}(\tau) \geq 1$.
At the right hand side of Eq. (\ref{monopole}) several forcing terms appear.  The 
term $\psi''$ dominates, if present, on super-horizon scales and causes 
a dilation effect on $\overline{\Delta}_{{\rm I}0}$. The term containing $k^2 \phi$ leads 
to the adiabatic growth of the photon-baryon fluctuations and becomes 
important for $k \tau \simeq 1$. 
In Eq. (\ref{monopole}) the damping term arises from the redshifting of the baryon momentum 
in an expanding Universe, while photon pressure provides the restoring force which is weakly suppressed by  the additional inertia of the baryons.  It is finally worth noticing that all the formalism developed in this section is nothing but 
an extension of the fluid treatment proposed in sections \ref{sec7} and \ref{sec8}. This 
aspect becomes immediately evident by comparing Eqs. (\ref{deltagamma}) and (\ref{monopole}). Equations 
(\ref{deltagamma}) and (\ref{monopole}) are indeed the same equation since $\overline{\Delta}_{{\rm I}0} =\delta_{\gamma}/4$.

\subsubsection{Solutions of the evolution of monopole and dipole}
Equation (\ref{monopole}) can be solved under different approximations (or even exactly \cite{HS1}).
The first brutal approximation would be to set $R_{\mathrm{b}}'= R_{\mathrm{b}} =0$, implying the the r\^ole of the baryons 
in the acoustic oscillations is totally neglected. As a consequence, in this case $c_{{\rm sb}}\equiv 1/\sqrt{3}$
which is nothing but the sound speed discussed in Eqs. (\ref{deltam})--(\ref{thetar}) for the fluid
analysis of the adiabatic mode. In the case of the adiabatic mode, neglecting 
neutrino anisotropic stress, $\psi = \phi = \psi_{\rm m}$ and $\psi' =0$. Hence, the solution 
for the monopole and the dipole to zeroth order in the tight coupling expansion 
follows by solving Eq. (\ref{monopole}) and by inserting the obtained result 
into Eqs. (\ref{zerothorder1}) and (\ref{zerothorder2}), i.e. 
\begin{eqnarray}
&& \overline{\Delta}_{{\rm I}0}(k,\tau) = \frac{\psi_{\rm }}{3} [ \cos{(k c_{{\rm sb}}\tau)} -3],
\nonumber\\
&& \overline{\Delta}_{{\rm I}1}(k,\tau) = - \frac{\psi_{\rm m}}{3} k c_{{\rm sb}} \sin{(k c_{\rm sb}\tau)},
\label{beta=0}
\end{eqnarray}
which is exactly the solution discussed in section 4 if we recall Eq. (\ref{zerothorder1}) and 
the definition (\ref{defvb}).

If $R_{\mathrm{b}}' =0$ but $R_{\mathrm{b}}\neq 0$, then the solution of Eqs. (\ref{zerothorder1})--
(\ref{zerothorder2}) and (\ref{monopole})  becomes, in the case of the adiabatic mode, 
\begin{eqnarray}
&& \overline{\Delta}_{{\rm I}0}(k,\tau) = \frac{\psi_{\rm m}}{3} (R_{\mathrm{b}} + 1) [ \cos{(k c_{\rm sb} \tau)} -3],
\nonumber\\
&& \overline{\Delta}_{{\rm I}1}(k,\tau) = \frac{\psi_{\rm m}}{3} \sqrt{\frac{R_{\mathrm{b}} + 1}{3}} \sin{(k c_{\rm sb} \tau)}.
\label{betaneq0}
\end{eqnarray}
Equation (\ref{betaneq0}) shows that the presence of the baryons increases the amplitude of the monopole 
by a factor $R_{\mathrm{b}}$. This phenomenon can be verified also in the case of generic time-dependent $R_{\mathrm{b}}$.
In the case of $R_{\mathrm{b}}\neq 0$ the shift in the monopole term is $(R_{\mathrm{b}} +1)$ with respect to the case $R_{\mathrm{b}} =0$.
This phenomenon produces a modulation of the height of the acoustic peak that depends on the baryon content 
of the model.

Consider now the possibility of setting directly initial conditions for the Boltzmann hierarchy during 
the radiation dominated epoch. During the radiation dominated epoch
and for modes which are outside the horizon, the initial conditions for the monopole 
and the dipole are fixed as 
\begin{eqnarray}
&&\Delta_{{\rm I}0}(k,\tau) =  - \frac{\phi_{0}}{2} - \frac{525 + 188 R_{\nu} + 16 R_{\nu}^2}{180( 25 + 2 R_{\nu}) } \phi_{0}
k^2 \tau^2 ,
\nonumber\\
&& \Delta_{{\rm I}1}(k,\tau)= \frac{\phi_{0}}{6} k\tau - \frac{65 + 16 R_{\nu}}{108( 25  + 2 R_{\nu})} \phi_{0} k^3 \tau^3
\label{RADBZ}
\end{eqnarray}
where $\phi_{0}$ is the constant value of $\phi$ during radiation. The constant value 
of $\psi$, i.e. $\psi_{0}$ will be related to $\phi_{0}$ through $R_{\nu}$, i.e. the fractional contribution
of the neutrinos to the total density.
It is useful to observe that in terms of the quantity $\Delta_{0} = (\overline{\Delta}_{{\rm I}0} -\psi)$,
Eq. (\ref{monopole})  becomes
\begin{equation}
\Delta_{0}'' + k^2 c_{{\rm sb}}^2 \Delta_{0} = - k^2 \biggl[ \frac{\phi}{3} + c_{{\rm sb}}^2 \psi\biggr].
\label{DEL0}
\end{equation}
The initial conditions for $\Delta_{0}$ are easily obtained from its definition in terms 
of $\Delta_{{\rm I}0}$ and $\psi$.
 
The same strategy can be applied to more realistic
 cases, such as the one where the scale factor interpolates between 
a radiation-dominated phase and a matter-dominated phase. In this case 
the solution of Eq. (\ref{monopole}) will be more complicated but always analytically tractable.
Equation (\ref{monopole}) can indeed be solved  in general terms. The general solution 
of the homogeneous equation is simply given, in the WKB 
approximation, as 
\begin{equation}
\overline{\Delta}_{{\rm I}0}  = \frac{1}{(R_{\mathrm{b}} + 1)^{1/4}} [ A \cos{k r_{\rm s}} + B \sin{k r_{\rm s}}].
\label{GEN}
\end{equation}
 For adiabatic fluctuations, $k^2 \phi$ contributes primarily to the cosine. The reason is that , in this 
 case, $\psi$ is constant until the moment of Jeans scale crossing at which moment it begins to decay.
 Non-adiabatic fluctuations, on the contrary, have vanishing gravitational potential at early 
 times and their monopole is dominated by sinusoidal harmonics.  Consequently, the peaks 
 in the temperature power spectrum will be located, for adiabatic fluctuations, 
 at a scale $k_{n}$ such that 
 $ k_{n} r_{\rm s} (\tau_{\ast}) = n \pi$.  Notice that, according  to Eq. (\ref{zerothorder2}) 
 the dipole, will be anticorrelated with the monopole. So if the monopole is 
 cosinusoidal , the dipole will be instead sinusoidal. Hence the ``zeros" of the cosine
 (as opposed to the maxima) will be filled by the monopole. 
  The solution of Eq. (\ref{monopole})  can then be obtained by supplementing the 
 general solution of the homogeneous equation (\ref{GEN}) with a particular 
 solution of the inhomogeneous equations that can be found easily with the 
 usual Green's function methods \cite{HS1}. 
 The amplitude of the monopole term shifts as $(1 + R_{\mathrm{b}})^{-1/4}$. Recalling 
the definition of $R_{\mathrm{b}}$ introduced in section \ref{sec8}, it can be argued that the height of the Doppler peak 
 is weakly sensitive to $h_{0}^2 \Omega_{\rm b0}$ in the $\Lambda$CDM model 
 where $\Omega_{\rm b0} \ll \Omega_{\rm M0}$ and $R_{\mathrm{b}}(\tau_{\rm dec}) < 1$.

\subsubsection{Simplistic estimate of the sound horizon at decoupling}
 
In $\ell$ space the position of the peaks for adiabatic and isocurvature modes is 
 given, respectively, by 
 \begin{eqnarray}
 && \ell^{(n)} = n\,\pi \frac{\overline{D}_{\mathrm{A}}(z_{\mathrm{dec}})}{ \,\,r_{\rm s}(\tau_{\rm dec})},
 \label{ADdop}\\
 && \ell^{(n)} = \biggl(n + \frac{1}{2}\biggr) \pi \frac{\overline{D}_{\mathrm{A}}(z_{\mathrm{dec}})}{ \,\,r_{\rm s}(\tau_{\rm dec})},
 \label{ISOdop}
 \end{eqnarray}
 where $\overline{D}_{\mathrm{A}}(z_{\mathrm{dec}})$ is the (comoving) angular diameter distance to decoupling 
 defined in Appendix \ref{APPA} (see in particular  Eq. (\ref{CANDD0}) and (\ref{defdec1})).
 We will be now interested in estimating (rather roughly) the position of the first peak, i.e. 
 \begin{equation}
 \ell_{\mathrm{A}} = \pi \frac{\overline{D}_{\mathrm{A}}(z_{\mathrm{dec}})}{ \,\,r_{\rm s}(\tau_{\rm dec})},
 \label{SEE1}
 \end{equation}
 where the subscript A stands for acoustic. The first thing we have to do is to estimate 
 the sound horizon at decoupling. From Eq. (\ref{Shor})  we have that 
 \begin{equation}
 r_\mathrm{s}(\tau_{\mathrm{dec}})= \int_{0}^{\tau_{\mathrm{dec}}} \frac{d\tau}{\sqrt{3(1 + R_{\mathrm{b}}(\tau))}}.
 \label{SEE2}
 \end{equation}
Equation (\ref{SEE2}) can also be written as
\begin{equation}
r_{\mathrm{s}}(\tau_{\mathrm{dec}})= \int_{0}^{\alpha_{\mathrm{dec}}} \frac{d\alpha}{\alpha \dot{\alpha}} c_{\mathrm{sb}}(\alpha),
\label{SEE3}
\end{equation}
where, following the notation of Eqs. (\ref{Eq1}) and (\ref{DEFEQ1}), $\alpha= (a/a_{0})$.
Indeed, recalling Eq. (\ref{Eq1}) we can write 
\begin{equation}
\alpha \dot{\alpha} = H_{0} \sqrt{\Omega_{\Lambda0} \alpha^4 + ( 1 - \Omega_{\Lambda0} - \Omega_{\mathrm{M0}}) \alpha^2 +
\Omega_{\mathrm{M0}} \alpha + \Omega_{\mathrm{R0}}}\simeq H_{0} \sqrt{\Omega_{\mathrm{M0}} \alpha + \Omega_{\mathrm{R0}}},
\label{SEE4}
\end{equation}
where, in the first equality we assumed that the spatial curvature vanishes; the second equality follows from the first 
since the contribution of matter and dark energy are subleading in the range of integration. 
Using Eq. (\ref{SEE4}) into Eq. (\ref{SEE3}) we can then write that 
\begin{equation}
\biggl(\frac{r_{\mathrm{s}}(\tau_{\mathrm{dec}})}{\mathrm{Mpc}}\biggr) = \frac{2998}{\sqrt{h_{0}^2 \Omega_{\mathrm{R0}}}} 
\int_{0}^{\alpha_{\mathrm{dec}}}\frac{d\alpha}{\sqrt{(1 + \beta_1\alpha)(1 + \beta_{2}\alpha)}},
\label{SEE5}
\end{equation}
where 
\begin{equation}
\beta_{1} = \frac{h_{0}^2 \Omega_{\mathrm{M0}}}{h_{0}^2 \Omega_{\mathrm{R0}}},\qquad 
\beta_{2} = \frac{3}{4}\frac{h_{0}^2 \Omega_{\mathrm{b0}}}{h_{0}^2 \Omega_{\gamma 0}}.
\label{SEE6}
\end{equation}
Recall that, according to Eqs. (\ref{omegaRval}), $h_{0}^2 \Omega_{\gamma 0} = 2.47\times 10^{-5}$ and 
$h_{0}^2 \Omega_{\mathrm{R0}}= 4.15 \times 10^{-5}$. 
From Eqs. (\ref{SEE5}) and (\ref{SEE6}) it is apparent that the sound horizon at decoupling depends both on $\Omega_{\mathrm{b0}}$ and $\Omega_{\mathrm{M0}}$. If we increase either $\Omega_{\mathrm{b0}}$ or $\Omega_{\mathrm{M0}}$
the sound horizon gets smaller. The integral appearing in Eq. (\ref{SEE5}) can be done analytically and the result 
is:
\begin{equation}
\biggl(\frac{r_{\mathrm{s}}(\tau_{\mathrm{dec}})}{\mathrm{Mpc}}\biggr) = \frac{2998}{\sqrt{1 + z_{\mathrm{dec}}}}
\frac{2}{\sqrt{3 \,h_{0}^2 \Omega_{\mathrm{M0}} c_1}} \ln{\biggl[ \frac{\sqrt{1 + c_1} + \sqrt{c_1 + c_1 c_{2}}}{1 + \sqrt{c_1 c_2}}\biggr]},
\label{SEE7}
\end{equation}
where
\begin{equation}
c_{1} = \beta_{2} \alpha_{\mathrm{dec}}= 27.6\,\, h_{0}^2 \Omega_{\mathrm{b0}}\,\,\biggl(\frac{1100}{1 + z_\mathrm{dec}}\biggr),\qquad 
c_{2} = \frac{1}{\beta_{1}\alpha_{\mathrm{dec}}} =\frac{0.045}{h_{0}^2 \Omega_{\mathrm{M0}}} \biggl(\frac{1 + z_{\mathrm{dec}}}{1100}\biggr).
\end{equation}
 With our fiducial values of the parameters, $r_{\mathrm{s}}(\tau_{\mathrm{dec}}) \simeq 150$ Mpc. 
 Recalling now that the comoving angular diameter distance to decoupling is estimated in Appendix \ref{APPA} (see 
 in particular Eq. (\ref{defdec3})) the sound, 
 $\ell_{A} \sim 300$. The sound horizon has been a bit underestimated with our approximation. For $r_{\mathrm{s}}(\tau_{\mathrm{dec}}) \sim 200$, $\ell_{A}\sim 220$, 
 which is around the measured value of the first Doppler peak in the temperature autocorrelation (see Fig. \ref{F4}).
 
It is difficult to obtain general analytic formulas for the position of the peaks. Degeneracies among the 
parameters may appear \cite{BE}. In \cite{W1s} a semi-analytical expression for the integral giving 
the angular diameter distance has been derived for various cases of practical interest.
Once the evolution of the lowest multipoles is known, the obtained expressions can be used in the 
integral solutions of the Boltzmann equation and the angular power spectrum can be 
computed analytically. Recently Weinberg in a series of papers \cite{Wbz1,Wbz2,Wbz3} 
computed the temperature fluctuations in terms of a pair of generalized form factors 
related, respectively, to the monopole and the dipole. This set of calculations were conducted in the synchronous 
gauge (see also \cite{old1,old2,old3} for earlier work on this subject; see also \cite{new1,new2}). Reference \cite{MUKa} also presents 
analytical estimates for the angular power spectrum exhibiting explicit dependence 
on the cosmological parameters in the case of the concordance model.

The results of the tight coupling expansion hold for $k \tau_{\rm c} \ll 1$. Thus 
the present approximation scheme breaks down, strictly speaking, for 
wave-numbers  $k > \tau_{\rm c}^{-1}$.
 Equation (\ref{monopole})  holds to zeroth-order in the tight coupling expansion, i.e. 
 it can be only applied on scales much larger than the photon mean free path. By comparing 
 the rate of the Universe expansion with the rate of dissipation we can estimate that 
 $\tau_{\rm c} k^2 \sim  \tau^{-1}$ defines approximately the scale above which 
 the wave-numbers will experience damping.  From these considerations the typical 
 damping scale can be approximated by 
 \begin{equation}
 k_{\rm d}^{-2} \simeq 0.3 ( \Omega_{\rm M} h^2)^{-1/2} 
 (\Omega_{\rm b0} h^2)^{-1} ( a/a_{\rm dec})^{5/2}\,\,\,{\rm Mpc}^2.
 \end{equation}
 The effect of diffusion is to damp the photon and baryon oscillations exponentially by the time
  of last scattering on comoving scales smaller than $3$ Mpc. For an experimental 
  evidence of this effect see \cite{DICK} and references therein.
 
 In order to have some qualitative estimate for the damping scale in the framework 
 of the tight coupling approximation, it is necessary to expand the temperature, polarization and 
 velocity fluctuations to second order in $\tau_{\rm c}$. Since for very small scales the r\^ole 
 of gravity is not important the longitudinal fluctuations 
 of the metric can be neglected. The result of this analysis \cite{HZ2} shows that 
 the monopole behaves approximately as 
 \begin{equation}
 \Delta_{{\rm I}0} \simeq e^{\pm i k r_{\rm s}} e^{ -(k/k_{\rm d})^2},
 \end{equation}
 where 
 \begin{equation}
 \frac{1}{k_{\rm d}^2} = \int_{0}^{\tau} \frac{\tau_{\rm c}}{6 (R_{\mathrm{b}} + 1)^2} \biggl[ R_{\mathrm{b}}^2 
+ \frac{16}{15} ( 1 + R_{\mathrm{b}})\biggr].
\label{KD}
\end{equation}
The  factor $16/15$ arises when the polarization fluctuations are taken consistently 
into account in the derivation \cite{HZ2}.

\subsubsection{First order in tight coupling expansion: polarization}

To first order in the tight-coupling limit, the relevant equations can be obtained by keeping all terms of order 
$\tau_{\rm c} $ and by using the 
first-order relations to simplify the expressions. From Eq. (\ref{L12}) the condition $\delta_{{\rm Q}1} =0$ can be derived. From Eqs. (\ref{L02}) and  (\ref{L21})--(\ref{L22}), the following remaining conditions are obtained respectively:
\begin{eqnarray}
&& - \delta_{{\rm Q}0} + \delta_{{\rm I} 2} + \delta_{{\rm Q} 2} =0,
\label{firstord1}\\
&& \frac{9}{10} \delta_{{\rm I}2} - \frac{1}{10} [ \delta_{{\rm Q}0} + \delta_{{\rm Q}2} ]= \frac{2}{5} k \overline{\Delta}_{{\rm I}1},
\label{firstord2}\\
&&  \frac{9}{10} \delta_{{\rm Q}2} - \frac{1}{10} [ \delta_{{\rm Q}0} + \delta_{{\rm I}2} ]=0.
\label{firstord3}
\end{eqnarray}
Equations  (\ref{firstord1})--(\ref{firstord3}) are a set of algebraic conditions 
implying that the  relations to be satisfied are:
\begin{eqnarray}
&& \delta_{{\rm Q}0} = \frac{5}{4} \delta_{{\rm I}2},
\label{Cond1}\\
&& \delta_{{\rm Q}2} = \frac{1}{4} \delta_{{\rm I}2},
\label{Cond2}\\
&& \delta_{{\rm I}2} = \frac{8}{15} k \overline{\Delta}_{{\rm I}1}.
\label{Cond3}
\end{eqnarray}
Recalling the original form of the expansion of the quadrupole as defined in Eq. (\ref{DEFEXP}),
 Eq. (\ref{Cond3}) can be also written 
\begin{equation}
\Delta_{{\rm I} 2} = \tau_{\rm c} \delta_{{\rm I}2}= \frac{8}{15} k\tau_{\rm c} \overline{\Delta}_{{\rm I}1},
\label{Cond4}
\end{equation}
since to zeroth order the quadrupole vanishes and the first non-vanishing effect comes 
from the first-order quadrupole whose value is determined from the zeroth-order monopole.

Now, from Eqs. (\ref{Cond1}) and (\ref{Cond2}), the quadrupole moment of $\Delta_{\rm Q}$ is proportional to the quadrupole of $\Delta_{\rm I}$, which is, in turn, 
proportional to the dipole evaluated to first order in $\tau_{\rm c}$. But $\Delta_{\rm Q}$ measures exactly 
the degree of  linear polarization of the radiation field. So, to first order in the tight-coupling expansion, the CMB  is linearly  polarized. Notice that the same derivation performed in the case of the equation for $\Delta_{\rm Q}$ 
can be more correctly performed in the case of the evolution equation of $\Delta_{\rm P}$ with the same 
result \cite{HZ2}.   Using the definition of $S_{\rm P}$ (i.e. Eq. (\ref{BRP})),
 and recalling Eqs. (\ref{Cond1})--(\ref{Cond3}),  we have that the source term 
 of Eq. (\ref{LSP}) can be approximated as
 \begin{equation}
 S_{\rm P} \simeq  \frac{4}{3} k \tau_{\rm c} \overline{\Delta}_{{\rm I}1}.
 \end{equation}
 Since $\tau_{\rm c}$ grows very rapidly during recombination, in order to have 
 quantitative estimates of the effect we have to know the evolution of $S_{\rm P}$ with 
 better accuracy. In order to achieve this goal, let us go back to the (exact) system 
 describing the coupled evolution of the various multipoles and, in particular, to Eqs. 
 (\ref{L02}) and (\ref{L21})--(\ref{L22}).  Taking the definition 
 of $S_{\rm P}$ (or $S_{\rm Q}$) and performing a first 
 time derivative we have 
 \begin{equation}
 S_{\rm P}' = \Delta_{{\rm I}2}' + \Delta_{{\rm P}2}' + \Delta_{{\rm P}0}'.
 \end{equation}
 Then, from Eqs. (\ref{L02}) and (\ref{L21})--(\ref{L22}), the time
 derivatives of the two quadrupoles and of the monopole can be expressed 
 in terms of the monopoles, quadrupoles and octupoles.  Simplifying 
 the obtained expression we get the following 
 evolution equation for $S_{\rm P}$, i.e.  \cite{HZ2}
 \begin{equation}
 S_{\rm P}' + \frac{3}{10} \epsilon' S_{\rm P} = k \biggl[ \frac{2}{5} \Delta_{{\rm I}1} - 
 \frac{3}{5}\biggl( \Delta_{{\rm P}1}+ \Delta_{{\rm P}3} + \Delta_{{\rm I}3}\biggr)\biggr].
 \end{equation}
This equation can be solved by evaluating the right hand side to zeroth-order 
in the tight coupling expansion, i.e. 
\begin{equation}
S_{\rm P}(\tau) =\frac{2}{5} k \int_{0}^{\tau}  d x \overline{\Delta}_{{\rm I}1} e^{-\frac{3}{10} \epsilon(\tau,x)}.
\end{equation}
This equation, giving the evolution of $S_{\rm P}$, can be inserted back into Eq. (\ref{LSP}) in order 
to obtain $\Delta_{\rm P}$. The result of this procedure is \cite{HZ2}
\begin{eqnarray}
&&\Delta_{\rm P} \simeq ( 1 - \mu^2) e^{i k\mu(\tau_{\rm dec} -\tau_{0})} {\cal D}(k),
\nonumber\\
&& {\cal D}(k)\simeq (0.51) \,\,k \,\,\sigma_{\rm dec}
 \overline{\Delta}_{{\rm I}1}(\tau_{\rm dec}),
 \label{IMPROVED}
\end{eqnarray}
where $\sigma_{\rm dec}$ is the width of the visibility function. This result allows to estimate 
with reasonable accuracy the angular power spectrum of the cross-correlation between 
temperature and polarization (see below), for instance, in the case of the adiabatic mode.

While the derivation of the polarization dependence of Thompson scattering 
has been conducted within the framework of the tight coupling approximation, it is useful to recall 
here that these properties follow dirrectly from the polarization dependence of Thompson 
scattering whose differential cross-section can be written as 
\begin{equation}
\frac{d\sigma}{d\Omega} = r_{0}^2 | \epsilon^{(\alpha)}\cdot \epsilon^{(\alpha')}|^2 \equiv 
 \frac{3 \sigma_{\rm T}}{8\pi} | \epsilon^{(\alpha)}\cdot \epsilon^{(\alpha')}|^2.
\end{equation}
where $\epsilon^{(\alpha)}$ is the incident polarization and $\epsilon^{(\alpha')}$ is the scattered polarization;
$r_{0}$ is the classical radius of the electron and $\sigma_{\rm T}$ is, as usual, the 
total Thompson cross-section.

Suppose that the incident radiation is  not polarized, i.e. $U = V = Q=0$; then we can write 
\begin{equation}
Q = {\cal I}_{x} - {\cal I}_{y}=0,\,\,\,\,\,\,\, {\cal I}_{x} = {\cal I}_{y} = \frac{{\cal I}}{2}.
\label{nonpol}
\end{equation}
Defining the incoming and outgoing polarization vectors as 
\begin{eqnarray}
&&\epsilon_{x} = (1,\,\,\,\,0,\,\,\,0),\,\,\,\,\,\,\,\,\,\,\epsilon_{y} = (0,\,\,\,\,1,\,\,\,0),\,\,\,\,\,\,\,\,\,\, \hat{k}=(0,\,\,\,\,0,\,\,\,1).
\nonumber\\
&& \epsilon_{x}' = (-\sin{\varphi},\,\,\,\,-\cos{\varphi},\,\,\,0),\,\,\,\,\,\,\,\,\,\,
\epsilon_{y}' = (\cos{\vartheta}\cos{\varphi},\,\,\,\,-\cos{\vartheta}\sin{\varphi},\,\,\,-\sin{\vartheta}),
\label{geom1}
\end{eqnarray}
the explicit form of the scattered amplitudes will be 
\begin{eqnarray}
&& {\cal I}_{x}'= \frac{3 \sigma_{T}}{8\pi} \biggl[ |\epsilon_{x}\cdot\epsilon_{x}'|^2 {\cal I}_{x} + |\epsilon_{y}\cdot\epsilon_{x}'|^2 {\cal I}_{y} \biggr] =\frac{3 \sigma_{T}}{16\pi} {\cal I},
\nonumber\\
&& {\cal I}_{y}'= \frac{3 \sigma_{T}}{8\pi} \biggl[ |\epsilon_{x}\cdot\epsilon_{y}'|^2 {\cal I}_{x} + |\epsilon_{y}\cdot\epsilon_{y}'|^2 {\cal I}_{y} \biggr]=\frac{3 \sigma_{T}}{16\pi} {\cal I} \cos^2{\vartheta}.
\label{geom2}
\end{eqnarray}

Recalling the definition of Stokes parameters:
\begin{eqnarray}
&& I'= {\cal I}_{x}' + {\cal I}_{y}' = \frac{3}{16 \pi } \sigma_{T}  {\cal I} ( 1 + \cos^2{\vartheta}),
\nonumber\\
&& Q' =  {\cal I}_{x}' - {\cal I}_{y}' = \frac{3}{16 \pi } \sigma_{T}  {\cal I}\sin^2{\vartheta}.
\end{eqnarray}
Even if $U'=0$ the obtained $Q$ and $U$ must be rotated 
to a common coordinate system:
\begin{equation}
Q'= \cos{2\varphi} Q,\,\,\,\,\,\,\,\,\,\,\,\,\,\,\,U'= -\sin{2\varphi} Q.
\end{equation}
So the final expressions  for the Stokes parameters of the scattered radiation are:
\begin{eqnarray}
&& I' =\frac{3}{16 \pi } \sigma_{\rm T}  {\cal I} ( 1 + \cos^2{\vartheta}),
\nonumber\\
&& Q'= \frac{3}{16 \pi } \sigma_{\rm T}  {\cal I} \sin^2{\vartheta} \cos{2\varphi},
\nonumber\\
&& U' = - \frac{3}{16 \pi } \sigma_{\rm T}  {\cal I} \sin^2{\vartheta} \sin{2\varphi}.
\end{eqnarray}
We can now expand the incident intensity in spherical harmonics:
\begin{equation}
{\cal I}(\theta,\varphi) = \sum_{\ell m} a_{\ell m} Y_{\ell m }(\vartheta,\varphi).
\label{CALI}
\end{equation}
So, for instance, $Q'$ will be
\begin{equation}
Q' = \frac{3}{16 \pi }\sigma_{\rm T} \int \sum_{\ell m} Y_{\ell m}(\vartheta,\varphi) a_{\ell m} \sin^2{\vartheta} \cos{2\varphi} d\Omega.
\label{HEU}
\end{equation}
By inserting the explicit form of the spherical harmonics into Eq. (\ref{HEU}) it can be easily shown that 
$Q'\neq0$ provided the term $a_{22}\neq 0$ in the expansion of Eq. (\ref{CALI}).  
\begin{figure}[tp]
\centering
\includegraphics[height=6cm]{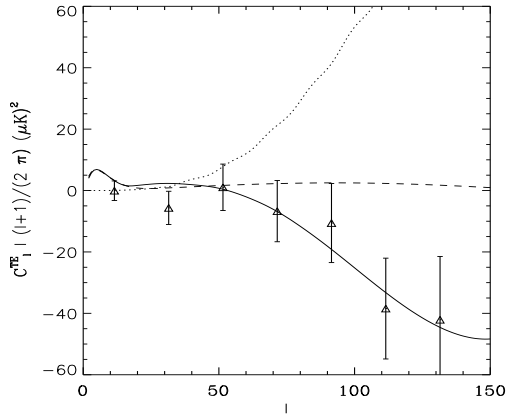}
\caption{The anticorrelation between the intensity fluctuations and the fluctuations 
in the degree of linear polarization is illustrated as presented in the WMAP one-year 
data.}
\label{POL}      
\end{figure}
Recall now that under clockwise rotations the Stokes parameters $Q$ and $U$ 
transform as in Eq. (\ref{clockwise}). As a consequence 
\begin{equation}
(Q \pm i U)^{\prime} = e^{\mp 2i \varphi}(Q \pm iU)
\end{equation} 
where $\varphi$ is the rotation angle.
From this observation it follows that the combinations 
\begin{equation}
(\Delta_{\rm Q} \pm i \Delta_{\rm U})(\hat{ n}) = \sum_{\ell m} a_{\pm 2,\ell m}
\,\,\,\,{_{\pm2}}{\cal Y}_{\ell m}(\hat{n})
\end{equation}
can be expanded in terms of the  spin-2 spherical
harmonics, i.e.   $_{\pm 2}{\cal Y}_{\ell}^{m}(\hat{n})$ \cite{sphharm1,sphharm2}.

The expansion coefficients are 
\begin{equation}
a_{\pm 2,\ell m}=\int d\Omega \,\,\,\,\,\,{_{\pm2}} {\cal Y}_{\ell m}^{*}(\Delta_{\rm Q} \pm i \Delta_{\rm U})(\hat{n}).
\end{equation}
In Ref. \cite{sphharm1,sphharm2}, the authors introduce the following 
linear combinations of 
$a_{\pm 2,  \ell m}$ to circumvent the impasse that the Stokes 
parameter are not invariant under rotations;
\begin{equation}
a_{E,\ell m}=-\frac{1}{2}(a_{2, \ell m}+a_{-2,\ell m})  \nonumber \\
a_{B, \ell m}= \frac{i}{2}(a_{2,\ell m}-a_{-2,\ell m}).
\end{equation}
These newly defined variables are expanded in terms of ordinary spherical
harmonics, $Y_{\ell m}(\hat{n})$,
\begin{equation}
E(\hat{n})= \sum_{lm} a_{E,\ell m} Y_{lm}(\hat{n}), \,\,\,\,\,\,\,\,\,\,\,\,\,
B(\hat{n})= \sum_{lm} a_{B,\ell m} Y_{lm}(\hat{ n}).
\end{equation}
The spin-zero spherical harmonics, $Y_{lm}(\hat{n})$, is free from
 the ambiguity with the rotation of the coordinate system,
and therefore $E$ and $B$ are rotationally invariant quantities.
The  $E$-mode  has $(-1)^{\ell}$ parity and the  $B$
mode $(-1)^{(\ell+1)}$ parity in analogy with electric and magnetic
fields.
Scalar perturbations generate only the $E$ mode \cite{kamionkowski}. 
While the scalar fluctuations only generate an $E$ mode, tensor fluctuations also 
generate a B mode. The Boltzmann equation for the tensor modes can be 
easily derived  following Refs. \cite{polnarev,crit} (see 
also \cite{kos2,knox}). 

Consider now, specifically, the adiabatic mode.
While the temperature fluctuation $\Delta_{\rm I}$ oscillates like $\cos{(kc_{\rm s,b}
\tau_{\rm dec})}$, the polarization is proportional to the dipole and oscillates like the sine 
of the same argument. The correlation function of the temperature and polarization, i.e. 
$\langle \Delta_{\rm I} \Delta_{\rm P}\rangle$ will then be 
proportional to $\sin{(k c_{\rm s,b} \tau_{\rm dec})} \cos{(kc_{\rm s,b} \tau_{\rm dec})}$.  
An analytical prediction for this quantity can be inferred from Eq. (\ref{IMPROVED})  (see \cite{HZ2}). 
The spectrum of the cross correlation must then have a peak for 
$k  c_{\rm s,b} \tau_{\rm dec} \sim \,\,3\pi/4$, corresponding to $\ell \sim 150$. This is the 
result suggested by Fig. \ref{POL} and taken from Ref. \cite{map1} reporting the 
measurement of the WMAP collaboration. In  Fig. \ref{POL} the temperature-polarization angular
power spectrum is reported for adiabatic models (solid line) and for isocurvature models 
(dashed line).  

We should mention here that a  rather effective method in order to treat on equal footing the scalar, vector 
and tensor radiative transfer equations is the total angular momentum method 
\cite{tot1,tot2,tot3}. Within this approach, the collision terms couple only the quadrupole moments 
of the distributions and each moment corresponds directly to observable patterns in the microwave sky. 
In this language the analysis of the polarization of the radiation field becomes somehow more transparent.

\newpage
\renewcommand{\theequation}{10.\arabic{equation}}
\setcounter{equation}{0}
\section{Early initial conditions?}
\label{sec10}

The presentation of the CMB anisotropies has been developed, in the last three sections, through a bottom-up approach.
It has been argued that the properties of the temperature and polarization autocorrelations 
are determined from a set of initial conditions that can include either an adiabatic mode, or a non-adiabatic mode, or both.
The most general set of initial conditions required to integrate the Einstein-Boltzmann hierarchy is formed by one 
adiabatic mode and by four non-adiabatic modes. In this approach, the initial conditions 
are set prior to equality but after neutrino decoupling. The wording {\em initial conditions} of CMB anisotropies 
may also have, in the present literature, a complementary (but crucially different) meaning.
If we believe to have a compelling and not erroneous picture of the thermodynamic history of the Universe 
also for temperatures much larger than $200$ GeV, then initial conditions for the fluctuations of the FRW 
metric may be assigned not prior to decoupling but much earlier, for instance during a stage of inflationary expansion.

While the logic summarized in the previous paragraph is the one followed in sections  \ref{sec7}, \ref{sec8} and \ref{sec9},
there is also a second (complementary) approach that has been partially explored already in section \ref{sec6} when 
computing, in a simplified situation, the spectrum of relic gravitons produced thanks to the sudden
transition form the de Sitter stage of expansion to the radiation-dominated stage of expansion. Moreover, 
in Fig. \ref{GRAV} the spectrum of relic gravitons has been discussed for diverse models making different assumptions 
both about the thermodynamic history of the Universe and about the nature of the laws of gravity at short distances.
The different curves appearing in Fig. \ref{GRAV} make different assumptions concerning the evolution of the Universe 
and, consequently, get to different experimental signatures. 
It is rather understandable, therefore, that this second approach necessarily demands the adoption of a specific model of evolution 
of the background geometry valid well before the moment when weak interactions fall out of thermal equilibrium. 
For this purpose the standard approach is to suppose that the Universe underwent a phase of accelerated 
expansion that was replaced by stage dominated by radiation. This oversimplified picture can be 
matched, below temperatures of the order of $1$ MeV, to the firmer model of evolution 
that has been discussed in the previous sections. In this sense the relevant plots 
that illustrate the overall evolution of the Hubble radius are the ones reported in Figs. \ref{mindur} and \ref{nonmindur}.
Concerning this general lore  various observations are in order:
\begin{itemize}
\item{} if the duration of inflation is minimal (see Fig \ref{mindur}) the initial conditions may not be necessarily 
quantum mechanical;
\item{} if the duration of inflation is non minimal (see Fig. \ref{nonmindur}) the initial conditions for the 
scalar and tensor modes are, most likely, of quantum mechanical origin;
\item{} the initial inflationary stage may be realized either through a single scalar degree of freedom or by means
of a collection of scalar fields;
\item{} in the case when not only one field is present isocurvature (i.e. non-adiabatic modes) will be present 
ab initio;
\item{} if inflation is realized through a single scalar field, still, there is the possibility that various other 
(spectator) fields are present during inflation and they may modify the evolution 
of curvature perturbations.
\end{itemize}
On top of the previous observations, it is also legitimate to stress that 
while we believe to have a clear and verified picture 
of the history of the Universe for $T< 1$ MeV, the same confidence may not be justified in the case of the early stages 
of the life of the Universe. If initial conditions for the scalar and tensor modes are set during the inflationary epoch, the 
tacit assumption is that we do know pretty well the evolution of our Universe between the $H_{\mathrm{inf}} \simeq 10^{-5} M_{\mathrm{P}} \simeq 10^{15}$ GeV  and $H_{\mathrm{BBN}} \simeq 10^{-31} \mathrm{GeV}$ (which is, according 
to Eq. (\ref{Hrho}) the curvature scale 
corresponding to a temperature $T\simeq $ MeV). 
  In this 
section we are going to {\em assume} that only a single scalar degree of freedom drives inflation and we will ask 
the question of what are the curvature fluctuations induced by the fluctuations of the inflaton. On a more technical ground, it is 
appropriate to mention that similar mathematical developments are required in the 
treatment of inhomogeneities in quintessence models driven by a single scalar field.

\subsection{Scalar modes induced by a minimally coupled scalar field}
In diverse situations it is important to compute the fluctuations induced by a single scalar degree
of freedom. This exercise is therefore technically relevant. 
The first step is this direction will be to write down the fluctuations of the energy-momentum tensor. We will 
do this in the conformally Newtonian gauge and we will then learn how to translate the obtained result 
in any other gauge needed for the resolution of physical problems. 
Consider therefore the energy-momentum tensor of a scalar field $\varphi$ characterized by a potential 
$V(\varphi)$. To first-order in the amplitude of the (scalar) metric fluctuations we will have:
\begin{equation}
\delta_{\mathrm{s}} T_{\mu}^{\nu} = \delta_{\mathrm{s}} g^{\nu\alpha} \partial_{\alpha}\varphi \partial_{\mu}\varphi
+ 2 \overline{g}^{\nu\alpha} \partial_{\alpha}\varphi \partial_{\mu}\chi 
- \delta_{\mu}^{\nu}\biggl[ \frac{1}{2} \delta_{\mathrm{s}} g^{\alpha\beta} \partial_{\alpha}\varphi \partial_{\beta}\varphi + \overline{g}^{\alpha\beta} \partial_{\alpha}\chi \partial_{\beta}\varphi -\frac{\partial V}{\partial\varphi}\biggr],
\label{deltatmunu}
\end{equation}
where $\chi= \delta\varphi$ is the first-order fluctuation of the scalar field in the conformally Newtonian gauge and 
$\overline{g}_{\alpha\beta}$ denotes the background metric while $\delta_{\mathrm{s}}g_{\alpha\beta}$ 
are its first-order fluctuations. In the conformally Newtonian gauge the only non-vanishing 
components are $\delta_{\mathrm{s}} g_{00} = 2 a^2 \phi$ and $\delta_{\mathrm{s}} g_{ij} = 2 a^2 \psi \delta_{ij}$.
Using the results of Appendix \ref{APPC} (in particular Eqs. (\ref{ENMOMS}), (\ref{enmomsc1}), (\ref{enmomsc2}) and 
(\ref{enmomsc3})), component by component, Eq. (\ref{deltatmunu}) will give
\begin{equation}
 \delta T_{0}^{0} = \delta_{\mathrm{s}}\rho_{\varphi},\qquad \delta T_{i}^{j} = - \delta_{\mathrm{s}} p_{\varphi} \delta_{i}^{j},\qquad \delta T_{0}^{i} = \varphi' \partial^{i} \chi,
 \label{dmunuex}
 \end{equation}
 where 
 \begin{equation}
 \delta \rho_{\varphi} = \frac{1}{a^2} \biggl[ - \phi {\varphi'}^2 +\chi' \varphi' + a^2\frac{\partial V}{\partial \varphi}\chi\biggr],
 \qquad \delta p_{\varphi} = \frac{1}{a^2} \biggl[ - \phi {\varphi'}^2 +\chi' \varphi' - a^2\frac{\partial V}{\partial \varphi}\chi\biggr].
 \label{dpphi}
 \end{equation}
 The anisotropic stress arises, in the case of a single scalar field, only to second order 
 in the amplitude of the fluctuations. This means, in our notations, that the anisotropic stress of a minimally coupled scalar field 
 will contain two derivatives of $\chi$ so it will be, in stenographic notation,  of ${\mathcal O}(|\partial\chi|^2)$ (see also Eqs. (\ref{CONVTS}) and (\ref{PPHI})). 
 Since we are here perturbing Einstein equations to first order, this contribution will 
 be neglected.
 The perturbed Einstein equations are then easily written. In particular, by using the explicit fluctuations of the 
 Einstein tensors reported in Eqs. (\ref{dg00}), (\ref{dgij}) and (\ref{dg0i}) of Appendix \ref{APPC} we have :
 \begin{eqnarray}
 &&\nabla^2 \psi - 3 {\cal H} ({\cal H}\phi + \psi') = 4\pi G  \biggl[ - \phi {\varphi'}^2 +\chi' \varphi' + a^2\frac{\partial V}{\partial \varphi}\chi\biggr],
\label{00S}\\
&& \psi'' + {\mathcal H}(\phi' + 2 \psi') + ( {\mathcal H}^2 + 2 {\mathcal H}')\phi + 
\frac{1}{3} \nabla^2 (\phi -\psi) 
\nonumber\\
&& = 4\pi G  \biggl[ - \phi {\varphi'}^2 +\chi' \varphi' - a^2\frac{\partial V}{\partial \varphi}\chi\biggr],
\label{ijS}\\
&& {\mathcal H}\phi +\psi' = 4\pi G \varphi' \chi,
\label{0iS}
\end{eqnarray}
together with the condition $\phi =\psi$ since, as already remarked,  the perturbed energy-momentum tensor of a single scalar 
degree of freedom does not possess, to first-order, an anisotropic stress.
To equations (\ref{00S}), (\ref{ijS}) and (\ref{0iS}) it is sometimes practical to add the perturbed 
Klein-Gordon equation (see Eqs. (\ref{KG1}), (\ref{KG1a}) and (\ref{KG2}) of the Appendix \ref{APPC}):
\begin{equation}
\chi'' + 2 {\mathcal H}\chi' -\nabla^2 \chi + \frac{\partial^2 V}{\partial\varphi^2}a^2 \chi + 2 \phi\frac{\partial V}{\partial\varphi} a^2 - \varphi' (\phi' + 3 \psi')=0.
\label{KGN}
\end{equation}

\subsubsection{From gauge-dependent to gauge-invariant descriptions}
Equations (\ref{00S})--(\ref{0iS}) are written in the conformally Newtonian gauge. Noticing  
that the gauge variation of the scalar field fluctuation reads
\begin{equation}
\chi \to \tilde{\chi} = \chi - \epsilon_{0} \varphi',
\label{chchtilde}
\end{equation}
the gauge-invariant generalization of $\phi$, $\psi$ and $\chi$ is given by the 
following three quantities:
\begin{eqnarray}
&& \Psi = \psi - {\mathcal H}(E' - B),
\label{gipsi}\\
&& \Phi = \phi + {\mathcal H}(E' - B) + (E' - B)',
\label{giphi}\\
&& X = \chi + (E' - B).
\label{gichi}
\end{eqnarray}
Sometimes $\Phi$ and $\Psi$ are called Bardeen potentials \cite{bardeen}. An interesting property of the 
conformally Newtonian gauge is that, in terms of $\Psi$, $\Phi$ and $X$, the evolution equations 
have the same form they would have in terms of $\psi$, $\phi$ and $\chi$. So the 
corresponding evolution equations for $\Psi$, $\Phi$ and $X$ can be obtained from Eqs. (\ref{00S}), 
(\ref{ijS}) and (\ref{0iS}) by replacing 
\begin{equation}
\psi \to \Psi, \qquad \phi\to \Phi,\qquad \chi\to X.
\end{equation}
The result of this trivial manipulation is given by 
\begin{eqnarray}
 &&\nabla^2 \Psi - 3 {\cal H} ({\cal H}\Phi + \Psi') = 4\pi G a^2 \delta^{(\mathrm{gi})} \rho_{\varphi} ,
\label{GI1}\\
&& \Psi'' + {\mathcal H}(\Phi' + 2 \Psi') + ( {\mathcal H}^2 + 2 {\mathcal H}')\Phi + 
\frac{1}{3} \nabla^2 (\Phi -\Psi) 
\nonumber\\
&& =  4\pi G a^2\delta^{(\mathrm{gi})} \rho_{\varphi}  ,
\label{GI2}\\
&& {\mathcal H}\Phi +\Psi' = 4\pi G \varphi' X,
\label{GI3}
\end{eqnarray}
where the gauge-invariant energy density and pressure fluctuations arise from  Eq. (\ref{dpphi}) 
\begin{equation}
\delta^{(\mathrm{gi})} \rho_{\varphi}= 
 \frac{1}{a^2}\biggl[ - \Phi {\varphi'}^2 +X' \varphi' + a^2\frac{\partial V}{\partial \varphi}X\biggr],\qquad \delta^{(\mathrm{gi})} p_{\varphi}=
 \frac{1}{a^2}\biggl[ - \Phi {\varphi'}^2 +X' \varphi' - a^2\frac{\partial V}{\partial \varphi}X\biggr].
 \label{GIRHO}
 \end{equation}
Suppose to be interested in the evolution equations of the fluctuations in a gauge which is totally different
from the conformally Newtonian gauge. The solution of this problem is very simple. 
Take, the evolution equations written in explicitly gauge-invariant terms. Then express the gauge-invariant 
quantities in the gauge you like. Finally substitute the expressions of the gauge-invariant quantities 
(now expressed in a specific gauge) back into the gauge-invariant equations. You will obtain 
swiftly the evolution equations in the gauge you like. Let us give an example 
of this procedure. Consider, for instance, the so-called uniform field gauge, i.e. the gauge where 
the scalar field is homogeneous. A possible choice of this gauge is 
\begin{equation}
\tilde{\chi} =0,\qquad \tilde{E} =0.
\label{gaugex}
\end{equation}
If we start from a generic gauge, we get to the uniform field gauge by fixing the relevant gauge 
parameters, i.e. $\epsilon$ and $\epsilon_{0}$ to the following values:
\begin{equation}
\epsilon_{0} = \frac{\chi}{\varphi'},\qquad \epsilon= E.
\label{gaugex2}
\end{equation}
These two conditions can be obtained from Eqs. (\ref{chchtilde}) and (\ref{E}) by imposing, respectively, 
$\tilde{\chi}=0$ and $\tilde{E}=0$. Notice that this gauge fixing eliminates completely the gauge freedom 
since $\epsilon_{0}$ and $\epsilon$ are not determined up to arbitrary constants.
In the uniform field gauge, the gauge-invariant quantities introduced in Eqs. (\ref{gipsi}), (\ref{giphi}) and (\ref{gichi}) 
assume the following form
\begin{equation}
\Phi = \phi + {\mathcal H}B + B',\qquad \Psi = \psi - {\mathcal H}B, \qquad X = \varphi' B.
\label{gaugex3}
\end{equation}
It is now easy to get all the evolution equations.  Consider, for instance, the following two equations, i.e.
\begin{equation}
{\mathcal H} ({\mathcal H} \Phi + \Psi') = 4\pi  G\varphi' X,\qquad \Phi =\Psi,
\label{gaugex4}
\end{equation}
which are, respectively, the gauge-invariant form of the momentum constraint and the gauge-invariant condition 
stemming from the off-diagonal terms of the perturbed Einstein equations.
Let us now use  Eq. (\ref{gaugex3}) into Eq. (\ref{gaugex4}). The following 
pair of equations can be swiftly obtained:
\begin{equation}
\psi' + {\mathcal H} \psi = 2{\mathcal H}( B' + {\mathcal H} B),\qquad \phi = \psi - (B' + 2 {\mathcal H}B).
\label{gaugex5}
\end{equation}
With similar procedure also the other relevant equations can be transformed in the uniform 
field gauge.  Notice that, depending on the problem at hand, gauge-dependent calculations may become
much shorter than fully gauge-invariant treatments. 

\subsubsection{Curvature perturbations and scalar normal modes}

Among all the gauge-invariant quantities some combinations have a special status. For instance 
the so-called curvature perturbations \cite{press,lyth} or the gauge-invariant density constrast \cite{bardeen2,press} (see 
also \cite{THTH} and references therein):
\begin{eqnarray}
&&{\mathcal R} = -\biggl( \Psi + \frac{{\mathcal H}({\mathcal H} \Phi + \Psi')}{{\mathcal H}^2 -{\mathcal H}'}\biggr),
\label{Rdefex}\\
&& \zeta= - \Psi + {\mathcal H} \frac{\delta^{(\mathrm{gi})}\rho_{\varphi}}{\rho_{\varphi}'},
\label{zetadefex}
\end{eqnarray}
where $\rho_{\varphi}$ denotes the (background) energy density of the scalar field
\begin{equation}
\rho_{\varphi} = \biggl( \frac{{\varphi'}^2}{2 a^2} + V\biggr),
\label{zetadefex2}
\end{equation}
and where $\delta^{(\mathrm{gi})}\rho_{\varphi}$ has been introduced in Eq. (\ref{GIRHO}).
 
The variable ${\mathcal R}$, already discussed in \ref{sec7}, has a simple interpretation 
in the comoving orthogonal gauge where it is exactly the fluctuation of the spatial 
curvature. In similar terms, $\zeta$ is nothing but the density contrast on hyeprsurfaces 
where the curvature is uniform. These two variables are clearly connected. In fact,
using Eqs. (\ref{Rdefex}) and (\ref{zetadefex}) into Eq. (\ref{GIRHO}) we get to the following 
(remarkably simple) expression:
\begin{equation}
{\mathcal R} =  \zeta - \frac{\nabla^2 \Psi}{12\pi G {\varphi'}^2}.
\label{message}
\end{equation}
The message of Eq. (\ref{message}) is very important. It tells us that when the wavelength of the 
fluctuations exceeds the Hubble radius, ${\mathcal R}\simeq \zeta + {\mathcal O}(k^2 \tau^2)$.
In other words, as long as $k\tau \ll 1$ (i.e. when the relevant wavelength is larger than the Hubble radius), the variables 
$\zeta$ and ${\mathcal R}$ have the same spectrum.

The variable ${\mathcal R}$ has also a special status since the scalar-tensor action perturbed to second 
order in the amplitude of the metric and scalar field fluctuations assumes the following simple form \cite{luk,muk} 
(see also \cite{THTH}):
\begin{equation}
S_{\mathrm{s}} = \delta_{\mathrm{s}}^{(2)} S = \frac{1}{2} \int d^4 x z^2 \eta^{\alpha\beta} \partial_{\alpha}{\mathcal R}
\partial_{\beta} {\mathcal R},
\label{secac1}
\end{equation} 
where
\begin{equation}
z= \frac{a \varphi'}{{\mathcal H}}.
\end{equation}
The Euler-Lagrange equations derived from Eq. (\ref{secac1}) imply 
\begin{equation}
{\mathcal R}'' + 2 \frac{z''}{z} {\mathcal R}' - \nabla^2 {\mathcal R} =0.
\label{secac2}
\end{equation}
Equation (\ref{secac2}) is derived in Appendix \ref{APPC} (see, in particular, the algebra prior to Eq.  (\ref{SA12})).
The canonical normal mode that can be easily read-off from Eqs. (\ref{secac1}) and (\ref{secac2}) 
is then $q = - z {\mathcal R}$. In terms of $q$ the action (\ref{secac1}) becomes then, up to total 
derivatives,
\begin{equation}
S_{\mathrm{s}} = \frac{1}{2} \int d^{4}x \biggl[ {q'}^2 + \frac{z''}{z} q^2 - \partial_{i} q\partial^{i} q\biggr].
\label{secac3}
\end{equation}
The canonical action (\ref{secac3}) is exactly of the same form of the 
one discussed, in Eq. (\ref{secon4}) for the tensor modes of the geometry. What changes is essentially 
the form of the pump field which is, in the case of the tensor modes, $a''/a = {\mathcal H}^2 + {\mathcal H}'$. 
In the case of the scalar modes the pump field is instead $z''/z$. 
From Eq. (\ref{secac3}) the quantum theory of the scalar modes can be easily 
developed in full analogy with waht has been done in the case of the tensor modes. In particular 
the canonical normal modes $q$and their conjugate momenta (i.e. $q'$) can be promoted to the status 
of quantum mechanical operators obeying equal-time commutation relations.
Recalling the notations of section \ref{sec6} (see, in particular, Eq. (\ref{expansiona})) the field 
operator will now be expanded as 
\begin{equation}
\hat{q}(\vec{x},\tau) = \frac{1}{2 (2\pi)^{3/2} } \int d^3 k \biggl[ \hat{q}_{\vk} e^{- i \vec{k} \cdot \vec{x} }
+ \hat{q}_{\vk}^{\dagger}  e^{ i \vec{k} \cdot \vec{x} }\biggr],
\label{qdec}
\end{equation}
and analogously for the conjugate momentum.
Also the phenomenon of super-adiabatic amplification (discussed in section \ref{sec6}) 
is simply translated in the context of the scalar modes since the operators  $\hat{q}$ obey now, in Fourier space, 
 a Schr\"odinger-like equation in the Heisenberg representation:
\begin{equation}
\hat{q}_{\vk}'' +\biggl[k^2 - \frac{z''}{z}\biggr]\hat{q}_{\vk}=0.
\label{secac4}
\end{equation}
Exactly as in the case of tensors (see, for instance Eqs. (\ref{secondform}) and (\ref{thirdform})) Eq. (\ref{secac4}) admits two physical regimes: the oscillating regime 
(i.e. $k^2\gg |z''/z|$) and the super-adiabatic regime (i.e. $k^2 \ll |z''/z|$) where the field operators 
are amplified and, in a more correct terminology, scalar phonons may be copiously produced.
The overall simplicity of these results must not be misunderstood. The perfect analogy between scalar 
and tensor modes only holds in the case of a {\em single} scalar field. Already in the case of two 
scalar degrees of freedom the generalization of these results is much more cumbersome. In the case 
of scalar fields and fluid variables, furthermore, the perfect mirroring between tensors and scalars is somehow lost.

\subsection{Exercise: Spectral relations}
\subsubsection{Some slow-roll algebra}
As a simple exercise the spectral relations (typical of single-field 
inflationary models) will now be derived.
The logic of the derivation will be to connect the spectral slopes and amplitudes of the scalar and of the 
tensor modes to the slow-roll parameters introduced in section \ref{sec5}:
\begin{eqnarray}
&& \epsilon = - \frac{\dot{H}}{H^2} = \frac{\overline{M}_{\mathrm{P}}^2}{2} \biggl(\frac{V_{,\varphi}}{V}\biggr)^2,
\nonumber\\
&& \eta = \frac{\ddot{\varphi}}{H \dot{\varphi}} = \epsilon - \overline{\eta},\qquad \overline{\eta} = 
\overline{M}_{\mathrm{P}}^2 \frac{V_{,\varphi\varphi}}{V},
\label{SRrec}
\end{eqnarray}
where the terms $V_{,\varphi}$ and $V_{,\varphi\varphi}$ denote, respectively, 
the first and second derivatives of the potential with respect to $\varphi$.
The slow-roll parameters affect the definition of the conformal time 
coordinate $\tau$. In fact, by definition
\begin{equation}
\tau = \int \frac{dt}{a(t)} = - \frac{1}{a H} + \epsilon \int \frac{d a}{a^2 H}
\end{equation}
where the second equality follows after integration by parts assuming that $\epsilon$ is 
constant (as it happens in the case when the potential, at least locally, can be 
approximated with a monomial in $\varphi$). Since
\begin{equation}
\int \frac{dt}{a} = \int \frac{da}{a^2 H},
\end{equation}
we will also have that 
\begin{equation}
a H = - \frac{1}{\tau ( 1 - \epsilon)}.
\label{RESR1}
\end{equation}
Using these observations, the pump fields of the scalar and tensor modes of the geometry 
can be expressed solely in terms of the slow-roll parameters. In particular, in the case 
of the tensor modes it is easy to derive the following chain of equality on the basis 
of the relation between cosmic and conformal time and using Eq. (\ref{SRrec}):
\begin{equation}
\frac{a''}{a} = {\mathcal H}^2 + {\mathcal H}' = a^2 H^2 (2 + \frac{\dot{H}}{H^2}) = a^2 H^2 ( 2 -\epsilon),
\label{RESR2}
\end{equation}
Inserting Eq. (\ref{RESR1}) into Eq. (\ref{RESR2}) we will also have, quite simply 
\begin{equation}
\frac{a''}{a} = \frac{ 2 -\epsilon}{4 \tau^2 (1 - \epsilon)^2}.
\label{RESR3}
\end{equation}
The evolution equation for the tensor mode functions is 
\begin{equation}
f_{k}'' +\biggl[ k^2 - \frac{a''}{a}\biggr] f_{k} =0,
\label{tensMF}
\end{equation}
whose solution is 
\begin{equation}
f_{k}(\tau) = \frac{{\mathcal N}}{\sqrt{2 k}} \sqrt{- k\tau} H_{\nu}^{(1)}(-k\tau),\qquad {\mathcal N} 
= \sqrt{\frac{\pi}{2}} e^{i \pi (2 \nu +1)/4},
\label{tensMF2}
\end{equation}
where $H_{\nu}^{(1)}(-k\tau)$ are the Hankel functions \cite{abr,tric} already encountered 
in section \ref{sec6}.
In Eq. (\ref{tensMF2}) the relation of $\nu$ tp $\epsilon$ is determined from the 
relation
\begin{equation}
\nu^2 - \frac{1}{4} = \frac{ 2 -\epsilon}{4  (1 - \epsilon)^2},
\end{equation}
which implies 
\begin{equation}
\nu = \frac{3 -\epsilon}{2 (1 -\epsilon)}.
\label{tensMF3}
\end{equation}
The same algebra allows to determine the relation of the scalar pump field with 
the slow-roll parameters. In particular, the scalar pump field is 
\begin{equation}
\frac{z''}{z} = \biggl(\frac{z'}{z}\biggr)^2  + \biggl(\frac{z'}{z}\biggr)'.
\label{tenspump1}
\end{equation}
and the corresponding evolution equation for the mode functions follows from Eq. (\ref{secac4})
and can be written as 
\begin{equation}
\tilde{f}_{k}'' + \biggl[ k^2 - \frac{z''}{z}\biggr] \tilde{f}_{k}=0.
\label{SCMF}
\end{equation}
Recalling now the explicit expression of $z$, i.e. 
\begin{equation}
z = \frac{a \varphi'}{{\mathcal H}} = \frac{a \dot{\varphi}}{H},
\label{tenspump2}
\end{equation}
we will have that 
\begin{equation}
\frac{\dot{z}}{z} = H + \frac{\ddot{\varphi}}{\dot{\varphi}} - \frac{\dot{H}}{H}.
\label{tenspump3}
\end{equation}
But using Eq. (\ref{tenspump2}), Eq. (\ref{tenspump1}) can be expressed as 
\begin{equation}
\frac{z''}{z} = a^2 \biggl[ \biggl(\frac{\dot{z}}{z}\biggr)^2 + H \frac{\dot{z}}{z} + 
\frac{\partial}{\partial t} \biggl( \frac{\dot{z}}{z} \biggr)\biggr]
\label{tenspump4}
\end{equation}
Using Eq. (\ref{tenspump3}) inside Eq. (\ref{tenspump4}) we get an expression 
that is the scalar analog of Eq. (\ref{RESR2}):
\begin{equation}
\frac{z''}{z} = a^2 H^2 (2 + 2 \epsilon + 3 \eta + \epsilon\eta +\eta^2)
\label{tenspump5}
\end{equation}
Notice that the explicit derivatives appearing in Eq. (\ref{tenspump4}) lead to two kinds of terms.
The terms of the first kind can be immediately written in terms of the slow-roll parameters. The 
second kind of terms involve three (time) derivatives either of the scalar field or of the Hubble 
parameter. In these two cases we can still say, from the definitions of $\epsilon$ and $\eta$,
that 
\begin{equation}
\frac{\partial^{3} \varphi}{\partial t^3} = \eta \dot{H} \dot{\varphi} + \eta H \ddot{\varphi},\qquad 
\ddot{H} = - 2 \epsilon H \dot{H}.
\end{equation}
Again, using Eq. (\ref{RESR1}) inside Eq. (\ref{tenspump5}) we do get 
\begin{equation}
\frac{z''}{z} = \frac{(2 + 2 \epsilon + 3 \eta + \epsilon\eta +\eta^2)}{(1-\epsilon)^2\tau^2},
\label{tenspump6}
\end{equation}
implying that the relation of the Bessel index $\tilde{\nu}$ to the slow-roll parameters 
is now determined from \cite{abr,tric}
\begin{equation}
\tilde{\nu}^2 - \frac{1}{4} = \frac{2 + 2 \epsilon + 3 \eta + \epsilon\eta +\eta^2}{(1 -\epsilon)^2}.
\label{nutildesq}
\end{equation}
By solving Eq. (\ref{nutildesq}) with respect to $\tilde{\nu}$ the following simple expression can be
readily obtained: 
\begin{equation}
\tilde{\nu}= \frac{3 + \epsilon + 2 \eta}{2( 1 -\epsilon)}.
\label{nutilde}
\end{equation}
Consequently, the solution of Eq. (\ref{SCMF}) will be, formally, the same as the one of the 
tensors (see Eq. (\ref{tensMF2})) but with a Bessel index $\tilde{\nu}$ instead of $\nu$:
\begin{equation}
\tilde{f}_{k}(\tau) = \frac{\tilde{{\mathcal N}}}{\sqrt{2 k}} \sqrt{- k\tau} H_{\tilde{\nu}}^{(1)}(-k\tau),\qquad \tilde{{\mathcal N}} 
= \sqrt{\frac{\pi}{2}} e^{i \pi (2 \tilde{\nu} +1)/4},
\label{SCMF2}
\end{equation}
Of course, this formal analogy should not be misunderstood: the difference in the Bessel index will entail, necessarily, a different 
behaviour in the small argument limit of Hankel functions \cite{abr,tric} and, ultimately,  slightly different spectra whose
essential features will be the subject of the remaining part of the present section.

\subsubsection{Tensor power spectra}

The tensor power-spectrum, in a given model, is the Fourier transform of the two-point function.
Consider, therefore, the two-point function of the tensor modes of the geometry
computed in the operator formalism:
\begin{equation}
\langle 0| \hat{h}_{i}^{j}(\vec{x},\tau) \hat{h}_{j}^{i}(\vec{y},\tau) |0\rangle = 
\frac{8\ell_{\mathrm{P}}^2}{a^2} \langle 0| \hat{\mu}(\vec{x},\tau) \hat{\mu}(\vec{y},\tau) |0\rangle.
\label{EXtens1}
\end{equation}
By now evaluating the expectation value we obtain 
\begin{equation}
\langle 0| \hat{h}_{i}^{j}(\vec{x},\tau) \hat{h}_{j}^{i}(\vec{y},\tau) |0\rangle =
\frac{8 \ell_{\mathrm{P}}^2}{a^2} \int \frac{d^3 k}{(2\pi)^3} |f_{k}(\tau)|^2 e^{ - i \vec{k}\cdot\vec{r}}.
\label{EXtens2}
\end{equation}
By making explicit the phase-space integral in Eq. (\ref{EXtens2}) we do get 
\begin{equation}
\langle 0| \hat{h}_{i}^{j}(\vec{x},\tau) \hat{h}_{j}^{i}(\vec{y},\tau) |0\rangle = 
\int d\ln{k} {\mathcal P}_{\mathrm{T}}(k) \frac{\sin{kr}}{kr},
\label{EXtens3}
\end{equation}
where ${\mathcal P}_{T}(k)$ is the tensor power spectrum, i.e. 
\begin{equation}
{\mathcal P}_{\mathrm{T}}(k) = \frac{4 \ell_{\mathrm{P}}^2 }{a^2 \pi^2} k^3 |f_{k}(\tau)|^2.
\label{EXtens4}
\end{equation}
But from Eq. (\ref{tensMF2}) we have that 
\begin{equation}
 |f_{k}(\tau)|^2 = \frac{|{\mathcal N}|^2}{2 k} (-x) H_{\nu}^{(1)}(-x) H_{\nu}^{(2)}(-x) \simeq \frac{\Gamma^2(\nu)}{4\pi k} 2^{2\nu}(-x)^{1-2\nu}.
 \label{EXtens4a}
 \end{equation}
 The second equality in Eq. (\ref{EXtens4a}) follows from the small argument limit 
 of Hankel functions \cite{abr,tric}:
 \begin{equation}
 H_{\nu}^{(1)}(-x) \simeq - \frac{i}{\pi} \Gamma(\nu) \biggl( - \frac{x}{2}\biggr)^{-\nu},
 \end{equation}
 for $|x|\ll 1$. The physical rationale for the small argument limit is that we are considering modes 
 whose wavelengths are larger than the Hubble radius, i.e. $|x|=k\tau \ll 1$.
 Equation (\ref{EXtens4}) gives then the {\em super-Hubble} tensor power spectrum, i.e. the spectrum  valid for those  modes whose wavelength 
is larger than the Hubble radius at the decoupling, i.e. 
\begin{equation}
{\mathcal P}_{\mathrm{T}}(k) =\ell_{\mathrm{P}} H^2 \frac{2^{2\nu}}{\pi^3} \Gamma^2(\nu) ( 1 - \epsilon)^{2\nu-1}  \biggl(\frac{k}{aH}\biggr)^{3 - 2 \nu}. 
\label{EXtens5}
\end{equation}
In Eq. (\ref{EXtens5}),  the term
$(1-\epsilon)^{2\nu -1}$ arises by eliminating $\tau$ is favour of $(aH)^{-1}$. There are now 
different (but equivalent) ways of expressing the result of Eq. (\ref{EXtens5}).
Recalling that $\ell_{\mathrm{P}} = \overline{M}_{\mathrm{P}}^{-1}$ the spectrum 
(\ref{EXtens5}) at horizon crossing (i.e. $k \simeq H a$) can be expressed as 
\begin{equation}
{\mathcal P}_{\mathrm{T}}(k)= \frac{2^{2\nu}}{\pi^3} \Gamma^2(\nu) ( 1 - \epsilon)^{2\nu-1}
\biggl(\frac{H^2}{\overline{M}_{\mathrm{P}}^2}\biggr)_{k\simeq a H}.
\label{EXtens6}
\end{equation}
The subscript arising in Eq. (\ref{EXtens6}) demands some simple explanations.
The moment at which a given wavelength crosses the Hubble radius is defined 
as the time at which 
\begin{equation}
 \frac{k}{{\mathcal H}} = \frac{k}{a H} \simeq k \tau \simeq  1.
\label{HCcond}
\end{equation}
 This condition is also called, somehow improperly, horizon crossing. Note that the equalities 
 in Eq. (\ref{HCcond}) simply follow from the relation between the cosmic and the 
 conformal time coordinate (see, for instance, Eq. (\ref{HtoHcal})).

As discussed in section \ref{sec5}, the slow-roll parameters are much smaller than one during 
inflation and become of order $1$ as inflation ends. Now the typical scales relevant for CMB anisotropies 
crossed the Hubble radius the first time (see Figs. \ref{mindur} and \ref{nonmindur})  about $60$ e-folds 
before the end of inflation (see Eqs. \ref{phi60} and (\ref{epsilon60}) when the slow-roll parameters 
had to be, for consistency sufficiently smaller than $1$. It is therefore legitimate to expand the Bessel 
indices in powers of the slow-roll parameters and, from Eq. (\ref{tensMF3}), we get:
\begin{equation}
\nu \simeq \frac{3}{2} + \epsilon + {\mathcal O}(\epsilon^2),
\label{EXtens7}
\end{equation}
Eq. (\ref{EXtens6}) can be also written as 
\begin{equation}
{\mathcal P}_{\mathrm{T}}(k) \simeq 
\frac{2}{3 \pi^2} \biggl(\frac{V}{\overline{M}_{\mathrm{P}}^4}\biggr)_{k\simeq a H},
\label{EXtens8}
\end{equation}
where we used the slow-roll relation $3 \overline{M}_{\mathrm{P}}^2 H^2 \simeq V$ (see Eqs. (\ref{SR1}) and (\ref{SR2})).
Finally, expressing $\overline{M}_{\mathrm{P}}$ in terms of $M_{\mathrm{P}}$ (see Eq. (\ref{PLdef})), 
\begin{equation}
{\mathcal P}_{\mathrm{T}}(k) \simeq 
\frac{128}{3 } \biggl(\frac{V}{M_{\mathrm{P}}^4}\biggr)_{k\simeq a H}.
\label{EXtens9}
\end{equation}
The tensor spectral index $n_{\mathrm{T}}$ is then defined as
\begin{equation}
{\mathcal P}_{\mathrm{T}}(k) \simeq k^{n_{\mathrm{T}}},\qquad n_{\mathrm{T}} = 
\frac{d \ln{{\mathcal P}_{\mathrm{T}}}}{d \ln{k}}.
\label{EXtens10}
\end{equation}
Taking now the spectrum in the parametrization of Eq. (\ref{EXtens9}) we will have that, 
from Eq. (\ref{EXtens10}), 
\begin{equation}
n_{\mathrm{T}} = \frac{V_{,\varphi}}{V} \frac{\partial \varphi}{\partial \ln{k}}.
\label{EXtens11}
\end{equation}
But since $k = a H$, we will also have that 
\begin{equation}
\frac{\partial \ln{k}}{\partial\varphi} = \frac{1}{a} \frac{\partial a}{\partial \varphi} + \frac{1}{H} 
\frac{\partial H}{\partial \varphi}.
\label{EXtens12}
\end{equation}
The right hand side of Eq. (\ref{EXtens12}) can then be rearranged by using the definitions of the 
slow-roll parameter $\epsilon$ and it is 
\begin{equation}
\frac{\partial \ln{k}}{\partial\varphi}= - \frac{V}{V_{,\varphi}} \biggl(\frac{1-\epsilon}{\overline{M}_{\mathrm{P}}^2}\biggr).
\label{EXtens13}
\end{equation}
Inserting Eq. (\ref{EXtens13}) into Eq. (\ref{EXtens11}) it is easy to obtain
\begin{equation}
n_{\mathrm{T}} = - \biggl(\frac{V_{,\varphi}}{V}\biggr)^2 \frac{\overline{M}_{\mathrm{P}}^2}{1 -\epsilon} \simeq - \frac{2\epsilon}{1 - \epsilon} \simeq - 2\epsilon + {\mathcal O}(\epsilon^2).
\label{EXtens14}
\end{equation}
It should be stressed that different definitions for the slow-roll parameters and for the spectral indices exist 
in the literature. At the very end the results obtained with different sets of conventions must necessarily all agree. 
In the present discussion the conventions adopted are, for practical reasons,  the same as the ones of the WMAP collaboration (see, for instance, 
\cite{map1,map2}).  
\subsubsection{Scalar power spectra}

The scalar power spectrum is computed by considering the two-point function 
of the curvature perturbations on comoving orthogonal hypersurfaces. 
This choice is, to some extent, conventional and also dictated by practical reasons since 
the relation of the curvature perturbations ${\mathcal R}$ to the scalar normal mode 
$q$ is rather simple, i.e. $z {\mathcal R} = -q$ (see Eqs. (\ref{secac1}) and (\ref{secac2}) as well as the algebra prior to Eq. (\ref{SA12}) 
in Appendix \ref{APPC}).
Having said this, the scalar modes of the geometry can be parametrized in terms 
of the two-point function of any other gauge-invariant operator. Eventually, after 
the calculation, the spectrum of the curvature perturbations can be always 
obtained by means of, sometimes long, algebraic manipulations.
In the operator formalism the quantity to be computed is 
\begin{equation}
\langle 0 | \hat{\mathcal R}(\vec{x},\tau) \hat{\mathcal R}(\vec{y},\tau) |0\rangle= 
\int {\mathcal P}_{{\mathcal R}}(k) \frac{\sin{k r}}{kr} d\ln{k},
\label{CURV}
\end{equation}
where 
\begin{equation}
{\mathcal P}_{{\mathcal R}}(k) = \frac{k^3}{2\pi^2 z^2} |\tilde{f}_{k}(\tau)|^2,
\label{EXscal1}
\end{equation}
where, now, the mode functions $\tilde{f}_{k}(\tau)$ are the functions given in Eq. (\ref{SCMF2}).
By repeating exactly the same steps outlined in the tensor case, the scalar 
power spectrum can be written as 
\begin{equation}
{\mathcal P}_{{\mathcal R}}(k) = \frac{2^{2\tilde{\nu}-3}}{\pi^3} \Gamma^2(\tilde{\nu}) (1 -\epsilon)^{1 -2\tilde{\nu}} 
\biggl(\frac{k}{aH}\biggr)^{3-2\tilde{\nu}} \biggl(\frac{H^2}{\dot{\varphi}}\biggr)^2.
\label{EXscal2}
\end{equation}
Recalling Eq. (\ref{nutilde}) and expanding in the limit $\epsilon \ll 1$ and $\eta\ll 1$ we have that, in full 
analogy with Eq. (\ref{EXtens7}),
\begin{equation}
\tilde{\nu} = \frac{3 +\epsilon + 2 \eta}{2(1-\epsilon)} \simeq \frac{3}{2} + 2\epsilon +\eta 
+ {\mathcal O}(\epsilon^2).
\label{EXscal3}
\end{equation}
By comparing Eqs. (\ref{EXtens7}) with (\ref{EXscal3}) it appears clearly that the difference 
between $\nu$ and $\tilde{\nu}$ arises as a first-order correction that depends upon (both) $\epsilon$ and 
$\eta$.
Equation (\ref{EXscal2}) can then be written in various (equivalent) forms, like, for instance,
evaluating the expression at horizon crossing and taking into account that $\tilde{\nu}$, to lowest order
is $3/2$. The result is:
\begin{equation}
{\mathcal P}_{{\mathcal R}}(k) = \frac{1}{2\pi^2} \biggl(\frac{H^2}{\dot{\varphi}}\biggr)^2_{k\simeq aH}.
\label{EXscal4}
\end{equation}
Since, from the slow-roll equations,
\begin{equation}
\dot{\varphi}^2 = \frac{V_{,\varphi}}{9 H^2},\qquad \frac{1}{2\pi^2} \frac{H^4}{\dot{\varphi}^2} = 
\frac{1}{12\pi^2} \frac{V}{\epsilon \overline{M}_{\mathrm{P}}}.
\label{EXscal5}
\end{equation}
Hence, Eq. (\ref{EXscal4}) becomes 
\begin{equation}
{\mathcal P}_{{\mathcal R}}(k) = \frac{8}{3\, M_{\mathrm{P}}^4} \biggl( \frac{V}{\epsilon}\biggr)_{k\simeq a H},
\label{EXscal6}
\end{equation}
where we used Eq. (\ref{EXscal5}) into Eq. (\ref{EXscal4}) and we recalled that $\overline{M}_{\mathrm{P}}^2 = M_{\mathrm{P}}^2/(8\pi)$.
The scalar spectral index is now defined as 
\begin{equation}
{\mathcal P}_{{\mathcal R}}(k) \simeq k^{n_{\mathrm s} -1},\qquad n_{\mathrm{s}} -1 = 
\frac{d \ln{{\mathcal P}_{\mathcal R}}}{d\ln{k}}
\label{EXscal7}
\end{equation}
It should be stressed that, again, the definition of the spectral index is conventional and, in particular, it appears 
that while in the scalar case the exponent of the wave-number has been parametrized as $n_{\mathrm{s}}-1$, in the 
tensor case, the analog quantity has been parametrized as $n_{\mathrm{T}}$.
Using the parametrization of the power spectrum given in Eq. (\ref{EXscal6}) and recalling Eq. (\ref{EXtens13}),
Eq. (\ref{EXscal7}) implies that
\begin{equation}
n_{\mathrm{s}} -1 = \frac{\dot{\varphi}}{H} \frac{1}{1-\epsilon} \biggl[ \frac{V_{,\varphi}}{V} - \frac{\epsilon_{,\varphi}}{\epsilon}\biggr].
\label{EXscal8}
\end{equation}
Since
\begin{equation}
\frac{\epsilon_{,\varphi}}{\epsilon} = 2 \frac{V_{,\varphi\varphi}}{V_{,\varphi}} - 2 \biggl(\frac{V_{,\varphi}}{V}\biggr),\qquad 
\frac{\dot{\varphi}}{H} = - \frac{V_{,\varphi}}{3 H^2},
\label{EXscal9}
\end{equation}
Eq. (\ref{EXscal8}) implies that 
\begin{equation}
n_{\mathrm{s}} = 1 + 6 \epsilon - 2 \overline{\eta}.
\label{Exscal10}
\end{equation}
Equation (\ref{Exscal10}) is the standard result for the scalar spectral index arising in single field 
inflationary models.
\subsubsection{Consistency relation}
Therefore, we will have, in summary,
\begin{equation}
{\mathcal P}_{\mathrm{T}} = \frac{128}{3 } \biggl(\frac{V}{M_{\mathrm{P}}^4}\biggr)_{k\sim aH},\qquad 
{\mathcal P}_{\mathcal R} = \frac{8}{3}\biggl(\frac{V}{\epsilon\,\, M_{\mathrm{P}}^4}\biggr)_{k\sim aH},
\label{EXscal11}
\end{equation}
with 
\begin{equation}
n_{\mathrm{T}} = - 2\epsilon,\qquad n_{\mathrm{s}} = 1 + 6\epsilon - 2 \overline{\eta}.
\label{EXscal12}
\end{equation}
For applications the ratio between the tensor and the scalar spectrum is also defined as 
\begin{equation}
r= \frac{{\mathcal P}_{\mathrm{T}}}{{\mathcal P}_{{\mathcal R}}}.
\label{EXscal13}
\end{equation}
Using Eq. (\ref{EXscal11}) into Eq. (\ref{EXscal13}) 
we obtain 
\begin{equation}
r = 16 \epsilon = - 8 n_{\mathrm{T}}
\label{EXscal14}
\end{equation}
which is also known as consistency condition. Again, as previously remarked, the conventions 
underlying Eq. (\ref{EXscal14}) are the same ones adopted from the WMAP collaboration 
\cite{map1,map2}.

The scalar and tensor power spectra computed here represent the (single filed) inflationary result 
for the CMB initial conditions. This exercise shows that the quantum-mechanically 
normalized inflationary perturbations lead, prior to decoupling, to a single adiabatic 
 mode. Few final comments are in order:
 \begin{itemize}
 \item{} while the situation is rather simple when only one scalar degree of freedom is present, 
 more complicated system can be easily imagined when more the one 
 scalar is present in the game;
 \item{} if many scalar fields are simultaneously present, the evolution of curvature 
 perturbations may be more complex.
 \end{itemize}
 The approximate conservation of curvature perturbations for typical wavelengths 
 larger than the Hubble radius holds, strictly speaking, only in the case of single field
 inflationary models. It is therefore the opinion of the author that the approach 
 of setting "early" initial conditions (i.e. during inflation) should always be 
 complemented and corroborated with a model-independent treatment of the 
 "late" initial conditions. In this way we will not only understand which is 
 the simplest model fitting the data but also, hopefully, the correct one.
   
\newpage
\begin{appendix}
\renewcommand{\theequation}{A.\arabic{equation}}
\setcounter{equation}{0}
\section{The concept of distance in cosmology}
\label{APPA}

In cosmology there are different concepts of distance. Galaxies emit electromagnetic 
radiation. Therefore we can ask what is the distance light travelled from the observed 
galaxy to the receiver (that we may take, for instance, located at the origin 
of the coordinate system). But we could also ask what was the actual distance 
of the galaxy at the time the signal was emitted. Or we can ask the distance of the galaxy now.
Furthermore, distances 
in FRW models depend upon the matter content. If we would know precisely 
the matter content we would also know accurately the distance. However, in practice, the distance 
of an object is used to infer a likely value of the cosmological parameters.
Different distance concepts can therefore be introduced such as:
\begin{itemize}
\item{} the proper coordinate distance;
\item{} the redshift;
\item{} the distance measure;
\item{} the angular diameter distance;
\item{} the luminosity distance.
\end{itemize}
In what follows these different concepts will be swiftly introduced and physically motivated.
\subsection{The proper coordinate distance}
The first idea coming to mind to define a distance is to look at a radial part of the FRW line 
element and define the coordinate distance to $r_{\mathrm{e}}$ as 
\begin{equation}
s(r_{\mathrm{e}}) = a(t) \int_{0}^{r_\mathrm{e}} \frac{ d r }{\sqrt{1 - k r^2}},
\label{pdist}
\end{equation}
where, conventionally, the origin of the coordinate system coincides with the position 
of the observer located while $r_{\mathrm{e}}$ defines the position of the emitter.
Since $k$ can be either positive, or negative or even zero, the integral at the right hand side of Eq. (\ref{pdist}) 
will change accordingly so that we will have:
\begin{eqnarray}
s(r_{\mathrm{e}}) &=&  a(t) r_{\mathrm{e}},\qquad k =0,
\nonumber\\
s(r_{\mathrm{e}}) &=& \frac{a(t)}{\sqrt{k}} \arcsin{[\sqrt{k} \,r_{\mathrm{e}}]}, \qquad k>0,
\nonumber\\
s(r_{\mathrm{e}}) &=& \frac{a(t)}{\sqrt{|k|}} {\mathrm{arcsinh}}{[\sqrt{|k|} \,r_{\mathrm{e}}]}, \qquad k<0,
 \label{pdist2}
 \end{eqnarray}
The only problem with the definition of Eq. (\ref{pdist}) is that it involves geometrical quantities 
that are not directly accessible through observations. A directly observable quantity, at least 
in principle, is the redshift and, therefore it will be important to substantiate the dependence 
of the distance upon the redshift and this will lead to complementary (and commonly used)
definitions of distance.

Before plunging into the discussion a lexical remark is in order.  
It should be clarified that the distance $r_{\mathrm{e}}$ is {\em fixed} and it 
is an example of {\em comoving coordinate} system. On the contrary, the distance $s(r_{\mathrm{e}}) = a(t) r_{\mathrm{e}}$ 
(we consider, for simplicity, the spatially flat case) is called {\em physical} distance since it gets stretched as the 
scale factor expands. It is a matter of convenience to use either comoving or physical systems of units. For instance, in the 
treatment of the inhomogeneities, it is practical to use comoving wavelengths and wave-numbers. In different 
situations physical frequencies are more appropriate. It is clear that physical distances make sense 
at a given moment in the life of the Universe. For instance, the Hubble radius at the electroweak 
epoch will be of the order of the cm. The same quantity, evaluated today, will be much larger (and of the order of the astronomical unit, i.e. $10^{13}$ cm) since 
the Universe expanded a lot between a temperature of $100$ GeV and the present temperature (of the order of $10^{-13}$ GeV).
In contrast, comoving distances are the same at any time in the life of the Universe. A possible convention (which is, though, not strictly necessary)
is to normalize to $1$ the present value of the scale factor so that, today, physical and comoving distances would coincide. 

\subsection{The redshift}
Suppose that a galaxy or a cloud of gas emits electromagnetic radiation of a given wavelength 
$ \lambda_{\mathrm{e}}$. The wavelength received by the observer will be denoted by 
$\lambda_{\mathrm{r}}$. If the Universe would not be expanding we would simply have that 
$\lambda_{\mathrm{e}}= \lambda_{\mathrm{r}}$. However since the Universe 
expands (i.e. $\dot{a} >0$), the observed wavelength will be more red (i.e. redshifted) in comparison
with the emitted frequency, i.e. 
\begin{equation}
\lambda_{\mathrm {r}} = \frac{a(t_{\mathrm{r}})}{a(t_{\mathrm{e}})} \lambda_{\mathrm{e}},\qquad 
1+ z = \frac{a_{\mathrm{r}}}{a_{\mathrm{e}}}.
\label{reds1}
\end{equation}
where $z$ is the redshift.  If the wavelength of the emitted radiation would be precisely known (for instance 
a known emission line of the hydrogen atom or of some other molecule or chemical compound)
the amount of expansion (i.e. the redshift) between the emission and the observation could be 
accurately determined. Equation (\ref{reds1}) can be directly justified from the FRW line 
element. Light rays follow null geodesics, i.e. $ds^2 =0$ in Eq. (\ref{FRW}). Suppose 
then that a signal is emitted at the time $t_{\mathrm{e}}$ (at a radial position $r_{\mathrm{e}}$) 
and received at the time $t_{\mathrm{r}}$ (at a radial position $r_{\mathrm{r}}=0$).
Then from Eq. (\ref{FRW}) with $ds^2 =0$ we will have \footnote{The position of the emitter 
is {\em fixed} in the comoving coordinate system. }
\begin{equation}
\int_{t_{{\mathrm{e}}}}^{t_{{\mathrm{r}}}}\frac{d t}{a(t)}= \int_{0}^{r_{\mathrm{e}}} \frac{d r}{\sqrt{1- kr^2}}.
\label{reds2}
\end{equation}
Suppose than that a second signal is emitted at a time  $t_{\mathrm{e}} +\delta t_{\mathrm{e}}$
and received at a time $t_{\mathrm{r}} +\delta t_{\mathrm{r}}$. It will take a time $\delta t_{\mathrm{r}} =
\lambda_{\mathrm{r}}/c$ to receive the signal and a time $\delta t_{\mathrm{e}} =
\lambda_{\mathrm{e}}/c$ to emit the signal:
\begin{equation}
\int_{t_{{\mathrm{e}}} + \delta t_{{\mathrm{e}}}}^{t_{{\mathrm{r}}} + \delta t_{{\mathrm{r}}}}\frac{d t}{a(t)}= \int_{0}^{r_{\mathrm{e}}} \frac{d r}{\sqrt{1- kr^2}}.
\label{reds3}
\end{equation}
By subtracting Eqs. (\ref{reds2}) and (\ref{reds3}) and by rearranging the limits of integration 
we do get:
\begin{equation}
\int_{t_{{\mathrm{e}}}}^{t_{{\mathrm{e}}} + \delta t_{{\mathrm{e}}}}\frac{d t}{a(t)} = 
\int_{t_{{\mathrm{r}}}}^{t_{{\mathrm{r}}} + \delta t_{{\mathrm{r}}}}\frac{d t}{a(t)}, 
\label{reds4}
\end{equation}
implying 
\begin{equation}
\frac{\delta t_{\mathrm{r}}}{\delta t_{\mathrm{e}}} = \frac{\lambda_{\mathrm{r}}}{\lambda_{\mathrm{e}}} = 
\frac{a(t_{\mathrm{r}})}{a(t_{\mathrm{e}})} = \frac{\nu_{\mathrm{e}}}{\nu_{\mathrm{r}}}.
\label{reds5}
\end{equation}
As far as the conventions are concerned we will remark that often (even if not always) the emission time 
will be generically denoted by $t$ while the observation time will be the present time $t_{0}$. Thus 
with this convention $z + 1 = a_0/a(t)$. 

Let us then suppose (incorrectly, as we shall see in a moment) that $t_{\mathrm{r}} = t_{\mathrm{e}} + r_{\mathrm{e}}$. Then, recalling the definition of the Hubble parameter evaluated at $t_{\mathrm{r}}= t_{0}$, i.e. 
$H_{\mathrm{r}} = H_{0} = \dot{a_{\mathrm{r}}}/{a_{\mathrm{r}}}$ we have: 
\begin{equation}
\lambda_{\mathrm{r}} = \frac{a(t_{\mathrm{r}})}{a(t_{\mathrm{e}})} \simeq \lambda_{\mathrm{e}}[ 1 + H_{0} r_{\mathrm{e}}],
\label{reds6}
\end{equation}
where the second equality in Eq. (\ref{reds6}) follows by expanding in Taylor series around $t_{\mathrm{r}}$  and 
by assuming $H_{0} r_{\mathrm{e}}<1$. Recalling the definition of Eqs. (\ref{reds1})  and (\ref{reds5}) 
we obtain an approximate form of the Hubble law, i.e. 
\begin{equation}
H_{0} r_{\mathrm{e}} \simeq z,\qquad z = \frac{\lambda_{\mathrm{r}} - \lambda_{\mathrm{e}}}{\lambda_{\mathrm{r}}}.
\label{reds7}
\end{equation}
This form of the Hubble law is approximate since it holds for small redshifts, i.e. $z< 1$. Indeed the 
assumption that $t_{\mathrm{r}} \simeq t_{\mathrm{e}} + r_{\mathrm{e}}$ is not strictly correct since it would 
imply, for a flat Universe, that the scale factor is approximately constant.  
To improve the situation it is natural to expand systematically the scale factor and the redshift. 
Using such a strategy we will have that:
\begin{eqnarray}
&&\frac{a(t)}{a_{0}} = 1 + H_{0} ( t - t_{0}) - \frac{q_{0}}{2} H_{0}^2 (t- t_{0})^2 +...
\nonumber\\
&& \frac{a_0}{a(t)} = 1 - H_{0} ( t-t_{0}) + \biggl(\frac{q_{0}}{2} + 1\biggr) H_{0}^2 ( t- t_{0})^2 +...
\label{exp1}
\end{eqnarray}
where $q_{0}$ is the deceleration parameter introduced in Eq. (\ref{decpar}).
Form the definition of redshift (see Eq. (\ref{reds1})) it is easy to obtain that 
\begin{equation}
z = H_{0} (t_{0} - t) + \biggl(\frac{q_{0}}{2} + 1\biggr) H_{0}^2 (t_{0} - t)^2,
\label{ztot}
\end{equation}
i.e. 
\begin{equation}
(t_{0} - t) = \frac{1}{H_{0}}\biggl[ z - \biggl(\frac{q_{0}}{2} + 1\biggr)^2 z^2\biggr].
\label{ttoz}
\end{equation}
Using then Eq. (\ref{exp1}) to express the integrand appearing at the left hand side 
of Eq. (\ref{reds2}) we obtain, in the limit $ |k r_{\mathrm{e}}^2|<1$, 
\begin{equation}
r_{\mathrm{e}} = \int_{t}^{t_{0}} \frac{dt'}{a(t')} = \frac{1}{H_{0} a_{0}} \biggl[ z - \frac{1}{2}(q_{0} +1) z^2\biggr],
\label{reds8}
\end{equation}
where, after performing the integral over $t'$, Eq. (\ref{ttoz}) has been used to eliminate 
$t$ in favour of $z$. Notice that while the leading result reproduces the one previously obtained in Eq. (\ref{reds7}), the 
correction involves the deceleration parameter.

\subsection{The distance measure}
As we saw in the previous subsection the term {\em distance} in cosmology can have different meanings.
Not all the distances we can imagine can be actually measured. A meaningful question can be, for instance,
the following: we see an object at a given redshift $z$; how far is the object we see at redshift $z$? 
Consider, again, Eq. (\ref{reds2}) where, now the integrand at the left hand side, can be expressed
as
\begin{equation}
\frac{dt}{a} = \frac{da}{H a^2} = - \frac{1}{a_{0}} \frac{dz}{H(z)}.
\end{equation}
From Eq. (\ref{FL1}), $H(z)$ can be expressed as 
\begin{equation}
H(z) = H_{0} \sqrt{ \Omega_{\mathrm{M}0} ( 1 + z)^3 + \Omega_{\Lambda0} + \Omega_{k} ( 1 + z)^2}\,,
\end{equation}
where 
\begin{eqnarray}
&&\Omega_{\mathrm{M}0} + \Omega_{\Lambda 0} + \Omega_{k} =1, \qquad\Omega_{k} = - \frac{k}{a_{0}^2 H_{0}^2}.
\label{curvcont}
\end{eqnarray}
In Eq. (\ref{curvcont}), $\Omega_{k}$ accounts for the curvature contribution to the total density. This parametrization may be confusing 
but will be adopted tom match the existing notations in the literature. Notice that if $k>0$ then $\Omega_{k}<0$ 
and vice-versa. Defining now the integral\footnote{It should be borne in mind that the inclusion of the radiation term (scaling, inside the squared root 
of the integrand, as $\Omega_{\mathrm{R}0} (1 + x)^4$) is unimportant for moderate redshifts (i.e. up to $z\simeq 20$). For even larger 
redshifts (i.e. of the order of $z_{\mathrm{eq}}$ or $z_{\mathrm{dec}}$) the proper 
inclusion of the radiation  contribution is clearly mandatory. See Eqs. (\ref{zeq}) and (\ref{phex}) for the derivation of $z_{\mathrm{eq}}$ and $z_{\mathrm{dec}}$.
See also the applications at the end  of this Appendix (in particular, Eqs. (\ref{defdec1}), (\ref{defdec2}) and (\ref{defdec3})).}
\begin{equation}
D_{0}(z) = \int_{0}^{z} \frac{dx}{\sqrt{\Omega_{\mathrm{M}0} ( 1 + x)^3 + \Omega_{\Lambda0} + \Omega_{k} ( 1 + x)^2}},
\label{D0}
\end{equation}
we have, from Eq. (\ref{reds2}), that 
\begin{eqnarray}
r_{\mathrm{e}}(z) &=& \frac{D_{0}(z)}{a_{0} H_{0}},\qquad k=0,
\nonumber\\
r_{\mathrm{e}}(z) &=& \frac{1}{a_{0} H_{0}\,\,\sqrt{|\Omega_{k}|}} \sin{\bigl[ \sqrt{|\Omega_{k}|} D_{0}(z)\bigr]},\qquad k>0,
\nonumber\\
r_{\mathrm{e}}(z) &=& \frac{1}{a_{0} H_{0}\,\,\sqrt{\Omega_{k}}} \sinh{\bigl[ \sqrt{\Omega_{k}} D_{0}(z)\bigr]},\qquad k<0.
\label{DM}
\end{eqnarray}
The quantity $r_{\mathrm{e}}(z)$ of Eq. (\ref{DM}) is also called, sometimes, {\em distance measure} and it 
is denoted by $d_{\mathrm{M}}(z)$. In the limit $|k r_{\mathrm{e}}^2| <1$ the open and closed expressions 
reproduce the flat case $k=0$. 

Some specific examples will now be given. Consider first the case $\Omega_{\mathrm{M}0}=1$, 
$\Omega_{\Lambda 0} =0$ and $\Omega_{k}=0$ (i.e. flat, mattter-dominated Universe). In such a case 
Eq. (\ref{DM}) gives 
\begin{equation}
r_{\mathrm{e}}(z) = \frac{1}{a_{0} H_{0}} \biggl[ 1 - \frac{1}{\sqrt{z +1 }}\biggr] \simeq \frac{1}{a_{0} H_{0}} \biggl[ z - \frac{3}{4} z^2 \biggr],
\label{MDdist}
\end{equation}
where the second equality has been obtained by expanding the exact result for $z,1$. Notice that the second equality 
of Eq. (\ref{MDdist}) reproduces Eq. (\ref{reds8}) since, for a flat matter-dominated Universe, 
$q_{0} = 1/2$. 
The other example we wish to recall is the one where $\Omega_{k} =1$, $\Omega_{\mathrm{M}0}=0$, and 
$\Omega_{\Lambda 0}=0$ (i.e. open Universe dominated by the spatial curvature). In this case Eqs. (\ref{D0}) 
and (\ref{DM}) give, respectively, 
\begin{equation}
D_{0}(z) = \ln{(z+1)},\qquad r_{\mathrm{e}}(z) = \frac{1}{2 a_{0} H_{0}} \biggl[ (z + 1) - \frac{1}{z+ 1}\biggr].
\label{CDdist}
\end{equation}
The considerations developed so far suggest the following chain of observations:
\begin{itemize}
\item{} in cosmology the distance measure to redshift $z$ depend upon the cosmological parameters;
\item{} while for $z<1$, $r_{\mathrm{e}} H_{0} \simeq z$, as soon as $z\simeq 1$ quadratic and
cubic terms start contributing to the distance measure;
\item{} in particular quadratic terms contain the deceleration parameter and are thus sensitive 
to the total matter content of the Universe.
\end{itemize}
If we would be able to scrutinize objects at high redshifts we may be able to get important clues not only on the 
expansion rate of the present Universe (i.e. $H_{0}$) but also on its present acceleration 
(i.e. $q_{0}<0$) or deceleration (i.e. $q_{0}>0$). 
This type of reasoning is the main rationale for the intensive study of type Ia supernovae. It has indeed 
been observed that type Ia supernovae are dimmer than expected.
This experimental study suggests that the present Universe is accelerating rather than decelerating, i.e. 
$q_{0} <0$ rather than $q_{0} >0$. Since 
\begin{equation}
q_{0} = - \frac{\ddot{a}_{0} a_{0}}{\dot{a}^2_{0}} = \frac{\Omega_{\mathrm{M}0}}{2} + (1 + 3 w_{\Lambda}) \Omega_{\Lambda 0},
\label{q0}
\end{equation}
where the dark-energy contribution as $p_{\Lambda} = w_{\lambda} \rho_{\Lambda}$ with $w_{\Lambda}\simeq -1$.
So, to summarize, suppose we know an object of given redshift $z$. Then we also know precisely the matter content of the Universe.
With these informations we can compute what is the distance to redshift $z$ by computing $r_{\mathrm{e}}(z)$.
The distance measure of Eq. (\ref{DM}) depends both on the redshift and on the precise
value of the cosmological parameters. But cosmological parameters are, in some sense, exactly 
what we would like to measure.
Astronomers, therefore, are interested in introducing more operational notions of distance like the angular 
diameter distance and the luminosity distance.

\subsection{Angular diameter distance}
Suppose to be in Eucledian (non-expanding) geometry. Then we do know that the arc of a curve $s$ 
is related to the diameter $d$ as $s = d \vartheta$ where $\vartheta$ is the angle subtended by $s$. Of course 
this is true in the situation where $\vartheta <1$. Suppose that $s$ is known, somehow. Then $d \simeq s/\vartheta$ 
can be determined by determining $\vartheta$, i.e. the angular size of the object. 
In FRW space-times the angular diameter distance can be defined from the angular part of the line 
element.
Since $ds_{\vartheta}^2 = a^2(t) r^2 d\vartheta^2$, the angular diameter distance to redshift $z$  will be 
\begin{equation}
D_{\mathrm{A}}(z) = \frac{s_{\vartheta}}{\vartheta} = a(t) r_{\mathrm{e}}  = \frac{a_{0} r_{\mathrm{e}}(z)}{1 + z},
\label{ADD0}
\end{equation}
where $r_{\mathrm{e}}(z)$ is exactly the one given, in the different cases, by Eq. (\ref{DM}).
The quantity introduced in Eq. (\ref{ADD0}) is the physical angular diameter distance. We 
can also introduce the comoving angular diameter distance $\overline{D}_{\mathrm{A}}(z)$:
\begin{equation}
D_{\mathrm{A}} = a(t) \overline{D}_{\mathrm{A}}= \frac{a_{0}}{1 + z} \overline{D}_{\mathrm{A}}
\label{CANDD0}
\end{equation}
which implies that the comoving angular diameter distance coincides with the distance measure 
defined in Eq. (\ref{DM}).

To determine the angular diameter distance we need, in practice, a set of standard rulers, i.e. objects that 
have the same size for different redshifts.  Then the observed angular sizes will give us the physical
angular diameter distance. Using the results of Eqs. (\ref{MDdist}) and (\ref{CDdist}) into Eq. (\ref{ADD0}) 
we obtain, respectively, that 
\begin{eqnarray}
D_{\mathrm{A}}(z) &=& \frac{1}{H_{0}}\frac{\sqrt{z +1} -1}{(z + 1)^{3/2}},
\label{ADDex1}\\
D_{\mathrm{A}}(z) &=& \frac{1}{2 H_{0}} \biggl[1 - \frac{1}{(z+1)^2}\biggr],
\label{ADDex2}
\end{eqnarray}
where Eq. (\ref{ADDex1}) applies in the case of a matter-dominated Universe and Eq. (\ref{ADDex2}) 
applies in the case of an open Universe dominated by the spatial curvature\footnote{It is clear that, in this case, from Eqs. (\ref{FL1}) and (\ref{FL2}) 
the scale factor expands linearly. In fact, suppose to take Eqs. (\ref{FL1}) and (\ref{FL2}) in the limit $\rho=0$ and $p=0$. In this case the spatial curvature 
must be negative for consistency with Eq. (\ref{FL1}) where the right hand side is positive semi-definite. By summing up Eqs. (\ref{FL1}) and (\ref{FL2}) 
we get to the condition $H^2 + \dot{H}=0$, i.e. $\ddot{a} =0$ which means $a(t) \sim t$. In this case, by definition, the deceleration parameter 
introduced in Eq. (\ref{decpar}) vanishes.} 
It is interesting to notice that the angular diameter distance in a flat (matter-dominated) Universe 
decreases for $z>1$: objects that are further away appear larger in the sky. In such a case $D_{\mathrm{A}}$ given by Eq. (\ref{ADDex1}) has a maximum and then decreases.
\subsection{Luminosity distance}

Suppose to be in Eucledian (transparent) space. Then, if the {\em absolute} luminosity (i.e. radiated energy 
per unit time) of an object 
at distance $D_{\mathrm{L}}$   is ${\mathcal L}_{\mathrm{abs}}$, the {\em apparent} luminosity ${\mathcal L}_{\mathrm{app}}$ will be 
\begin{equation}
{\mathcal L}_{\mathrm{app}} = \frac{\mathcal{L}_{\mathrm{abs}}}{4\pi \,D_{\mathrm{L}}^2},\qquad 
D_{\mathrm{L}} = \sqrt{\frac{{\mathcal L}_{\mathrm{abs}}}{4\pi {\mathcal L}_{\mathrm{app}}}}.
\label{DL1}
\end{equation}

If the observer is located at a position $r_{\mathrm{e}}$ from the source, the detected photons will be spread 
over an area ${\mathcal A} = 4 \pi\, a_{0}^2 r_{\mathrm{e}}^2$. Thus, from the emission time 
$t_{\mathrm{e}}$ to the observation time $t_{0}$  we will have that the energy density of radiation 
evolves as:
\begin{equation}
\rho(t_{0}) = \rho(t_{\mathrm{e}}) \biggl(\frac{a_{\mathrm{e}}}{a_{0}}\biggr)^4,\qquad 
\sqrt{\frac{\rho_{\mathrm{e}}}{\rho_{0}}} = (1 + z)^{-2}.
\label{DL2}
\end{equation}
But now,
\begin{equation}
{\mathcal L}_{\mathrm{abs}} \propto \sqrt{\rho_{\mathrm{e}}},\qquad {\mathcal L}_{\mathrm{app}} \propto \frac{\sqrt{\rho_{0}}}{4\pi a_{0}^2 r_{\mathrm{e}}^2}.
\label{DL3}
\end{equation}
Thus  the luminosity distance as a function of the redshift becomes:
\begin{equation}
D_{\mathrm{L}}(z) =(1 + z) a_{0} r_{\mathrm{e}}(z),
\label{DL4}
\end{equation}
By comparing Eq. (\ref{ADD0}) to Eq. (\ref{DL4}) we also have that
\begin{equation}
D_{\mathrm{A}}(z) = \frac{D_{\mathrm{L}}(z)}{(1 + z)^2}.
\label{DL5}
\end{equation}

To give two examples consider, as usual, the cases of a (flat) matter-dominated Universe and the 
case of an open (curvature-dominated) Universe.  In these two cases, Eq. (\ref{DL4}) in 
combination with Eqs. (\ref{MDdist}) and (\ref{CDdist}) leads to
\begin{equation}
D_{\mathrm{L}}(z) = \frac{1}{H_{0}} \sqrt{z + 1}[ \sqrt{z + 1} -1],
\qquad D_{\mathrm{L}}(z) = \frac{1}{2 H_{0}}[ ( z + 1)^2 -1].
\end{equation}

As in the case of the angular diameter distance there it is mandatory to have a set of standard 
rulers, in the case of the luminosity distance there is the need of a set of standard candles, i.e. 
a set of objects that are known to have all the same absolute luminosity ${\mathcal L}_{\mathrm{abs}}$ .
Then by measuring the apparent luminosity ${\mathcal L}_{\mathrm{app}}$ the luminosity distance 
can be obtained at a given redshift. The observed $D_{\mathrm{L}}(z)$ can then be compared with various theoretical models and precious informations on the underlying cosmological parameters can be obtained.

\subsection{Horizon distances} 

In the discussion of the kinematical features of FRW models, a key r\^ole is played by the concept of 
{\em event horizon} and of {\em particle horizon}. 

The physical distance of the event horizon is defined as 
\begin{equation}
d_{\mathrm{e}}  = a(t)  \int_{t}^{t_{\mathrm{max}}} \frac{dt'}{a(t')}.
\label{eventh}
\end{equation}
The quantity defined in Eq. (\ref{eventh}) measures the maximal distance over which we can admit, even 
in the future, a causal connection. If $d_{\mathrm{e}}(t)$ is finite in the limit $t_{\mathrm{max}}\to \infty$ 
(for finite $t$) we can conclude that the event horizon exist. In the opposite case, i.e. 
$d_{\mathrm{e}}(t)\to \infty$ for $t_{\mathrm{max}}\to \infty$  the event horizon does not exist. 

The physical distance  of the particle horizon is defined as 
\begin{equation}
d_{\mathrm{p}}(t)  = a(t) \int_{t_{\mathrm{min}}}^{t} \frac{dt'}{a(t')}.
\label{parth}
\end{equation}
Equation (\ref{parth}) measures the extension of the regions admitting a causal connection at time $t$.
If the integral converges in the limit $t_{\mathrm{min}}\to 0$ we say that there exist a particle 
horizon. 

 \subsection{Few simple applications}
In CMB studies it is often useful to compute the (comoving) angular diameter distance 
to decoupling or to equality (see Eq. (\ref{CANDD0}) for a definition of the 
comoving angular diameter distance). As already discussed, the (comoving) angular diameter distance 
coincides with the distance measure. So suppose, for instance, to be interested in the model where
$\Omega_{\mathrm{M}0} =1$. In this case we have to compute 
\begin{equation}
\overline{D}_{\mathrm{A}} (z)= \frac{D_{\mathrm{A}}(z)}{a(t)} = \frac{a_{0} r_{\mathrm{e}}(z)}{a ( 1 + z)} =
r_{\mathrm{e}}(z), 
\label{defdec1}
\end{equation}
where the last equality follows from the definition of redshift and where $\overline{D}_{\mathrm{A}}$ 
denotes the (comoving) angular diameter distance. 
If $\Omega_{\mathrm{M}0}=1$, the comoving angular diameter distance to decoupling and to 
equality is given, respectively, by:
\begin{eqnarray}
&& \overline{D}_{\mathrm{A}}(z_{\mathrm{dec}}) = \frac{2}{H_{0}} \biggl(1 - \frac{1}{\sqrt{1 + z_{\mathrm{dec}}}}\biggr)
\simeq \frac{1.939}{H_{0}},
\nonumber\\
&&  \overline{D}_{\mathrm{A}}(z_{\mathrm{eq}}) =  \frac{2}{H_{0}} \biggl(1 - \frac{1}{\sqrt{1 + z_{\mathrm{eq}}}}\biggr)
\simeq \frac{2}{H_{0}},
\label{defdec2}
\end{eqnarray}
where we assumed that the Universe was always matter-dominated from equality onwards (which is not 
an extremely good approximation since radiation may modify a bit the estimate). The comoving 
angular diameter distance enters crucially in the determination of the multipole number
on the last scattering sphere. Suppose indeed to be interested in the following questions:
\begin{itemize}
\item{} what is the multipole number corresponding to a wavelength comparable with the Hubble radius at 
decoupling (or at equality)?
\item{} what is the angle subtended by such a wavelength?
\end{itemize}
Before answering these questions let us just remark that when the spatial curvature vanishes 
(and when $\Omega_{\mathrm{M}0} + \Omega_{\Lambda0} =1$) the comoving angular diameter 
distance must be computed numerically but a useful approximate expression is 
\begin{equation}
\overline{D}_{\mathrm{A}}(z_{\mathrm{dec}}) = \frac{2}{\Omega_{\mathrm{M}0}^{0.4}} H_{0}^{-1}.
\label{defdec3}
\end{equation}
When the (comoving) wave-number $k_{\mathrm{dec}}$ is comparable with the Hubble radius, the following chain of equalities holds:
\begin{equation}
k_{\mathrm{dec}} = {\mathcal H}_{\mathrm{dec}} = a_{\mathrm{dec}} H_{\mathrm{dec}}  =
 \sqrt{\Omega_{\mathrm{M}0}} H_{0} \sqrt{ 1 + z_{\mathrm{dec}}},
 \label{defdec4}
 \end{equation}
 where the second equality follows by assuming that the Hubble radius at decoupling 
 is (predominantly) determined by the matter contribution, i.e. 
 \begin{equation}
 H_{\mathrm{dec}}^2 \simeq \frac{8\pi G}{3} \rho_{\mathrm{M}} = H_{0}^2 \Omega_{\mathrm{M}0} ( 1 + z_{\mathrm{dec}})^3.
\label{defdec5} 
 \end{equation}
The corresponding multipole number on the last scattering sphere will then be given 
by 
\begin{equation}
\ell_{\mathrm{dec}} = k_{\mathrm{dec}} \overline{D}_{\mathrm{A}}(z_{\mathrm{dec}})  = \sqrt{ 1 + z_{\mathrm{dec}}} 
\Omega_{\mathrm{M}0}^{0.1} \simeq 66.3 \,\,\Omega_{\mathrm{M}0}^{0.1},
\label{defddec6}
\end{equation}
where $z_{\mathrm{dec}} \simeq 1100$. The angle subtended by $\pi/k_{\mathrm{dec}}$ will then be 
\begin{equation}
\theta_{\mathrm{dec}} = \frac{180^{0}}{\ell_{\mathrm{dec}}} = 2.7^{0}\,\, \Omega_{\mathrm{M}0}^{-0.1}.
\label{defdec7}
\end{equation}
Following analog steps, it is possible to show that 
\begin{equation}
k_{\mathrm{eq}} = \frac{h_{0}^2\Omega_{\mathrm{M}0} }{14\,\, \mathrm{Mpc}},\qquad 
 \ell_{\mathrm{eq}} = k_{\mathrm{eq}} \overline{D}_{\mathrm{A}}(z_{\mathrm{eq}})  = 430\,\,h_{0}\Omega_{\mathrm{M}0}^{0.6}.
\label{defdec8}
\end{equation}
\renewcommand{\theequation}{B.\arabic{equation}}
\setcounter{equation}{0}
\section{Kinetic description of hot plasmas}
\label{APPB}
\subsection{Generalities on thermodynamic systems}
In thermodynamics we distinguish, usually, intensive variables (like 
the pressure and the temperature) which do not depend upon the 
total matter content of the system and extensive variables (like 
internal energy, volume, entropy, number of particles).
The first principle of thermodynamics tells us that 
\begin{equation}
d {\mathcal E} = T d S - p d V + \mu dN,
\label{firstprinc}
\end{equation}
where $S$ is the entropy, $p$ is the pressure, $V$ is the volume, 
$\mu$ is the chemical potential, $N$ the number of 
particles and ${\mathcal E}$ the internal energy.
From Eq. (\ref{firstprinc}) it can be easily deduced that 
\begin{equation}
T = \biggl( \frac{\partial{\mathcal E}}{\partial S} \biggr)_{V,N},\qquad 
p = -\biggl( \frac{\partial{\mathcal E}}{\partial V} \biggr)_{S,N},\qquad 
\mu = \biggl( \frac{\partial{\mathcal E}}{\partial N} \biggr)_{V,S},
\label{firstprinc2}
\end{equation}
where the subscripts indicate that each partial derivation is done 
by holding fixed the remaining two variables.
Suppose now that the system is described only by an appropriate 
set of extensive variables. In this situation we can think that, for instance, 
the internal energy is a function of the remaining extensive 
variables, i.e. ${\mathcal E} = {\mathcal E}(S, V, N)$. Let us then perform 
a scale transformation of all the variables, i.e.
\begin{equation}
{\mathcal E}\to \sigma {\mathcal E},\qquad S\to \sigma S,\qquad 
V\to \sigma V, \qquad N \to \sigma N.
\label{rescale}
\end{equation}
We will have, consequently, that $\sigma {\mathcal E} = {\mathcal E}(\sigma S, ,\sigma V, \sigma N)$. By taking the derivative of the latter relation with respect to $\sigma$ and by then fixing $\sigma =1$ we do get the following relation: 
\begin{equation}
{\mathcal E} = \biggl(\frac{\partial {\mathcal E}}{\partial S}\biggr)_{V,N} S + 
\biggl(\frac{\partial {\mathcal E}}{\partial V}\biggr)_{S,N} V +
\biggl(\frac{\partial {\mathcal E}}{\partial N}\biggr)_{V,S} N.
\label{rescale2}
\end{equation}
Using now Eqs. (\ref{firstprinc2}) into Eq. (\ref{rescale2}) we do get the 
following important relation called, sometimes, {\em fundamental 
relation of thermodynamics}:
\begin{equation}
{\mathcal E }= T S - p V + \mu N.
\label{fund}
\end{equation}
If the system is formed by different particle species, a chemical potential
for each species is introduced and, consequently, $\mu N = \sum_{\mathrm{i}}
\mu_{\mathrm{i}} N_{\mathrm{i}}$. Equation (\ref{fund}) tells us that if 
the chemical potential vanishes (as in the case of a gas of photons) 
the entropy will be simply given by $S= ({\mathcal E} + p V)/T$.
In statistical mechanics it is sometimes useful to introduce 
different {\em potentials} such as the free energy $F$, the Gibbs free energy 
$G$ and the so-called thermodynamic potential $\Omega$:
\begin{equation}
F= {\mathcal E} - T S,\qquad G = F + p V, \qquad \Omega = F - \mu N.
\label{Tpotentials}
\end{equation}
The free energies $F$ and $G$ or the thermodynamic potential $\Omega$ 
allow to reduce the number of extensive variables employed for the 
description of a given system in favour of one or more intensive variables\footnote{To avoid ambiguities 
in the notations we did not mention the enthalpy, customarily defined as $H = {\mathcal E} + pV$ (this 
nomenclature may clash with the notation employed for the Hubble parameter $H$). Note, however, that 
the enthalpy density is exactly what appears in the second of the Friedmann-Lema\^itre equations (see Eq. (\ref{FL2})).}.
So the description provided via the potentials is always semi-extensive 
in the sense that it includes always one or more intensive variables. 
Notice, for instance, that $\Omega(T, V,\mu)$ and that, using Eqs. (\ref{Tpotentials}) and (\ref{firstprinc}) 
\begin{equation}
d \Omega = - p d V - S dT - N d\mu.
\label{Tpot2}
\end{equation}
Equation (\ref{Tpot2}) implies that $\Omega = \Omega(V, T, \mu)$ and
\begin{equation}
S = - \biggl( \frac{\partial \Omega}{\partial T}\biggr)_{V,\mu},\qquad 
p = - \biggl( \frac{\partial \Omega}{\partial V}\biggr)_{T,\mu},\qquad
N = - \biggl( \frac{\partial \Omega}{\partial \mu}\biggr)_{T,V}.
\label{Tpot3}
\end{equation}
In the case of a gas of photons $\mu=0$, and $\Omega = F$. This 
implies, using Eq. (\ref{Tpotentials}) and (\ref{firstprinc}) that 
$d F = (\mu dN - p dV - S dT)$. Hence, 
the condition of equilibrium of a photon gas is given by 
\begin{equation}
\mu = \biggl( \frac{\partial F}{\partial N}\biggr)_{V, T}=0.
\end{equation}
For a boson gas and for a fermion gas we have that $\Omega$ can be written, respectively, as 
\begin{eqnarray}
&& \Omega^{\mathrm{B}} = \sum_{\vec{k}}  \Omega^{\mathrm{B}}_{\vec{k}},
\qquad \Omega^{\mathrm{F}} = \sum_{\vec{k}}  \Omega^{\mathrm{F}}_{\vec{k}},
\label{defFB}\\
&&  \Omega^{\mathrm{B}}_{\vec{k}} = T \sum_{\vec{k}} \ln{\biggl[ 1 
- e^{\frac{\mu - E_{k}}{T}}\biggr]},
\label{defOMB}\\
&& \Omega^{\mathrm{F}}_{\vec{k}} = -T \sum_{\vec{k}} \ln{\biggl[ 1 
+ e^{\frac{\mu - E_{k}}{T}}\biggr]}.
\label{defOMF}
\end{eqnarray}
Recalling that the Bose-Einstein and Fermi-Dirac occupation 
numbers are defined as 
\begin{equation}
N^{\mathrm{B}} = \sum_{\vec{k}} \overline{n}^{\mathrm{B}}_{k},\qquad 
N^{\mathrm{F}} = \sum_{\vec{k}} \overline{n}^{\mathrm{F}}_{k},
\label{occnum}
\end{equation}
 the third relation reported in Eq. (\ref{Tpot3}) allows to determine 
 $\overline{n}^{\mathrm{B}}_{k}$ and $\overline{n}^{\mathrm{F}}_{k}$:
 \begin{eqnarray}
 && \overline{n}^{\mathrm{B}}_{k} = - \biggl( \frac{\partial
  \Omega_{\vec{k}}^{\mathrm{B}}}{\partial \mu}\biggr)_{T,V} = \frac{1}{e^{(E_{k} - \mu)/T} -1},
  \label{BEoc}\\
  && \overline{n}^{\mathrm{F}}_{k} = - \biggl( \frac{\partial
  \Omega_{\vec{k}}^{\mathrm{F}}}{\partial \mu}\biggr)_{T,V} = \frac{1}{e^{(E_{k} - \mu)/T} +1}.
\label{FDoc}
\end{eqnarray}

\subsection{Fermions and bosons}
To determine the concentration, the energy density, the pressure and the entropy we can now follow two complementary procedures. For instance 
the entropy and the pressure can be deduced from Eq. (\ref{Tpot3}). 
Then Eq. (\ref{fund}) allows to determine the internal energy ${\mathcal E}$.
It is also possible to write the energy density, the pressure and the concentration in terms of the occupation numbers:
\begin{eqnarray}
&& n^{\mathrm{B/F}} = \frac{g}{(2\pi)^3} \int d^{3}k\,\, \overline{n}_{k}^{\mathrm{B/F}},
\label{concBF}\\
&& \rho^{\mathrm{B/F}} = \frac{g}{(2\pi)^3} \int d^{3}k\,\,E_{k}\,\, \overline{n}_{k}^{\mathrm{B/F}},
\label{rhoBF}\\
&& p^{\mathrm{B/F}} = \frac{g}{(2\pi)^3} \int d^{3}k\,\,\frac{|\vec{k}|^2}{3 E_{k}}\,\, \overline{n}_{k}^{\mathrm{B/F}},
\label{pBF}
\end{eqnarray}
where $g$ denotes the effective number of relativistic degrees of freedom and where the superscripts indicate that each relation holds, independently 
for the Bose-Einstein or Fermi-Dirac occupation number.
Then, Eq. (\ref{fund}) can be used to determine the entropy or the entropy density.

Consider, for instance, the ultra-relativistic case when the temperature 
is much larger than the masses and than the chemical potential:
\begin{equation}
T\gg m, \qquad T\gg |\mu|, \qquad E_{k} = \sqrt{k^2 + m^2} \simeq k.
\label{ultrarel}
\end{equation}
In this case we will have that 
\begin{eqnarray}
&&n^{\mathrm{B}} = \frac{\zeta(3)}{\pi^2} g T^3,\qquad n^{\mathrm{F}} = \frac{3}{4} n^{\mathrm{B}},
\label{concBF2}\\
&&\rho^{\mathrm{B}} = \frac{\pi^2}{30} g T^4, \qquad \rho^{\mathrm{F}} = \frac{7}{8} \rho^{\mathrm{B}},
\label{rhoBF2}\\
&& p^{\mathrm{B}} = \frac{\pi^2}{90} g T^4, \qquad p^{\mathrm{F}} = \frac{7}{8} p^{\mathrm{B}},
\label{pBF2}
\end{eqnarray}
From Eq. (\ref{fund}) the entropy density will then be swiftly determined as 
\begin{equation}
s^{\mathrm{B}} = \frac{2\pi^2}{45} g T^3,\qquad s^{\mathrm{F}} =\frac{7}{8} s^{\mathrm{B}} 
\label{entrdens}
\end{equation} 
To perform the integrations implied by the above results it is useful to recall that 
\begin{equation}
{\mathcal I}_{\mathrm{B}} =\int_{0}^{\infty} \frac{x^3 dx}{e^{x} -1} = 
\frac{\pi^4}{15}, \qquad {\mathcal I}_{\mathrm{F}} = \int_{0}^{\infty} \frac{x^3 dx}{e^{x} +1} = \frac{7}{8} {\mathcal I}_{\mathrm{B}}= \frac{7\pi^4}{120}.
\label{integrals}
\end{equation}
Notice that the value of ${\mathcal I}_{\mathrm{F}}$ can be swiftly 
determined once the value of ${\mathcal I}_{\mathrm{B}}$ is known. In fact 
\begin{equation} 
{\mathcal I}_{\mathrm{F}} - {\mathcal I}_{\mathrm{B}} = - 2 \int_{0}^{\infty} 
\frac{x^3 dx }{e^{2 x} -1} = - \frac{1}{8} {\mathcal I}_{\mathrm{B}},
\label{integrals2}
\end{equation}
where the second equality follows by changing the integration variable 
from $x$ to $y = 2 x$.
Furthermore, the following pair of integrals is useful to obtain the explicit 
expressions for the concentrations:
\begin{equation}
\int_{0}^{\infty} dx \frac{x^{s -1}}{e^{a x} -1} = \frac{1}{a^{s}}
 \Gamma(s) \zeta(s), \qquad \int_{0}^{\infty} dx \frac{x^{s -1}}{e^{a x} +1} = \frac{1}{a^{s}}
 \Gamma(s)( 1 - 2^{1 - s})\zeta(s),
\label{integrals3}
\end{equation}
where $\Gamma(s)$ is the Euler Gamma function and $\zeta(s)$ is the Riemann $\zeta$ function (recall $\zeta(3) = 1.20206$).

As an example of the first procedure described consider the calculation 
of the entropy density of a boson gas directly from the first relation of 
Eq. (\ref{Tpot3}):
\begin{equation}
S^{\mathrm{B}} = \frac{g V}{2\pi^2} \biggl\{ {\mathcal I}_{\mathrm{B}} - 
\int_{0}^{\infty} x^2 dx \ln{[ 1 - e^{-x}]} \biggr\} = \frac{2 g V}{45 \pi^2} T^3,
\label{entrex}
\end{equation}
where the second integral can be evaluated by parts and
where the sum has been transformed into an integral according to
\begin{equation}
\sum_{\vec{k}} \to \frac{g}{(2\pi)^3} V \int d^{3} k.
\end{equation}
The result of Eq. (\ref{entrex}) clearly coincides with the one of Eq. (\ref{entrdens}) recalling that, by definition, $s^{\mathrm{B}}= S^{\mathrm{B}}/V$.
In similar terms, the wanted thermodynamic variables can be obtained 
directly from the thermodinamic potentials.

In the ultra-relativistic limit the boson and fermion gases have a radiative 
equation of state, i.e. 
\begin{equation}
p^{\mathrm{B}} = \frac{\rho^{\mathrm{B}}}{3},\qquad p^{\mathrm{F}} = \frac{\rho^{\mathrm{F}}}{3}.
\end{equation}
In the non-relativistic limit the equation of state is instead $p =0$ both for 
bosons and fermions. In fact, in the non-relativistic limit,
\begin{equation}
|m - \mu| > T,\qquad E_{k} = \sqrt{k^2 + m^2}\simeq m + \frac{k^2}{2 m}.
\end{equation}
Then, in this limit
\begin{equation}
\overline{n}_{k}^{\mathrm{B/F}} = \overline{n}_{k} = e^{(\mu - E_{k})/T}.
\end{equation}
Then, from the definitions (\ref{concBF}), (\ref{rhoBF}) and (\ref{pBF}) 
it can be easily obtained, after Gaussian integration
\begin{eqnarray}
&& n= \frac{g}{(2\pi)^{3/2}} (m T)^{3/2} e^{(\mu - m)/T},
\nonumber\\
&& \rho = m n,\qquad p = n T = \rho \biggl(\frac{T}{m}\biggr) \ll \rho,
\label{NReq}
\end{eqnarray}
which shows that, indeed $p =0$ in the non-relativistic limit. Note that 
to derive Eq. (\ref{NReq}) the well known result
\begin{eqnarray}
&&{\mathcal I}(\alpha) = \int_{0}^{\infty} dk\, e^{- \alpha k^2} = \frac{1}{2} 
\sqrt{\frac{\pi}{\alpha}}, 
\nonumber\\
&& \int_{0}^{\infty}  dk\,k^2\, e^{- \alpha k^2} = 
- \frac{\partial}{\partial\alpha} {\mathcal I}(\alpha) = \frac{\sqrt{\pi}}{4} \alpha^{-3/2}.
\end{eqnarray}

The energy density in the ultra-relativistic limit (i.e. Eq. (\ref{rhoBF2})) goes 
as $T^{4}$. The energy density in the non-relativistic limit (i.e. Eq. (\ref{NReq})) is exponentially suppressed as $e^{-m/T}$. Therefore, as soon as 
the temperature drops below the threshold of pair production, the energy 
density and the concentration are exponentially suppressed. This is 
the result of particle-antiparticle annihilations. At very high temperatures 
$T\gg m$ particles annihilate with anti-particles but the energy-momentum 
supply of the thermal bath balances the annihilations with the 
production of particles-antiparticles pairs. At lower temperatures 
(i.e. $T< m$) the thermal energy of the particles is not sufficient for a 
copious production of pairs.

\subsection{Thermal, kinetic and chemical equilibrium}

Let us now suppose that the primordial plasma is formed by a mixture 
of $N_{b}$ bosons and $N_{f}$ fermions. Suppose also that  
the ultra-relativistic limit holds so that the masses and the chemical potentials 
of the different species can be safely neglected. Suppose 
finally that, in general, each bosonic species carries $g_{b}$ degrees of freedom at a temperature and that each fermionic species carries $g_{f}$ degrees of freedom. Each of the bosonic degrees of freedom will be in thermal 
equilibrium at a temperature $T_{b}$ and; similarly each of the fermionic 
degrees of freedom will be in thermal equilibrium at a temperature $T_{f}$.
Under the aforementioned assumptions, Eqs. (\ref{rhoBF2}) and 
(\ref{entrdens}) imply  that the total energy density and the total entropy density of the system are given by 
\begin{equation}
\rho(T) = g_{\rho}(T) \frac{\pi^2}{30} T^4,\qquad s(T) = g_{s}(T) \frac{2 \pi^2}{45} T^3,
\label{ultra1}
\end{equation}
where 
\begin{eqnarray}
g_{\rho}(T) &=& \sum_{b=1}^{N_{b}} g_{b} \biggl(\frac{T_{b}}{T}\biggr)^4 + 
\frac{7}{8} \sum_{f=1}^{N_{f}} g_{f} \biggl(\frac{T_{f}}{T}\biggr)^4,
\label{grho}\\
g_{s}(T) &=& \sum_{b=1}^{N_{b}} g_{b} \biggl(\frac{T_{b}}{T}\biggr)^3 + 
\frac{7}{8} \sum_{f=1}^{N_{f}} g_{f} \biggl(\frac{T_{f}}{T}\biggr)^3,
\label{gs}
\end{eqnarray}
Equations (\ref{grho}) and (\ref{gs}) are clearly different. 
If all the fermionic and 
bosonic species are in thermal equilibrium at a common temperature 
$T$, then
\begin{equation}
T_{b}  = T_{f} = T, \qquad g_{\rho} = g_{s}.
\label{theq}
\end{equation}
However, if at least one of the various species has a different 
temperature, then $g_{\rho} \neq g_{s}$.
In more general terms we can say that:
\begin{itemize}
\item{} if all the species are in equilibrium at a common 
temperature $T$, then the system is in thermodynamic 
equlibrium;
\item{} if some species are in equilibrium at a temperature 
different from $T$, then the system is said to be in kinetic 
equilibrium.
\end{itemize}
There is a third important notion of equilibrium, i.e. the chemical 
equilibrium. Consider, indeed, the situation where 
$2 H + 0 \to H_{2}0$. In chemical equilibrium the latter reaction 
is balanced by $H_{2} 0\to 2 H + 0$. We can attribute to each 
of the reactants of a chemical process a coefficient. For instance, in the 
aforementioned naive example we will have $\alpha_{H} = 1$, 
$\alpha_{O}=1$, $\alpha_{H_{2}O} = -1$ satisfying the sum rule 
$\sum_{R} \alpha_{R} R=0$ where $R$ denotes each of the reactants.
Such a sum rule simply means that the disappearance of a water molecule 
implies the appearance of two atoms of hydrogen and one atom of oxygen 
and vice versa.  This concept of chemical equilibrium can be generalized 
to more general reactions, like $e^{+} + e^{-} \to \gamma$ 
or $e + p \to H + \gamma$ and so on. 
By always bearing in mind the chemical analogy, let us suppose 
to conduct a chemical reaction at fixed temperature and fixed pressure 
(as it sometimes the case for practical applications). Then the Gibbs free 
energy is the appropriate quantity to use since 
\begin{equation}
dG = \sum_{R} \mu_{R} d N_{R} - S dT + V dp.
\label{gibb1}
\end{equation}
If $dT=0$ and $d p=0$ the condition of chemical equilibrium is expressed 
by $d G=0$ which can be expressed as 
\begin{equation}
\sum_{R} \mu_{R} \biggl(\frac{\partial N_{R}}{\partial \lambda}\biggr) d \lambda 
\equiv \sum_{R} \alpha_{R} \mu_{R} =0,
\end{equation}
where the second equality follows from the first one since $d N_{R}$ are not 
independent and are all connected by the fact that $d N_{R}/\alpha_{R}$ must have the same value $d\lambda$ for all the reactants.
Thus, in the case of $e^{+} + e^{-} \to \gamma$ (and vice versa) the condition 
of chemical equilibrium implies $\mu_{e^{+}} + \mu_{e^{-}} = \mu_{\gamma}$, i.e. $\mu_{e^{+}} =- \mu_{e^{-}}$ since $\mu_{\gamma}=0$. Similarly, for 
the reactions $p + e \to H + \gamma$ (Hydrogen formation) and 
$ \gamma + H \to p + e$ (Hydrogen photo-dissociation) the condition 
of chemical equilibrium implies that $\mu_{e} + \mu_{p} = \mu_{H}$.

\subsection{An example of primordial plasma}
The considerations presented in this Appendix will now be applied to 
few specific examples that are useful in the treatment of the hot Universe.
Consider first the case when the primordial soup is formed by all the 
degrees of freedom of the Glashow-Weinberg-Salam (GWS) model 
$SU_{L}(2)\otimes U_{Y}(1) \otimes SU_{c}(3)$. 
Suppose that the plasma is at a temperature $T$ larger than $175$ GeV, i.e. 
a temperature larger than the top mass which is the most massive 
species of the model (the Higgs mass, still unknown, in such that $m_{H} > 115$ GeV). In this situation all the species of the GWS model are in thermodynamic equilibrium at the temperature $T$. 
In this situation $g_{\rho}$ and $g_{s}$ are simply given by 
\begin{equation}
g_{\rho} = g_{s} = \sum_{b} g_{b} + \frac{7}{8} \sum_{f} g_{f},
\end{equation}
where the sum now extends over all the fermionic and bosonic 
species of the GWS model. 
In the GWS the quarks come in six flavours 
\begin{equation}
(m_{t},\,\, m_{b},\,\,m_{c},\,\,m_{s},\,\,m_{u},\,\, m_{d}) = (175,\,\,4,\,\,1,\,\,
0.1,\,\,5\times10^{-3},\,\,1.5 \times10^{-3}) \mathrm{GeV}
\end{equation}
where the quark masses have been listed in GeV.
The lepton masses are, roughly,
\begin{equation}
(m_{\tau},\,\,m_{\mu},\,\,m_{e}) = (1.7,\,\,0.105,0.0005)\mathrm{GeV}.
\end{equation}
Finally, the $W^{\pm}$ and $Z^{0}$, are, respectively, $80.42$ and $91.18$
GeV.

Let us then compute, separately, the bosonic and the fermionic 
contributions. In the GWS model the fermionic species are constituted 
by the six quarks, by the three massive leptons and by the neutrinos 
(that we will take massless). For the quarks the number of relativistic 
degrees of freedom is given by $(6 \times 2\times 2\times 3)= 72$, i.e. 
$6$ particles times a factor $2$ (for the corresponding antipartcles) 
times another factor $2$ (the spin) times a factor $3$ (since each quark 
may come in three different colors). Leptons do not carry color so the 
effective number of relativistic degrees of freedom of $e$, $\mu$ 
and $\tau$ (and of the corresponding neutrinos) is $18$. 
Globally, the fermions carry $90$ degrees of freedom.
The eight massless gluons each of them with two physical polarizations 
amount to $8\times 2=16$ bosonic degrees of freedom. The $SU_{L}(2)$ 
(massless) gauge bosons and the $U_{Y}(1)$ (massless) 
gauge boson lead to $3\times 2 + 2 =8$
bosonic degrees of freedom. Finally the Higgs field (an $SU_{L}(2)$ complex doublet ) carries $4$ degrees of freedom. Globally the bosons carry $28$ 
degrees of freedom. Therefore, we will have that, overall, in the GWS 
model
\begin{equation}
g_{\rho} =g_{s} = 28 + \frac{7}{8}\times 90 =106.75.
\label{DFGWS}
\end{equation}
Notice, as a side remark, that the counting given above assumes indeed 
that the electroweak symmetry is restored (as it is probably the case for 
$T> 175$ GeV). When the electroweak symmetry is broken down to the 
$U_{\mathrm{em}}(1)$, the vector bosons acquire a mass and the Higgs 
field looses three of its degrees of freedom. It is therefore easy to understand 
that the gauge bosons of the electroweak sector lead to the same 
number of degrees of freedom obtained in the symmetric phase: the three 
massive gauge bosons will carry $3\times3$ degrees of freedom, the photons 
$2$ degrees of freedom and the Higgs field (a real scalar, after symmetry 
breaking) $1$ degree of freedom. 

When the temperature drops below $170$ GeV, the top quarks 
start annihilating and $g_{\rho} \to 96.25$. When the temperature 
drops below $80$ GeV the gauge bosons annihilate and
 $g_{\rho}\to 86.25$. When the temperature drops below $4$ GeV, the bottom
quarks start annihilating and $g_{\rho}\to 75.75$ and so on.
While the electroweak phase transition takes place around 
$100$ GeV the quark-hadron phase transition takes place 
around $150$ MeV.  At the quark-hadron phase transition the quarks are not 
free anymore and start combining forming colorless hadrons. In particular 
bound states of three quarks are called baryons while bound states of quark-antiquark 
pairs are mesons. The massless degrees of freedom around the quark-hadron 
phase transition are $\pi^{0}$, $\pi^{\pm}$, $e^{\pm}$, $\mu^{\pm}$ and the 
neutrinos. This implies that $g_{\rho} = 17.25$. When the temperature drops even more 
(i.e. $T < 20 \,\mathrm{MeV}$) muons and also pions annihilate and for $T\sim {\mathcal O}(\mathrm{MeV})$
$g_{\rho} = 10.75$.

It is relevant to notice that the informations on the temperature and on the number of relativistic degrees of freedom can be usefully converted into informations on the Hubble expansion rate and on the Hubble radius at each 
corresponding epoch.
In fact, using Eq. (\ref{FL1}) the temperature of the Universe determines directly the Hubble 
rate and, consequently, the Hubble radius. According to Eq. (\ref{FL1}) and taking into account 
Eq. (\ref{ultra1}) we will have, 
\begin{equation}
H = 1.66 \sqrt{g_{\rho}} \frac{T^2}{M_{\mathrm{P}}},
\label{Hrho}
\end{equation}
or in Planck units:
\begin{equation}
\biggl(\frac{H}{M_{\mathrm{P}}}\biggr) = 1.15\times10^{-37} \,\,\biggl(\frac{g_{\rho}}{106.75}\biggr)^{1/2}  
\biggl(\frac{T}{{\mathrm{GeV}}}\biggr)^2.
\label{Hplanck}
\end{equation}
Equation (\ref{Hplanck}) measures the curvature scale at each cosmological epoch. For instance, at 
the time of the electroweak phase transition $ H\simeq 10^{-34} M_{\mathrm{P}}$ while 
for $T\sim {\mathrm{MeV}}$, $H \simeq 10^{-44} \,\,M_{\mathrm{P}}$.
To get an idea of the size of the Hubble radius we we can express $H^{-1}$ in centimeter:
\begin{equation}
r_{H} = 1.4 \biggl(\frac{106.75}{g_{\rho}}\biggr)^{1/2} \biggl(\frac{100\,{\mathrm{GeV}}}{T}\biggr)^2 \,\,{\mathrm{cm}},
\label{Hradius}
\end{equation}
which shows that the Hubble radius is of the order of the centimeter at the electroweak epoch while 
it is of the order of $10^{4} \,\,\mathrm{Mpc}$ at the present epoch. Finally, recalling that during the radiation 
phase $H^{-1} = 2 t$, 
\begin{equation}
t_{H} = 23\,\,\biggl(\frac{106.75}{g_{\rho}}\biggr)^{1/2} \biggl(\frac{100\,{\mathrm{GeV}}}{T}\biggr)^2 \,\,{\mathrm{psec}},
\label{Htime}
\end{equation}
which shows that the Hubble time is of the order of $20$ psec at the electroweak time while it is $t\simeq 0.73 \,\mathrm{sec}$ right before electron positron annihilation and for $T\simeq\mathrm{MeV}$.

\subsection{Electron-positron annihilation and neutrino decoupling}

As soon as the Universe is old of one second two important phenomena take place: the annihilation 
of electrons and positrons and the decoupling of neutrinos. 
For sufficiently high temperatures the weak interactions are in equilibrium and the reactions
\begin{eqnarray}
e^{-} + p \to n+ \nu_{e},\qquad e^{+} + n \to p + \overline{\nu}_{e},
\label{weint}
\end{eqnarray}
are balanced by their inverse. However, at some point, the rate of the weak interactions 
equals the Hubble expansion rate and, eventually, it becomes smaller than $H$.
Recalling that, roughly, 
\begin{equation}
\Gamma_{\mathrm{weak}} \simeq \sigma_{\mathrm{F}} T^3, \qquad \sigma_{\mathrm F} = G_{\mathrm{F}}^2 T^2,
\qquad G_{\mathrm{F}} =1.16 \times
10^{-5} \,\, \mathrm{GeV}^{-2},
\label{wrate}
\end{equation}
 it is immediate to show, neglecting numerical factors that 
\begin{equation}
\frac{\Gamma_{\mathrm{weak}}}{H} \simeq 
\biggl(\frac{T}{\mathrm{MeV}}\biggr)^3, \qquad H \simeq \frac{T^2}{M_{\mathrm{P}}}.
\label{wrate2}
\end{equation}
Thus, as soon as $T$ drops below a temperature of the order of the MeV the weak interactions 
are not in equilibrium anymore. This temperature scale is also of upmost importance 
for the formation of the light nuclei since, below the MeV, the neutron to proton ratio is depleted via 
free neutron decay.

It is important to appreciate, at this point, that the neutrino distribution is preserved by the expansion and, therefore,
we may still assume that $a(t) T$ is constant. However, around the MeV the electrons and positrons 
start annihilating. This occurrence entails the sudden heating of the photons since the annihilations 
of electrons and positrons ends up in photons. The net result of this observation is an increase 
of the temperature of the photons with respect to the kinetic temperature of the neutrinos. 
In equivalent terms, after electron-positron annihilation, 
the temperature of the neutrinos will be systematically smaller than the temperature 
of the photons. To describe this phenomenon let us consider an initial time $t_{\mathrm{i}}$ before $e^{+}$--$e^{-}$
annihilation and a final time $t_{\mathrm{f}}$ after $e^{+}$--$e^{-}$ annihilation. 
Using the conservation of the entropy density we can say that 
\begin{eqnarray}
&& g_{s}(t_{\mathrm{i}})\, a^3(t_{\mathrm{i}})\,T_{\gamma}^3(t_{\mathrm{i}}) = g_{s}(t_{\mathrm{f}})\, a^3(t_{\mathrm{f}})\,T_{\gamma}^3(t_{\mathrm{f}}),
\label{consnugamma}\\
&& g_{s}(t_{\mathrm{i}}) = 10.75,\qquad g_{s}(t_{\mathrm{f}})= 2 + 5.25 
\biggl[\frac{T_{\nu}(t_{\mathrm{f}})}{T_{\gamma}(t_{\mathrm{f}})}\biggr]^3,
\label{effdf}
\end{eqnarray}
where it has been taken into account that while before $e^{+}$--$e^{-}$
annihilation the temperature of the photons coincides with the temperature of the neutrinos, it may not be the case 
after  $e^{+}$--$e^{-}$ annihilation. Using the same observation, we can also say that 
\begin{equation}
a^3(t_{\mathrm{i}})\,T_{\nu}^3(t_{\mathrm{i}}) = a^3(t_{\mathrm{f}})\,T_{\nu}^3(t_{\mathrm{f}}).
\label{effdf2}
\end{equation}
Dividing then Eq. (\ref{effdf}) by Eq. (\ref{effdf2}) the anticipated result is that 
\begin{equation}
T_{\gamma}(t_{\mathrm{f}}) = \biggl(\frac{11}{4}\biggr)^{1/3} T_{\nu}(t_{\mathrm{f}}),
\label{nuT}
\end{equation}
i.e. the temperature of the photons gets larger than the kinetic temperature 
of the neutrinos. This result also implies that the energy density of a massless 
neutrino background would be today 
\begin{equation}
\rho_{\nu0} = \frac{21}{8} \biggl(\frac{4}{11}\biggr)^{4/3} \rho_{\gamma0},\qquad h_{0}^2 \Omega_{\nu0} = 
1.68\times 10^{-5}.
\label{OMnu}
\end{equation}
It is worth noticing that, according to the considerations related to the phenomenon of neutrino 
decoupling, the effective number of relativistic degrees of freedom around the eV is given by 
\begin{equation}
g_{\rho} = 2 + \frac{7}{8} \times 6 \times \biggl(\frac{4}{11}\biggr)^{4/3} = 3.36,
\label{grev}
\end{equation}
where the two refers to the photon and the neutrinos contribute weighted by their 
kinetic temperature.
\subsection{Big-bang nucleosynthesis (BBN)}

In this subsection the salient aspects of BBN will be summarized. BBN is the process where light nuclei 
are formed. Approximately one quarter of the baryonic matter in the Universe is in the 
form of $^{4}\mathrm{He}$. The remaining part is made, predominantly, by Hydrogen in its different 
incarnations (atomic, molecular and ionized). During BBN the protons and neutrons 
(formed during the quark-hadron phase transition) combine to form nuclei. According 
to Eq. (\ref{Htime})   the quark-hadron phase transition takes place when the Universe 
is old of $20 \mu\mathrm{sec}$ to be compared with $t_{\mathrm{BBN}}\simeq \mathrm{sec}$.
During BBN only light nuclei are formed and, more specifically, $^{4}\mathrm{He}$, $^{3}\mathrm{He}$,
$D$, $^{7}\mathrm{Li}$. The reason why the  $^{4}\mathrm{He}$ is the most abundant element 
has to do with the fact that,  $^{4}\mathrm{He}$ has the largest binding energy for nuclei with 
atomic number $A<12$ (corresponding to carbon). Light nuclei provide stars with the 
initial set of reactions necessary to turn on the synthesis of heavier elements (iron, cobalt and so on).

In short, the logic of BBN is the following:
\begin{itemize}
\item{} after the quark-hadron phase transition the antinucleons annihilate with the nucleons, thus 
the total baryon concentration will be given by $n_{\mathrm{B}} = n_{\mathrm{n}} + n_{\mathrm{p}}$;
\item{} if the baryon number is conserved (as it is rather plausible) the baryon concentration 
will stay constant and, in particular, it will be $n_{\mathrm{B}} \simeq 10^{-10} n_{\gamma}$;
\item{} for temperatures lower than  $T\simeq {\mathcal O}(\mathrm{MeV})$ weak interactions 
fall out of equilibrium; at this stage the concentrations of neutrons and protons are determined 
from their equilibrium values and, approximately, 
\begin{equation}
\frac{n_{\mathrm{n}}}{n_{\mathrm{p}}} = e^{- \frac{Q}{T}}\simeq \frac{1}{6},\qquad Q = m_{\mathrm{n}} -m_{\mathrm{p}};
\label{1/6}
\end{equation}
for $T\simeq 0.73$ MeV;
\item{} if nothing else would happen, the neutron concentration would be progressively 
depleted by free neutron decay, i.e. $n \to p + e^{-} + \overline{\nu}_{e}$;
\item{} however, when $T\simeq 0.1$ MeV the reactions for the formation of the deuterium (D) are 
in equilibrium, i.e. $p + n \to D+ \gamma$ and $D + \gamma \to p + n$;
\item{} the reactions for the formation of deuterium fall out of thermal equilibrium only at a much 
lower temperature (i.e. $T_{D} \simeq 0.06$ MeV);
\item{} as soon as deuterium is formed, $^{4}\mathrm{He}$ and $^{3}\mathrm{He}$ can arise
according to the following chain of reactions:
\begin{eqnarray}
&& D + n \to T +\gamma,\qquad T + p \to ^{4}\mathrm{He} + \gamma,
\label{He1}\\
&& D + p \to ^{3}\mathrm{He} + \gamma,\qquad ^{3}\mathrm{He}+ n \to ^{4}\mathrm{He} +\gamma;
\label{He2}
\end{eqnarray}
\end{itemize}
As soon as the helium is formed a little miracle happens: since the temperature of equilibration 
of the helium is rather large (i.e. $T\simeq 0.3$ MeV), the helium is not in equilibrium at the moment 
when it is formed. In fact the helium can only be formed when deuterium is already present. 
As a consequence the reactions (\ref{He1}) and (\ref{He2}) only take place from right to left.
When the deuterium starts being formed, free neutron decay has already depleted a bit the 
neutron to proton ratio which is equal to $1/7$. 
Since each $^{4}\mathrm{He}$ has two neutrons, we have that $n_{\mathrm{n}}/2$ nuclei of 
 $^{4}\mathrm{He}$ can be formed per unit volume. Therefore, the  $^{4}\mathrm{He}$ mass fraction will be 
 \begin{equation}
 Y_{p} = \frac{4(n_{\mathrm{n}}/2)}{n_{\mathrm{n}} + n_{\mathrm{p}}} \simeq \frac{2(1/7)}{1 + 1/7} \simeq 0.25.
\label{heab}
\end{equation}
The abundances of the other light elements are comparatively smaller than the one of  $^{4}\mathrm{He}$ and, in
particular:
\begin{equation}
D/H \simeq 10^{-5},\qquad ^{3}\mathrm{He}/H \simeq 10^{-3},\qquad ^{7}\mathrm{Li}/H \simeq 10^{-10}.
\end{equation}
The abundances of the light elements computed from BBN calculations agree with the observations.
The simplest BBN scenario implies that the only two free parameters are the temperature and what 
has been called $\eta_{\mathrm{b}}$ i.e. the baryon to photon ratio which must be of the order 
of $10^{-10}$ to agree with experimental data. BBN represents, therefore, one of the successes 
of the standard cosmological model.
\newpage
\renewcommand{\theequation}{C.\arabic{equation}}
\setcounter{equation}{0}
\section{Scalar modes of the geometry}
\label{APPC}
In this appendix we are going to derive the results that are the starting point for the study of the
evolution of the scalar modes both around equality and during the primeval inflationary phase.
As exemplified in section \ref{sec10} it will be always possible to pass from a gauge-dependent 
description to a fully gauge-invariant set of evolution equations. In this appendix the 
longitudinal gauge will be consistently followed. Generalizations of the results derived in this 
appendix can be found in \cite{THTH}. 

\subsection{Scalar fluctuations of the Einstein tensor}
The scalar fluctuations of the geometry with covariant and controvariant indices can be written, to first order 
and in the longitudinal gauge, as 
\begin{eqnarray}
&& \delta_{\rm s} g_{00} = 2 a^2 \phi,\,\,\,\,\,\,\,\,\,\,\,\,\,\,\,\,\,\,\,\,\,\,\,\, 
\delta_{\rm s} g^{00} = - \frac{2}{a^2} \phi,
\nonumber\\
&&  \delta_{\rm s} g_{ij} = 
2 a^2 \psi \delta_{i j} ,\,\,\,\,\,
\delta_{\rm s} g^{ij} = 
-\frac{2}{ a^2} \psi \delta^{i j},
\label{MFLONG}
\end{eqnarray}
where the notations of section \ref{sec6} have been followed.
Since the fluctuations of the Christoffel connections can be expressed as 
\begin{equation}
\delta_{\mathrm{s}} \Gamma_{\alpha\beta}^{\mu} = \frac{1}{2}  \overline{g}^{\mu\nu}( 
-\partial_{\nu} \delta_{\mathrm{s}} g_{\alpha\beta} + \partial_{\beta} \delta_{\mathrm{s}} g_{\nu\alpha} 
+ \partial_{\alpha} \delta_{\mathrm{s}} g_{\beta\nu} ) + 
\frac{1}{2} \delta_{\mathrm{s}} g^{\mu\nu} ( -\partial_{\nu}  
\overline{g}_{\alpha\beta} + \partial_{\beta}  \overline{g}_{\nu\alpha} 
+ \partial_{\alpha}  \overline{g}_{\beta\nu} ),
\label{christ}
\end{equation}
where $\overline{g}_{\mu\nu} = a^2(\tau)\eta_{\mu\nu}$ denotes the background metric and 
$\delta_{\mathrm{s}} g_{\mu\nu}$ the first-order fluctuations in the longitudinal gauge 
which are given, in explicit terms, by Eq. (\ref{MFLONG}).
Inserting Eq. (\ref{MFLONG}) into  Eq. (\ref{christ}) the explicit form of the fluctuations 
of the Christoffel connections can be obtained:
\begin{eqnarray}
&& \delta_{\rm s} \Gamma^{0}_{00} = \phi' ,
\nonumber\\
&& \delta_{\rm s} \Gamma_{i j}^{0} = - 
[ \psi' + 2 {\cal H}( \phi + \psi)]\delta_{ij} \nonumber\\
&& \delta_{\rm s} \Gamma_{i0}^{0} = \delta_{\rm s} \Gamma_{0i}^{0}= 
\partial_{i} \phi,
\nonumber\\
&&\delta_{\rm s} \Gamma^{i}_{00} = \partial^{i}\phi,
\nonumber\\
&& \delta_{\rm s} \Gamma_{ij}^{k} = ( \partial^{k} \psi \delta_{ij} - 
\partial_{i} \psi \delta^{k}_{j} - \partial_{j}\psi \delta^{k}_{i}),
\nonumber\\
&& \delta_{\rm s} \Gamma_{0 i}^{j} = - \psi' \delta_{i}^{j}.
\label{christ2}
\end{eqnarray}
We remark, incidentally, that the fluctuations of the Christoffel connections 
in an inhomogeneous Minkowski metric (used in section \ref{sec7} for the derivation of the scalar 
Sachs-Wolfe effect) are simply obtained from Eqs. (\ref{christ2}) by setting 
${\mathcal H} =0$.

The fluctuations of the Ricci tensor can be now expressed, as 
\begin{equation}
\delta_{\mathrm{s}} R_{\mu\nu} = \partial_{\alpha} \delta_{\mathrm{s}} \Gamma_{\mu\nu}^{\alpha}
- \partial_{\nu} \delta_{\mathrm{s}} \Gamma_{\mu \beta}^{\beta} + 
\delta_{\mathrm{s}} \Gamma_{\mu\nu}^{\alpha}\, \overline{\Gamma}_{\alpha\beta}^{\beta} 
+ \overline{\Gamma}_{\mu\nu}^{\alpha}\,\delta_{\mathrm{s}} \Gamma_{\alpha\beta}^{\beta}
- \delta_{\mathrm{s}}\Gamma_{\alpha\mu}^{\beta} \overline{\Gamma}_{\beta\nu}^{\alpha} 
- \overline{\Gamma}_{\alpha\mu}^{\beta} \delta_{\mathrm{s}} \Gamma_{\beta\nu}^{\alpha}.
\label{riccisf}
\end{equation}
where, as usual, $\overline{\Gamma}_{\alpha\beta}^{\mu}$ are the background values of the 
Christoffel connections, i.e., as already obtained:
\begin{equation}
\overline{\Gamma}_{00}^{0} = {\cal H},\,\,\,\,\,\, \overline{\Gamma}_{i j}^{0} 
={\cal H} \delta_{i j}, \,\,\,\,\,\,\, \overline{\Gamma}_{0i}^{j} = 
{\cal H} \delta_{i}^{j}.
\label{christ3}
\end{equation}
Using Eqs. (\ref{christ2}) into Eq. (\ref{riccisf}) and taking into account Eq. (\ref{christ3}) the explicit 
form of the components of the (perturbed) Ricci tensors can be easily obtained and they are:
\begin{eqnarray}
 \delta_{\rm s} R_{00} &=& 3 [ \psi'' + {\cal H}( \phi' + \psi') ],
\nonumber\\
 \delta_{\rm s} R_{0 i} &=&2  \partial_{i} ( \psi' + {\cal H} \phi),
\nonumber\\
 \delta_{\rm s} R_{i j} &=& - \delta_{i j}\{ [\psi'' + 2 ( {\cal H}' + 2 {\cal H}^2) 
(\psi + \phi)  + {\cal H} ( \phi' + 5 \psi') - \nabla^2 \psi ] \} +\partial_{i} \partial_{j}( \psi - \phi) 
\label{SRICCI}
\end{eqnarray}
The fluctuations of the Ricci tensor with mixed (i.e. one covariant the other controvariant) indices 
can be easily obtained since 
\begin{equation}
\delta_{\mathrm{s}} R_{\alpha}^{\beta} = \delta_{\mathrm{s}}( g^{\alpha\mu} R_{\beta\mu}) = 
 \delta_{\mathrm{s}}g^{\alpha\mu} \overline{R}_{\beta\mu} +  \overline{g}^{\alpha\mu}  \delta_{\mathrm{s}}R_{\beta\mu},
\label{SRICCI2}
\end{equation}
where $\overline{R}_{\alpha\beta}$ are the Ricci tensors evaluated on the background, i.e. 
\begin{eqnarray}
&& \overline{R}_{00} = - 3 {\cal H}', \,\,\,\,\,\,\, \overline{R}_{0}^{0} = - 
\frac{3}{a^2} {\cal H}',
\nonumber\\
&& \overline{R}_{i j} = ( {\cal H}' + 2 {\cal H}^2) \delta_{i j},
\,\,\,\,\,\,\, \overline{R}_{i}^{j} = - \frac{1}{a^2} ( {\cal H}' + 2 {\cal H}^2) \delta_{i}^{j},
\nonumber\\
&& \overline{R} = - \frac{6}{a^2}( {\cal H}^2 + {\cal H}').
\label{SRICCI3}
\end{eqnarray}
 Using Eqs. (\ref{SRICCI}) into Eq. (\ref{SRICCI2}) and recalling Eq. (\ref{SRICCI3}) we get 
 \begin{eqnarray}
\delta_{\rm s} R_{0}^{0} &=& \frac{1}{a^2} \{ \nabla^2 \phi +
3 [\psi'' + {\cal H} ( \phi' + \psi') + 2 {\cal H}' \phi]\},
\nonumber\\
\delta_{\rm s} 
 R_{i}^{j} &=& \frac{1}{a^2} [ \psi'' + 2 ( {\cal H}' + 2 {\cal H}^2)
 \phi + {\cal H} ( \phi' + 5 \psi') - \nabla^2 \psi ] \delta_{i}^{j} - \frac{1}{a^2}\partial_{i}\partial^{j}  ( \psi -\phi),
\nonumber\\
\delta_{\rm s} R_{i}^{0} &=& \frac{2}{a^2} \partial_{i}( \psi' + {\cal H} \phi),
\nonumber\\
\delta_{\rm s} R_{0}^{i} &=&- \frac{2}{a^2} \partial^{i} ( \psi' + {\cal H} \phi) .
\label{mixedR}
\end{eqnarray}
Finally the fluctuations of the components of the Einstein tensor
with mixed indices are computed to be 
\begin{eqnarray}
\delta_{\rm s} {\cal G}_{0}^{0} &=& \frac{2}{a^2} \{ \nabla^2 \psi  - 3 {\cal H} ( \psi' + {\cal H} \phi)\},
\label{dg00}\\
\delta_{\rm s} {\cal G}_{i}^{j} &=& \frac{1}{a^2} \{ [- 2 \psi'' - 
2 ({\cal H}^2 + 2 {\cal H}') \phi - 2 {\cal H} \phi' - 4 {\cal H}\psi'
-\nabla^2 ( \phi - \psi)] \delta_{i}^{j}
- \partial_{i}\partial^{j} (\psi - \phi) \},
\label{dgij}\\
\delta_{\rm s} {\cal G}_{i}^{0} &=& \delta_{\mathrm{s}} R_{i}^{0}.
\label{dg0i}
\end{eqnarray} 
Equations (\ref{dg00}), (\ref{dgij}) and (\ref{dg0i}) are extensively used in section \ref{sec7} and \ref{sec10}.

\subsection{Scalar fluctuations of the energy-momentum tensor(s)}
All along the present lectures two relevant energy-momentum tensors 
have been extensively used, namely the energy-momentum tensor 
of a minimally coupled scalar field and the energy-momentum tensor 
of a mixture of perfect fluids. For the applications discussed in sections 
\ref{sec7}, \ref{sec8}, \ref{sec9} and \ref{sec10} it is relevant to derive the 
first-order form of the energy-momentum tensor. Needless to say that 
since the inverse metric appears in several places in the explicit form 
of the energy-momentum tensor(s), an explicit dependence upon 
the scalar fluctuations of the metric may enter the various (perturbed) components
of $T_{\mu}^{\nu}$.

Let us start with the case of a fluid source.  The energy-momentum tensor of a perfect 
fluid is 
\begin{equation}
T_{\mu \nu} = ( p + \rho) u_{\mu} u_{\nu} - p g_{\mu \nu}.
\label{PFEMT}
\end{equation}
By perturbing to first-order the normalization condition $g_{\mu\nu} u^{\mu} u^{\nu} = 1$ we have that 
\begin{equation}
\delta_{\mathrm{s}}  g^{\mu\nu} \overline{u}_{\mu} \overline{u}_{\nu} + \overline{g}^{\mu\nu} ( \overline{u}_{\mu} 
\delta_{\mathrm{s}} u_{\nu} + \delta_{\mathrm{s}} u_{\mu} \overline{u}_{\nu}) =0,
\label{NORMfirst}
\end{equation}
implying, together with Eq. (\ref{MFLONG}) 
\begin{equation}
\overline{u}_0 = a,\qquad \delta_{\mathrm{s}} u^{0} = - \phi/a.
\label{NORMfirst2}
\end{equation}
It is important to stress that, on the background, the spatial component of $\overline{u}_{\mu}$, i.e. 
$\overline{u}_{i}$, vanish exactly. The contribution to the peculiar velocity arises instead 
to first-order since, in the longitudinal gauge, $\delta_{\mathrm{s}} u_{i} \neq 0$.
By taking the first-order fluctuation of Eq. (\ref{PFEMT}) the result is 
\begin{equation}
\delta_{\mathrm{s}} T_{\mu\nu} = (\delta p + \delta \rho) \overline{u}_{\mu} \overline{u}_{\nu} + 
(p+\rho) (\overline{u}_{\mu} \delta_{\mathrm{s}}u_{\nu} + \delta_{\mathrm{s}} u_{\mu} \overline{u}_{\nu})
- \delta p \overline{g}_{\mu\nu} - p \delta_{\mathrm{s}} g_{\mu\nu}.
\label{FFEMT}
\end{equation}
The perturbed components of energy-momentum tensor with mixed indices can 
be of course obtained from the expression 
\begin{equation}
\delta_{\mathrm{s}} T_{\alpha}^{\beta} = \delta_{\mathrm{s}}( g^{\alpha\mu} T_{\beta\mu}) = 
 \delta_{\mathrm{s}}g^{\alpha\mu} \overline{T}_{\beta\mu} +  \overline{g}^{\alpha\mu}  \delta_{\mathrm{s}}T_{\beta\mu}
 \label{FFEMT2}
 \end{equation}
 where $\overline{T}_{\mu\nu}$ denote the background components 
 of the energy-momentum tensor, i.e.
 \begin{equation}
 \overline{T}_{00} = a^2 \rho,\qquad \overline{T}_{ij} = a^2 p.
 \label{FFEMT3}
 \end{equation}
Inserting Eqs. (\ref{MFLONG}) into Eq. (\ref{FFEMT}) and taking into account Eqs. (\ref{FFEMT2}) and (\ref{FFEMT3}) 
we obtain:
\begin{equation}
\delta_{\rm s} T_{0}^{0} = \delta\rho, \,\,\,\,\,\,\,\,\,\,\,\,\,
\delta_{\rm s} T^{00} = \frac{1}{a^2} ( \delta \rho - 2 \rho \phi),
\label{dT00}
\end{equation}
and 
\begin{eqnarray}
&&\delta_{\rm s} T_{i}^{j} = - \delta p \delta_{i}^{j},
\label{dTij1}\\
&&\delta_{\rm s} T^{i j} = \frac{1}{a^2} [ \delta p \delta^{i j} + 2 p \psi \delta^{i j} ],
\label{dTij2}\\
&& \delta_{\rm s} T_{0}^{i} = ( p + \rho) v^{i},
\label{dT0i}\\
&&  \delta_{\rm s} T^{0i} 
= \frac{1}{a^2} ( p + \rho) v^{i}.
\label{dT0i2}
\end{eqnarray}
where we have chosen to define $ \delta u^{i} = v^{i}/a$.

Let us now consider the fluctuations of the energy-momentum tensor of 
a scalar field $ \varphi$ characterized by a potential $V(\varphi)$:
\begin{equation}
T_{\mu\nu} = \partial_{\mu} \varphi \partial_{\nu} \varphi - g_{\mu\nu} 
\biggl[\frac{1}{2} g^{\alpha\beta} \partial_{\alpha} \varphi \partial_{\beta} 
\varphi - V(\varphi)\biggr].
\end{equation}
Denoting with $ \chi$ the first-order fluctuation of the scalar field 
$\varphi$ we will have
\begin{eqnarray}
&& \delta_{\rm s} T_{\mu\nu} = 
\partial_{\mu}\chi \partial_{\nu} \varphi + 
\partial_{\mu} \varphi \partial_{\nu} \chi
\nonumber\\
&& - 
\delta_{\rm s} 
g_{\mu\nu} \biggl[ \frac{1}{2} g^{\alpha\beta} \partial_{\alpha} 
\varphi \partial_{\beta} \varphi 
- V\biggr] - g_{\mu\nu} \biggl[ \frac{1}{2} \delta_{\rm s} g^{\alpha\beta} 
\partial_{\alpha} \varphi \partial_{\beta} \varphi
+ g^{\alpha\beta} \partial_{\alpha} \chi \partial_{\beta}
 \varphi - \frac{\partial V}{\partial\varphi} \chi\biggr].
\label{ENMOMS} 
\end{eqnarray}
Inserting Eqs. (\ref{MFLONG}) into Eq. (\ref{ENMOMS}) we obtain, in explicit terms:
\begin{eqnarray}
&& \delta_{\rm s} T_{00} = \chi' \varphi' + 2 a^2 \phi V + a^2 
\frac{\partial V}{\partial \varphi} \chi,
\nonumber\\
&& \delta_{\rm s} T_{0i} = \varphi' \partial_{i} \chi,
\nonumber\\
&& \delta_{\rm s} T_{i j} = 
\delta_{ij} \biggl[ \varphi' \chi' - 
\frac{\partial V}{\partial\varphi} \chi a^2 - ( \phi + \psi) {\varphi'}^2 + 
2 a^2 V \psi\biggr].
\end{eqnarray}
Recalling that 
\begin{equation}
\overline{T}_{00} = \frac{{\varphi'}^2}{2 } +a^2 V, \,\,\,\,\,\,\,\,\,\,\,\,
\overline{T}_{ij} = \biggl[\frac{{\varphi'}^2}{2 } -a^2 V \biggr] 
\delta_{ij},
\end{equation}
the perturbed components of the scalar field energy-momentum tensor 
with mixed (one covariant the other controvariant) indices 
can be written, following Eq. (\ref{FFEMT2}), as 
\begin{eqnarray}
&& \delta_{\rm s} T_{0}^{0}= \frac{1}{a^2} \biggl( - \phi {\varphi'}^2 
+ \frac{\partial V}{\partial \varphi} a^2 \chi+ \chi' \varphi'\biggr),
\label{enmomsc1}\\
&& \delta_{\rm s} T_{i}^{j}= \frac{1}{a^2}\biggl( \phi {\varphi'}^2 
+ \frac{\partial V}{\partial \varphi} a^2\chi - \chi' \varphi'\biggr) 
\delta_{i}^{j}, 
\label{enmomsc2}\\
&& \delta_{\rm s} T_{0}^{i} = - \frac{1}{a^2} \varphi' \partial^{i} \chi. 
\label{enmomsc3}
\end{eqnarray}
These equations have been extensively used in section \ref{sec10}.

\subsection{Scalar fluctuations of the covariant conservation equations}
The perturbed Einstein equations are sufficient to determine the evolution
of the perturbations. However, for practical purposes, it is often useful 
to employ the equations stemming from the first-order counterpart 
of the covariant conservation equations. 
Consider, first, the case of a fluid, then the perturbation of the covariant 
conservation equation can be written as:
\begin{equation}
\partial_{\mu} \delta_{\mathrm{s}} T^{\mu\nu} + 
\overline{\Gamma}_{\mu\alpha}^{\mu} \delta_{\mathrm{s}} T^{\alpha\nu} 
+ \delta_{\mathrm{s}} \Gamma_{\mu\alpha}^{\mu}\overline{T}^{\alpha\nu} 
+ \overline{\Gamma}_{\alpha\beta}^{\nu} \delta_{\mathrm{s}} T^{\alpha\beta} 
+ \delta_{\mathrm{s}} \Gamma_{\alpha\beta}^{\nu} \overline{T}^{\alpha\beta} =0.
\label{pertconsgen1}
\end{equation}
Recalling Eqs. (\ref{MFLONG}) and (\ref{dTij1})--(\ref{dT0i}) the 
 $(0)$ and $(i)$ components of  Eq. (\ref{pertconsgen1}) can be written, 
 \begin{eqnarray}
&&\partial_{0} \delta_{\rm s} T^{00} + \partial_{j} \delta_{\rm s} T^{j0} + 
( 2 \delta_{\rm s} \Gamma_{00}^{0} + \delta_{\rm s} \Gamma_{k_0}^{k} ) 
\overline{T}^{00} 
\nonumber\\
&& + 
(2 \overline{\Gamma}_{00}^{0}+ \overline{\Gamma}_{k 0}^{k}) \delta_{\rm s} 
 T^{00}
+ \overline{\Gamma}_{ij}^{0} \delta_{\rm s} T^{ij} + 
\delta_{\rm s} \Gamma_{ij}^{0} \overline{T}^{ij} =0,
\label{T0a}\\
&& \partial_{0} \delta_{\rm s} T^{0 j} + \partial_{k} \delta_{\rm s} T^{ k j}
+ ( \delta_{\rm s} \Gamma_{0 k}^{0} + \delta_{\rm s} \Gamma_{m k}^{m}) 
\overline{T}^{ k j} 
\nonumber\\
&& + ( \overline{\Gamma}_{00}^{0} + \overline{\Gamma}_{k 0}^{k}) 
\delta_{\rm s} T^{0 j} 
+ \delta_{\rm s} \Gamma_{00}^{j} \overline{T}^{00}  + \delta_{\rm s} 
\Gamma_{k m}^{j} \overline{T}^{k m} + 
2 \overline{\Gamma}_{0 k}^{j} \delta_{\rm s}  T^{0 k} =0.
\label{T0b}
\end{eqnarray}
Inserting now the specific form of the perturbed connections of Eqs. (\ref{christ2}) 
into Eqs. (\ref{T0a}) and (\ref{T0b}) the following 
result can be, respectively, obtained:
\begin{equation}
 \delta \rho' - 3 \psi' ( p + \rho) + ( p + \rho) \theta + 
3 {\cal H} ( \delta \rho + \delta p) =0,
\label{gencov1}
\end{equation}
for the  $(0)$ component, and  
\begin{eqnarray}
&& ( p + \rho ) \theta' + \theta [ ( p' + \rho') + 4 {\cal H}( p + \rho)]
 + \nabla^2 \delta p + ( p + \rho) \nabla^2 \phi =0.
\label{gencov2}
\end{eqnarray}
for the $(i)$ component.
In the above equations, as explained in the text, the 
divergence of the velocity field, i.e. $ \theta = \partial_{i}v^{i} = \partial_{i}\partial^{i} v$, has been 
directly introduced. Notice that the possible anisotropic stress (arising, for instance, in the case 
of neutrinos has been neglected). Its inclusion modifies the left hand side of Eq. (\ref{gencov2}) 
by the term $-(p + \rho) \nabla^2 \sigma$. In section \ref{sec7}, Eqs. (\ref{gencov1}) and (\ref{gencov2}) 
have been written in the case of  a mixture of fluids.

Finally, to conclude this appendix, it is relevant to compute the fluctuation of the Klein-Gordon equation 
which is equivalent to the covariant conservation of the energy-momentum tensor 
of the the (minimally coupled) scalar degree of freedom that has been already extensively discussed.
The  Klein-Gordon equation in curved spaces can be written as (see section \ref{sec5})
\begin{equation}
g^{\alpha\beta} \nabla_{\alpha} \nabla_{\beta} \varphi + 
\frac{\partial W}{\partial \varphi} a^2 =0. 
\label{KG1}
\end{equation}
From Eq. (\ref{KG1}) the perturbed Klein-Gordon equation can be 
written as 
\begin{equation}
\delta_{\rm s} g^{\alpha\beta} [ \partial_{\alpha} \partial_{\beta} \varphi 
- \overline{\Gamma}_{\alpha\beta}^{\sigma} \partial_{\sigma} \varphi] 
+ \overline{g}^{\alpha\beta}[ \partial_{\alpha}\partial_{\beta} \chi - 
\delta_{\rm s} \Gamma^{\sigma}_{\alpha\beta} \partial_{\sigma} \varphi - 
\overline{\Gamma}_{\alpha\beta}^{\sigma} \partial_{\sigma} \chi] 
+ \frac{\partial^2 V}{\partial\varphi^2} =0.
\end{equation}
Using Eqs. (\ref{MFLONG}) and (\ref{christ2}) we have:
\begin{eqnarray}
&&\delta_{\mathrm{s}}g^{00}[ \varphi'' - {\mathcal H} \varphi'] - 
\delta_{\mathrm{s}}g^{ij} \overline{\Gamma}_{ij}^{0} \varphi' 
\nonumber\\
&& + \overline{g}^{00} [ \chi'' - \delta_{\mathrm{s}}\Gamma_{00}^{0} \varphi' - \overline{\Gamma}_{00}^{0} \chi']
+ \overline{g}^{ij}[\partial_{i}\partial_{j} \chi - \delta_{\mathrm{s}} \Gamma_{ij}^{0} \varphi' - 
\overline{\Gamma}_{ij}^{0} \chi' ] + \frac{\partial V}{\partial\varphi^2} \chi=0.
\label{KG1a}
\end{eqnarray}
Finally, recalling the explicit forms of the Christoffel connections and of the metric fluctuations 
the perturbed Klein-Gordon equation becomes:
\begin{equation}
\chi'' + 2 {\cal H} \chi' - \nabla^2 \chi +
\frac{\partial^2 V}{\partial\varphi^2}a^2 \chi + 
2 \phi \frac{\partial V}{\partial\varphi}a^2 - 
\varphi'( \phi' + 3 \psi')  =0.
\label{KG2}
\end{equation}
It should be appreciated that the perturbed Klein-Gordon equation also contains 
a contribution arising from the metric fluctuations and it is not only 
sensitive to the fluctuations of the scalar field.
\subsection{Some algebra with the scalar modes}
We will now develop some algebra that is rather useful when dealing with the scalar modes 
induced by a minimally coupled scalar degree of freedom. We will assume Eqs. (\ref{00S}),
(\ref{ijS}) and (\ref{0iS}) valid in the longitudinal gauge. Subtracting Eq. (\ref{00S}) from Eq. (\ref{ijS}) 
the following equation can be obtained (recall that $\phi =\psi$ since the scalar 
field, to first-order, does not produce an anisotropic stress):
\begin{equation}
\psi'' + 6 {\mathcal H} \psi' + 2 ({\mathcal H}' + 2 {\mathcal H}^2) \psi - \nabla^2 \psi = - 8\pi G \, \frac{\partial V}{\partial\varphi} a^2 \chi.
\label{SA1}
\end{equation}
From Eq. (\ref{0iS}) it follows easily that 
\begin{equation}
\chi = \frac{\psi'  +{\mathcal H}\psi}{4\pi\, G \, \varphi'}.
\label{SA2}
\end{equation}
Using then Eq. (\ref{SA2}) inside Eq. (\ref{SA1}) (to eliminate $\chi$) and recalling Eq. (\ref{FLS3}) 
(to eliminate the derivative of the scalar potential with respect to $\varphi$) we obtain the following decoupled equation:
\begin{equation}
\psi'' + 2 \biggl[ {\mathcal H} - \frac{\varphi''}{\varphi'}\biggr] \psi' + 2 \biggl[ {\mathcal H}' - {\mathcal H} \frac{\varphi''}{\varphi'}\biggr] \psi - \nabla^2 \psi =0.
\label{SA3}
\end{equation}
Note that Eq. (\ref{FLS3}) is written in the cosmic time coordinate. Here we need its conformal time counterpart which is easily obtained:
\begin{equation}
\varphi'' + 2 {\mathcal H} \varphi' + a^2 \frac{\partial V}{\partial \varphi}=0.
\end{equation}
It is appropriate to mention, incidentally, that Eq. (\ref{SA3}) can be also written in a slightly simpler form that may be of some use in specific applications namely:
\begin{equation}
f'' - \nabla^2 f - z \biggl(\frac{1}{z}\biggr)'' f =0, \qquad f = \frac{a}{\varphi'} \psi,\qquad z = \frac{a \varphi'}{{\mathcal H}}.
\label{SA3a}
\end{equation}
It could be naively thought that the variable defined in Eq. (\ref{SA3a}) is the canonical normal mode of the system. This is 
not correct since, as we see, Eq. (\ref{SA3a}) does not contain any information on the fluctuation of the scalar field. The correct 
normal mode of the system will now be derived.

Recall now the definition of the curvature perturbations introduced in section \ref{sec7} (see Eq. (\ref{defR2})):
\begin{equation}
{\mathcal R} = - \psi - \frac{{\mathcal H}}{{\mathcal H}^2 - {\mathcal H}'} (\psi' + {\mathcal H} \phi).
\label{SA4}
\end{equation}
By setting $\phi = \psi$ in Eq. (\ref{SA4}) we can express the first (conformal) time derivative of ${\mathcal R}$ as:
\begin{equation}
\frac{\partial  {\mathcal R}}{\partial \tau} = - \psi' - \frac{{\mathcal H}}{{\mathcal H}^2 - {\mathcal H}'}[\psi'' + {\mathcal H}' \psi + {\mathcal H}\psi] - [\psi' + {\mathcal H}\psi] \frac{\partial }{\partial\tau}\biggl(\frac{{\mathcal H}}{{\mathcal H}^2 - {\mathcal H}'}\biggr).
\label{SA5}
\end{equation}
Recalling the conformal time analog of Eq. (\ref{FLS2}), i.e. 
\begin{equation}
{\mathcal H}^2 - {\mathcal H}' = 4\pi\, G\, {\varphi'}^2,
\label{SA6}
\end{equation}
the derivation appearing in the second term of Eq. (\ref{SA5}) can be made explicit. Using then Eq. (\ref{SA3}) 
inside the obtained expression, all the terms can be eliminated except the one containing 
the Laplacian. The final result will be:
\begin{equation}
{\mathcal R}' = - \frac{{\mathcal H}}{4\pi\, G\, {\varphi'}^2} \nabla^2 \psi.
\label{SA7}
\end{equation}
Equation (\ref{SA7}) can be used to obtain a decoupled equation for ${\mathcal R}$ that has been quoted and used 
in section \ref{sec10} (see, in particular, Eq. (\ref{secac2})). 
From Eqs. (\ref{SA4}) and (\ref{SA6}) we can write:
\begin{equation}
\frac{{\mathcal H}}{ 4\pi \, G \, {\varphi'}^2}(\psi' + {\mathcal H} \psi) = - ({\mathcal R} + \psi).
\label{SA8}
\end{equation}
Taking the Laplacian of both sides of Eq. (\ref{SA8}) and recalling Eq. (\ref{SA7}) 
the following relation can be derived:
\begin{equation}
\frac{{\mathcal H}}{ 4\pi \, G \, {\varphi'}^2}\nabla^2\psi' = - \nabla^2{\mathcal R} + \biggl[2 {\mathcal H} - 
\frac{{\mathcal H}'}{{\mathcal H}}\biggr]{\mathcal R}'.
\label{SA9}
\end{equation}
By now taking the derivative of Eq. (\ref{SA7}) we do obtain 
\begin{equation}
{\mathcal R}'' = - \frac{{\mathcal H}}{4\pi\, G\, {\varphi'}^2} \nabla^2 \psi' - \frac{{\mathcal H}}{4\pi\, G\, {\varphi'}^2}
\biggl[ \frac{{\mathcal H}'}{{\mathcal H}} - 2 \frac{\varphi''}{\varphi'}\biggr] \nabla^2 \psi.
\label{SA10}
\end{equation}
Using now Eqs. (\ref{SA7}) and (\ref{SA9}) inside Eq. (\ref{SA10})  (to eliminate, respectively, $\nabla^2 \psi$ and 
 $\nabla^2 \psi'$) the following equation is readily derived:
 \begin{equation}
 {\mathcal R}'' + 2 \biggl[ {\mathcal H} - \frac{{\mathcal H}'}{{\mathcal H}} + \frac{\varphi''}{\varphi'}\biggr] {\mathcal R}' 
 - \nabla^2 {\mathcal R}=0.
 \label{SA11}
 \end{equation}
Equation (\ref{SA11}) can be finally rewritten as 
\begin{equation}
{\mathcal R}'' + 2 \frac{z'}{z} {\mathcal R}' 
 - \nabla^2 {\mathcal R}=0,\qquad z = \frac{a \varphi'}{{\mathcal H}},
 \label{SA12}
 \end{equation}
which is exactly Eq. (\ref{secac2}). As discussed in Eq. (\ref{secac1}), ${\mathcal R}$ is related with 
the scalar normal mode as $q = - {\mathcal R} z$. Recalling Eq. (\ref{SA3}) and Eq. (\ref{SA6}), 
we have that 
\begin{equation}
q = a \chi + z \psi.
\end{equation}
The derivation presented in this appendix is gauge-dependent. However, since ${\mathcal R}$ and $q$ 
are both gauge-invariant, their equations will also be gauge-invariant.  
Finally, it should be stressed that the same result obtained here by working with the evolution equations of the 
fluctuations can be obtained by perturbing (to second-order) the (non-gauge-fixed) scalar tensor action. This 
procedure is rather lengthy and the final results (already quoted in the bulk of the paper) are the ones 
of Eqs. (\ref{secac1}) and (\ref{secac3}).
\end{appendix}

\end{document}